\documentclass[english, hidelinks]{article}
\usepackage{geometry}
\usepackage{setspace}
\usepackage{authblk}
\usepackage[utf8]{inputenc}
\usepackage{color,soul,yhmath}
\usepackage{xcolor}
\usepackage{bm}
\usepackage{graphicx}
\graphicspath{ {.} }
\usepackage{hyperref}
\usepackage{natbib}
\usepackage{bbm}
\usepackage{amsmath}
\usepackage{amssymb}
\usepackage{booktabs}
\usepackage{graphicx}
\usepackage{multirow}
\usepackage{tabularx}
\usepackage{enumitem}

\usepackage[makeroom]{cancel}

\DeclareMathOperator*{\argmin}{arg\,min}
\hypersetup{
    colorlinks=true,
    linkcolor=blue,
    citecolor=blue,
    filecolor=magenta,
    urlcolor=blue
}

\newtheorem{theorem}{Theorem}

\newcounter{mmalgorithm}

\newtheorem{corollary}[theorem]{Corollary}

\newtheorem{lemma}{Lemma}

\newtheorem{proposition}{Proposition}
\newtheorem{remark}{Remark}

\numberwithin{equation}{section}

\def\bb#1{\mathbbm{#1}}
\def\bf#1{\mathbf{#1}}

\def\wt#1{\widetilde{#1}}
\def\wh#1{\widehat{#1}}
\def\vec{\textnormal{\textbf{vec}}}

\def\MSE{\mathrm{MSE}}

\DeclareFontFamily{U}{mathx}{\hyphenchar\font45}
\DeclareFontShape{U}{mathx}{m}{n}{
      <5> <6> <7> <8> <9> <10>
      <10.95> <12> <14.4> <17.28> <20.74> <24.88>
      mathx10
      }{}
\DeclareSymbolFont{mathx}{U}{mathx}{m}{n}
\DeclareFontSubstitution{U}{mathx}{m}{n}
\DeclareMathAccent{\widecheck}{0}{mathx}{"71}
\def\check#1{\widecheck{#1}}

\def\cA{\mathcal{A}}
\def\cB{\mathcal{B}}
\def\cC{\mathcal{C}}
\def\cD{\mathcal{D}}

\def\cG{\mathcal{G}}

\def\cI{\mathcal{I}} 
\def\cII{\mathcal{II}} \def\cIII{\mathcal{III}} \def\cIV{\mathcal{IV}} \def\cV{\mathcal{V}} \def\cVI{\mathcal{VI}} \def\cVII{\mathcal{VII}} \def\cVIII{\mathcal{VIII}} \def\cIX{\mathcal{IX}}
\def\cJ{\mathcal{J}}
\def\cK{\mathcal{K}}
\def\cL{\mathcal{L}}

\def\cR{\mathcal{R}}

\def\e{\mathbf{e}}

\def\u{\mathbf{u}}
\def\v{\mathbf{v}}
\def\x{\mathbf{x}}

\def\A{\mathbf{A}}
\def\B{\mathbf{B}}
\def\C{\mathbf{C}}
\def\G{\mathbf{G}}
\def\Q{\mathbf{Q}}
\def\S{\mathbf{S}}
\def\D{\mathbf{D}}
\def\Y{\mathbf{Y}}
\def\M{\mathbf{M}}
\def\F{\mathbf{F}}
\def\E{\mathbf{E}}
\def\H{\mathbf{H}}
\def\U{\mathbf{U}}
\def\V{\mathbf{V}}
\def\J{\mathbf{J}}
\def\R{\mathbf{R}}
\def\I{\mathbf{I}}
\def\L{\mathbf{L}}
\def\O{\mathbf{O}}
\def\W{\mathbf{W}}
\def\X{\mathbf{X}}
\def\Z{\mathbf{Z}}
\def\z{\mathbf{z}}
\def\a{\mathbf{a}}
\def\b{\mathbf{b}}
\def\y{\mathbf{y}}
\def\w{\mathbf{w}}
\def\P{\mathbf{P}}
\def\K{\mathbf{K}}
\def\0{\mathbf{0}}
\def\1{\mathbf{1}}
\def\diag{\textnormal{diag}}

\def\bSigma{\boldsymbol{\Sigma}}

\def\balpha{\boldsymbol{\alpha}}
\def\bbeta{\boldsymbol{\beta}}

\def\btheta{\boldsymbol{\theta}}
\def\bdelta{\boldsymbol{\delta}}
\def\bGamma{\boldsymbol{\Gamma}}
\def\bOmega{\boldsymbol{\Omega}}
\def\bDelta{\boldsymbol{\Delta}}
\def\bPi{\boldsymbol{\Pi}}
\def\bPsi{\boldsymbol{\Psi}}
\def\bXi{\boldsymbol{\Xi}}
\def\bepsilon{\boldsymbol{\epsilon}}
\def\bzeta{\boldsymbol{\zeta}}
\def\var{\textnormal{Var}}
\def\cov{\textnormal{Cov}}
\def\tr{\text{tr}}

\newcommand{\norm}[1]{\big\| #1 \big\|}

\allowdisplaybreaks[1]

\begin{document}
\setlength{\parindent}{18pt}
	\doublespacing
	\begin{titlepage}
		
		\title{Multilevel Main Effects Matrix Factor Model}
                \author{Kaixin Liu\thanks{Kaixin Liu is PhD student, Department of Statistics, London School of Economics. Email: K.Liu31@lse.ac.uk}}
                \author{Clifford Lam\thanks{Clifford Lam is Professor, Department of Statistics, London School of Economics. Email: C.Lam2@lse.ac.uk}}
                \author{Zetai Cen\thanks{Zetai Cen is Senior Research Associate, School of Mathematics, University of Bristol, and Visiting Fellow, Department of Statistics, London School of Economics. Email: zetai.cen@bristol.ac.uk. Supported by Engineering and Physical Sciences Research Council (EP/Z531327/1).}}

		\affil{Department of Statistics, London School of Economics and Political Science\\
        School of Mathematics, University of Bristol}
		
		\date{}
		
		\maketitle

\begin{abstract}
We propose a multilevel main effects matrix factor model (MMEFM) to extract meaningful row and column effects from groups of matrix data. As a generalization of a multilevel matrix factor model, we provide rigorous definitions of MMEFM and identifications, together with iterative algorithms for the estimators of all global and local components in the main effects and common component, including global/local main effects and core ranks, with rates of convergence spelt out. An important feature of the model is that all main effects can be nonstationary and still be consistently estimated at each time point. At the same time, our estimated global common factors are alignment independent, and hence are truly global, and can be presented and given interpretations more easily. We propose a Bootstrap procedure for detecting the existence and identifying the groups of data that are driven by the global common factors. The power of the test, together with the accuracy of various estimators, are demonstrated in  extensive simulation settings. Estimation performance is also compared to other state-of-the-art methods for matrix time series by simulations and a set of U.S. equity data. All numerical works are done using the R package \texttt{MMEFM}, which is available at \texttt{CRAN}.
\end{abstract}
		
		\bigskip
		\bigskip


\noindent

	\end{titlepage}
	
	\setcounter{page}{2}

\newpage
\section{Introduction}\label{sec:introduction}

With rapid advances in data collection and storage technologies, large numbers of measurements are now routinely recorded over time in matrix form. Such matrix-valued time series arise, for example, as asset-by-indicator panels in finance, country-by-indicator panels in economics, and origin-by-destination traffic matrices. For such data, matrix factor models, which extend factor models for vector time series \citep{Bai2003,Lametal2011,LamYao2012}, provide a natural approach to dimension reduction by preserving the intrinsic row and column structure, rather than treating the vectorized observations through an unrestricted loading structure; see, among others, \cite{Wangetal2019}, \cite{chen2020constrained}, \cite{Heetal2021}, \cite{Yuetal2022a}, \cite{chen2023statistical} and \cite{He2023}. More recently, \cite{lam2024matrix} introduce a main effects matrix factor model (MEFM) that explicitly separates a time-varying grand mean and row and column main effects from the interaction (common) component. This distinction allows different forms of systematic variation within a matrix to be modeled and interpreted separately.

In some applications, several related high-dimensional time series are observed, and their dimensions need not agree across groups. Such data may exhibit another layer of structure: some latent dynamics are shared across groups, while others are group-specific. This distinction has been well studied in the literature on multilevel factor models for vector time series; see, among others, \cite{moench2013dynamic}, \cite{choi2018multilevel}, \cite{han2021shrinkage} and \cite{choi2023canonical}. Related global--local factor structures have also been considered for higher-dimensional data. \cite{lu2025three} study a three-dimensional factor model with global and local factors, while \cite{he2025factor} consider common and cluster-specific factors for biclustering matrix-valued time series. More generally, \cite{Cen2026_multiorder} develops a multi-order tensor factor model that separates global and local components across multiple tensor time series. Most closely related to our setting, \cite{Zhangetal2025} propose a multilevel matrix factor model (MFM) with global and local bilinear components. A global matrix factor process is shared among groups through group-specific row and column loadings, while local matrix factors capture group-specific variation.

The two forms of structure, multilevel variation across groups and additive main effects within each matrix, are complementary and can arise simultaneously in grouped matrix-valued time series. As pointed out by \cite{lam2024matrix}, additive main effects can be absorbed into an enlarged bilinear representation, but treating such a representation as unstructured obscures their distinct interpretation and rank structure. We therefore introduce a \textbf{M}ultilevel \textbf{M}ain \textbf{E}ffects matrix \textbf{F}actor \textbf{M}odel (MMEFM), which extends the MFM of \citet{Zhangetal2025} by incorporating time-varying grand means and row and column main effects, each with global and local components. While \cite{lam2024matrix} discuss low-rank factor representations for the main effects, we develop such structures in a multilevel setting by decomposing the row and column main effects into global and local components, respectively. This exploits cross-group information and provides a parsimonious representation while retaining group-specific dynamics. At the model level, the MFM of \cite{Zhangetal2025} is a restricted case of the MMEFM, so that the latter provides a more general representation by allowing global and local main-effect structures; see Remark~\ref{remark2}.

Using pre-averaging ideas from \cite{ChenLam2024}, our estimation procedure utilizes cross-group information for the global main effects as well as the global common factors in the common component through projected cross-group moments. Local components in each data group are then estimated and removed from the data, with global factor processes still expressed under different row and column transformations.  We therefore align and pool them through a nearest-Kronecker approximation to obtain a global common factor series $\wh\G_t$, followed by a refitting the group-specific global loadings. This leads to one of our contributions distinguished from \citet{Zhangetal2025}, that unlike their group-specific representations $\wh\S_{mt}$ of the global common factors which are still group-dependent, our global main effects and common factor series are truly common across groups. They can then be easier visualized and compared with other estimated group-specific components to provide further insight.

We further establish convergence rates for the proposed estimators under a class of piecewise-stationary linear factor processes. In particular, the class $\cG_T$ accommodates both short- and long-memory dependence as well as changes in dynamics; see Section~\ref{sec:assumption_theories} for details. Within the same setting, we also provide rank estimators of all loading matrices in Section~\ref{subsec:rank_estimation}, with theoretical results presented in Section~\ref{sec:assumption_theories}.

Another important contribution is to determine the presence of a global common component. Imposing such a component without first assessing its presence may give a false impression of shared dynamics when no cross-group common factor exists. We construct group-level statistics and show that the order of the statistics for groups driven by the global components are larger than those for the remaining groups. Building on this separation, we develop a practical existence test for the global common components together with a screening rule for identifying the groups that share them.

The remainder of the paper is organized as follows. Section~\ref{sec:notations} introduces notation. Section~\ref{sec:model_estimation} presents the MMEFM and its estimation. Section~\ref{sec:assumption_theories} introduces the assumptions for the MMEFM and establishes its theoretical properties, while Section~\ref{sec:practical_global_screening} presents the practical detection and screening procedure for global common factors. Section~\ref{sec: numerical_results} reports simulation results, including comparisons with other state-of-the-art matrix factor models, and an empirical analysis of the U.S.\ stock market. Proofs and additional results are included in the Supplementary Materials.

\section{Notations}\label{sec:notations}
Throughout this paper, we use a letter, bold lower-case letter, and bold capital letter, i.e., $a$ or $A$, $\bf{a}$, and $\bf{A}$, to denote a scalar, a vector, and a matrix respectively. We also use $a_i, A_{i,j}, \bf{A}_{i\cdot}, \bf{A}_{\cdot i}$ to denote, respectively, the $i$th element of $\bf{a}$, the $(i,j)$th element of $\bf{A}$, the $i$th row vector (as a column vector) of $\bf{A}$, and the $i$th column vector of $\bf{A}$. For any $a,b\in\bb{R}$, we write $a\vee b:=\max\{a,b\}$ and $a\wedge b:=\min\{a,b\}$. We use $a\asymp b$ to denote $a=O(b)$ and $b=O(a)$, while $a\asymp_P b$ to denote $a=O_P(b)$ and $b = O_P(a)$. We also use $a \gtrsim b$ and $a\lesssim b$ to denote $b = O(a)$ and $a = O(b)$ respectively, and $\gtrsim_P$ and $\lesssim_P$ their respective asymptotic probabilistic versions. We use $\circ$ to denote the Hadamard product and $\otimes$ the Kronecker product.

Given a positive integer $m$, define $[m]:=\{1,2,\dots,m\}$.  The vector $\1_m$ denotes a vector of ones with length $m$. The $i$th largest eigenvalue of a matrix $\A$ is denoted by $\lambda_i(\bf{A})$. The notation $\bf{A}\succcurlyeq 0$ (\text{resp.} $\bf{A} \succ 0$) means that $\bf{A}$ is positive semi-definite (\text{resp.} positive definite). We use $\bf{A}'$ to denote the transpose of $\bf{A}$, and $\diag(\A)$ to denote a diagonal matrix with the diagonal elements of $\A$, while $\diag(\{a_1,\dots,a_n\})$ represents the diagonal matrix with $\{a_1,\dots,a_n\}$ on the diagonal. 
The matrix $\A^\perp$ has orthonormal columns spanning the orthogonal complement of the column space of $\A$, so that $\A'\A^{\perp} = \0$ and $\A^{\perp'}\A^{\perp}=\I$. We use $\A^{\otimes 2}$ to denote $\A\A'$.

Some norm notations.
For a given set, we denote by $|\cdot|$ its cardinality. We use $\|\bf{\cdot}\|$ to denote the spectral norm of a matrix or the $L_2$ norm of a vector, and $\|\bf{\cdot}\|_F$ to denote the Frobenius norm of a matrix. We use $\|\cdot\|_{\max}$ to denote the maximum absolute value of the elements in a vector or a matrix. The notations $\|\cdot\|_1$ and $\|\cdot\|_{\infty}$ denote the $L_1$ and $L_{\infty}$-norm of a matrix respectively, defined by $\|\A\|_{1} := \max_{j}\sum_{i}|(\A)_{ij}|$ and $\|\A\|_{\infty} := \max_{i}\sum_{j}|(\A)_{ij}|$. Without loss of generality, we always assume the eigenvalues of a matrix are arranged by descending orders, and so are their corresponding eigenvectors.

\section{Model and Estimation}\label{sec:model_estimation}
We propose the MMEFM as follows. For each data group $m\in[M]$ and $t\in[T]$,
\begin{align}\label{eqn:MMEFM}
\Y_t^{(m)} &= (\mu_t + \mu_t^{(m)})\1_{p_m}\1_{q_m}' + [(\A_{1m}\balpha_t) + (\A_{2m}\balpha_t^{(m)})]\1_{q_m}' + \1_{p_m}[(\B_{1m}\bbeta_t) + (\B_{2m}\bbeta_t^{(m)})]'\notag\\ 
&\;\;\;\; + \Q_m\G_t\J_m' + \R_m\F_t^{(m)}\C_m' + \E_t^{(m)}\notag\\
&= \mu_t\1_{p_m}\1_{q_m}' + (\A_{1m}\balpha_t)\1_{q_m}' + \1_{p_m}(\B_{1m}\bbeta_t)' + \Q_m\G_t\J_m'\notag\\
&\;\;\;\; + \mu_t^{(m)}\1_{p_m}\1_{q_m}' + (\A_{2m}\balpha_t^{(m)})\1_{q_m}' + \1_{p_m}(\B_{2m}\bbeta_t^{(m)})' + \R_m\F_t^{(m)}\C_m' + \E_t^{(m)},
\end{align}
where $\Y_t^{(m)}$ is the $m$th group observed matrix of dimension $p_m\times q_m$ at time $t$, $\mu_t$, $\balpha_t$, $\bbeta_t$, $\mu_t^{(m)}$, $\balpha_t^{(m)}$ and $\bbeta_t^{(m)}$ are the global grand mean, row main effects, column main effects, and their corresponding local versions at group $m$ respectively at time $t$. The matrices $\A_{1m} \in \mathbb{R}^{p_m\times r_1}$ (\text{resp.} $\A_{2m} \in \mathbb{R}^{p_m\times r_2^{(m)}}$) and $\B_{1m} \in \mathbb{R}^{q_m\times \ell_1}$ (resp.\ $\B_{2m} \in \mathbb{R}^{q_m \times \ell_2^{(m)}}$) are the global (\text{resp.} local) row main effect and the global (\text{resp.} local) column main effect loading matrix respectively at group $m$. The matrices $\Q_m\in\mathbb{R}^{p_m\times k_r}$, $\J_m \in \mathbb{R}^{q_m\times k_c}$, $\R_m\in\mathbb{R}^{p_m\times k_r^{(m)}}$ and $\C_m \in \mathbb{R}^{q_m\times k_c^{(m)}}$ are the global row, global column, local row and local column factor loading matrix respectively at group $m$. Finally, $\G_t$ and $\F_t^{(m)}$ are the global factor and local factor matrices (at group $m$) respectively. 

The upper equality of model (\ref{eqn:MMEFM}) lists the grand means and the main effects first before the factor components, whereas the lower equality lists the global components first before the local components. This model can be considered a generalization of the one in \cite{Zhangetal2025}, with global and local main effects included in our model. It is clear that the global time series $\{\balpha_t\}$, $\{\bbeta_t\}$ and $\{\G_t\}$ are all group-independent, whereas all other components of the model are dependent on the group $m$. It is also clear that at time $t$, the grand mean, row and column main effects for the group $m$ data are given respectively by
\begin{equation}\label{eqn:combined_main_effects}
c_t^{(m)} := \mu_t + \mu_t^{(m)}, \;\;\; \a_t^{(m)} := \A_{1m}\balpha_t + \A_{2m}\balpha_t^{(m)}, \;\;\; \b_t^{(m)} := \B_{1m}\bbeta_t + \B_{2m}\bbeta_t^{(m)} ,
\end{equation}
which are all an affine transformation of the global and local grand means/main effects and are naturally unidentifiable without further restrictions. The same applies to the global and local common components $\Q_m\G_t\J_m'$ and $\R_m\F_t^{(m)}\C_m'$, which are themselves unidentifiable because they are the product of unknown matrices.

With (\ref{eqn:combined_main_effects}), model (\ref{eqn:MMEFM}) can also be written as, for each $t\in[T]$, 
\begin{align*}
\Y_t^{(m)} = c_t^{(m)}\1_{p_m}\1_{q_m}' + \a_t^{(m)}\1_{q_m}' + \1_{p_m}\b_t^{(m) \prime} + (\Q_m,\R_m)\Bigg(
  \begin{array}{cc}
    \G_t & \0 \\
    \0 & \F_t^{(m)} \\
  \end{array}
\Bigg)\Bigg(
  \begin{array}{c}
    \J_m' \\
    \C_m' \\
  \end{array}
\Bigg) + \E_t^{(m)}, \;\; m\in[M], 
\end{align*}
which is a main effect matrix factor model in \cite{lam2024matrix} for each group $m$, with a special block diagonal structure for the factor matrix. If we ignore this structure, the total number of factor series to be estimated is then $\sum_{m=1}^M(k_r + k_r^{(m)})(k_c + k_c^{(m)})$, which is much larger than $k_rk_c + \sum_{m=1}^Mk_r^{(m)}k_c^{(m)}$, the number of factor series in model \eqref{eqn:MMEFM}.
More importantly, the interplay between these groups of time series and thus their interpretation would be lost.

\begin{remark}\label{remark1}
Model (\ref{eqn:MMEFM}) has the main effects split into ``global'' and ``local'', apparently increasing the number of parameters and main effect series to estimate. However, as is evident in Section \ref{subsec:estimation}, the split is mainly there to promote interpretation rather than to increase the difficulty of estimation, since $c_t^{(m)}$, $\a_t^{(m)}$ and $\b_t^{(m)}$ can all be estimated consistently for each $t\in[T]$ from Theorem \ref{thm:main_effect}.  
\end{remark}

\subsection{Examples}\label{subsec:Example}
Consider the case where each data group represents an industry sector. For instance, $M=11$ and each data matrix from group $m$, $\Y_t^{(m)}$, records data at time $t$ from one of the eleven sectors classified by the Global Industry Classification Standard (GICS). Within each group, we may consider countries from the G20 as the rows of $\Y_t^{(m)}$, and various economic indicators of each industry/sub-industry related to the sector at time $t$ as the columns. Hence in this example we have $p_m = 20$ countries, but $q_m$ can be different for each $m\in[11]$, depending on what economic indicators and what industry/sub-industry classifications we go for in each sector. 

As for interpretations, the term $(\A_{1m}\balpha_t)\1_{q_m}'$ from model (\ref{eqn:MMEFM}) represents the country main effects which are \textit{the same} across different economic indicators/ industries/sub-industries. The same  vector $\balpha_t$ continues to drive the country main effects across different sectors. Any genuine variation for the country main effects arising from differences of sectors are captured by the terms $(\A_{2m}\balpha_t^{(m)})\1_{q_m}'$ for $m\in[11]$. Reversing the roles of country and economic indicators/industries/sub-industries while replacing $\balpha_t$ with $\bbeta_t$ above give interpretations of $\1_{p_m}(\B_{1m}\bbeta_t)'$ and $\1_{p_m}(\B_{2m}\bbeta_t^{(m)})'$ respectively. Finally, the term $\Q_m\G_t\J_m'$ represents any \textit{global} interaction effects remaining after accounting for all the global and local main effects, which are all driven by some global factor time series in $\G_t$. Any genuine differences in interaction effects across sectors are captured by the terms $\R_m\F_t^{(m)}\C_m'$.

Another example would be the financial data set used in \cite{Zhangetal2025}. Same as the example described above, each data group still represents an industry sector. However, each row now represents a stock within the sector, and there are $p_m=20$ stocks included for each sector. For the columns, financial data like return, volume of trade and turnover rate are included, so that $q_m$ are the same for each $m$.

\subsection{Identification}\label{subsec:Identification}

We first present (IC1) for identifying the main effects from the common component and (IC2) for identifying the global effects from the local ones. Condition (IC3) is needed for separating the loading matrices from the time series, so that both can be identified.

\begin{itemize}
    \item[(IC1)] (Main effects v.s.\ common component) {For $m\in[M]$ and $j=1,2$, we have
    \[\1_{p_m}'\A_{jm}=\0, \; \1_{q_m}'\B_{jm} = \0, \; \1_{p_m}'\Q_m = \0, \; \1_{q_m}'\J_m = \0, \; \1_{p_m}'\R_m = \0, \; \1_{q_m}'\C_m = \0.\]}

    \item[(IC2)] (Global v.s.\ local) {For each $m\in[M]$, the series $\{\G_t\}$ is uncorrelated with $\{\F_t^{(m)}\}$}. The series $\{\balpha_t^{(m)}\}$, $\{\bbeta_t^{(m)}\}$ and $\{\F_t^{(m)}\}$ are uncorrelated to $\{\balpha_t^{(n)}\}$, $\{\bbeta_t^{(n)}\}$ and $\{\F_t^{(n)}\}$ respectively for all $m\neq n$. Also $\sum_{m=1}^M\mu_t^{(m)}=0$, $\sum_{t=1}^T\balpha_t\balpha_t^{(m)'} = \0$ and $\sum_{t=1}^T\bbeta_t\bbeta_t^{(m)'} = \0$ for each $m\in[M]$. The elements of error series $\{\E_t^{(m)}\}$ are uncorrelated with those of $\{\E_t^{(n)}\}$ for $m\neq n$, and are also uncorrelated with those of $\{\balpha_t\}$, $\{\bbeta_t\}$, $\{\G_t\}$, $\{\balpha_t^{(s)}\}$, $\{\bbeta_t^{(s)}\}$ and $\{\F_t^{(s)}\}$ for all $s\in[M]$.

    \item[(IC3)] (Separation of loading matrices and time series)
    All loading matrices involved in model (\ref{eqn:MMEFM}) are of full rank; of full rank are also the matrices $\bb{E}(\balpha_t^{\otimes 2})$, $\bb{E}(\bbeta_t^{\otimes 2})$, $\bb{E}(\balpha_t^{(m) \otimes 2})$, $\bb{E}(\bbeta_t^{(m) \otimes2})$, $\bb{E}(\vec(\G_t)^{\otimes 2})$,  $\bb{E}(\vec(\F_t^{(m)})^{\otimes 2})$, $\A_{1m}^{\perp'}\A_{2m}$, $\B_{1m}^{\perp '}\B_{2m}$, $\Q_m^{\perp '}\R_m$ and $\J_m^{\perp '}\C_m$.
\end{itemize}
The assumptions $\sum_{t=1}^T\balpha_t\balpha_t^{(m)'} = \0$ and $\sum_{t=1}^T\bbeta_t\bbeta_t^{(m)'} = \0$ in (IC2) are not restrictive at all and are only for convenience in relevant proofs. For instance, 
\begin{align*}
\A_{1m}\balpha_t + \A_{2m}\balpha_t^{(m)} &= \Bigg\{\A_{1m} + \A_{2m}\Bigg(\sum_{s=1}^T\balpha_s^{(m)}\balpha_s'\Bigg)\Bigg(\sum_{s=1}^T\balpha_s\balpha_s'\Bigg)^{-1}\Bigg\}
\balpha_t \\
&+ \A_{2m}\Bigg\{\balpha_t^{(m)} - \Bigg(\sum_{s=1}^T\balpha_s^{(m)}\balpha_s'\Bigg)\Bigg(\sum_{s=1}^T\balpha_s\balpha_s'\Bigg)^{-1}\balpha_t\Bigg\} =: \wt\A_{1m}\balpha_t + \A_{2m}\wt\balpha_t^{(m)},  
\end{align*}
now with $\sum_{t=1}^T\balpha_t\wt\balpha_t^{(m)'} = \0$. The (asymptotic) invertibility of $\sum_{t=1}^T\balpha_t\balpha_t'$ comes from Assumption~(IC3), and (S1) in Section~\ref{subsec:assumption}. Meanwhile, if $\{\balpha_t\}$ is uncorrelated with $\{\balpha_t^{(m)}\}$ for $m\in[M]$, then elements in $T^{-1}\sum_{t=1}^T\wt\balpha_t^{(m)}\wt\balpha_t^{(n)'}$ are going to 0 at the same rate as those in $T^{-1}\sum_{t=1}^T\balpha_t^{(m)}\balpha_t^{(n)'}$ for $m\neq n$, rendering $\{\wt\balpha_t^{(m)}\}$ and $\{\wt\balpha_t^{(n)}\}$ asymptotically uncorrelated for $m\neq n$. For simplicity, we do not engage in asymptotic identification in the proof of Theorem \ref{thm:identification} and only use (IC2) as above for exact identification.

\begin{theorem}[Identification]\label{thm:identification}
Under (IC1), (IC2) and (IC3), all main effects, loading matrices, global and local means and common factors in model (\ref{eqn:MMEFM}) are identified.
\end{theorem}

\begin{remark}\label{remark2}
Following a demonstration in \cite{lam2024matrix}, using the notations in (\ref{eqn:combined_main_effects}), model (\ref{eqn:MMEFM}) can be written as a multilevel matrix FM as follows:
\begin{align*}
    \Y_t^{(m)} &= \wt\Q_m\wt\W_t^{(m)}\wt\J_m' + \E_t^{(m)}, \;\;\; \text{ where}\\
    \wt\Q_m &:= (\I_{p_m}, \Q_m, \R_m, \1_{p_m}), \;\;\; \wt\J_m := (\1_{q_m}, \J_m, \C_m, \I_{q_m}),  \;\;\;
    \wt\W_t^{(m)} :=  \left(
  \begin{array}{cccc}
    \a_t^{(m)} & \0 & \0 & \0 \\
    \0 & \G_t & \0 & \0 \\
    \0 & \0 & \F_t^{(m)} & \0 \\
    c_t^{(m)} & \0 & \0 & \b_t^{(m)'} \\
  \end{array}
\right).
\end{align*}
Ignoring the above structure means that we are estimating $(p_m + k_r + k_r^{(m)} + 1)(q_m + k_c + k_c^{(m)} + 1)$ time series in $\wt\W_t^{(m)}$, which tends to overfit the data because $p_mq_m$ is usually too large relative to $T$.

On the other hand, consider a multilevel matrix FM given by $\Acute{\Y}_t^{(m)} = \Acute{\C}_t + \Acute{\E}_t$ for $m\in[M]$,  where \mbox{$\Acute{\C}_t := \Q_m\G_t\J_m' + \R_m\F_t^{(m)}\C_m'$}. Defining $\M_n := \I_n - n^{-1}\1_n\1_n'$, we construct
\[c_t^{(m)} := (p_mq_m)^{-1}\1_{p_m}'\Acute{\C}_t\1_{q_m}, \;\; \a_t^{(m)} := q_m^{-1}\M_{p_m}\Acute{\C}_t\1_{q_m}, \;\;\; \b_t^{(m)} := p_m^{-1}\M_{q_m}\Acute{\C}'_t\1_{p_m}.\]
Then $\Y_t^{(m)}$ can be expressed as
\begin{align*}
    \Y_t^{(m)} &= c_t^{(m)}\1_{p_m}\1_{q_m}' + \a_t^{(m)}\1_{q_m}' + \1_{p_m}\b_t^{(m)'} + (\M_{p_m}\Q_m)\G_t(\M_{q_m}\J_m)' +  (\M_{p_m}\R_m)\F_t^{(m)}(\M_{q_m}\C_m)' + \Acute{\E}_t,
\end{align*}
which is a multilevel MEFM in (\ref{eqn:MMEFM}), with grand means and main effects given by
\begin{align*}
    \mu_t &= (p_mq_m)^{-1}\1_{p_m}'\Q_m\G_t\J_m'\1_{q_m}, \; \mu_t^{(m)} = (p_mq_m)^{-1}\1_{p_m}'\R_m\F_t^{(m)}\C_m'\1_{q_m}, \; \balpha_t = \G_t\J_m'\1_{q_m}, \\ \bbeta_t&=\G_t'\Q_m'\1_{p_m}, \;    \balpha_t^{(m)} = \F_t^{(m)}\C_m'\1_{q_m}, \; \bbeta_t^{(m)} = \F_t^{(m)'}\R_m'\1_{p_m},
\end{align*}
and the associated main effect loading matrices for $\balpha_t$, $\bbeta_t$, $\balpha_t^{(m)}$ and $\bbeta_t^{(m)}$ given respectively by
\[\A_{1m} = q_m^{-1}\M_{p_m}\Q_m, \;\; \B_{1m} = p_m^{-1}\M_{q_m}\J_m, \;\;  \A_{2m} = q_m^{-1}\M_{p_m}\R_m, \;\; \B_{2m} = p_m^{-1}\M_{q_m}\C_m.\]
Two important observations are that all factor loading matrices are now one rank less because of the multiplication of $\M_{p_m}$ and $\M_{q_m}$, and $\balpha_t^{(m)}$ and $\bbeta_t^{(m)}$ are now of length $k_r^{(m)}$ and $k_c^{(m)}$ rather than being $r_2^{(m)}$ and $\ell_2^{(m)}$ respectively in model~\eqref{eqn:MMEFM}. Since $r_2^{(m)}$ and $\ell_2^{(m)}$ can be larger than $k_r^{(m)}$ and $k_c^{(m)}$, a multilevel MEFM is certainly more flexible than a multilevel matrix FM. 
\end{remark}

\subsection{Estimation}\label{subsec:estimation}
The proof of Theorem~\ref{thm:identification} is instrumental to the estimators to be presented below. We first present the estimators for the grand mean and main effects as a whole. Then we present all global and local estimators for the main effects, before all common component estimators. 

Hereafter, whenever an estimated loading matrix is formed from the first $k$ eigenvectors of a symmetric matrix $\V$, these eigenvectors are chosen to be orthonormal and to correspond to the $k$ largest eigenvalues of $\V$.

\begin{itemize}
    \item[1.] (Grand mean and main effects) The estimators of $c_t^{(m)}$, $\a_t^{(m)}$ and $\b_t^{(m)}$ are given respectively by
    \[\wh{c}_t^{(m)} := (p_mq_m)^{-1}\1_{p_m}'\Y_t^{(m)}\1_{q_m}, \;\;\; \wh\a_t^{(m)} := q_m^{-1}\M_{p_m}\Y_t^{(m)}\1_{q_m}, \;\;\; \wh\b_t^{(m)} := p_m^{-1}\M_{q_m}\Y_t^{(m)'}\1_{p_m}.\]
    We estimate the global and local grand means by respectively
    \[\wh\mu_t := M^{-1}\sum_{m=1}^M\wh{c}_t^{(m)}, \;\;\; \wh{\mu}_t^{(m)} := \wh{c}_t^{(m)} - \wh{\mu}_t.\]

    \item[2.] (Global and local main effects) 
    Define
    \begin{align}
        \wh{\bOmega}_{A_1,m} &:= (M-1)^{-1}\sum_{n=1;n\neq m}^M\Big\{(Tq_mq_n)^{-1}\M_{p_m}\sum_{t=1}^T\Y_t^{(m)}\1_{q_m}\1_{q_n}'\Y_t^{(n)'}\M_{p_n}\Big\}^{\otimes 2},\label{eqn:Omega_A1hat}\\
        \wh{\bOmega}_{B_1,m} &:= (M-1)^{-1}\sum_{n=1;n\neq m}^M\Big\{(Tp_mp_n)^{-1}\M_{q_m}\sum_{t=1}^T\Y_t^{(m)'}\1_{p_m}\1_{p_n}'\Y_t^{(n)}\M_{q_n}\Big\}^{\otimes 2}. \label{eqn:Omega_B1}
    \end{align}
    We set $\wh\A_{1m}$ and $\wh\B_{1m}$ respectively as the first $r_1$ eigenvectors of $\wh{\bOmega}_{A_1,m}$, and the first $\ell_1$ eigenvectors of $\wh\bOmega_{B_1,m}$.
    We take $\wh\A_{2m}$ and $\wh\B_{2m}$ respectively as the first $r_2^{(m)}$ and $\ell_2^{(m)}$ eigenvectors of 
    \begin{align}
        \wh\bOmega_{A_2,m} := \bigg\{T^{-1}\sum_{t=1}^T\wh\a_t^{(m)\otimes 2}\wh\A_{1m}^\perp\bigg\}^{\otimes 2} \; \text{ and } \; \wh\bOmega_{B_2,m} := \bigg\{T^{-1}\sum_{t=1}^T\wh\b_t^{(m)\otimes 2}\wh\B_{1m}^\perp\bigg\}^{\otimes 2} . \label{eqn:Omega_A2B2}
    \end{align}

    The local main effect time series are estimated as
    \begin{equation}\label{eqn:maineffects_time_series}
    \begin{split}
        \wh\balpha_t^{(m)} &:= q_m^{-1}(\wh\A_{2m}'\wh\A_{1m}^{\perp}\wh\A_{1m}^{\perp'}\wh\A_{2m})^{-1}\wh\A_{2m}'\wh\A_{1m}^{\perp}\wh\A_{1m}^{\perp'}\M_{p_m}\Y_t^{(m)}\1_{q_m},\\
        \wh\bbeta_t^{(m)} &:= p_m^{-1}(\wh\B_{2m}'\wh\B_{1m}^{\perp}\wh\B_{1m}^{\perp'}\wh\B_{2m})^{-1}\wh\B_{2m}'\wh\B_{1m}^{\perp}\wh\B_{1m}^{\perp'}\M_{q_m}\Y_t^{(m)'}\1_{p_m}.
    \end{split}
    \end{equation}

    Define
    \begin{alignat*}{2}
        \wh\Z_{\alpha,m}&:=\wh\A_{1m}'(\wh\y_1^{(m)},\ldots,\wh\y_T^{(m)}),
        \quad&&\text{where}\quad
        \wh\y_{t}^{(m)} := q_m^{-1}\M_{p_m}\Y_t^{(m)}\1_{q_m} - \wh\A_{2m}\wh\balpha_t^{(m)}, \\
        \wh\Z_{\beta,m}&:=\wh\B_{1m}'(\wh\w_1^{(m)},\ldots,\wh\w_T^{(m)}),
        \quad&&\text{where}\quad
        \wh\w_t^{(m)} := p_m^{-1}\M_{q_m}\Y_t^{(m)'}\1_{p_m} - \wh\B_{2m}\wh\bbeta_t^{(m)}.
    \end{alignat*}
    We estimate $\wh\A_\alpha:=(\wh\balpha_1,\ldots,\wh\balpha_T)$ and $\wh\B_\beta:=(\wh\bbeta_1,\ldots,\wh\bbeta_T)$ as $\sqrt{T}$ times the matrices whose rows are, respectively, the first $r_1$ and $\ell_1$ eigenvectors of
    \begin{align*}
        M^{-1}\sum_{m=1}^M\wh\Z_{\alpha,m}'(\wh\Z_{\alpha,m}\wh\Z_{\alpha,m}')^{-1}\wh\Z_{\alpha,m}
        \quad\text{and}\quad
        M^{-1}\sum_{m=1}^M\wh\Z_{\beta,m}'(\wh\Z_{\beta,m}\wh\Z_{\beta,m}')^{-1}\wh\Z_{\beta,m}.
    \end{align*}

    We then re-estimate $\A_{1m}$ and $\B_{1m}$ respectively as 
    \begin{align}
        \wt\A_{1m} := T^{-1}(\wh\y_1^{(m)},\ldots,\wh\y_T^{(m)})\wh\A_\alpha'
        \quad\text{and}\quad
        \wt\B_{1m} := T^{-1}(\wh\w_1^{(m)},\ldots,\wh\w_T^{(m)})\wh\B_\beta'.
        \label{eqn:A1mtilde}
    \end{align}

    \item[3.] (Common component) Define
    \[\check\Y_t^{(m)} := \Y_t^{(m)} - \wh{c}_t^{(m)}\1
    _{p_m}\1_{q_m}' - \wh{\a}_t^{(m)}\1_{q_m}' - \1_{p_m}\wh\b_t^{(m)'}.\]
    Using the pre-averaging idea in \cite{ChenLam2024}, we find from a family of randomly generated directions the one that aligns the most with the direction of the strongest factors from $\Q_m$, say $\v_a^{(m)}$. Do the same for $\J_m$, say $\v_b^{(m)}$. See Section~\ref{app:finding_directions} for all the details of this procedure.     

    Let $\v_a^{(m)}$ and $\v_b^{(m)}$ be standardized as unit vectors.
    Define 
    \begin{align}
    \X_{Q,m} &:= T^{-2}(M-1)^{-1}\sum_{n=1;n\neq m}^M\bigg\{\sum_{t=1}^T\check\Y_t^{(m)}\v_b^{(m)}\v_b^{(n)'}\check\Y_t^{(n)'}\bigg\}^{\otimes 2}, \label{eqn:X_A}\\
    \X_{J,m} &:= T^{-2}(M-1)^{-1}\sum_{n=1;n\neq m}^M\bigg\{\sum_{t=1}^T\check\Y_t^{(m)'}\v_a^{(m)}\v_a^{(n)'}\check\Y_t^{(n)}\bigg\}^{\otimes 2}. \label{eqn:X_B}
    \end{align}
    We take $\wh\Q_m$ as the first $k_r$ eigenvectors of $\X_{Q,m}$, and $\wh\J_m$ as the first $k_c$ eigenvectors of $\X_{J,m}$. 
    
    We then define
    \begin{align}
    \X_{R,m} &:= T^{-1}\sum_{t=1}^T\{\check\Y_t^{(m)}\wh{\J}_m^{\perp}\}^{\otimes 2}, \;\;\; \X_{C,m} := T^{-1}\sum_{t=1}^T\{\check\Y_t^{(m)'}\wh\Q_m^{\perp}\}^{\otimes 2}. \label{eqn:XR_XC}
    \end{align}
    We take $\wh\R_m$ as the first $k_r^{(m)}$ eigenvectors of $\X_{R,m}$, and $\wh\C_m$ as the first $k_c^{(m)}$ eigenvectors of $\X_{C,m}$.

    The local common factor $\F_t^{(m)}$ is estimated using least square method:
    \begin{align*}
        \wh\F_t^{(m)} &:= (\wh\R_m'\wh\Q_m^{\perp}\wh\Q_m^{\perp'}\wh\R_m)^{-1}\wh\R_m'\wh\Q_m^{\perp}\wh\Q_m^{\perp'}\check\Y_t^{(m)}\wh\C_m.
    \end{align*}

    Define $\check\Y_{G,t}^{(m)}:=\check\Y_t^{(m)}-\wh\R_m\wh\F_t^{(m)}\wh\C_m'$, and
    \begin{align*}
        \wh\Z_{G,m}
        :=\left(\vec(\wh\Q_m'\check{\Y}_{G,1}^{(m)}\wh\J_m), \dots, \vec(\wh\Q_m'\check{\Y}_{G,T}^{(m)}\wh\J_m)\right).
    \end{align*}
    We first choose a reference group $m_0\in[M]$. For each $m\ne m_0$, define
    \begin{align*}
        \wh\bXi_{m}:=\wh\Z_{G,m_0}\wh\Z_{G,m}'(\wh\Z_{G,m}\wh\Z_{G,m}')^{-1}.
    \end{align*}
    We obtain a two-sided alignment by solving  $\min_{\bXi_{Q,m}\in \bb{R}^{k_r\times k_r}, \bXi_{J,m}\in \bb{R}^{k_c\times k_c}} \norm{\bXi_{J,m}\otimes \bXi_{Q,m}-\wh\bXi_m}_F^2$  (see e.g., Section~6 of \citeauthor{Loan2000}, \citeyear{Loan2000}). We estimate $\bXi_{Q,m}$ and $\bXi_{J,m}$ using $\vec(\wh\bXi_{Q,m})=\wh\sigma_{\Xi,m}^{1/2}\wh\u_{\Xi,m}$ and $\vec(\wh\bXi_{J,m})=\wh\sigma_{\Xi,m}^{1/2}\wh\v_{\Xi,m}$ respectively, where $\wh\sigma_{\Xi,m}$, $\wh\u_{\Xi,m}$ and $\wh\v_{\Xi,m}$ are the largest singular value, the corresponding left and right singular vectors of $\cR(\wh\bXi_m)$, respectively;
    the matrix $\cR(\W)$ for a block matrix $\W = (\W_{ij})_{i\in[k_c], j\in[k_c]}$, with each $\W_{ij}$ of size $k_r\times k_r$, is defined as 
    \begin{align*}
        \cR(\W) := (\vec(\W_{11}),\ldots,\vec(\W_{k_c1}),\ldots,\vec(\W_{1k_c}),\ldots,\vec(\W_{k_ck_c})).
    \end{align*}
    For the reference group, set $\wh\bXi_{Q,m_0}:=\I_{k_r}$ and $\wh\bXi_{J,m_0}:=\I_{k_c}$. The global common factor series is then estimated as
    \begin{align*}
        \wh\G_t:=M^{-1}\sum_{m=1}^M \wh\bXi_{Q,m}\wh\Q_m'\check{\Y}_{G,t}^{(m)}\wh\J_m\wh\bXi_{J,m}'.
    \end{align*}
    We then re-estimate $\J_m\otimes \Q_m$ by least squares:
    \begin{align*} 
        \wh{\J_m\otimes\Q_m}
        := \big(\vec(\check\Y_{G,1}^{(m)}),\ldots, \vec(\check\Y_{G,T}^{(m)})\big)
        (\vec(\wh\G_1),\ldots,\vec(\wh\G_T))' [(\vec(\wh\G_1),\ldots,\vec(\wh\G_T))^{\otimes 2}]^{-1}. 
    \end{align*}
    Finally, to solve $\min_{\Q_m,\J_m}\norm{\J_m\otimes\Q_m-\wh{\J_m\otimes\Q_m}}_F^2$, we estimate $\Q_m$ and $\J_m$ by setting $\vec(\wt\Q_m) = \wh\sigma_{Q,m}^{1/2}\wh\u_{Q,m}$ and $\vec(\wt\J_m) = \wh\sigma_{Q,m}^{1/2}\wh\v_{J,m}$ respectively, where $\wh\sigma_{Q,m}$, $\wh\u_{Q,m}$ and $\wh\v_{J,m}$ are the largest singular value, and the corresponding left and right singular vectors of $\cR(\wh{\J_m \otimes \Q_m})$ respectively.

\end{itemize}
Different from the global time series estimators proposed in \cite{Zhangetal2025}, our proposed $\wh\balpha_t$, $\wh\bbeta_t$ in Step~2 and $\wh\G_t$ in Step~3 are all independent of the index $m$, with all associated loading matrices re-estimated based on these truly global time series estimators.

\begin{remark}\label{remark3}
We stress that the grand mean time series $\{c_t^{(m)}\}$ and the main effects time series $\{\a_t^{(m)}\}$ and $\{\b_t^{(m)}\}$ themselves can be nonstationary, which is an important feature for MEFM in \cite{lam2024matrix}.
The global or local grand means and main effects can also be nonstationary, as long as Conditions (ME1) and (ME2) in Section \ref{subsec:assumption} are satisfied. 
\end{remark}

\begin{remark}\label{remark:global_common_alignment}
The two-sided alignment step is needed because, for each $m$, $\wh\Q_m'\check{\Y}_{G,t}^{(m)}\wh\J_m$ is expressed in group-specific row and column coordinates. In particular, if $\wh\Q_m'\check{\Y}_{G,t}^{(m)}\wh\J_m\approx\L_m\G_t\S_m'$, then direct averaging yields $M^{-1}\sum_{m=1}^M\L_m\G_t\S_m'$, while $M^{-1}\sum_{m=1}^M(\S_m\otimes\L_m)$ need not be a Kronecker product. Hence, the average generally cannot be written as $\L\G_t\S'$ for a common pair $(\L,\S)$, and the resulting coordinate change cannot in general be absorbed by the subsequent two-sided least-squares refitting.

At the population level, $\G_t$ is represented in the coordinate systems of group $m$ and $m_0$ as $\L_m\G_t\S_m'$ and $\L_{m_0}\G_t\S_{m_0}'$, respectively. The vectorized coordinate change from group $m$ to group $m_0$ is then $(\S_{m_0}\S_m^{-1})\otimes(\L_{m_0}\L_m^{-1})$, since applying this transformation to $\vec(\L_m\G_t\S_m')$ yields $\vec(\L_{m_0}\G_t\S_{m_0}')$. Under the rearrangement operator,
\begin{align*}
    \cR\left((\S_{m_0}\S_m^{-1})\otimes(\L_{m_0}\L_m^{-1})\right)
    =\vec(\L_{m_0}\L_m^{-1})\vec(\S_{m_0}\S_m^{-1})'.
\end{align*}
Hence, the leading left and right singular vectors of $\cR(\wh\bXi_m)$ estimate the row and column alignment matrices, bringing the representation from group $m$ into the coordinate system of group $m_0$ before averaging.
\end{remark}

\begin{remark}\label{remark:reference_group}
The reference group in the global alignment can be chosen arbitrarily in principle. In numerical implementation, we choose the group with the largest projected-core norm, for instance the largest value of $\sum_{t=1}^T\left\|\wh\Q_m'\check\Y_{G,t}^{(m)}\wh\J_m\right\|_F^2$ to improve numerical stability and to avoid using a group whose global common component is nearly degenerate as the reference. A more formal procedure for detecting whether a group contains nonzero global common factors is given in Section~\ref{subsec:detect_global_cm}.
\end{remark}

\subsection{Rank estimation for all loading matrices}\label{subsec:rank_estimation}
The rank of all loading matrices in model (\ref{eqn:MMEFM}) are assumed known so far. We propose the following eigenvalue-ratio based estimators for $r_1$, $r_2^{(m)}$, $\ell_1$, $\ell_2^{(m)}$, $k_r$, $k_c$,  $k_r^{(m)}$ and $k_c^{(m)}$ respectively:
\begin{align*}
   \wh{r}_1 &:= \argmin_{j\in[p_m]}\frac{\max_{m\in[M]}\lambda_{j+1}(\wh\bOmega_{A_1,m})}{\max_{m\in[M]}\lambda_{j}(\wh\bOmega_{A_1,m})}, \;\; \wh{r}_2^{(m)} :=  \argmin_{j\in[p_m]}\frac{\lambda_{j+1}(\wh\bOmega_{A_2,m})}{\lambda_{j}(\wh\bOmega_{A_2,m})}, \;\; \wh{\ell}_1 := \argmin_{j\in[q_m]}\frac{\max_{m\in[M]}\lambda_{j+1}(\wh\bOmega_{B_1,m})}{\max_{m\in[M]}\lambda_{j}(\wh\bOmega_{B_1,m})},\\
   \wh{\ell}_2^{(m)} &:=  \argmin_{j\in[q_m]}\frac{\lambda_{j+1}(\wh\bOmega_{B_2,m})}{\lambda_{j}(\wh\bOmega_{B_2,m})}, \;\; \wh{k}_r := \argmin_{j\in[p_m]}\frac{\max_{m\in[M]}\lambda_{j+1}(\X_{Q,m})}{\max_{m\in[M]}\lambda_j(\X_{Q,m})}, \;\; \wh{k}_c := \argmin_{j\in[q_m]} \frac{\max_{m\in[M]}\lambda_{j+1}(\X_{J,m})}{\max_{m\in[M]}\lambda_j(\X_{J,m})}, \\
   \wh{k}_r^{(m)} &:= \argmin_{j\in[p_m]} \frac{\lambda_{j+1}(\X_{R,m})}{\lambda_j(\X_{R,m})}, \;\;  \wh{k}_c^{(m)} := \argmin_{j\in[q_m]} \frac{\lambda_{j+1}(\X_{C,m})}{\lambda_j(\X_{C,m})},
\end{align*}
where $\wh\bOmega_{A_1,m}$, $\wh\bOmega_{B_1,m}$, $\wh\bOmega_{A_2,m}$, $\wh\bOmega_{B_2,m}$, $\X_{Q,m}$, $\X_{J,m}$, $\X_{R,m}$ and $\X_{C,m}$
are defined in (\ref{eqn:Omega_A1hat}), (\ref{eqn:Omega_B1}), (\ref{eqn:Omega_A2B2}), (\ref{eqn:X_A}), (\ref{eqn:X_B}) and (\ref{eqn:XR_XC}), respectively.

\section{Assumptions and Theoretical Results}\label{sec:assumption_theories}

\subsection{Assumptions}\label{subsec:assumption}

We say $\{\w_t\}_{t\in[T]} \in \cG_T(\{d_1,\ldots,d_m\}|\{\x_i,i\in\mathbb{Z}\})$ if the process $\{\w_t\}_{t\in[T]}$ is piecewise-stationary with $m$ stationary segments, each with length of order $T$. In particular, for $t$ within the $j$-th segment, $\w_t$ can be expressed as
\[\w_t = \sum_{w\geq 0}\psi_j(w)\x_{t-w}, \;\;\; \psi_j(0)=1, \;\;\; \psi_j(w) := w^{d_j-1}l_j(w)\text{ for }w\ge 1,\]
where $d_j<1/2$, and $l_j(\cdot): \mathbb{R} \mapsto \mathbb{R}$ is a function such that $l_j(w)$ converges to a constant as $w\rightarrow \infty$.
All elements in $\{\x_w\}$ have mean 0 and are independent and identically distributed with bounded 4th order moments. To facilitate notations under the regime of $\cG_T(\cdot)$, define
\begin{equation}\label{eqn:R(T,d)}
    R(T,d):=
    \begin{cases}
        T^{-1}, \quad\;&d < 1/4, \\
        T^{-1}\log T, &d = 1/4, \\
        T^{4d-2}, &1/4<d<1/2.
    \end{cases}
\end{equation}

\begin{itemize}
    \item[(ME1)] (Main effect (uniform) factor strength).
    We assume that for each $m\in[M]$, as $p_m,q_m\to\infty$, 
    \begin{alignat*}{2}
        &\A_{1m}'\A_{1m}/p_m^{\gamma_r} \to \bSigma_{A_1,m},\qquad
        &&\B_{1m}'\B_{1m}/q_m^{\gamma_c} \to \bSigma_{B_1,m},\\
        &\A_{2m}'\A_{2m}/p_m^{\gamma_r^{(m)}} \to \bSigma_{A_2,m},\qquad
        &&\B_{2m}'\B_{2m}/q_m^{\gamma_c^{(m)}} \to \bSigma_{B_2,m}.%
    \end{alignat*}
    where $\gamma_r, \gamma_r^{(m)}, \gamma_c, \gamma_c^{(m)}$ are all in $(1/2,1]$. The matrices $\bSigma_{A_1,m}$, $\bSigma_{B_1,m}$, $\bSigma_{A_2,m}$, and $\bSigma_{B_2,m}$ 
    are positive definite with eigenvalues bounded away from zero and infinity for all $m\in[M]$.

    \item[(ME2)] (Time series in main effects).
    {
    The time series $\{\balpha_t\}, \{\balpha_t^{(m)}\}, \{\bbeta_t\}$ and $\{\bbeta_t^{(m)}\}$ for $t\in[T]$ are all in $\cG_T$ with potentially different set and number of piecewise-stationary segments, different set of $d_j$'s and $l_j$'s, and defined with time series $\{\x_{\alpha,t}\}, \{\x_{\alpha,t}^{(m)}\}, \{\x_{\beta,t}\}$ and $\{\x_{\beta,t}^{(m)}\}$ respectively. Also, $\{\x_{\alpha,t}^{(m)}\}$ and $\{\x_{\beta,t}^{(m)}\}$ are independent of $\{\x_{\alpha,t}^{(n)}\}$ and $\{\x_{\beta,t}^{(n)}\}$ respectively for $m\neq n$. 
    }

    \item[(E1)] (Decomposition of error).
    { 
    We assume that
    \begin{align*}
        \E_t^{(m)} = \Q_{e,m}\G_{e,t}\J_{e,m}' + \R_{e,m}\F_{e,t}^{(m)}\C_{e,m}' + \bSigma_{\epsilon,m}\circ \bepsilon_t^{(m)}, 
    \end{align*}
    where $\G_{e,t}$ and $\F_{e,t}^{(m)}$ are $k_{e,r}\times k_{e,c}$ and $k_{e,r}^{(m)}\times k_{e,c}^{(m)}$ matrices containing independent elements with mean 0 and variance 1, respectively. The matrix $\bepsilon_t^{(m)} \in \bb{R}^{p_m\times q_m}$ consists of independent entries with mean 0 and variance 1. The series $\{\G_{e,t}\}$, $\{\F_{e,t}^{(m)}\}$ and $\{\bepsilon_t^{(m)}\}$ are mutually independent. The matrix $\bSigma_{\epsilon,m}$ contains standard deviations, uniformly bounded away from 0 and infinity, of the corresponding elements in $\bepsilon_{t}^{(m)}$. Moreover, the error loadings are all (approximately) sparse matrices uniformly across all groups, i.e., for all $m\in[M]$, $\|\Q_{e,m}\|_1 = \|\J_{e,m}\|_1 = \|\R_{e,m}\|_1 = \|\C_{e,m}\|_1 = O(1)$, with $k_{e,r}, k_{e,c}, k_{e,r}^{(m)}, k_{e,c}^{(m)} = O(1)$. Furthermore, assume $\Q_{e,m}\Q_{e,n}'=\0$ or $\J_{e,m}\J_{e,n}'=\0$ for each $m\ne n$. 
    }

    \item[(E2)] (Time series in $\E_t^{(m)}$).
    {
    There is $\X_{e,t}$ the same dimension as $\G_{e,t}$, $\X_{e, t}^{(m)}$ the same dimension as $\F_{e,t}^{(m)}$ and $\X_{\epsilon,t}^{(m)}$ the same dimension as $\bepsilon_t^{(m)}$ such that $\G_{e,t} = \sum_{w\geq 0}a_{e,w}\X_{e,t-w}$, $\F_{e,t}^{(m)} = \sum_{w\geq 0}a_{e,w}^{(m)}\X_{e,t-w}^{(m)}$ and $\bepsilon_t^{(m)} = \sum_{w\geq 0} a_{\epsilon,w}^{(m)}\X_{\epsilon,t-w}^{(m)}$, with $\{\X_{e,t}\}$, $\{\X_{e,t}^{(m_1)}\}$, $\{\X_{e,t}^{(m_2)}\}$, $\{\X_{\epsilon,t}^{(n_1)}\}$ and $\{\X_{\epsilon,t}^{(n_2)}\}$ independent of each other for all $m_1,m_2,n_1,n_2\in[M]$. The series $\{\X_{e,t}\}$ and $\{\X_{e,t}^{(m)}\}$ have independent elements for each $m\in[M]$, while $\{\X_{\epsilon,t}^{(m)}\}$ has i.i.d.\ elements for each $m\in[M]$, all with mean 0 and variance 1 with uniformly bounded fourth order moments. The time series $\{\X_{e,t}\}$ $\{\X_{e,t}^{(m)}\}$, and $\{\X_{\epsilon,t}^{(m)}\}$ are independent of all main effect and factor series in model (\ref{eqn:MMEFM}) for each $m\in[M]$.

    The coefficients $a_{e,w}$, $a_{e,w}^{(m)}$ and $a_{\epsilon,w}^{(m)}$ are such that $\sum_{w\geq 0}a_{e,w}^2 = \sum_{w\geq 0}a_{e,w}^{(m)2} = \sum_{w\geq 0}a_{\epsilon,w}^{(m)2}=1$ and $\sum_{w\geq 0}|a_{e,w}|,\sum_{w\geq 0}|a_{e,w}^{(m)}|, \sum_{w\geq 0}|a_{\epsilon,w}^{(m)}| \leq c$ for some constant $c>0$.
    }

    \item[(F1)] (Time series in $\G_{t}$ and $\F_{t}^{(m)}$).
    {
    For $t\in[T]$, define $\z_{G,t} := \vec(\G_t')$, and $\z_{F,t}^{(m)} := \vec(\F_t^{(m)'})$ for each $m\in[M]$.
    The time series $\{\z_{G,t}\}$ and $\{\z_{F,t}^{(m)}\}$ are in $\cG_T$ with potentially different sets and numbers of piecewise-stationary segments, different sets of $d_j$'s and $l_j$'s, and defined with independent time series $\{\x_{G,t}\}$ and $\{\x_{F,t}^{(m)}\}$ respectively.
    Also, $\{\x_{F,t}^{(m)}\}$ is independent of $\{\x_{F,t}^{(n)}\}$ for $m\neq n$. Both $\{\G_t\}$ and $\{\F_t^{(m)}\}$ are independent of $\{\E_t^{(n)}\}$ for $m,n\in[M]$.
    }
    
    \item[(L1)] (Factor loading matrices)  
    {
    For each $m\in[M]$, define $\Z_q^{(m)} := \diag(\Q_m'\Q_m)$, $\Z_j^{(m)} := \diag(\J_m'\J_m)$, $\Z_r^{(m)} := \diag(\R_m'\R_m)$ and $\Z_c^{(m)} := \diag(\C_m'\C_m)$. We assume that 
    \begin{align*}(\Z_q^{(m)})^{-1/2}\Q_m'\Q_m(\Z_q^{(m)})^{-1/2} &\rightarrow \bSigma_{Q,m}, \;\;\; (\Z_j^{(m)})^{-1/2}\J_m'\J_m(\Z_j^{(m)})^{-1/2} \rightarrow \bSigma_{J,m},\\(\Z_r^{(m)})^{-1/2}\R_m'\R_m(\Z_r^{(m)})^{-1/2} &\rightarrow \bSigma_{R,m}, \;\;\; (\Z_c^{(m)})^{-1/2}\C_m'\C_m(\Z_c^{(m)})^{-1/2} \rightarrow \bSigma_{C,m},
    \end{align*}
    where $\bSigma_{Q,m}$, $\bSigma_{J,m}$, $\bSigma_{R,m}$ and $\bSigma_{C,m}$ have eigenvalues uniformly bounded away from 0 and infinity for all $m\in[M]$. Also, we assume
    \begin{align*}
    (\Z_q^{(m)})_{ii} &\asymp p_m^{\delta_{q,i}^{(m)}} \text{ for} \;\; i\in[k_r],\;\;\;\;\;\;\;\;
    (\Z_j^{(m)})_{ii} \asymp q_m^{\delta_{j,i}^{(m)}} \text{ for} \;\; i\in[k_c],\\
        (\Z_r^{(m)})_{ii} &\asymp p_m^{\delta_{r,i}^{(m)}} \text{ for} \;\; i\in[k_r^{(m)}],\;\;\;\;\;
    (\Z_c^{(m)})_{ii} \asymp q_m^{\delta_{c,i}^{(m)}} \text{ for} \;\; i\in[k_c^{(m)}],
    \end{align*}
    where $\delta_{q,i}^{(m)}, \delta_{j,i}^{(m)}, \delta_{r,i}^{(m)}$ and $\delta_{c,i}^{(m)}$ are all in $(1/2, 1]$, ordered non-increasingly.
    }

    \item[(S1)] (Non-degeneracy and separation conditions)
    {
    As $T\to\infty$, the eigenvalues of
    $T^{-1}\sum_{t=1}^T\bb{E}(\balpha_t^{\otimes 2})$,
    $T^{-1}\sum_{t=1}^T\bb{E}(\bbeta_t^{\otimes 2})$,
    $T^{-1}\sum_{t=1}^T\bb{E}(\balpha_t^{(m)\otimes 2})$,
    $T^{-1}\sum_{t=1}^T\bb{E}(\bbeta_t^{(m)\otimes 2})$,
    $T^{-1}\sum_{t=1}^T\bb{E}(\vec(\G_t)^{\otimes 2})$, and \\
    $T^{-1}\sum_{t=1}^T\bb{E}(\vec(\F_t^{(m)})^{\otimes 2})$
    are all bounded away from zero and infinity over $m\in[M]$. 
    In addition, for sufficiently large $p_m$ and $q_m$, the singular values of
    \begin{alignat*}{2}
        &\A_{1m}^{\perp'}\A_{2m}(\A_{2m}'\A_{2m})^{-1/2},\qquad
        &&\B_{1m}^{\perp'}\B_{2m}(\B_{2m}'\B_{2m})^{-1/2}, \\
        &\Q_m^{\perp'}\R_m(\R_m'\R_m)^{-1/2},\qquad
        &&\J_m^{\perp'}\C_m(\C_m'\C_m)^{-1/2}.
    \end{alignat*}
    are bounded away from zero and infinity for each $m\in[M]$.
    }

\end{itemize}

The overall structure of Assumptions~(ME1)--(L1) broadly follows \cite{lam2024matrix}, with modifications to accommodate the multilevel structure and the broader temporal class $\cG_T$. Assumption~(ME1) allows global and local main effects to differ in strength, but assumes equal strength within each loading matrix for simplicity. By contrast, (L1) allows heterogeneous strengths across loading directions in the common components. Assumptions~(ME2) and (F1) place the main-effect and common-component factors in $\cG_T$, allowing changes in dynamics as well as short- and long-memory dependence. Under (F1), the entries of $\{\z_{G,t}\}$ and $\{\z_{F,t}^{(m)}\}$ are independent, while the cross-sectional dependence is accommodated through $\bSigma_{Q,m}$, $\bSigma_{J,m}$, $\bSigma_{R,m}$ and $\bSigma_{C,m}$ in (L1). Assumptions~(E1)--(E2) allow cross-sectional and serial dependence in the errors, while the cross-group orthogonality condition in (E1) prevents the errors from generating spurious global signals. Similar to Conditions~2 and 3 in \cite{Zhangetal2025}, Assumption~(S1) ensures non-degenerate factor processes and sufficient separation between the local and global loading spaces, so that the local signals are preserved after projecting out the global loading spaces.

\begin{remark}\label{remark:E1}
    The orthogonality condition $\Q_{e,m}\Q_{e,n}'=\0$ or $\J_{e,m}\J_{e,n}'=\0$ for each $m\ne n$ in (E1) is imposed to ensure that the error matrices are uncorrelated across groups, as required by (IC2). However, this condition could be relaxed, in which case the model would be asymptotically identified. Such a relaxation would introduce additional terms into the cross-group error products and lead to more complicated proofs and convergence rates. We do not pursue the corresponding theoretical extension here, but in simulation in subsection~\ref{subsec: simulation}, this condition is not imposed. 
\end{remark}

\subsection{Rate of convergence for estimators}

We establish the asymptotic rates of convergence for the estimators depicted in Section~\ref{subsec:estimation}.
First, we consider the estimators of grand means and main effects.

\begin{theorem}[Consistency of grand mean and main effect estimators]\label{thm:main_effect}
    Let Assumptions~(IC1), (IC2), (E1) and (E2) hold. Then, for each $m\in[M]$ and $t\in[T]$, 
    \begin{align*}
        (\wh{c}_t^{(m)} - c_t^{(m)})^2 = O_P\Big(\frac{1}{p_mq_m}\Big), \quad\;
        \frac{1}{p_m}\norm{\wh\a_t^{(m)} - \a_t^{(m)}}^2 = O_P\Big(\frac{1}{q_m}\Big), \quad\;
        \frac{1}{q_m}\norm{\wh\b_t^{(m)} - \b_t^{(m)}}^2 = O_P\Big(\frac{1}{p_m}\Big).
    \end{align*}
    Moreover, for each $m\in[M]$ and $t\in[T]$, 
    \begin{align*}
        \big(\wh\mu_t - \mu_t\big)^2 
        = O_P\Big(\max_{m\in[M]}\frac{1}{p_mq_m}\Big), \qquad
        \big(\wh\mu_t^{(m)} - \mu_t^{(m)}\big)^2 
        = O_P\Big(\max_{m\in[M]}\frac{1}{p_mq_m}\Big).
    \end{align*}
\end{theorem}

Theorem~\ref{thm:main_effect} demonstrates the performance of direct moment estimation and reflects the gain from cross-sectional averaging.

\begin{itemize}
    \item[(R1)] (Rate conditions for main effects)
    {
    For each $m\in[M]$, define $\nu_{r,m}:=p_m^{\gamma_r^{(m)}-\gamma_r}$ and $\nu_{c,m}:=q_m^{\gamma_c^{(m)}-\gamma_c}$. Let $d_{r,\star}$ be the largest memory parameter across all elements of $\balpha_t$ and $\balpha_t^{(m)}$, $m\in[M]$, with $d_{c,\star}$ defined analogously for $\bbeta_t$ and $\bbeta_t^{(m)}$.
    We assume that
    \begin{alignat*}{2}
        &\max_{m\in[M]}\left\{\frac{p_m^{1-\gamma_r}}{q_m}, \frac{p_m^{1-\gamma_r^{(m)}}}{q_m}\right\}=o(1),
        &&\qquad R(T,d_{r,\star})\max_{\substack{m,n\in[M] \\ m\neq n}}\left\{(\nu_{r,m}\vee\frac{1}{\nu_{r,m}})(1\vee \nu_{r,n})\right\}=o(1), \\
        &\max_{m\in[M]}\left\{\frac{q_m^{1-\gamma_c}}{p_m}, \frac{q_m^{1-\gamma_c^{(m)}}}{p_m}\right\}=o(1),
        &&\qquad R(T,d_{c,\star})\max_{\substack{m,n\in[M] \\ m\neq n}}\left\{(\nu_{c,m}\vee\frac{1}{\nu_{c,m}})(1\vee \nu_{c,n})\right\}=o(1).
    \end{alignat*}
    }
\end{itemize}

The quantities $\nu_{r,m}$ and $\nu_{c,m}$ measure the local main effect strength relative to the corresponding global effect. Assumption~(R1) controls how this relative strength can vary with the matrix dimensions and temporal dependence. In particular, (R1) automatically holds when all main effect factors are pervasive. 

\begin{theorem}[Consistency of initial global main effect loading estimators]\label{thm:initial_A1m}
    Let the assumptions of Theorem~\ref{thm:main_effect} hold. Further assume Assumptions~(ME1), (ME2), (R1) and (S1). For each $m\in[M]$, define $\check{\A}_{1m}:=\A_{1m}(\A_{1m}'\A_{1m})^{-1/2}$ and $\check{\B}_{1m}:=\B_{1m}(\B_{1m}'\B_{1m})^{-1/2}$. Then there exist asymptotically orthogonal matrices $\H_{A_1,m}$ and $\H_{B_1,m}$ with $\H_{A_1,m}\H_{A_1,m}'=\I+o_P(1)$ and $\H_{B_1,m}\H_{B_1,m}'=\I+o_P(1)$, such that
    \begin{align*}
        \norm{\wh\A_{1m}-\check{\A}_{1m}\H_{A_1,m}}_F^2
        &= O_P(\eta_{A_{1m}}), \; \eta_{A_{1m}} := R(T,d_{r,\star})\nu_{r,m}
        \max_{n\in[M]\setminus\{m\}}\nu_{r,n}
        +\frac{p_m^{1-\gamma_r}}{Tq_m}
        \big(1\vee \max_{n\in[M]\setminus\{m\}}\nu_{r,n}\big), \\
        \norm{\wh\B_{1m}-\check{\B}_{1m}\H_{B_1,m}}_F^2
        &= O_P(\eta_{B_{1m}}), \;  \eta_{B_{1m}} := R(T,d_{c,\star})\nu_{c,m}
        \max_{n\in[M]\setminus\{m\}}\nu_{c,n}
        +\frac{q_m^{1-\gamma_c}}{Tp_m}
        \big(1\vee \max_{n\in[M]\setminus\{m\}}\nu_{c,n}\big).
    \end{align*}
\end{theorem}

\begin{theorem}[Consistency of local main effect loading and factor estimators]\label{thm:A2m_alphatm}
    Let the assumptions of Theorem~\ref{thm:initial_A1m} hold. For each $m\in[M]$, define $\check{\A}_{2m}:=\A_{2m}(\A_{2m}'\A_{2m})^{-1/2}$ and $\check{\B}_{2m}:=\B_{2m}(\B_{2m}'\B_{2m})^{-1/2}$. Then there exist asymptotically orthogonal matrices $\H_{A_2,m}$ and $\H_{B_2,m}$ such that
    \begin{align*}
        \norm{\wh\A_{2m}-\check{\A}_{2m}\H_{A_2,m}}_F^2
        =O_P\Big(&R(T,d_{r,\star})\frac{\max_{n\in[M]\setminus\{m\}}\nu_{r,n}}{\nu_{r,m}}
        +\frac{p_m^{1-\gamma_r}}{Tq_m}
        \frac{\big(1\vee \max_{n\in[M]}\nu_{r,n}\big)}{\nu_{r,m}^2}
        +\frac{p_m^{1-2\gamma_r^{(m)}}}{q_m^2}\Big), \\
        \norm{\wh\B_{2m}-\check{\B}_{2m}\H_{B_2,m}}_F^2
        =O_P\Big(&R(T,d_{c,\star})\frac{\max_{n\in[M]\setminus\{m\}}\nu_{c,n}}{\nu_{c,m}}
        +\frac{q_m^{1-\gamma_c}}{Tp_m}
        \frac{\big(1\vee \max_{n\in[M]}\nu_{c,n}\big)}{\nu_{c,m}^2}
        +\frac{q_m^{1-2\gamma_c^{(m)}}}{p_m^2}\Big).
    \end{align*}
    Moreover, define $\check\balpha_t^{(m)}:=(\A_{2m}'\A_{2m})^{1/2}\balpha_t^{(m)}$ and $\check\bbeta_t^{(m)}:=(\B_{2m}'\B_{2m})^{1/2}\bbeta_t^{(m)}$. Then,
    \begin{align*}
        \frac{1}{p_m^{\gamma_r^{(m)}}}
        \norm{\wh\balpha_t^{(m)}-\H_{A_2,m}^{-1}\check\balpha_t^{(m)}}^2
        =O_P\Big(&\frac{1}{p_m^{\gamma_r^{(m)}}q_m}
        +R(T,d_{r,\star})\frac{\max_{n\in[M]\setminus\{m\}}\nu_{r,n}}{1\wedge \nu_{r,m}}
        +\frac{p_m^{1-\gamma_r^{(m)}}}{Tq_m}
        \frac{1\vee \max_{n\in[M]}\nu_{r,n}}{1\wedge \nu_{r,m}} \Big), \\
        \frac{1}{q_m^{\gamma_c^{(m)}}}
        \norm{\wh\bbeta_t^{(m)}-\H_{B_2,m}^{-1}\check\bbeta_t^{(m)}}^2
        =O_P\Big(&\frac{1}{q_m^{\gamma_c^{(m)}}p_m}
        +R(T,d_{c,\star})\frac{\max_{n\in[M]\setminus\{m\}}\nu_{c,n}}{1\wedge \nu_{c,m}}
        +\frac{q_m^{1-\gamma_c^{(m)}}}{Tp_m}
        \frac{1\vee \max_{n\in[M]}\nu_{c,n}}{1\wedge \nu_{c,m}} \Big).
    \end{align*}
\end{theorem}

\begin{theorem}[Consistency of global main effect factor and refitted loading estimators]\label{thm:A1m_alphat}
    Let the assumptions in Theorem~\ref{thm:A2m_alphatm} hold. Define $\check\balpha_t:=(T^{-1}\sum_{s=1}^T\balpha_s\balpha_s')^{-1/2}\balpha_t$ and $\check\bbeta_t:=(T^{-1}\sum_{s=1}^T\bbeta_s\bbeta_s')^{-1/2}\bbeta_t$. Then there exist asymptotically orthogonal matrices $\H_\alpha$ and $\H_\beta$ such that, for each $t\in[T]$,
    \begin{align*}
        \norm{\wh\balpha_t-\H_\alpha^{-1}\check\balpha_t}^2
        =O_P\Big(\max_{m\in[M]}\Big\{
        &\frac{1}{p_m^{\gamma_r}q_m}
        +R(T,d_{r,\star})(1\vee\nu_{r,m})\max_{n\in[M]\setminus\{m\}}\nu_{r,n}
        +\frac{p_m^{1-\gamma_r}}{Tq_m}
        \frac{1\vee\max_{n\in[M]}\nu_{r,n}}{1\wedge\nu_{r,m}} \Big\}\Big), \\
        \norm{\wh\bbeta_t-\H_\beta^{-1}\check\bbeta_t}^2
        =O_P\Big(\max_{m\in[M]}\Big\{
        &\frac{1}{q_m^{\gamma_c}p_m}
        +R(T,d_{c,\star})(1\vee\nu_{c,m})\max_{n\in[M]\setminus\{m\}}\nu_{c,n}
        +\frac{q_m^{1-\gamma_c}}{Tp_m}
        \frac{1\vee\max_{n\in[M]}\nu_{c,n}}{1\wedge\nu_{c,m}}\Big\}\Big).
    \end{align*}
    Moreover, let $\eta_{A_{1m}}$ and $\eta_{B_{1m}}$ denote the rates in Theorem~\ref{thm:initial_A1m}, and $\eta_\alpha$ and $\eta_\beta$ denote the rates in the two factor bounds above. Define $\check{\check{\A}}_{1m}:=\A_{1m}(T^{-1}\sum_{t=1}^T\balpha_t\balpha_t')^{1/2}$ and $\check{\check{\B}}_{1m}:=\B_{1m}(T^{-1}\sum_{t=1}^T\bbeta_t\bbeta_t')^{1/2}$. Then,
    \begin{align*}
        \frac{1}{p_m^{\gamma_r}}
        \norm{\wt\A_{1m}-\check{\check\A}_{1m}\H_\alpha}_F^2
        =O_P\Big(\eta_{A_{1m}}
        +\Big\{\frac{p_m^{1-\gamma_r}}{q_m}
        +R(T,d_{r,\star})(1\vee\nu_{r,m})
        \max_{n\in[M]\setminus\{m\}}\nu_{r,n}\Big\}\eta_\alpha\Big), \\
        \frac{1}{q_m^{\gamma_c}}
        \norm{\wt\B_{1m}-\check{\check\B}_{1m}\H_\beta}_F^2
        =O_P\Big(\eta_{B_{1m}}
        +\Big\{\frac{q_m^{1-\gamma_c}}{p_m}
        +R(T,d_{c,\star})(1\vee\nu_{c,m})
        \max_{n\in[M]\setminus\{m\}}\nu_{c,n}\Big\}\eta_\beta\Big).
    \end{align*}
\end{theorem}

Theorems~\ref{thm:initial_A1m}--\ref{thm:A1m_alphat} reveal how the relative strengths $\nu_{r,m}$ and $\nu_{c,m}$ induce a trade-off between global and local estimation. The refitted loading $\wt\A_{1m}$ inherits both the error of $\wh\A_{1m}$ and that of the estimated global main effect factor series. Although refitting does not necessarily improve the loading convergence rate, it aligns the group-specific global loadings to a common temporal coordinate. This alignment substantially improves the estimation of the global main effect products, as demonstrated in subsection~\ref{subsubsec:accuracy_of_main_effects}.

A useful benchmark is obtained when all main effect factors are pervasive and their memory parameters are below $1/4$, as summarized in the following corollary.

\begin{corollary}[Simplified rates of main effect estimators]\label{corollary:simplified_main_effect_rates}
    Let Assumptions~(IC1), (IC2), (E1), (E2) and (S1) hold. Suppose that (ME1) and (ME2) hold with pervasive main effect factors, i.e., $\gamma_r=\gamma_c=\gamma_r^{(m)}=\gamma_c^{(m)}=1$ for all $m\in[M]$, and that all main effect memory parameters are less than $1/4$. Then, for each $m\in[M]$ and $t\in[T]$, 
    \begin{itemize}[leftmargin=1.5em, itemsep=1pt, parsep=0pt, topsep=2pt]
        \item $\norm{\wh\A_{1m}-\check{\A}_{1m}\H_{A_1,m}}_F^2=O_P\big(1/T\big)$, $\norm{\wh\B_{1m}-\check{\B}_{1m}\H_{B_1,m}}_F^2=O_P\big(1/T\big)$.
        \item $\norm{\wh\A_{2m}-\check{\A}_{2m}\H_{A_2,m}}_F^2=O_P\big(1/T+1/(p_mq_m^2)\big)$, $\norm{\wh\B_{2m}-\check{\B}_{2m}\H_{B_2,m}}_F^2=O_P\big(1/T+1/(q_mp_m^2)\big)$.
        \item $1/p_m\norm{\wh\balpha_t^{(m)}-\H_{A_2,m}^{-1}\check\balpha_t^{(m)}}^2=O_P\big(1/T+1/(p_mq_m)\big)$, $1/q_m\norm{\wh\bbeta_t^{(m)}-\H_{B_2,m}^{-1}\check\bbeta_t^{(m)}}^2=O_P\big(1/T+1/(p_mq_m)\big)$.
        \item $\norm{\wh\balpha_t-\H_\alpha^{-1}\check\balpha_t}^2=O_P\big(1/T+\max_{n\in[M]}1/(p_nq_n)\big)$, $\norm{\wh\bbeta_t-\H_\beta^{-1}\check\bbeta_t}^2=O_P\big(1/T+\max_{n\in[M]}1/(p_nq_n)\big)$.
        \item $1/p_m\norm{\wt\A_{1m}-\check{\check\A}_{1m}\H_\alpha}_F^2=O_P\big(1/T+1/q_m\max_{n\in[M]}1/(p_nq_n)\big)$, $1/q_m\norm{\wt\B_{1m}-\check{\check\B}_{1m}\H_\beta}_F^2=O_P\big(1/T+1/p_m\max_{n\in[M]}1/(p_nq_n)\big)$.
    \end{itemize}
\end{corollary}

Before imposing some regularity conditions for Step~3 in Section~\ref{subsec:estimation},
recall $\v_a^{(m)}$ and $\v_b^{(m)}$ therein.
Then define $s_{Q,m} := \big\| \Q_m' \v_a^{(m)} \big\|$ and $s_{J,m} := \big\| \J_m' \v_b^{(m)} \big\|$,
which are similarly seen in \cite{ChenLam2024}. In particular, we would have $s_{Q,m} \asymp p_m^{\delta_{q,1}/2}$ when $\v_a^{(m)}$ aligns with the column of $\Q_m$ corresponding to the strongest factor.
In the following, we present some assumptions on the projection vectors $\v_a^{(m)}$ and $\v_b^{(m)}$.

\begin{itemize}
    \item[(L2)] (Cross-group alignment of projection directions)
    There exists an absolute constant $c>0$ such that for all $m,n\in[M]$,
    with probability tending to one,
    \begin{align*}
        \big|\v_b^{(m)'} \J_m \J_n' \v_b^{(n)} \big| \geq c\,\|\J_m' \v_b^{(m)}\|\,\|\J_n' \v_b^{(n)}\| ,  \qquad 
        \big|\v_a^{(m)'} \Q_m \Q_n' \v_a^{(n)} \big| \geq c\,\|\Q_m' \v_a^{(m)}\|\,\|\Q_n' \v_a^{(n)}\| .
    \end{align*}
\end{itemize}
\begin{itemize}
    \item[(R2)] (Signal dominance for global common component)
    Let $d_{\text{fac},\star}$ be the largest long-memory exponent among all elements of $\G_t$ and $\F_t^{(m)}$ for $m\in[M]$.
    Then for all $m,n\in[M]$ with $m\neq n$, assume that
    \begin{align*}
        & p_m^{\delta_{r,1}^{(m)} - \delta_{q,1}^{(m)}} q_m^{\delta_{c,1}^{(m)}} = O(s_{J,m}^2) , \;\;
        p_n^{\delta_{r,1}^{(n)} - \delta_{q,1}^{(n)}} q_n^{\delta_{c,1}^{(n)}} = O(s_{J,n}^2) ,  \;\;
         p_m^{1 - \delta_{q,1}^{(m)}} = O(s_{J,m}^2), \;\;
        p_n^{1 - \delta_{q,1}^{(n)}} = O(s_{J,n}^2) , \\
        & q_m^{\delta_{c,1}^{(m)} - \delta_{j,1}^{(m)}} p_m^{\delta_{r,1}^{(m)}} = O(s_{Q,m}^2) ,  \;\;
        q_n^{\delta_{c,1}^{(n)} - \delta_{j,1}^{(n)}} p_n^{\delta_{r,1}^{(n)}} = O(s_{Q,n}^2) , \;\;
        q_m^{1 - \delta_{j,1}^{(m)}} = O(s_{Q,m}^2), \;\;
        q_n^{1 - \delta_{j,1}^{(n)}} = O(s_{Q,n}^2)  .
    \end{align*}
\end{itemize}

Assumption~(L2) formalizes that the selected projection directions align with the same latent global column and row factor directions across different groups. 
This assumption would hold if $\v_b^{(m)}$ and $\v_b^{(n)}$ are the leading singular directions of $\J_m$ and $\J_n$ correspond to the same strongest global factor. Our numerical results in Section~\ref{subsubsec:accuracy_of_common_components} show that the iterative projection procedure in Section~\ref{app:finding_directions} estimate those vectors appropriately.

\begin{theorem}[Consistency of local loading and factor estimators]
\label{thm:common_comp_global_local_init}
Let Assumptions~(IC1), (E1), (E2), (F1), (L1), (L2), and (R2) hold.
Define $\check{\R}_m := \R_m (\R_m'\R_m)^{-1/2}$, $\check{\C}_m := \C_m (\C_m'\C_m)^{-1/2}$,
and
\begin{align*}
    \eta_{Q,m} &:= 
    R(T,d_{\text{fac},\star}) \;
    p_m^{2 (\delta_{q,1}^{(m)} -\delta_{q,k_r}^{(m)})}
    \Bigg( \frac{  \max_{n\in[M] \setminus \{m\}} s_{J,n}^2 p_n^{\delta_{q,1}^{(n)}}}{\max_{n\in[M] \setminus\{m\}} s_{J,n}^2 p_n^{\delta_{q,k_r}^{(n)}}} \Bigg)^2 , \\
    \eta_{J,m} &:=
    R(T,d_{\text{fac},\star}) \;
    q_m^{2 (\delta_{j,1}^{(m)} -\delta_{j,k_c}^{(m)})}
    \Bigg( \frac{  \max_{n\in[M] \setminus \{m\}} s_{Q,n}^2 q_n^{\delta_{j,1}^{(n)}}}{\max_{n\in[M] \setminus\{m\}} s_{Q,n}^2 q_n^{\delta_{j,k_c}^{(n)}}} \Bigg)^2 .
\end{align*}
Further assume $\eta_{Q,m}, \eta_{J,m} = o(1)$.
Then there exist asymptotically orthogonal matrices $\H_{R,m}$ and $\H_{C,m}$ with $\H_{R,m} \H_{R,m}'=\I+o_P(1)$ and $\H_{C,m} \H_{C,m}'=\I+o_P(1)$,
such that
\begin{align*}
    &\quad\quad\quad
    \big\| \wh\R_m - \check{\R}_m \H_{R,m} \big\|_F^2 = O_P(\Delta_{R,m}) , \;\quad
    \big\| \wh\C_m - \check{\C}_m \H_{C,m} \big\|_F^2 = O_P(\Delta_{C,m}) ,
    \quad
    \text{where}  \\
    \Delta_{R,m} &:=
    p_m^{-2 \delta_{r,k_r^{(m)}}^{(m)}}
    q_m^{-2 \delta_{c,1}^{(m)}} 
    \Bigg\{
    \big(p_m^{2 \delta_{r,1}^{(m)}} q_m^{2 \delta_{c,1}^{(m)}} 
    + p_m^{1 + \delta_{q,1}^{(m)}} q_m^{1 + \delta_{j,1}^{(m)}} \big) \eta_{J,m}
    + \frac{p_m^{1 + \delta_{r,1}^{(m)}} q_m^{1 + \delta_{c,1}^{(m)}}}{T} 
    + p_m^{2(1 - \delta_{r,1}^{(m)})}
    + \frac{p_m^2 q_m}{T} 
    \Bigg\} , 
    \\
    \Delta_{C,m} &:=
    q_m^{-2 \delta_{c,k_c^{(m)}}^{(m)}}
    p_m^{-2 \delta_{r,1}^{(m)}} 
    \Bigg\{
    \big(q_m^{2 \delta_{c,1}^{(m)}} p_m^{2 \delta_{r,1}^{(m)}} 
    + q_m^{1 + \delta_{j,1}^{(m)}} p_m^{1 + \delta_{q,1}^{(m)}} \big) \eta_{Q,m}
    + \frac{q_m^{1 + \delta_{c,1}^{(m)}} p_m^{1 + \delta_{r,1}^{(m)}}}{T} 
    + q_m^{2(1 - \delta_{c,1}^{(m)})}
    + \frac{q_m^2 p_m}{T} 
    \Bigg\} .
\end{align*}
Further let Assumption~(S1) hold,
and define $\check{\F}_t^{(m)} := \big( \R_m'\R_m \big)^{1/2} \F_t^{(m)} \big( \C_m'\C_m \big)^{1/2}$.
Then
\begin{align*}
    &\quad
    \frac{1}{
    p_m^{\delta_{r,1}^{(m)}} 
    q_m^{\delta_{c,1}^{(m)}}
    }
    \big\| \wh\F_t^{(m)} - \H_{R,m}^{-1} \check{\F}_t^{(m)} \H_{C,m} \big\|_F^2
    = 
    O_P( \Delta_{R,m} + \Delta_{C,m} 
    + \Delta_{F,m} ) ,
    \quad \text{where} 
    \\
    \Delta_{F,m} &:=
    \min\Big( 
    q_m^{1 - 2\delta_{c,1}^{(m)}}
    p_m^{1 - \delta_{r,1}^{(m)}}
    \Delta_{R,m}
    + 
    q_m^{1 - 2\delta_{c,1}^{(m)}}
    p_m^{ - \delta_{r,1}^{(m)}} 
    ,
    p_m^{1 - 2\delta_{r,1}^{(m)}}
    q_m^{1 - \delta_{c,1}^{(m)}}
    \Delta_{C,m}
    +
    p_m^{1 - 2\delta_{r,1}^{(m)}}
    q_m^{ - \delta_{c,1}^{(m)}}
    \Big) .
\end{align*}
\end{theorem}

Theorem~\ref{thm:common_comp_global_local_init} establishes the consistency of the local factor loading and factor estimators. The terms $\eta_{Q,m}$ and $\eta_{J,m}$ measure the error inherited from the estimated global loading directions, and the condition $\eta_{Q,m},\eta_{J,m}=o(1)$ ensures that the projected cross-group moment matrices $\X_{C,m}$ and $\X_{R,m}$ recover the local factor spaces consistently. 
When all common factors are strong, i.e., all $\delta$'s equal one, and dimensions are comparable, then $\Delta_{R,m}$ and $\Delta_{C,m}$ become $O_P(T^{-1})$. The additional Assumption~(S1) then yields the rate for the local factor series $\wh\F_t^{(m)}$, whose leading term combines $\Delta_{R,m}$, $\Delta_{C,m}$, and an additional term $\Delta_{F,m}$ accounting for the asymmetry between the row and column sides in the refitting.

\begin{theorem}[Consistency of global loading and factor estimators]
\label{thm:common_comp_global}
Let all assumptions in Theorem~\ref{thm:common_comp_global_local_init} hold.
Let $m_0\in[M]$ be a fixed reference group, and define $\check\G_t := \big( \Q_{m_0}'\Q_{m_0} \big)^{1/2} \G_t \big( \J_{m_0}'\J_{m_0} \big)^{1/2}$.
For any $m\in[M]$, define
\begin{align*}
    \Delta_{G,m} &:= 
    \min(p_m,q_m)
    + p_m^{\delta_{r,1}^{(m)}}
    q_m^{\delta_{c,1}^{(m)}}
    \big( \Delta_{R,m} + \Delta_{C,m} + \Delta_{F,m} \big)
    + \big( \eta_{Q,m} + \eta_{J,m} \big) p_m q_m , \\
    S_m &:= \lambda_{\min}\big( \Q_m'\Q_m \big) \lambda_{\min}\big( \J_m'\J_m \big) 
    \asymp 
    p_m^{\delta_{q,k_r}^{(m)}} q_m^{\delta_{j,k_c}^{(m)}} .
\end{align*}
and assume $\Delta_{G,m} = o\big\{ \lambda_{\min}\big( \Q_m'\Q_m \big) \lambda_{\min}\big( \J_m'\J_m \big) \big\}$.
Further assume that the reference group $m_0$ is chosen such that for any $m\in[M]$,
$\lambda_{\min}\big( \Q_{m_0}'\Q_{m_0} \big) \lambda_{\min}\big( \J_{m_0}'\J_{m_0} \big)
= O\big\{ \lambda_{\min}\big( \Q_m'\Q_m \big) \lambda_{\min}\big( \J_m'\J_m \big) \big\}$.
Then there exist asymptotically orthogonal matrices $\H_{Q,m_0}$ and $\H_{J,m_0}$ such that
\begin{align*}
    \big\| \wh\G_t - \H_{Q,m_0}' \check\G_t \H_{J,m_0} \big\|_F^2
    = O_P(\Delta_G) ,
    \quad \text{where} 
    \quad
    \Delta_G :=
    \max_{m\in[M]} \Delta_{G,m}
    + \Big( \max_{m'\in[M]} \frac{\Delta_{G,m'}}{S_{m'}} \Big) \max_{m\in[M]} S_m .
\end{align*}
Also, we have
\begin{align*}
    \big\| \wt\Q_m - \Q_m \big( \Q_{m_0}' \Q_{m_0} \big)^{-1/2} \H_{Q,m_0} \big\|_F^2
    =  O_P(\Delta_{K,m}),
    \quad \text{where} 
    \quad
    \Delta_{K,m} := 
    \Delta_G \frac{ p_m^{\delta_{q,1}^{(m)}}
    q_m^{\delta_{j,1}^{(m)}} }{ p_{m_0}^{2 \delta_{q,k_r}^{(m_0)}}
    q_{m_0}^{2 \delta_{j,k_c}^{(m_0)}} }
    + \frac{ p_m q_m}{T S_{m_0}} ,
\end{align*}
and similarly $\big\| \wt\J_m - \J_m \big( \J_{m_0}' \J_{m_0} \big)^{-1/2} \H_{J,m_0} \big\|_F^2 = O_P(\Delta_{K,m})$.
\end{theorem}

Built upon Theorem~\ref{thm:common_comp_global_local_init},
Theorem~\ref{thm:common_comp_global} addresses the global common component. 
The quantity $\Delta_G$ has two parts. The first, $\max_{m\in[M]} \Delta_{G,m}$, aggregates the local-component estimation errors across groups. The second term in $\Delta_G$ arises from the two-sided alignment and the refitting step, where $S_m$ measures the smallest squared signal in the $m$-th group's projected global loading matrices. 
The signal-dominance condition $\Delta_{G,m}=o(S_m)$ ensures that the perturbed Gram matrices in the alignment step remain well-conditioned, and the reference-group condition guarantees the alignment from any group to the reference group does not blow up. 
Under these conditions, the first part of the theorem shows that $\wh\G_t$ is consistent for the aligned version $\H_{Q,m_0}'\check\G_t\H_{J,m_0}$ of the global factor, which is infeasible by merely using the group-wise ``global'' part in \cite{Zhangetal2025}.
Then the theorem gives the rates for the refitted loadings $\wt\Q_m$ and $\wt\J_m$. In the strong-factor setting where all $\delta$'s equal one and all groups have comparable dimensions, both terms reduce to $O_P(T^{-1})$, delivering consistency of the refitted global loadings up to the asymptotic orthogonal matrices $\H_{Q,m_0}$ and $\H_{J,m_0}$.

\subsection{Detecting the existence of global common factors}\label{subsec:detect_global_cm}
So far we have assumed that the global factors in $\G_t$ drive the dynamics of each data set $\Y_t^{(m)}$ for $m\in[M]$. However, if we are exploring new data sets which may not share common dynamics with others, blindly applying Algorithm~\ref{alg:projection} in Section \ref{app:finding_directions} will give us a false impression that there are global factors driving the dynamics of these new data sets, when there may be none. It is important to first investigate if there are global factors driving the dynamics of any data set at all.

To this end, suppose we have $\wh\Q_{m}$ and $\wh\J_{m}$ as the output of Algorithm~\ref{alg:projection} in Section \ref{app:finding_directions}. Also, for $m\in[M]$, let $\U_m = (\Q_m,\R_m)$ and $\V_m = (\J_m, \C_m)$, and $\wh\U_{m,\cdot c_m}$ and $\wh\V_{m,\cdot d_m}$ be the $c_m$th and $d_m$th eigenvectors of 
\[T^{-1}\sum_{t=1}^T\M_{p_m}\Y_t^{(m)}\M_{q_m}\Y_t^{(m)'}\M_{p_m} \;\text{ and }\; T^{-1}\sum_{t=1}^T\M_{q_m}\Y_t^{(m)'}\M_{p_m}\Y_t^{(m)}\M_{q_m} ,\]
respectively \citep{lam2024matrix}. This means that we are using the data $\{\Y_t^{(m)}\}_{t\in[T]}$ only to estimate the MMEFM %
in (\ref{eqn:MMEFM}) without distinguishing global and local factors.

\begin{theorem}\label{thm:global_factor_detection}
Let all the conditions in Theorem \ref{thm:common_comp_global_local_init} hold. Define the global active set as 
\begin{align}
    \cA_G
    :=\left\{m\in[M]:\text{there exists } n\neq m\in[M] \text{ such that } \{\Q_m\G_t\J_m'\}_{t\in[T]} \text{ and } \{\Q_n\G_t\J_n'\}_{t\in[T]} \text{ are nonzero}\right\},
    \label{eqn:active_set}
\end{align}
and the statistics
\begin{equation}\label{eqn:S_G}
S_{G,m} := \frac{\sum_{n=1;n\neq m}^M \max_{c_m, c_n\in[k_r], d_m,d_n\in[k_c]}\Big\{T^{-1}\sum_{t=1}^T\wh\Q_{m,\cdot c_m}'\check\Y_t^{(m)}\wh\J_{m,\cdot d_m}\wh\J_{n,\cdot d_n}'\check\Y_t^{(n)'}\wh\Q_{n,\cdot c_n}\Big\}^2}{\max_{c_m\in[k_r+k_r^{(m)}], d_m\in[k_c + k_c^{(m)}]}\Big(T^{-1}\sum_{t=1}^T\wh\U_{m,\cdot c_m}'\check\Y_t^{(m)}\wh\V_{m,\cdot d_m}\wh\V_{m,\cdot d_m}'\check\Y_t^{(m)'}\wh\U_{m,\cdot c_m}\Big)^2}, \;\;\; m\in[M].
\end{equation}
Then for any $m \in \cA_G$ and $k\notin \cA_G$, assuming 
\[\max_{n\neq m,n\in\cA_G}p_n^{\delta_{q,k_r}^{(n)}}q_n^{\delta_{j,k_c}^{(n)}}\gtrsim \max(\max_{n\neq m,n\in\cA_G}p_n^{\delta_{q,1}^{(n)}}q_n^{\delta_{j,1}^{(n)}}, \max_{n\notin \cA_G}p_n^{\delta_{r,1}^{(n)}}q_n^{\delta_{c,1}^{(n)}})R(T,d_{\text{fac},\star}),\]
we have
\begin{align*}
S_{G,m} &\gtrsim_P \frac{\max_{n\neq m,n\in\cA_G}p_n^{\delta_{q,k_r}^{(n)}}q_n^{\delta_{j,k_c}^{(n)}}}{p_m^{\delta_{q,1}^{(m)}}q_m^{\delta_{j,1}^{(m)}}}, \;\text{ and }\;\\
S_{G,k} &\lesssim_P \frac{\max\big(\max_{n\in\cA_G}p_n^{\delta_{q,1}^{(n)}}q_n^{\delta_{j,1}^{(n)}}, \max_{n\neq k,n\notin\cA_G}p_n^{\delta_{r,1}^{(n)}}q_n^{\delta_{c,1}^{(n)}}\big)}{p_k^{\delta_{r,1}^{(k)}}q_k^{\delta_{c,1}^{(k)}}}R(T,d_{\text{fac},\star}),\end{align*}
where $R(\cdot,\cdot)$ is defined in (\ref{eqn:R(T,d)}) and $d_{\text{fac},\star}$ in Assumption {\em (R2)}. In particular, if $p_m^{\delta_{q,1}^{(m)}}q_m^{\delta_{j,1}^{(m)}} \asymp p_n^{\delta_{q,1}^{(n)}}q_n^{\delta_{j,1} ^{(n)}} \asymp p_n^{\delta_{q,k_r}^{(n)}}q_n^{\delta_{j,k_c}^{(n)}}$ for any $n, m \in \cA_G$, and $\max_{n\in[M]}(p_n^{\delta_{q,1}^{(n)}}q_n^{\delta_{j,1}^{(n)}}, p_n^{\delta_{r,1}^{(n)}}q_n^{\delta_{c,1}^{(n)}})R(T,d_{\text{fac},\star}) = o(\min_{k\notin\cA_G}p_k^{\delta_{r,1}^{(k)}}q_k^{\delta_{c,1}^{(k)}})$, then
\[S_{G,m} \asymp_P 1, \;\;\; S_{G,k} = o_P(1).\]
\end{theorem}
The last conditions in the above theorem translate to that all signals from the global common factors from different groups are of the same order, while the maximum global or local signals are of a smaller order than the minimum local ones after multiplying a factor of $R(T,d_{\text{fac},\star})$.
In practice, we need to replace $k_r$, $k_r^{(m)}$, $k_c$, $k_c^{(m)}$ in the definition of $S_{G,m}$ by their estimators described in Section \ref{subsec:rank_estimation}. We stress that over-estimating these numbers of factors is not much of a problem, but under-estimating them can lead to failure in detecting the existence of global common factors potentially. Below we provide a theorem for the rank estimators for all the loading matrices presented in Section \ref{subsec:rank_estimation}.

\begin{theorem}\label{thm:rank_estimation}
Let all the conditions in Theorem \ref{thm:common_comp_global_local_init} hold. Then for each $m\in[M]$, as $\min(T,p_m,q_m) \rightarrow \infty$,
\[P(\wh{r}_1 \geq r_1) \rightarrow 1, \;\; P(\wh{\ell}_1 \geq \ell_1) \rightarrow 1, \;\; P(\wh{r}_2^{(m)} \geq r_2^{(m)}) \rightarrow 1, \;\; P(\wh{\ell}_2^{(m)} \geq \ell_2^{(m)}) \rightarrow 1,\]
if $\eta_{A_{1m}} = o(\nu_{r,m}^2)$ and $\eta_{B_{1m}} = o(\nu_{c,m}^2)$, where $\eta_{A_{1m}}$ and $\eta_{B_{1m}}$ are the rates defined in Theorem \ref{thm:initial_A1m}.
Moreover, we also have $P(\wh{k}_r \geq k_r) \rightarrow 1$ and $P(\wh{k}_c \geq k_c) \rightarrow 1$ if
\begin{align*}
&\max_{m,n\in[M],n\neq m}p_m^{\delta_{q,1}^{(m)}}q_m^{\delta_{j,1}^{(m)}}p_n^{\delta_{q,1}^{(n)}}q_n^{\delta_{j,1}^{(n)}}\cdot\max_{m,n\in[M],n\neq m}\big[T^{-1}p_mq_mp_nq_n + p_m^{\delta_{q,1}^{(m)}}q_m^{\delta_{j,1}^{(m)}}p_n^{\delta_{q,1}^{(n)}}q_n^{\delta_{j,1}^{(n)}}R(T,d_{\text{fac},\star})\big] \\
&=o\Big(\min\big(\max_{m,n\in[M],n\neq m} p_m^{2\delta_{q,k_r}^{(m)}}q_m^{2\delta_{j,1}^{(m)}}p_n^{2\delta_{q,k_r}^{(n)}}q_n^{2\delta_{j,1}^{(n)}}, 
\max_{m,n\in[M],n\neq m} p_m^{2\delta_{q,1}^{(m)}}q_m^{2\delta_{j,k_c}^{(m)}}p_n^{2\delta_{q,1}^{(n)}}q_n^{2\delta_{j,k_c}^{(n)}}\big)\Big).
\end{align*}
Finally, we also have $P(\wh{k}_r^{(m)} \geq k_r^{(m)}) \rightarrow 1$ and $P(\wh{k}_c^{(m)} \geq k_c^{(m)}) \rightarrow 1$ if
\begin{align*}
    q_m^{2(1-\delta_{c,1}^{(m)})} + T^{-1}q_m^2p_m + q_m^{1+\delta_{j,1}^{(m)}}p_m^{1+\delta_{q,1}^{(m)}}\eta_{Q,m} &= o(q_m^{4\delta_{c,k_c^{(m)}}^{(m)}-2\delta_{c,1}^{(m)}}p_m^{2\delta_{r,1}^{(m)}}),\\
    p_m^{2(1-\delta_{r,1}^{(m)})} + T^{-1}p_m^2q_m + p_m^{1+\delta_{q,1}^{(m)}}q_m^{1+\delta_{j,1}^{(m)}}\eta_{J,m} &= o(p_m^{4\delta_{r,k_r^{(m)}}^{(m)}-2\delta_{r,1}^{(m)}}q_m^{2\delta_{c,1}^{(m)}}).
\end{align*}
If all factors are strong and all processes have short memory, the greater than or equal sign in all asymptotic probabilities above will become the equal sign, if we add the following terms to the numerator and denominator of the eigenvalue ratios defining $\wh{r}_1$, $\wh{r}_2^{(m)}$, $\wh{\ell}_1$, $\wh{\ell}_2^{(m)}$, $\wh{k}_r$, $\wh{k}_c$, $\wh{k}_r^{(m)}$ and $\wh{k}_c^{(m)}$ respectively ($C$ is a generic constant):
\begin{align*}
&\max_{m\neq n\in[M]}CT^{-1}p_mp_n, \; CT^{-1}p_m^2, \; \max_{m\neq n\in[M]}CT^{-1}q_mq_n, \;  CT^{-1}q_m^2\;\\
&\max_{m\neq n\in[M]} CT^{-1}p_mq_mp_nq_n, \; \max_{m\neq n\in[M]} CT^{-1}p_mq_mp_nq_n, \;  CT^{-1/2}p_mq_m, \; CT^{-1/2}p_mq_m.
\end{align*}
\end{theorem}
The addition of the rates in the last part of Theorem \ref{thm:rank_estimation} stabilizes the eigenvalue ratios to constant order after the dip to $o_P(1)$ at the true number of factors, hence proving the consistency of the rank estimators.

\section{Practical Detection and Screening of Global Common Factors}\label{sec:practical_global_screening}

Theorem~\ref{thm:global_factor_detection} provides a theoretical grounding for detecting global common factors through the group-level statistics $S_{G,m}$, $m\in[M]$. From this, we derive two practical procedures for detecting which group of data, $\{\Y_t^{(m)}\}_{t\in[T]}$ for $m\in[M]$, share the global common factors. The first one investigates whether any nontrivial global common factors exist, and the second one screens which groups are driven by these factors.

For practical implementation, we  standardize $\check\Y_t^{(m)}$ and define
\begin{align}
    \check\Z_t^{(m)}=\check\Y_t^{(m)} / \wh\sigma_m,
    \qquad
    \wh\sigma_m^2=(Tp_mq_m)^{-1}\sum_{t=1}^T\norm{\check\Y_t^{(m)}}_F^2.
    \label{eqn:standardize_Yt_check}
\end{align}
The statistic $S_{G,m}$ in \eqref{eqn:S_G} is then computed using $\check\Z_t^{(m)}$ in place of $\check\Y_t^{(m)}$. This normalization reduces the influence of purely marginal scale differences between groups, preventing signal domination from groups just for their larger scales.

Let $S_{G,(1)}\ge S_{G,(2)}\ge \cdots \ge S_{G,(M)}$ denote the decreasing order statistics of $S_{G,1},\ldots,S_{G,M}$. Testing the existence of global factors  is equivalent to testing the following hypotheses:
\begin{align*}
    H_{0,G}:\cA_G=\varnothing
    \qquad\text{v.s.}\qquad
    H_{1,G}:|\cA_G|\ge 2 .
\end{align*}
Theorem~\ref{thm:global_factor_detection} implies that $S_{G,(1)}$ could be used for this test, since all $S_{G,m}$ decay under $H_{0,G}$, while at least two are of constant order under $H_{1,G}$. 
Nevertheless, we use $S_{G,(2)}$ because $H_{1,G}$ requires the global common component to be shared by at least two groups, so second-largest statistic directly reflects the alternative.
Given an existence cutoff $\tau_1$, we reject $H_{0,G}$ when $S_{G,(2)}>\tau_1$. 
After rejecting the null, we may further screen which groups share the global common factors. Let $\tau_2$ be a screening cutoff and define the initial selected set $\wh\cA_G^{\circ}=\{m\in[M]:S_{G,m}>\tau_2\}$. The final selected set is then
\begin{align*}
    \wh\cA_G
    =
    \begin{cases}
        \wh\cA_G^{\circ},
        & S_{G,(2)}>\tau_1 \text{ and } |\wh\cA_G^{\circ}|\ge 2,\\
        \varnothing,
        & \text{otherwise}.
    \end{cases}
\end{align*}
The definition of $\wh\cA_G$ enforces the convention that a global common component must involve at least two groups, cf.\ Remark~3 of \cite{Cen2026_multiorder}. It also prevents reporting isolated selected groups when the evidence for a global factor is insufficient.

We next discuss how to choose the cutoffs $\tau_1$ and $\tau_2$. Although Theorem~\ref{thm:global_factor_detection} suggests that inactive statistics decay at the rate $R(T,d_{\text{fac},\star})$, we do not directly use a deterministic rate threshold in the main procedure, since the finite-sample magnitude of $S_{G,m}$ can still depend on constants related to $M$, the dimensions $p_m,q_m$, factor strengths, among others. Instead, we use a circular-shift bootstrap to choose the existence and screening cutoffs $\tau_1$ and $\tau_2$.

The bootstrap is designed to preserve the within-group temporal dependence and marginal residual structure, while breaking the synchronous cross-group alignment that would be induced by global common factors. The procedure is summarized as follows.

\noindent\hrulefill
\vspace{-9pt}

\refstepcounter{mmalgorithm}\label{alg:bootstrap}
\noindent\textbf{Algorithm~\themmalgorithm} Circular-shift bootstrap calibration for global common factor detection and screening

\vspace{-13pt}
\noindent\hrulefill
\begin{itemize}
\vspace{-12pt}

    \item[1.] Given the standardized arrays $\{\check\Z_t^{(m)}\}$ from \eqref{eqn:standardize_Yt_check}, choose the bootstrap size $B$ and calibration level $\alpha$. Fix one reference group $m_0\in[M]$.

    \item[2.] For each bootstrap replication $b=1,\ldots,B$, keep the reference group fixed: $\check\Z_t^{*(m_0,b)}=\check\Z_t^{(m_0)}$ for $t\in[T]$. For each $m\ne m_0$, draw an independent circular shift $\ell_m^{(b)}$ uniformly from $\{0,1,\ldots,T-1\}$, and define $\check\Z_t^{*(m,b)}=\check\Z_{1+\{(t+\ell_m^{(b)}-1)\bmod T\}}^{(m)}$ for $t\in[T]$.

    \item[3.] Reapply the detection procedure used to compute $S_{G,m}$ to the shifted arrays $\{\check\Z_t^{*(m,b)}\}$, and obtain $S_{G,1}^{*(b)},\ldots,S_{G,M}^{*(b)}$.

    \item[4.] For each $b\in[B]$, let $S_{G,(1)}^{*(b)}\ge S_{G,(2)}^{*(b)}\ge\cdots\ge S_{G,(M)}^{*(b)}$ denote the decreasing order statistics of $S_{G,1}^{*(b)},\ldots,S_{G,M}^{*(b)}$. After $B$ bootstrap replications, set
    \begin{align*}
        \tau_1=\tau_2=\wh q_{1-\alpha}\left\{S_{G,(2)}^{*(b)}:b=1,\ldots,B\right\},
    \end{align*}
    where $\wh q_{1-\alpha}$ denotes the empirical $(1-\alpha)$-quantile.
\end{itemize}
\vspace{-12pt}
\noindent\hrulefill

We use $S_{G,(2)}^{*(b)}$ to calibrate both $\tau_1$ and $\tau_2$, aligning the screening rule with the existence test. Comparisons in Section~\ref{app:orderstat_comparison} of the Supplementary Materials show that the analogous $S_{G,(1)}$-based procedure has similar existence-test performance but is more conservative in screening and can substantially reduce recovery.

\begin{remark}\label{remark:bootstrap}
The circular-shift bootstrap in Algorithm~\ref{alg:bootstrap} is formulated under the assumption that the factor series have a single stationary segment. For data with known or unknown structural breaks, or more generally with multiple regimes, whole-sample circular shifting may mix observations from different regimes. A possible extension is to first divide the sample into regimes, either using known break points or estimated break points from change point detection methods, and then apply circular shifting within each regime.
\end{remark}

\section{Numerical Results}\label{sec: numerical_results}
\subsection{Simulation}\label{subsec: simulation}
\subsubsection{Data generating process}\label{subsubsec: DGP}

We generate a collection of matrix time series $\{\Y_t^{(m)}\}_{t\in[T]}$, where $\Y_t^{(m)}\in\bb{R}^{p_m\times q_m}$, from the MMEFM. 

\textbf{Loadings}: For each group $m$, the entries of $\A_{1m}$, $\A_{2m}$, $\B_{1m}$, $\B_{2m}$, $\Q_m$, $\J_m$, $\R_m$, and $\C_m$ are first generated independently from $N(0,1)$. We impose the desired factor strengths by columnwise rescaling. Specifically, if $\L$ is a loading matrix with $p$ rows and its $j$th column has strength parameter $\xi_j\in(1/2,1]$, we set $\L_{\cdot,j}=p^{-(1-\xi_j)/2}\Z_{\cdot,j}$ with $\Z_{\cdot,j}\sim N(\0,\I_p)$, so that $\norm{\L_{\cdot,j}}_2^2\asymp p^{\xi_j}$. Each column is then demeaned to satisfy (IC1). We set $r_1=r_2^{(m)}=\ell_1=\ell_2^{(m)}=2$ for the main effects, $k_r=k_c=3$ for the global common component, and $k_r^{(m)}=k_c^{(m)}=2$ for the local common components.

\textbf{Factor dynamics}: All latent factor series are generated according to the piecewise-stationary construction described in subsection~\ref{subsec:assumption}. This applies to $\mu_t$, $\mu_t^{(m)}$, and the entries of $\balpha_t$, $\bbeta_t$, $\balpha_t^{(m)}$, $\bbeta_t^{(m)}$, $\G_t$, and $\F_t^{(m)}$. We use two equal-length segments. For a generic scalar series, we first generate $x_t=\sum_{w\geq 0}\phi^w\eta_{t-w}$ with $\phi=0.6$ and $\eta_t\sim N(0,1)$ independently over $t$. Within segment $s$, we then set $y_t=(1-B)^{-d_s}x_t$, where $B$ denotes the backshift operator and $(1-B)^{-d_s}=\sum_{w\ge 0}\pi_w(d_s)B^w$, with $\pi_0(d_s)=1$ and $\pi_w(d_s)=\pi_{w-1}(d_s)(w-1+d_s)/w$ for $w\ge 1$. When $d_s=0$, this reduces to the short-memory AR(1) case. This yields segmentwise stationary dynamics with polynomially decaying dependence when $d_s>0$, while allowing the structure to change at the midpoint. To enforce the sample orthogonality in (IC2), let $\A=[\balpha_1,\ldots,\balpha_T]'$ and $\A^{(m)}=[\balpha_1^{(m)},\ldots,\balpha_T^{(m)}]'$. We replace $\A^{(m)}$ by $(\I-\P_\alpha)\A^{(m)}$, where $\P_\alpha=\A(\A'\A)^{-1}\A'$. Thus $\A'\A^{(m)}=0$. We apply the same construction to $\bbeta_t$ and $\bbeta_t^{(m)}$. The local grand means are demeaned across groups at each $t$ so that $\sum_{m=1}^M\mu_t^{(m)}=0$.

\textbf{Idiosyncratic noise}: For $\Q_{e,m}$, $\J_{e,m}$, $\R_{e,m}$, and $\C_{e,m}$, entries are first generated independently from $N(0,1)$ and are then set to zero independently with probability $0.95$. The entries of $\G_{e,t}$ and $\F_{e,t}^{(m)}$ follow independent AR(1) processes with coefficient $0.2$ and standard normal innovations. The entries of $\bepsilon_t^{(m)}$ are generated in the same way. Finally, the entries of $\bSigma_{\epsilon,m}$ are generated independently from $|N(0,1)|$.

Unless otherwise stated, we take $M=3$ groups and use 500 Monte Carlo replications in each setting. To evaluate the accuracy of our estimators for time-varying quantities, given any series of parameters $\btheta=\{\btheta_t\}_{t\in[T]}$, where $\btheta_t$ can be a scalar, vector, or matrix, and its estimator $\wh\btheta=\{\wh\btheta_t\}_{t\in[T]}$, we define the (relative) mean squared error as
$$
\MSE(\wh\btheta):=\frac{\sum_{t=1}^{T}\|\btheta_t-\wh\btheta_t\|_F^2}{\sum_{t=1}^T\|\btheta_t\|_F^2}.
$$
Moreover, for any given loading matrix $\Q$ and its estimator $\wh\Q$, we use the column-space distance
$$
\cD(\Q,\wh\Q):=\Big\|\Q(\Q'\Q)^{-1}\Q' - \wh\Q(\wh\Q'\wh\Q)^{-1}\wh\Q' \Big\|,
$$
which is widely used in the literature such as \cite{chen2022factor}, among others. For group-specific quantities, $\MSE$ or $\cD$ is first computed within each group and then averaged over $m$. 

\subsubsection{Accuracy of main effects}\label{subsubsec:accuracy_of_main_effects}

We first study estimation of the grand means and main effects under four designs (Ia)--(Id). Each design is considered at $T\in\{100,400\}$. The common component part of the DGP follows the baseline design (IIa) in subsection~\ref{subsubsec:accuracy_of_common_components}.

\begin{itemize}
    \item[(Ia)] \textbf{Main effect baseline.} Set $(p_m,q_m)=(30,30)$ for each $m\in[M]$. All loading strength parameters in (ME1) are set to $1$. For all latent series, we set memory parameters $d_1=d_2=0$.

    \item[(Ib)] \textbf{Long-memory main effects.} Same as (Ia), except that $d_1=d_2=0.1$ for $\mu_t$, $\mu_t^{(m)}$, and each entry of $\balpha_t$, $\bbeta_t$, $\balpha_t^{(m)}$, and $\bbeta_t^{(m)}$.

    \item[(Ic)] \textbf{Weaker main effect factors.} Same as (Ia), except that $\gamma_r=\gamma_c=\gamma_r^{(m)}=\gamma_c^{(m)}=0.8$ for each $m\in[M]$.

    \item[(Id)] \textbf{Larger imbalanced dimensions.} Same as (Ia), except that $(p_m,q_m)=(30,60)$ for each $m\in[M]$.
\end{itemize}

Table~\ref{tab:main-effect-accuracy} reports the results. The direct moment estimators change little with $T$, since their pointwise accuracy is driven by cross-sectional averaging; the slightly smaller relative MSEs under long memory in (Ib) mainly reflect the larger signal magnitude in the denominator rather than improved estimation accuracy. In contrast, the main effect loading space errors decrease clearly with $T$, consistent with Theorems~\ref{thm:initial_A1m}--\ref{thm:A1m_alphat}. Long memory mainly increases dispersion, while weaker loadings moderately reduce accuracy. Increasing only $q_m$ in (Id) gives a clearer improvement on the column side. The initial and refitted global loading estimates have nearly identical errors, indicating that refitting mainly links the loadings to their corresponding estimated global main effect factor series rather than changing their estimated spaces. Accordingly, the global main effect products improve sharply after refitting, while the local products also become more accurate as $T$ increases. Long memory and weaker loadings mainly affect small-$T$ accuracy, while the imbalanced design remains stable.

\begin{table}[ht!]
\centering
\caption{Main effect estimation accuracy. All entries are Monte Carlo means multiplied by $10^2$, with standard deviations in parentheses.}
\label{tab:main-effect-accuracy}
\setlength{\tabcolsep}{3.5pt}
\resizebox{\textwidth}{!}{%
\begin{tabular}{l|cc|cc|cc|cc}
\toprule
Settings & \multicolumn{2}{c}{Ia} & \multicolumn{2}{c}{Ib} & \multicolumn{2}{c}{Ic} & \multicolumn{2}{c}{Id}\\
 & $T=100$ & $T=400$ & $T=100$ & $T=400$ & $T=100$ & $T=400$ & $T=100$ & $T=400$\\
\midrule
$\MSE(\wh c_t^{(m)})$ & 0.05(0.01) & 0.05(0.00) & 0.04(0.01) & 0.04(0.00) & 0.05(0.01) & 0.05(0.00) & 0.02(0.00) & 0.02(0.00)\\
$\MSE(\wh\a_t^{(m)})$ & 0.59(0.07) & 0.58(0.06) & 0.48(0.07) & 0.47(0.04) & 1.17(0.13) & 1.14(0.11) & 0.30(0.04) & 0.29(0.03)\\
$\MSE(\wh\b_t^{(m)})$ & 0.60(0.07) & 0.58(0.05) & 0.49(0.07) & 0.47(0.05) & 1.17(0.14) & 1.16(0.11) & 0.59(0.06) & 0.58(0.04)\\
$\MSE(\wh\mu_t)$ & 0.03(0.01) & 0.03(0.00) & 0.02(0.01) & 0.02(0.00) & 0.03(0.01) & 0.03(0.00) & 0.01(0.00) & 0.01(0.00)\\
$\MSE(\wh\mu_t^{(m)})$ & 0.08(0.01) & 0.08(0.01) & 0.06(0.01) & 0.06(0.01) & 0.08(0.02) & 0.08(0.01) & 0.04(0.01) & 0.04(0.00)\\
\addlinespace[0.45em]
$\cD(\A_{1m},\wh\A_{1m})$ & 6.53(4.16) & 2.77(1.11) & 8.27(7.02) & 3.19(1.63) & 6.82(4.71) & 2.89(1.07) & 6.14(4.46) & 2.57(1.02)\\
$\cD(\A_{1m},\wt\A_{1m})$ & 6.51(4.16) & 2.76(1.11) & 8.26(7.02) & 3.19(1.63) & 6.79(4.71) & 2.88(1.07) & 6.14(4.46) & 2.57(1.02)\\
$\cD(\A_{2m},\wh\A_{2m})$ & 5.91(2.37) & 2.72(0.86) & 7.55(4.45) & 3.12(1.21) & 6.43(2.71) & 2.89(0.91) & 5.69(2.72) & 2.57(0.88)\\
$\cD(\B_{1m},\wh\B_{1m})$ & 6.78(5.27) & 2.76(1.07) & 9.64(9.27) & 3.26(1.46) & 7.07(4.26) & 2.82(0.95) & 4.92(4.10) & 2.01(0.71)\\
$\cD(\B_{1m},\wt\B_{1m})$ & 6.76(5.27) & 2.76(1.07) & 9.62(9.27) & 3.26(1.46) & 7.05(4.26) & 2.80(0.95) & 4.91(4.10) & 2.00(0.71)\\
$\cD(\B_{2m},\wh\B_{2m})$ & 6.28(3.54) & 2.68(0.86) & 8.38(6.53) & 3.16(1.15) & 6.59(2.57) & 2.82(0.78) & 4.57(3.22) & 1.98(0.59)\\
\addlinespace[0.45em]
$\MSE(\wh\A_{1m}\wh\balpha_t)$ & 101.74(10.48) & 102.27(10.74) & 102.17(9.27) & 101.31(8.93) & 104.18(14.56) & 104.66(14.85) & 102.22(9.85) & 102.09(10.05)\\
$\MSE(\wt\A_{1m}\wh\balpha_t)$ & 0.44(0.60) & 0.10(0.06) & 0.89(2.41) & 0.12(0.13) & 0.60(1.85) & 0.13(0.07) & 0.45(1.99) & 0.07(0.05)\\
$\MSE(\wh\A_{2m}\wh\balpha_t^{(m)})$ & 0.59(0.58) & 0.18(0.07) & 1.11(2.49) & 0.20(0.13) & 0.82(1.59) & 0.28(0.08) & 0.56(1.46) & 0.13(0.06)\\
$\MSE(\wh\B_{1m}\wh\bbeta_t)$ & 101.86(9.91) & 102.09(10.26) & 102.14(10.31) & 101.64(9.40) & 103.32(15.62) & 104.52(15.73) & 101.31(7.17) & 101.09(7.34)\\
$\MSE(\wt\B_{1m}\wh\bbeta_t)$ & 0.62(2.25) & 0.10(0.05) & 1.41(4.77) & 0.12(0.10) & 0.56(0.90) & 0.13(0.06) & 0.31(1.30) & 0.05(0.03)\\
$\MSE(\wh\B_{2m}\wh\bbeta_t^{(m)})$ & 0.76(1.97) & 0.18(0.07) & 1.60(5.09) & 0.21(0.11) & 0.78(0.80) & 0.27(0.07) & 0.43(2.26) & 0.09(0.03)\\
\bottomrule
\end{tabular}%
}
\end{table}

\subsubsection{Accuracy of common components}
\label{subsubsec:accuracy_of_common_components}

We next study estimation of the global and local common components under four designs (IIa)--(IId). Each design is considered at $T\in\{100,400\}$. The main effect part of the DGP follows the baseline design (Ia) in subsection~\ref{subsubsec:accuracy_of_main_effects}.

\begin{itemize}
    \item[(IIa)] \textbf{Common component baseline.} Set $(p_m,q_m)=(30,30)$ for each $m\in[M]$. All loading strength parameters in (L1) are set to $1$. For all latent series, we set memory parameters $d_1=d_2=0$.

    \item[(IIb)] \textbf{Long-memory common factors.} Same as (IIa), except that $d_1=d_2=0.1$ for each entry of $\G_t$ and $\F_t^{(m)}$.

    \item[(IIc)] \textbf{Weaker common factors.} Same as (IIa), except that for each $m\in[M]$, $\delta_{q,i}^{(m)}=\delta_{j,i}^{(m)}=0.8$ for $i=1,2,3$ and $\delta_{r,i}^{(m)}=\delta_{c,i}^{(m)}=0.8$ for $i=1,2$.

    \item[(IId)] \textbf{Larger imbalanced dimensions.} Same as (IIa), except that $(p_m,q_m)=(30,60)$ for each $m\in[M]$.
\end{itemize}

Table~\ref{tab:common-component-accuracy} reports the results. The refitted global loading estimates $\wt\Q_m$ and $\wt\J_m$ are substantially more accurate than the initial estimates $\wh\Q_m$ and $\wh\J_m$ in every design, showing the benefit of cross-group alignment and refitting. The loading-space errors also fall clearly as $T$ increases; weaker loadings in (IIc) make estimation more difficult, while the larger imbalanced dimensions in (IId) remain stable and generally improve accuracy. The improvement is even more pronounced for the global common component product, whose error drops sharply after alignment and refitting; the local product also improves with $T$. Under long memory, several $T=100$ entries have large Monte Carlo variation, but this largely disappears at $T=400$.

\begin{table}[ht!]
\centering
\caption{Common component estimation accuracy. All entries are Monte Carlo means multiplied by $10^2$, with standard deviations in parentheses.}
\label{tab:common-component-accuracy}
\setlength{\tabcolsep}{3.5pt}
\resizebox{\textwidth}{!}{%
\begin{tabular}{l|cc|cc|cc|cc}
\toprule
Settings & \multicolumn{2}{c}{IIa} & \multicolumn{2}{c}{IIb} & \multicolumn{2}{c}{IIc} & \multicolumn{2}{c}{IId}\\
 & $T=100$ & $T=400$ & $T=100$ & $T=400$ & $T=100$ & $T=400$ & $T=100$ & $T=400$\\
\midrule
$\cD(\Q_m,\wh\Q_m)$ & 7.98(3.09) & 3.90(1.45) & 9.84(5.71) & 4.52(1.77) & 9.21(3.03) & 4.46(1.41) & 5.75(2.21) & 2.73(1.05)\\
$\cD(\Q_m,\wt\Q_m)$ & 1.77(0.40) & 0.79(0.15) & 2.13(4.02) & 0.79(0.18) & 3.17(0.46) & 1.44(0.18) & 1.13(0.20) & 0.50(0.07)\\
$\cD(\J_m,\wh\J_m)$ & 8.22(3.15) & 3.93(1.57) & 9.69(5.65) & 4.51(1.74) & 9.27(3.16) & 4.24(1.25) & 8.24(3.12) & 3.73(1.32)\\
$\cD(\J_m,\wt\J_m)$ & 1.56(0.20) & 0.71(0.08) & 1.75(3.54) & 0.67(0.08) & 3.01(0.37) & 1.38(0.16) & 1.46(0.15) & 0.66(0.06)\\
\addlinespace[0.45em]
$\cD(\R_m,\wh\R_m)$ & 2.53(0.64) & 1.01(0.18) & 3.28(4.47) & 1.05(0.22) & 4.23(0.79) & 2.42(0.55) & 2.23(0.72) & 0.76(0.15)\\
$\cD(\C_m,\wh\C_m)$ & 2.50(0.66) & 1.01(0.16) & 3.33(4.54) & 1.04(0.22) & 4.23(0.90) & 2.41(0.56) & 2.09(0.48) & 0.84(0.10)\\
\addlinespace[0.45em]
$\MSE(\wh\Q_m\wh\G_t\wh\J_m')$ & 155.20(39.46) & 153.60(34.79) & 155.72(38.31) & 155.31(36.50) & 156.11(38.47) & 156.13(34.55) & 152.32(37.03) & 153.54(36.18)\\
$\MSE(\wt\Q_m\wh\G_t\wt\J_m')$ & 0.07(0.02) & 0.04(0.01) & 0.49(9.09) & 0.03(0.01) & 0.25(0.05) & 0.16(0.02) & 0.04(0.01) & 0.02(0.00)\\
$\MSE(\wh\R_m\wh\F_t^{(m)}\wh\C_m')$ & 0.25(0.11) & 0.13(0.03) & 0.82(10.24) & 0.12(0.03) & 0.71(0.20) & 0.48(0.10) & 0.14(0.06) & 0.06(0.01)\\
\bottomrule
\end{tabular}%
}
\end{table}

To assess our common component estimation procedure, we compare it with the basic and augmented MFM estimators of \cite{Zhangetal2025}. For comparability, all three estimators use the same $\check\Y_t^{(m)}$ for common component estimation. In addition to setting~(IIa), we consider setting~(IIc') which is the same as (IIa), except that the global loading strengths are $(1,0.8,0.8)$ in both modes, while the local loading strengths remain $(1,1)$. The row and column designs are symmetric, so we report only the row-side results. Table~\ref{tab:zhang-comparison} shows that the augmented MFM improves substantially on the basic estimator, while our estimator yields smaller errors throughout. The advantage is more pronounced when the global loadings are weaker.

\begin{table}[ht!]
\centering
\caption{Global row loading-space estimation accuracy for MMEFM and MFM estimators.
MFM Augmented uses 15 iterations. All entries are Monte Carlo means multiplied
by $10^2$, with standard deviations in parentheses.}
\label{tab:zhang-comparison}
\small
\setlength{\tabcolsep}{5pt}
\renewcommand{\arraystretch}{0.95}
\begin{tabular}{l|cc|cc}
\toprule
Settings & \multicolumn{2}{c}{IIa} & \multicolumn{2}{c}{IIc$'$} \\
 & $T=100$ & $T=400$ & $T=100$ & $T=400$ \\
\midrule
MMEFM          & 1.78(0.42)  & 0.80(0.14) & 5.15(13.91)  & 1.30(0.22) \\
MFM Basic      & 14.96(8.95) & 3.84(1.34) & 69.33(21.23) & 17.08(11.93) \\
MFM Augmented  & 2.71(2.08)  & 1.22(0.16) & 10.07(13.33) & 2.05(0.30) \\
\bottomrule
\end{tabular}
\end{table}

\subsubsection{Behavior of the statistic
  \texorpdfstring{$S_{G,m}$}{S(G,m)}}\label{subsubsec:sgm_behavior}

We next examine the finite-sample behavior of the statistic $S_{G,m}$ in Theorem~\ref{thm:global_factor_detection}. Let $\cA_G$ denote the active set of groups containing a shared global common component, and let $\cA_G^c=[M]\setminus\cA_G$. Since a global common component must be shared by at least two groups, we consider $|\cA_G|\in\{0,2,3,4,5\}$ with $M=5$. The active groups are randomly selected in each Monte Carlo replication. We use 400 replications and set $T=200$ unless otherwise stated.

We consider six settings. The first four settings are inherited from (IIa)--(IId) and are relabelled as (IIIa)--(IIId), respectively. We additionally consider

\begin{itemize}
    \item[(IIIe)] \textbf{Weaker local common component.} Same as (IIIa), except that $\delta_{r,i}^{(m)}=\delta_{c,i}^{(m)}=0.8$ for $i=1,2$ and each $m\in[M]$.

    \item[(IIIf)] \textbf{Weaker global common component.} Same as (IIIa), except that $\delta_{q,i}^{(m)}=\delta_{j,i}^{(m)}=0.8$ for $i=1,2,3$ and each $m\in[M]$.
\end{itemize}

\begin{figure}[ht!]
\centering
\includegraphics[width=.95\textwidth]{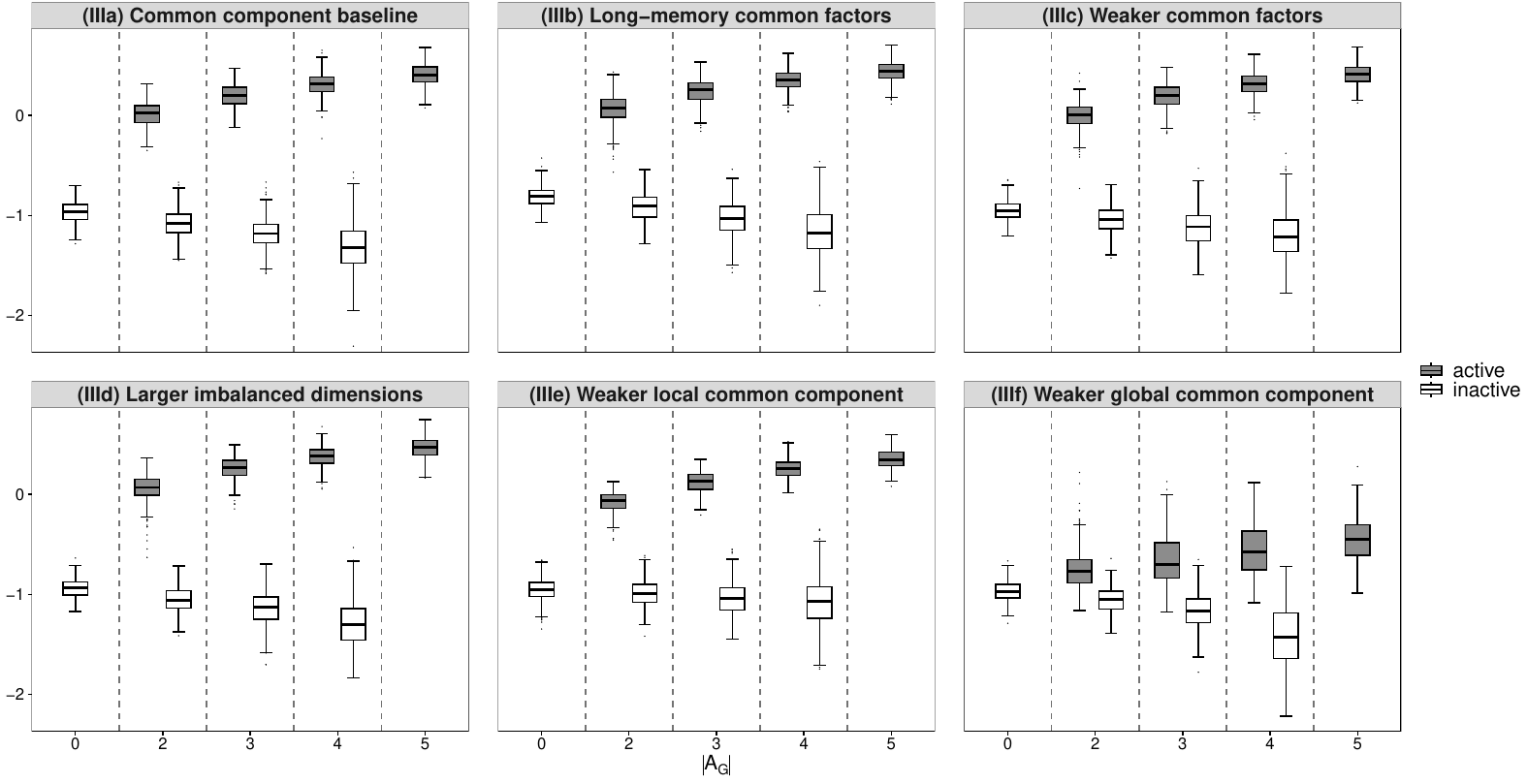}
\caption{Averages of $S_{G,m}$ over the active and inactive groups at $T=200$. The y-axis is $\log_{10}$-transformed.}
\label{fig:sgm_boxplot}
\end{figure}

Before computing $S_{G,m}$, we standardize $\check{\Y}_t^{(m)}$ using \eqref{eqn:standardize_Yt_check}. Figure~\ref{fig:sgm_boxplot} reports the within-replication averages of $S_{G,m}$ over the active and inactive groups at $T=200$. The active and inactive averages are clearly separated in settings (IIIa)--(IIIe). The weaker-global setting (IIIf) is more difficult, especially when only a few groups are active. The separation also tends to increase with $|\cA_G|$, as each active group then shares the global component with more groups.

\begin{figure}[ht!]
\centering
\includegraphics[width=.95\textwidth]{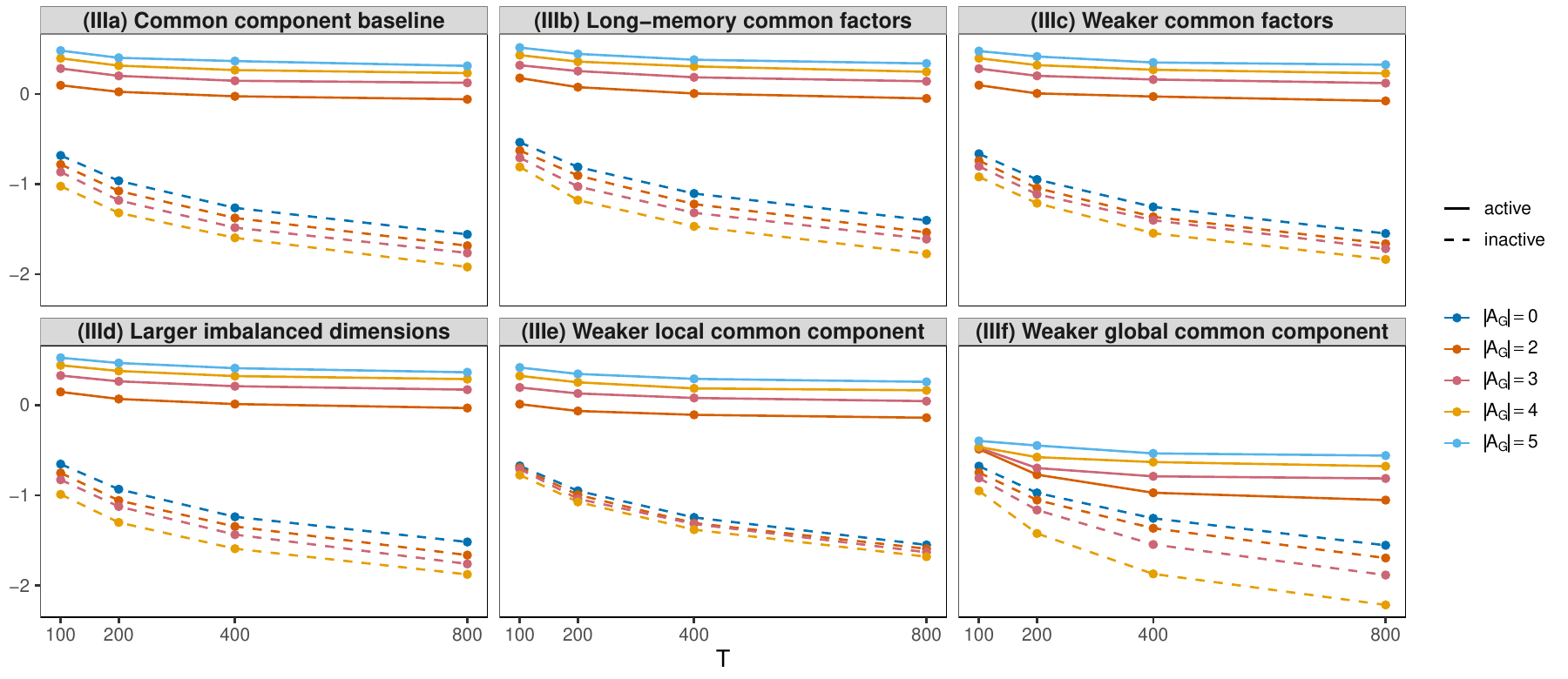}
\caption{Medians of the averages of $S_{G,m}$ over the active and inactive groups, respectively. The y-axis is $\log_{10}$-transformed. Solid and dashed lines correspond to active and inactive groups, respectively.}
\label{fig:sgm_curve}
\end{figure}

Figure~\ref{fig:sgm_curve} plots the Monte Carlo medians of these active and inactive averages as $T$ increases. Across all settings, the active averages remain relatively stable, while the inactive averages decrease. The gap therefore widens with $T$, consistent with Theorem~\ref{thm:global_factor_detection}. In the weaker-global setting (IIIf), the active averages are lower in finite samples but remain much more stable than the inactive averages.

\subsubsection{Accuracy of detection and screening}

We next examine whether the separation of $S_{G,m}$ translates into reliable detection and screening using the procedure in Section~\ref{sec:practical_global_screening}. For these experiments, we use 400 Monte Carlo replications and true factor ranks, set the nominal level to $5\%$, and use $B=199$ bootstrap replications.

Tables~\ref{tab:bootstrap-existence-baseline} and \ref{tab:bootstrap-existence-settings} report the performance of the bootstrap existence test. In the baseline setting, the empirical size remains close to the nominal $5\%$ level across sample sizes, while power increases rapidly with both $T$ and $|\cA_G|$. At $T=200$, the test continues to control size well across settings. Power remains high in settings (IIIa)--(IIIe), with some loss under long memory when only a few groups are active. The weaker-global setting (IIIf) is substantially more difficult, consistent with the lower active statistics observed in Section~\ref{subsubsec:sgm_behavior}. Power nevertheless improves as more groups share the global component.

\begin{table}[ht!]
\centering
\caption{Bootstrap existence test in the baseline setting (IIIa). Entries are empirical rejection rates with Monte Carlo standard errors in parentheses; $|\cA_G|=0$ reports empirical size and $|\cA_G|\ge 2$ reports power.}
\label{tab:bootstrap-existence-baseline}
\small
\setlength{\tabcolsep}{5pt}
\renewcommand{\arraystretch}{0.95}
\begin{tabular}{@{}lccccc@{}}
\toprule
$T$ & $|\cA_G|=0$ & $|\cA_G|=2$ & $|\cA_G|=3$ & $|\cA_G|=4$ & $|\cA_G|=5$ \\
\midrule
100 & 0.043 (0.010) & 0.828 (0.019) & 0.993 (0.004) & 1.000 (0.000) & 1.000 (0.000) \\
200 & 0.058 (0.012) & 0.995 (0.004) & 1.000 (0.000) & 1.000 (0.000) & 1.000 (0.000) \\
400 & 0.050 (0.011) & 0.998 (0.002) & 1.000 (0.000) & 1.000 (0.000) & 1.000 (0.000) \\
\bottomrule
\end{tabular}
\end{table}

\begin{table}[ht!]
\centering
\caption{Bootstrap existence test across settings at $T=200$. Entries are described in Table~\ref{tab:bootstrap-existence-baseline}.}
\label{tab:bootstrap-existence-settings}
\small
\setlength{\tabcolsep}{5pt}
\renewcommand{\arraystretch}{0.95}
\begin{tabular}{@{}lcccccc@{}}
\toprule
$|\cA_G|$ & (IIIa) & (IIIb) & (IIIc) & (IIId) & (IIIe) & (IIIf) \\
\midrule
0 & 0.058 (0.012) & 0.068 (0.013) & 0.050 (0.011) & 0.055 (0.011) & 0.058 (0.012) & 0.053 (0.011) \\
2 & 0.995 (0.004) & 0.930 (0.013) & 0.993 (0.004) & 0.985 (0.006) & 1.000 (0.000) & 0.085 (0.014) \\
4 & 1.000 (0.000) & 1.000 (0.000) & 1.000 (0.000) & 1.000 (0.000) & 1.000 (0.000) & 0.648 (0.024) \\
\bottomrule
\end{tabular}
\end{table}

Tables~\ref{tab:bootstrap-screening-baseline} and \ref{tab:bootstrap-screening-settings} report the screening performance, where $\text{TPR}=|\wh\cA_G\cap\cA_G|/|\cA_G|$, $\text{FPR}=|\wh\cA_G\cap\cA_G^c|/|\cA_G^c|$, and Exact denotes $\mathbbm{1}\{\wh\cA_G=\cA_G\}$, averaged over replications. In the baseline setting, screening becomes nearly exact as $T$ increases when global common factors are present, while false positives remain rare. The screening procedure therefore becomes increasingly accurate as the separation in $S_{G,m}$ strengthens. At $T=200$, screening remains accurate across settings (IIIa)--(IIIe), with high true positive rates and low false positive rates. The weaker-global setting (IIIf) is again the most challenging, where the reduced shared signal leads to substantially lower true positive and exact-selection rates. This pattern closely follows the existence-test results, since the final selected set is gated by the existence decision.

\begin{table}[ht!]
\centering
\caption{Bootstrap screening in the baseline setting (IIIa). Entries are empirical averages computed from the final gated selected set $\wh{\cA}_G$.}
\label{tab:bootstrap-screening-baseline}
\small
\renewcommand{\arraystretch}{0.95}
\setlength{\tabcolsep}{3.0pt}
\begin{tabular}{@{}l*{15}{c}@{}}
\toprule
\multirow{2}{*}{Metric}
& \multicolumn{3}{c}{$|\cA_G|=0$}
& \multicolumn{3}{c}{$|\cA_G|=2$}
& \multicolumn{3}{c}{$|\cA_G|=3$}
& \multicolumn{3}{c}{$|\cA_G|=4$}
& \multicolumn{3}{c}{$|\cA_G|=5$} \\
\cmidrule(lr){2-4}\cmidrule(lr){5-7}\cmidrule(lr){8-10}\cmidrule(lr){11-13}\cmidrule(l){14-16}
& $100$ & $200$ & $400$ & $100$ & $200$ & $400$ & $100$ & $200$ & $400$ & $100$ & $200$ & $400$ & $100$ & $200$ & $400$ \\
\midrule
TPR & -- & -- & -- & 0.826 & 0.995 & 0.998 & 0.951 & 0.998 & 1.000 & 0.971 & 0.996 & 0.999 & 0.985 & 0.995 & 1.000 \\
FPR & 0.019 & 0.026 & 0.021 & 0.001 & 0.002 & 0.000 & 0.000 & 0.000 & 0.000 & 0.000 & 0.000 & 0.000 & -- & -- & -- \\
Exact & 0.958 & 0.943 & 0.950 & 0.825 & 0.990 & 0.998 & 0.868 & 0.993 & 1.000 & 0.888 & 0.983 & 0.998 & 0.928 & 0.975 & 1.000 \\
\bottomrule
\end{tabular}
\end{table}

\begin{table}[ht!]
\centering
\caption{Bootstrap screening across settings at $T=200$. Entries are described in Table~\ref{tab:bootstrap-screening-baseline}.}
\label{tab:bootstrap-screening-settings}
\small
\setlength{\tabcolsep}{5pt}
\renewcommand{\arraystretch}{0.95}
\begin{tabular}{@{}llcccccc@{}}
\toprule
$|\cA_G|$ & Metric & (IIIa) & (IIIb) & (IIIc) & (IIId) & (IIIe) & (IIIf) \\
\midrule
\multirow{2}{*}{$0$} & FPR & 0.026 & 0.029 & 0.023 & 0.024 & 0.025 & 0.022 \\
 & Exact & 0.943 & 0.933 & 0.950 & 0.945 & 0.943 & 0.948 \\
\midrule
\multirow{3}{*}{$2$} & TPR & 0.995 & 0.930 & 0.993 & 0.985 & 1.000 & 0.073 \\
 & FPR & 0.002 & 0.000 & 0.001 & 0.002 & 0.029 & 0.009 \\
 & Exact & 0.990 & 0.930 & 0.990 & 0.980 & 0.915 & 0.060 \\
\midrule
\multirow{3}{*}{$4$} & TPR & 0.996 & 0.986 & 0.995 & 0.999 & 0.999 & 0.425 \\
 & FPR & 0.000 & 0.000 & 0.000 & 0.000 & 0.005 & 0.003 \\
 & Exact & 0.983 & 0.950 & 0.980 & 0.998 & 0.990 & 0.085 \\
\bottomrule
\end{tabular}
\end{table}

Finally, as a robustness check, we repeat the estimation experiments and the baseline statistic-separation experiment under Student $t(6)$ innovations, with the corresponding results reported in Section~\ref{app:heavy_tail_robustness} of the Supplementary Materials. The qualitative patterns remain largely unchanged. Heavy tails mainly increase finite-sample variation, especially for smaller $T$, while having little effect on the overall estimation accuracy or the separation between active and inactive groups.

\subsection{Real data analysis}\label{subsec: real_data}

We study daily U.S.\ equity data from CRSP, grouping stocks into the 11 GICS sectors. The set of stocks is fixed on December 30, 2022 by selecting the 30 largest U.S.\ companies by market capitalization within each sector. Six stocks with structural exits during 2023--2024 are excluded from the primary sample, leaving 324 stocks in total. We consider nine indicators: daily return, 20-day cumulative log return, log dollar volume, log turnover, 20-day relative turnover, 20-day log realized volatility, high-low range, log bid-ask spread, and 20-day log Amihud illiquidity \citep{AMIHUD200231}. The sample runs from January 3, 2023 to December 31, 2024, with $T=502$ trading days, giving 11 grouped matrix time series $\{\Y_t^{(m)}\}_{t=1}^{T}$. Observations in 2025 are reserved for out-of-sample forecasting.

For each indicator, the mean and standard deviation are computed from all observed entries in the 2023--2024 primary sample before imputation and used for standardization. Ordinary missing values are imputed using same-day sector medians, and each series is then demeaned over 2023--2024. We continue to write $\{\Y_t^{(m)}\}$ for the resulting cell-demeaned standardized matrices. Indicator construction and preprocessing are described in subsection~\ref{app:data_preprocessing} of the Supplementary Materials.

\paragraph{Why a MMEFM structure?}
We first motivate the two features of the MMEFM that matter here: explicit main effects and a shared global structure.

First, under (IC1), we have the decomposition
$\norm{\Y_t^{(m)}}_F^2=p_mq_m\big(\wh c_t^{(m)}\big)^2+q_m\norm{\wh\a_t^{(m)}}^2+p_m\norm{\wh\b_t^{(m)}}^2+\norm{\check\Y_t^{(m)}}_F^2$.
The interaction component accounts only for
$\sum_m\sum_t\norm{\check\Y_t^{(m)}}_F^2/\sum_m\sum_t\norm{\Y_t^{(m)}}_F^2\approx56.77\%$
of the total squared magnitude, leaving 43.23\% attributable to the additive components. This sizable contribution motivates modeling the main effects explicitly. 

Second, we apply the global-common-factor detection and screening procedure developed in Section~\ref{sec:practical_global_screening}. The two largest statistics are $S_{G,(1)}=8.254$ and $S_{G,(2)}=7.571$, both exceeding the 95\% circular-shift bootstrap cutoff of $4.589$, providing evidence for a nontrivial global common component. Sector-level screening retains Industrials, Communication Services, and Real Estate. We nevertheless estimate the MMEFM using all 11 sectors, since the remaining $S_{G,m}$'s are also away from zero and may reflect weaker shared market dynamics. For comparison, we also consider a screened specification in which the global common component is estimated using only the three retained sectors, while local components are retained for all sectors. Further evidence on the contributions of the main effects and global component, together with the complete set of sector-level $S_{G,m}$ statistics, is provided in the Supplementary Materials.

\paragraph{Model estimation and interpretation.}
We next estimate the MMEFM on our stock data. The factor dimensions are selected using the procedure in subsection~\ref{subsec:rank_estimation}. The global main-effect and common-component ranks are both $(1,1)$, while the local ranks vary across sectors, with the complete selection provided in Supplementary Table~\ref{tab:stock_rank_selection}.

\begin{figure}[ht!]
\centering
\includegraphics[width=.7\textwidth]{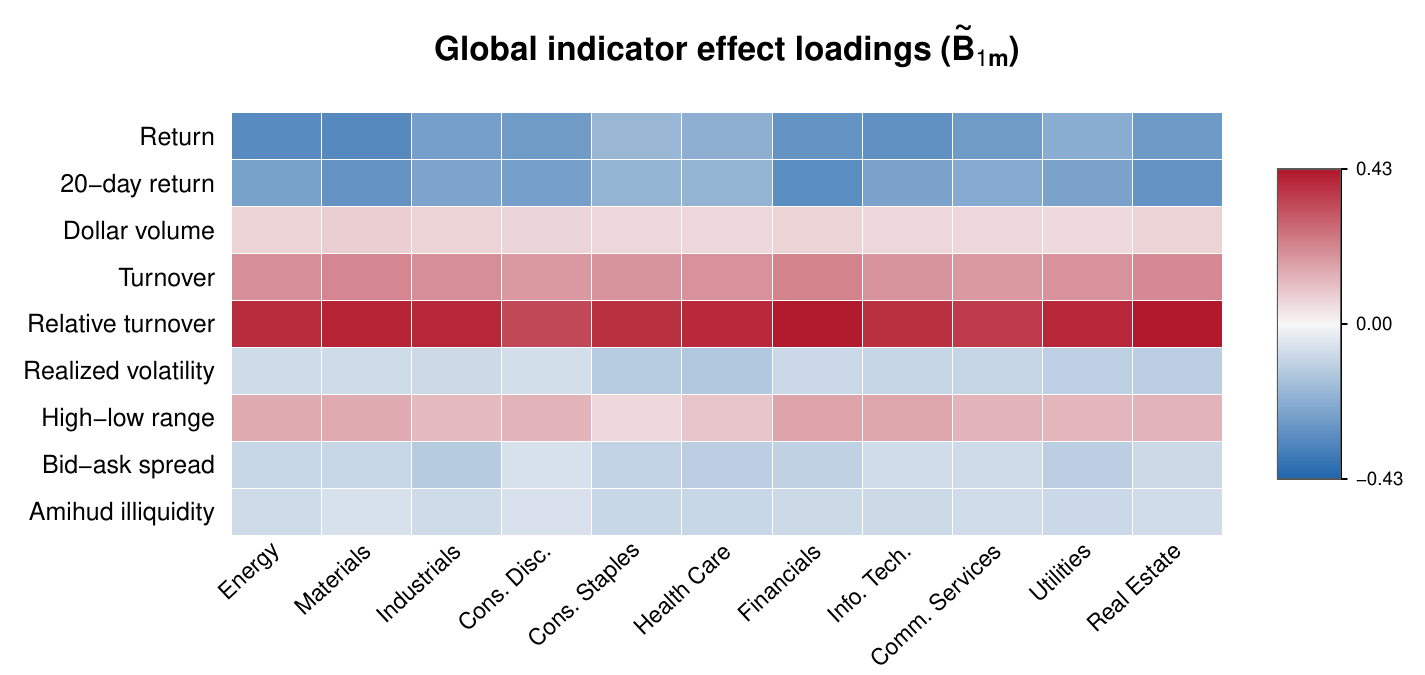}
\caption{Global indicator main-effect estimates.}
\label{fig:stock_global_indicator_main}
\end{figure}

\begin{figure}[ht!]
\centering
\includegraphics[width=.95\textwidth]{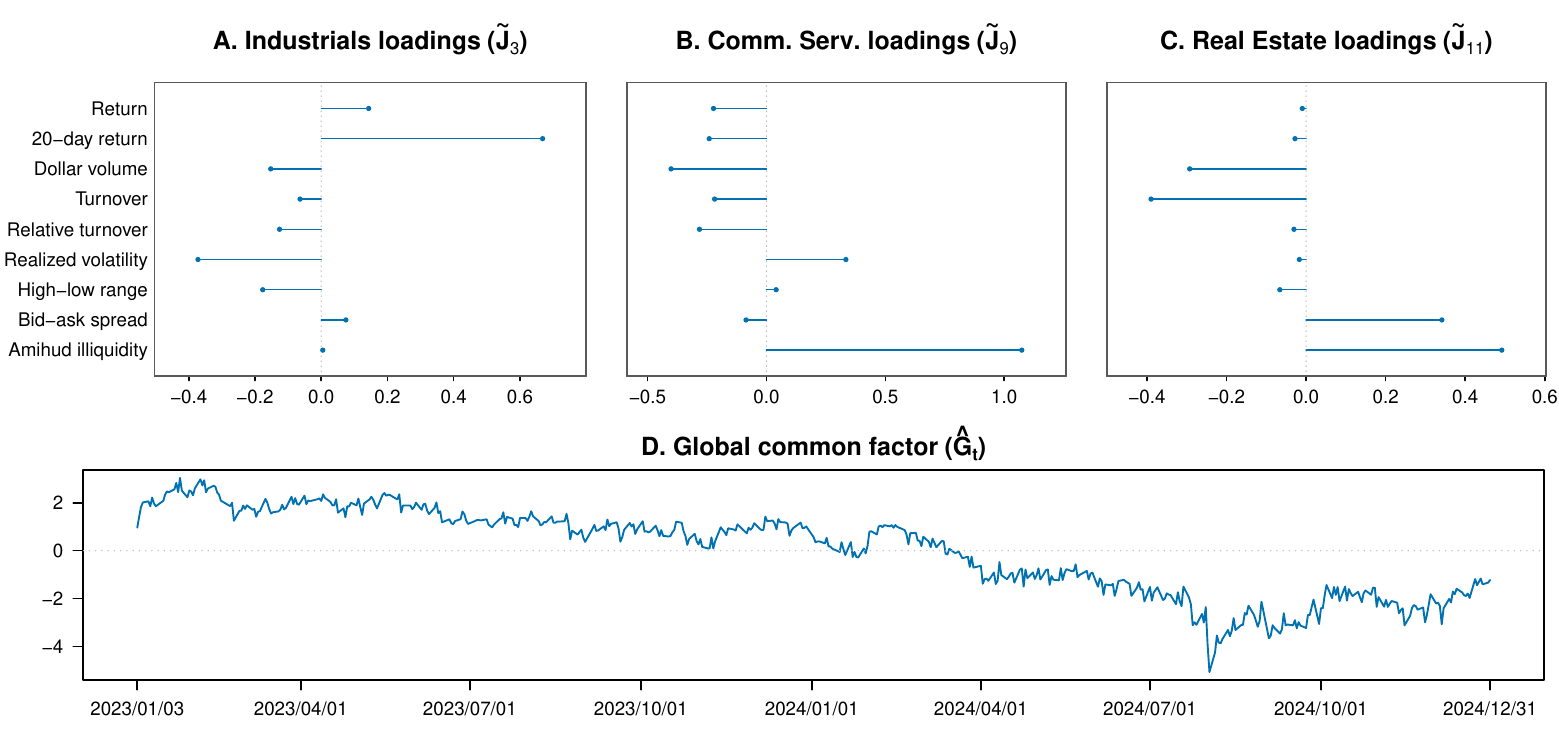}
\caption{Global common component estimates.}
\label{fig:stock_global_interaction_screened}
\end{figure}

We first examine the global structure. Figure~\ref{fig:stock_global_indicator_main} shows that the global indicator-main loadings $\wt\B_{1m}$ are broadly similar across sectors, with turnover and relative turnover loading in the opposite direction to daily and 20-day returns, suggesting a market-wide contrast between trading activity and recent price performance. Figure~\ref{fig:stock_global_interaction_screened} further shows that the global common component has more heterogeneous indicator-loading profiles across the three selected sectors: 20-day return is most prominent in Industrials, whereas liquidity and trading measures are more prominent in Communication Services and Real Estate. The shared global factor also shows a large movement around the late-July--early-August 2024 market dislocation, pointing to a market-wide change rather than a sector-specific episode.

\begin{figure}[ht!]
\centering
\includegraphics[width=.85\textwidth]{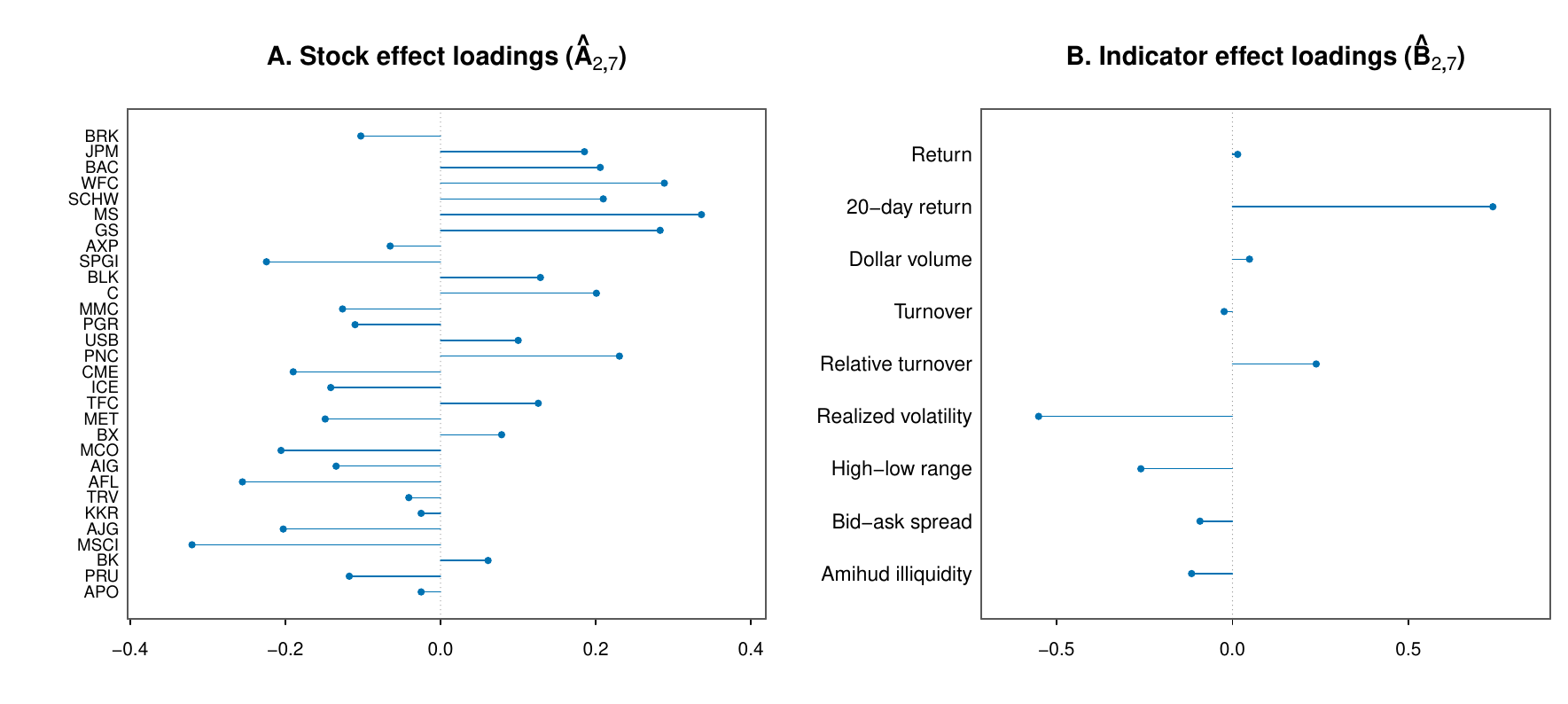}
\caption{Local main-effect estimates for Financials.}
\label{fig:stock_financials_local_main_effects}
\end{figure}

\begin{figure}[ht!]
\centering
\includegraphics[width=.85\textwidth]{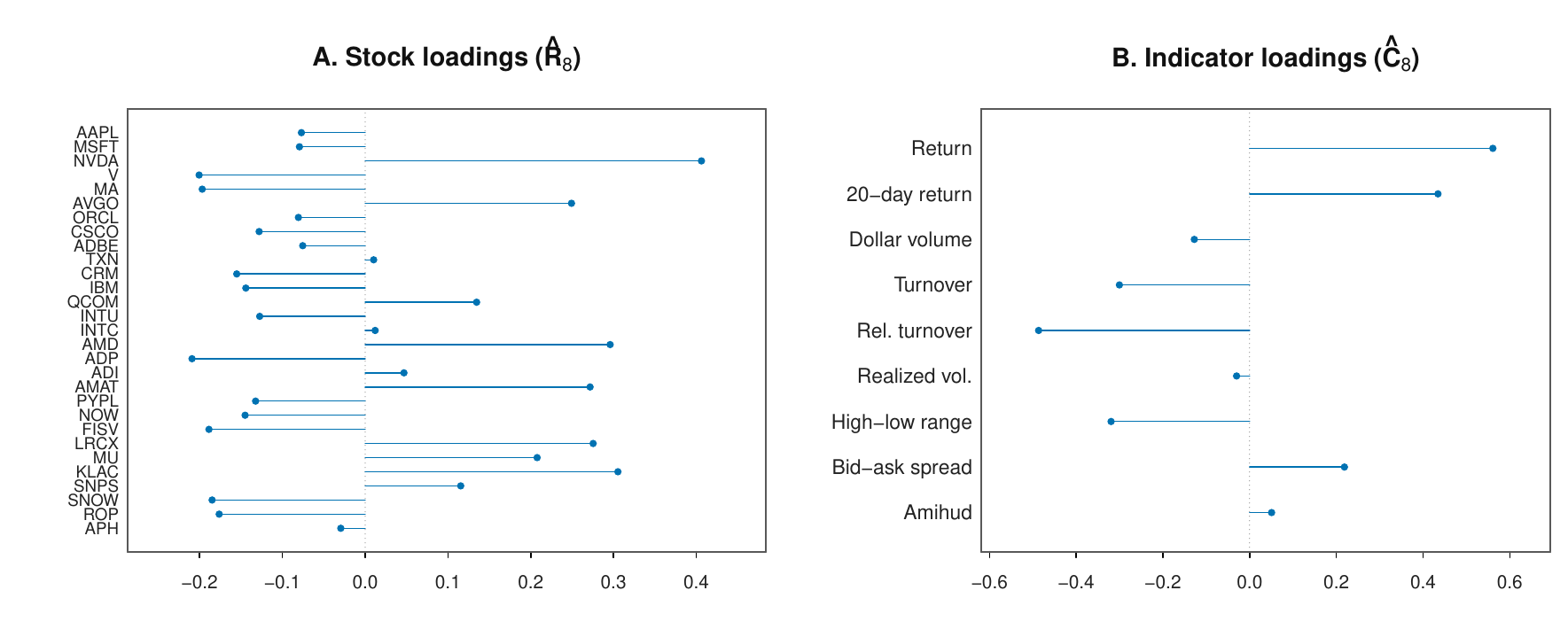}
\caption{Local common-component estimates for Information Technology.}
\label{fig:stock_IT_local_interaction}
\end{figure}

We next illustrate the local structure using Financials and Information Technology. Figure~\ref{fig:stock_financials_local_main_effects} shows that the Financials stock-main loadings broadly separate large banks and broker-dealers from insurers and several market-infrastructure firms. Figure~\ref{fig:stock_IT_local_interaction} shows that the Information Technology common stock loadings distinguish semiconductor and chip-equipment firms from software, payment, and other large technology firms, while returns load in the opposite direction to trading activity and the high-low range.

Additional estimates are reported in the Supplementary Materials.

\paragraph{Model comparison.}
We next compare the in-sample fit of several MMEFM specifications and MFM. Table~\ref{tab:stock_model_comparison} reports $RSS/TSS$ together with a simple parameter-count adjustment, $(RSS/TSS)\cdot N/(N-\nu)$, where $N:=\sum_{m=1}^M Tp_mq_m$ and $\nu:= T\cdot \#\{\text{latent series per date}\} + \#\{\text{loading coefficients}\}$. 

The local-only and global-only specifications retain only local and global dynamics, respectively. MFM Basic uses the baseline estimator of \cite{Zhangetal2025}, whereas MFM Augmented uses their Section~3.5 iterative estimator with 15 iterations. The MMEFM specifications fit better than the two single-level structured benchmarks, while the screened-global specification achieves essentially the same fit as the all-sector model with fewer loading coefficients. MFM Augmented improves markedly over MFM Basic, but both fit less well than the MMEFM specifications.

\begin{table}[ht!]
\centering
\caption{In-sample reconstruction across MMEFM specifications and MFM benchmarks. The MFM rank estimator selects global rank $(1,1)$; the local ranks are $(3,3)$ for Energy, $(1,2)$ for Information Technology, $(2,3)$ for Consumer Discretionary, Consumer Staples, and Communication Services, and $(1,3)$ otherwise.}
\label{tab:stock_model_comparison}
\small
\setlength{\tabcolsep}{5pt}
\renewcommand{\arraystretch}{0.95}
\begin{tabular}{lrrrr}
\toprule
Model & Per-date dim. & Loading coeffs. & $RSS/TSS$ & Adjusted \\
\midrule
MMEFM, all sectors       & 68 & 2,074 & 0.712 & 0.730 \\
MMEFM, screened global   & 68 & 1,737 & \textbf{0.708} & \textbf{0.726} \\
MMEFM, local only        & 64 & 1,228 & 0.742 & 0.760 \\
MMEFM, global only       &  4 &  846  & 0.869 & 0.871 \\
MFM Basic                & 58 & 1,180 & 1.001 & 1.022 \\
MFM Augmented            & 58 & 1,180 & 0.771 & 0.787 \\
\bottomrule
\end{tabular}
\end{table}

We complement the comparison with daily rolling forecasts for 2025. At each origin, ranks and loading matrices are re-estimated on a 252-day window advanced by one trading day; the recovered time-varying quantities are forecast entrywise by AR$(p)$ models with $p\in[5]$ chosen by BIC from the latest 250 observations. We consider $h=1,3,5,10$, keep the three screened sectors fixed throughout 2025, and define MSPE as the total squared Frobenius forecast error divided by the number of evaluated entries.

Table~\ref{tab:stock_forecast} shows that both MMEFM specifications outperform MFM Basic at every horizon. Relative to MFM Augmented, the all-sector MMEFM reduces MSPE by around 10\%, and the screened-global specification performs similarly.

\begin{table}[ht!]
\centering
\caption{Daily rolling forecasting MSPEs for 2025.}
\label{tab:stock_forecast}
\small
\setlength{\tabcolsep}{5pt}
\renewcommand{\arraystretch}{0.95}
\begin{tabular}{lrrrr}
\toprule
Model & $h=1$ & $h=3$ & $h=5$ & $h=10$ \\
\midrule
MMEFM, all sectors
& \textbf{0.802} & \textbf{0.870} & \textbf{0.894} & \textbf{0.940} \\
MMEFM, screened global
& 0.810 & 0.875 & 0.897 & \textbf{0.940} \\
MFM, Basic
& 0.983 & 1.022 & 1.028 & 1.038 \\
MFM, Augmented
& 0.908 & 0.953 & 0.973 & 1.003 \\
\bottomrule
\end{tabular}
\end{table}

\bibliographystyle{apalike}
\bibliography{ref, ref_supp}

\clearpage

\begin{center}
{\Large\bfseries Supplementary Materials for the paper ``Multilevel Main Effects Matrix Factor Model''}\\[1em]
\end{center}

\appendix

\numberwithin{figure}{section}
\numberwithin{table}{section}
\numberwithin{mmalgorithm}{section}

\section{Iterative Projection for Estimating
  \texorpdfstring{$\Q_m$}{Q(m)} and
  \texorpdfstring{$\J_m$}{J(m)}}\label{app:finding_directions}
For finding the common components in Step~3 in Section~\ref{subsec:estimation}, we assume that the directions that align more to the strongest factors of $\Q_m$ and $\J_m$ are available for all $m\in[M]$, namely $\v_a^{(m)}$ and $\v_b^{(m)}$ for $m\in[M]$ in (\ref{eqn:X_A}) and (\ref{eqn:X_B}), which are of course not the case in reality.
In the following, we present an iterative algorithm for finding $\v_a^{(m)}$ and $\v_b^{(m)}$, and more accurate estimators of $\Q_m$ and $\J_m$ in Step 3 in Section \ref{subsec:estimation}.

\noindent\hrulefill
\vspace{-9pt}

\refstepcounter{mmalgorithm}\label{alg:projection}
\noindent\textbf{Algorithm~\themmalgorithm} Algorithm for choosing good projection directions for estimating $\Q_m$ and $\J_m$

\vspace{-13pt}
\noindent\hrulefill
\begin{itemize}
\vspace{-12pt}
    \item[1.] Preset $K_0$ (recommend $K_0=20$). Let $\{S_{m,k}\}_{m\in[M],k\in[K_0]}$ be a sequence (in $k$) of subsets of $[q_m]$ for each $m\in[M]$, which chooses elements in $[q_m]$ uniformly, with $|S_{m,k}| = \lfloor q_m/2\rfloor$. Denote by $\1_{d,S}$ a vector of ones and zeros of length $d$, where the positions in $S\subseteq [d]$ are 1 and 0 otherwise.  

    \item[2.] Let $m=1$, and let $k_n\in[K_0]$ for each $n\neq m$. Define
    \begin{equation*}
        \X_{Q,m,k|k_n} := T^{-2}(M-1)^{-1}\sum_{n\neq m;n=1}^M\Big\{\sum_{t=1}^T\check\Y_t^{(m)}\1_{q_m,S_{m,k}}\1_{q_n,S_{n,k_n}}'\check\Y_t^{(n)'}\Big\}^{\otimes 2}.
    \end{equation*}
    Find $\wt{k}_m$ such that $\X_{Q,m,\wt{k}_m|k_n}$ has the largest eigenvalue. Take $\v_b^{(m)} = \1_{q_m,S_{m,\wt{k}_m}}$.

    \item[3.] For $m\in[M]$ larger than 1, repeat Step 2, replacing $k_n$ by $\wt{k}_n$ found in previous steps in the definition of $\X_{Q,m,k|k_n}$. 

    \item[4.] With $\v_b^{(m)} = \1_{q_m,S_{m,\wt{k}_m}}$ for each $m\in[M]$, take $\wh\Q_m$ as the first $k_r$ eigenvectors of $\X_{Q,m}$ defined in (\ref{eqn:X_A}).

    \item[5.] Set $\v_a^{(m)} = \wh\Q_{m,\cdot 1}$ for each $m\in[M]$. 
    
    Fix  $m=1$. Replace $\v_a^{(m)} = \wh\Q_{m, \cdot k}$ and find the eigenvalue of $\X_{J,m}$ defined in (\ref{eqn:X_B}), $k\in[k_r]$. Set $\v_a^{(m)} = \wh\Q_{m, \cdot k}$ so that $\X_{J,m}$ has the largest eigenvalue. Repeat for $m=2,\ldots,M$, resulting in a new set of $\v_a^{(m)}$ for $m\in[M]$. 

    Estimate $\wh\J_m$ as the first $k_c$ eigenvectors of $\X_{J,m}$.

    \item[6.] Set $\v_b^{(m)} = \wh\J_{m,\cdot 1}$ for each $m\in[M]$. 
    
    Fix $m=1$. Replace $\v_b^{(m)} = \wh\J_{m, \cdot k}$ and find the eigenvalue of $\X_{Q,m}$ defined in (\ref{eqn:X_A}), $k\in[k_c]$. Set $\v_b^{(m)} = \wh\J_{m, \cdot k}$ so that $\X_{Q,m}$ has the largest eigenvalue. Repeat for $m=2,\ldots,M$, resulting in a new set of $\v_b^{(m)}$ for $m\in[M]$. 

    Estimate $\wh\Q_m$ as the first $k_r$ eigenvectors of $\X_{Q,m}$. 

    \item[7.] Iterate Step 5 and 6 until some kind of convergence is reached.
\end{itemize}
\vspace{-12pt}
\noindent\hrulefill

One major difference between Algorithm~\ref{alg:projection} and the iterative algorithm in \cite{Zhangetal2025} is that we only use the directions that align more to the strongest factors for iterative projection in Steps 5 and 6 above. This is to avoid projection using weak factors when it is likely that in real data such weak factors exist, which
can numerically lead to less accurate estimators if used as projection directions. Our final two-step aligned estimators $\wt\Q_m$ and $\wt\J_m$ in step 3 in Section \ref{subsec:estimation} are constructed based on $\wh\Q_m$ and $\wh\J_m$ and indeed perform better than the augmented estimators from \cite{Zhangetal2025}. See Table \ref{tab:zhang-comparison} in Section \ref{sec: numerical_results} for more details.

\section{Alternative Procedures for Global Common Component Estimation}
\label{app:alternative_global_common_component}

The estimation procedure in Step~3 of Section~\ref{subsec:estimation} uses the nearest Kronecker product approximation, described in Section~6 of \cite{Loan2000}, implemented through a matrix rearrangement and singular value decomposition, for both the alignment procedure and the re-estimation procedure. For completeness, we describe two alternative iterative procedures. Subsection~\ref{app:procrustes_alignment} presents a two-sided Procrustes alternative for the alignment step, while Subsection~\ref{app:ALS_refitting} presents an alternating least-squares alternative. These two alternatives concern separate stages of the estimation procedure and may be applied independently.

\subsection{Two-sided Procrustes alignment}
\label{app:procrustes_alignment}
To align the coordinate systems of the $\wh\Q_m$ and $\wh\J_m$ across $m\in[M]$, we first choose a reference group $m_0\in[M]$. For each $m\ne m_0$, we find orthogonal alignment matrices $\wh\O_{Q,m}$ and $\wh\O_{J,m}$ by the two-sided Procrustes criterion, namely
\begin{align}
    (\wh\O_{Q,m}, \wh\O_{J,m}) 
    = \argmin_{\O_Q\in\bb{O}(k_r),\O_J\in\bb{O}(k_c)} \sum_{t=1}^T
    \left\| \wh\Q_{m_0}'\check\Y_{G,t}^{(m_0)}\wh\J_{m_0} - \O_Q'\wh\Q_m'\check\Y_{G,t}^{(m)}\wh\J_m\O_J \right\|_F^2, \label{eqn:Procrustes_problem}
\end{align}
where $\bb{O}(a)$ denotes the set of $a\times a$ orthogonal matrices. To this end, given $\O_{J,m}$, define 
\begin{align*}
    \bPsi_{Q,m}:=\sum_{t=1}^T \wh\Q_m'\check\Y_{G,t}^{(m)}\wh\J_m\O_{J,m}\wh\J_{m_0}'\check\Y_{G,t}^{(m_0)'}\wh\Q_{m_0}.
\end{align*}
The corresponding Procrustes update is
\begin{align}
    \O_{Q,m} = \U_{Q,m}\V_{Q,m}', \label{eqn:Procrustes_update_Q}
\end{align}
where $\U_{Q,m}$ and $\V_{Q,m}$ are the left- and right-singular vectors of $\bPsi_{Q,m}$. Similarly, given $\O_{Q,m}$, define 
\begin{align*}
    \bPsi_{J,m}:=\sum_{t=1}^T \wh\J_m'\check\Y_{G,t}^{(m)'}\wh\Q_m\O_{Q,m}\wh\Q_{m_0}'\check\Y_{G,t}^{(m_0)}\wh\J_{m_0},
\end{align*}
and its corresponding Procrustes update is
\begin{align}
    \O_{J,m} = \U_{J,m}\V_{J,m}', \label{eqn:Procrustes_update_J}
\end{align}
where $\U_{J,m}$ and $\V_{J,m}$ are the left- and right-singular vectors of $\bPsi_{J,m}$. Starting from $\O_{Q,m}=\I_{k_r}$ and $\O_{J,m}=\I_{k_c}$, we iterate \eqref{eqn:Procrustes_update_Q} and \eqref{eqn:Procrustes_update_J} to obtain $\wh\O_{Q,m}$ and $\wh\O_{J,m}$. We then replace $\wh\Q_m$ and $\wh\J_m$ with $\wh\Q_m\wh\O_{Q,m}$ and $\wh\J_m\wh\O_{J,m}$ respectively for $m\ne m_0$. After the alignment, the global common factor $\G_t$ is estimated by least square as
\begin{align}
    \wh\G_t &:= M^{-1}\sum_{m=1}^M\wh\Q_m'(\check\Y_t^{(m)} - \wh\R_m\wh\F_t^{(m)}\wh\C_m')\wh\J_m. \label{eqn:estimation_Gt}
\end{align}

\subsection{Alternating least-squares refitting}
\label{app:ALS_refitting}

For $m\in[M]$, we re-estimate $\Q_m$ and $\J_m$ by minimizing $\sum_{t=1}^T\|\check\Y_t^{(m)} - \wh\R_m\wh\F_t^{(m)}\wh\C_m' - \Q_m\wh\G_t\J_m'\|_F^2$ over $\Q_m$ and $\J_m$. To this end, given $\J_m$, we have
\begin{equation}\label{eqn:Q_iterate}
\wt\Q_m = \bigg(\sum_{t=1}^T(\check\Y_t^{(m)} - \wh\R_m\wh\F_t^{(m)}\wh\C_m')\J_m\wh\G_t'\bigg)
\bigg(\sum_{t=1}^T\wh\G_t\J_m'\J_m\wh\G_t'\bigg)^{-1},
\end{equation}
and given $\Q_m$, we have
\begin{equation}\label{eqn:J_iterate}
\wt\J_m = \bigg(\sum_{t=1}^T(\check\Y_t^{(m)} - \wh\R_m\wh\F_t^{(m)}\wh\C_m')'\Q_m\wh\G_t\bigg)
\bigg(\sum_{t=1}^T\wh\G_t'\Q_m'\Q_m\wh\G_t\bigg)^{-1}.
\end{equation}
Using estimators $\check\Q_m$ and $\check\J_m$, we can iterate (\ref{eqn:Q_iterate}) and (\ref{eqn:J_iterate}) to obtain the final estimators $\wt\Q_m$ and $\wt\J_m$ for $\Q_m$ and $\J_m$, respectively. This is in parallel to the algorithm in Section 2.2 of \cite{Caietal2025}.

\begin{table}[ht!]
\centering
\caption{Common component estimation accuracy using Procrustes alignment and ALS refitting. All entries are Monte Carlo means multiplied by $10^2$, with standard deviations in parentheses.}
\label{tab:common-component-accuracy-procrustes-als}
\resizebox{\textwidth}{!}{%
\begin{tabular}{l|cc|cc|cc|cc}
\toprule
Settings & \multicolumn{2}{c}{IIa} & \multicolumn{2}{c}{IIb} & \multicolumn{2}{c}{IIc} & \multicolumn{2}{c}{IId}\\
 & $T=100$ & $T=400$ & $T=100$ & $T=400$ & $T=100$ & $T=400$ & $T=100$ & $T=400$\\
\midrule
$\cD(\Q_m,\wh\Q_m)$ & 7.98(3.09) & 3.90(1.45) & 9.84(5.71) & 4.52(1.77) & 9.21(3.03) & 4.46(1.41) & 5.75(2.21) & 2.73(1.05)\\
$\cD(\Q_m,\wt\Q_m)$ & 1.67(0.42) & 0.75(0.15) & 2.09(4.45) & 0.76(0.20) & 2.85(0.46) & 1.35(0.17) & 1.06(0.22) & 0.48(0.07)\\
$\cD(\J_m,\wh\J_m)$ & 8.22(3.15) & 3.93(1.57) & 9.69(5.65) & 4.51(1.74) & 9.27(3.16) & 4.24(1.25) & 8.24(3.12) & 3.73(1.32)\\
$\cD(\J_m,\wt\J_m)$ & 1.42(0.20) & 0.66(0.07) & 1.66(4.15) & 0.62(0.07) & 2.69(0.34) & 1.28(0.13) & 1.31(0.15) & 0.61(0.05)\\
\addlinespace[0.45em]
$\cD(\R_m,\wh\R_m)$ & 2.53(0.64) & 1.01(0.18) & 3.28(4.47) & 1.05(0.22) & 4.23(0.79) & 2.42(0.55) & 2.23(0.72) & 0.76(0.15)\\
$\cD(\C_m,\wh\C_m)$ & 2.50(0.66) & 1.01(0.16) & 3.33(4.54) & 1.04(0.22) & 4.23(0.90) & 2.41(0.56) & 2.09(0.48) & 0.84(0.10)\\
\addlinespace[0.45em]
$\MSE(\wt\Q_m\wh\G_t\wt\J_m')$ & 0.52(2.44) & 0.06(0.64) & 1.23(9.15) & 0.03(0.00) & 0.58(2.27) & 0.15(0.57) & 0.30(1.87) & 0.02(0.00)\\
$\MSE(\wh\R_m\wh\F_t^{(m)}\wh\C_m')$ & 0.25(0.11) & 0.13(0.03) & 0.82(10.24) & 0.12(0.03) & 0.71(0.20) & 0.48(0.10) & 0.14(0.06) & 0.06(0.01)\\
\bottomrule
\end{tabular}%
}
\end{table}

\clearpage

\section{Proof of Theorems}\label{app:proof_of_theorems}

\textbf{\textit{Proof of Theorem \ref{thm:identification}.}}
The grand mean $c_t^{(m)}$ (as defined in \eqref{eqn:combined_main_effects}) is identified when we pre- and post-multiply by $\1_{p_m}'$ and $\1_{q_m}$ respectively on both sides of model (\ref{eqn:MMEFM}) under (IC1). Then for $\{\a_t^{(m)}\}$ and $\{\b_t^{(m)}\}$ we can post-multiply by $\1_{q_m}$ and pre-multiply by $\1_{p_m}'$ on both sides of (\ref{eqn:MMEFM}) respectively. The remaining common component is then identified as well.

For global and local effects, since $c_t^{(m)} = \mu_t + \mu_t^{(m)}$ is identified, we have $\mu_t = M^{-1}\sum_{m=1}^M c_t^{(m)}$ using (IC2), and hence both $\{\mu_t\}$ and $\{\mu_t^{(m)}\}$ are identified. Next, we in fact identify the row and column main effects at group $m$ respectively as
\begin{align}\label{eqn:main_effects}
    \a_t^{(m)} = q_m^{-1}\M_{p_m}(\Y_t^{(m)} - \E_t^{(m)})\1_{q_m}, \;\;\; \b_t^{(m)} = p_m^{-1}\M_{q_m}(\Y_t^{(m)} - \E_t^{(m)})'\1_{p_m}.
\end{align}
Hence for $m\neq n$, using (\ref{eqn:combined_main_effects}) and (IC2), we have
\begin{align}
    \A_{1m}\bb{E}\bigg(T^{-1}\sum_{t=1}^T\balpha_t^{\otimes 2}\bigg)\A_{1n}' &= \bb{E}\bigg(T^{-1}\sum_{t=1}^T\a_t^{(m)}\a_t^{(n)'}\bigg) \nonumber \\
    &= q_m^{-1}q_n^{-1}\M_{p_m}\bb{E}\bigg\{T^{-1}\sum_{t=1}^T(\Y_t^{(m)} - \E_t^{(m)})\1_{q_m}\1_{q_n}'(\Y_t^{(n)} - \E_t^{(n)})'\bigg\}\M_{p_n} \nonumber \\
    &= T^{-1}q_m^{-1}q_n^{-1}\M_{p_m}\sum_{t=1}^T\bb{E}\{\Y_t^{(m)}\1_{q_m}\1_{q_n}'\Y_t^{(n)'}\}\M_{p_n}, \label{eqn: Omega_population} \\
    \B_{1m}\bb{E}\bigg(T^{-1}\sum_{t=1}^T\bbeta_t^{\otimes 2}\bigg)\B_{1n}' &= \bb{E}\bigg(T^{-1}\sum_{t=1}^T\b_t^{(m)}\b_t^{(n)'}\bigg) \nonumber \\ 
    &= p_m^{-1}p_n^{-1}\M_{q_m}\bb{E}\bigg\{T^{-1}\sum_{t=1}^T(\Y_t^{(m)}-\E_t^{(m)})'\1_{p_m}\1_{p_n}'(\Y_t^{(n)} - \E_t^{(n)})\bigg\}\M_{q_n} \nonumber \\
    &= T^{-1}p_m^{-1}p_n^{-1}\M_{q_m}\sum_{t=1}^T\bb{E}\{\Y_t^{(m)'}\1_{p_m}\1_{p_n}'\Y_t^{(n)}\}\M_{q_n}. \nonumber 
\end{align}
If we define 
\begin{align*}
    \bOmega_{A_1,m} &:= (M-1)^{-1}\sum_{n=1;n\neq m}^M \bigg\{T^{-1}q_m^{-1}q_n^{-1}\M_{p_m}\sum_{t=1}^T\bb{E}\{\Y_t^{(m)}\1_{q_m}\1_{q_n}'\Y_t^{(n)'}\}\M_{p_n}\bigg\}^{\otimes 2}, \\ 
    \bOmega_{B_1,m} &:= (M-1)^{-1}\sum_{n=1;n\neq m}^M \bigg\{T^{-1}p_m^{-1}p_n^{-1}\M_{q_m}\sum_{t=1}^T\bb{E}\{\Y_t^{(m)'}\1_{p_m}\1_{p_n}'\Y_t^{(n)}\}\M_{q_n}\bigg\}^{\otimes 2},
\end{align*}
then a version of $\A_{1m}$ (resp.\ $\B_{1m}$) can be identified as the first $r_1$ eigenvectors of $\bOmega_{A_1,m}$  (resp.\ first $\ell_1$ eigenvectors of $\bOmega_{B_1,m}$). 

As for $\A_{2m}$, since $\A_{1m}^\perp$ have orthonormal columns spanning the orthogonal complement of the column space of $\A_{1m}$, from \eqref{eqn:combined_main_effects}, we have
\[\A_{1m}^{\perp'}\a_t^{(m)} = \A_{1m}^{\perp'}\A_{2m}\balpha_t^{(m)}.\]
Hence 
\[T^{-1}\sum_{t=1}^T\a_t^{(m)}\a_t^{(m)'}\A_{1m}^\perp = \A_{1m}\bigg(T^{-1}\sum_{t=1}^T\balpha_t\balpha_t^{(m)'}\bigg)\A_{2m}'\A_{1m}^\perp + \A_{2m}\bigg(T^{-1}\sum_{t=1}^T\balpha_t^{(m)}\balpha_{t}^{(m)'}\bigg)\A_{2m}'\A_{1m}^\perp.\]
Hence (IC2) gives
\begin{align}
    \bigg\{T^{-1}\sum_{t=1}^T(\a_t^{(m)\otimes 2})\A_{1m}^\perp\bigg\}^{\otimes 2} = \A_{2m}\bigg\{T^{-1}\sum_{t=1}^T(\balpha_t^{(m)\otimes 2})\A_{2m}'\A_{1m}^\perp\bigg\}^{\otimes 2}\A_{2m}'.
    \label{eqn:S_A2m}
\end{align}
Therefore, we can take the first $r_2^{(m)}$ eigenvectors of $\Big\{T^{-1}\sum_{t=1}^T\a_t^{(m)\otimes 2}\A_{1m}^\perp\Big\}^{\otimes 2}$ as a version of $\A_{2m}$. Similarly, we can take the first $\ell_{2}^{(m)}$ eigenvectors of $\Big\{T^{-1}\sum_{t=1}^T\b_t^{(m)\otimes 2}\B_{1m}^\perp\Big\}^{\otimes 2}$ as a version of $\B_{2m}$.

To identify $\balpha_t^{(m)}$, since
$\A_{1m}^{\perp '}\A_{2m}\balpha_t^{(m)} = q_m^{-1}\A_{1m}^{\perp'}\M_{p_m}(\Y_t^{(m)} - \E_t^{(m)})\1_{q_m}$, we have
\[\balpha_t^{(m)} = q_m^{-1}(\A_{2m}'\A_{1m}^{\perp}\A_{1m}^{\perp '}\A_{2m})^{-1}\A_{2m}'\A_{1m}^{\perp}\A_{1m}^{\perp'}\M_{p_m}(\Y_t^{(m)} - \E_t^{(m)})\1_{q_m}.\]
Similarly,
\[\bbeta_t^{(m)} = p_m^{-1}(\B_{2m}'\B_{1m}^{\perp}\B_{1m}^{\perp'}\B_{2m})^{-1}\B_{2m}'\B_{1m}^{\perp}\B_{1m}^{\perp'}\M_{q_m}(\Y_t^{(m)} - \E_t^{(m)})'\1_{p_m}.\]
We can then identify $\balpha_t$ and $\bbeta_t$ as
\begin{align*}
\balpha_t &= (\A_{1m}'\A_{1m})^{-1}\A_{1m}'\{q_m^{-1}\M_{p_m}(\Y_t^{(m)} - \E_t^{(m)})\1_{q_m} - \A_{2m}\balpha_t^{(m)}\},\\
\bbeta_t &= (\B_{1m}'\B_{1m})^{-1}\B_{1m}'\{p_m^{-1}\M_{q_m}(\Y_t^{(m)} - \E_t^{(m)})'\1_{p_m} - \B_{2m}\bbeta_t^{(m)}\}.
\end{align*}
(A refinement procedure can be carried out to estimate $\balpha_t$ and $\bbeta_t$ as an average over all $m\in[M]$, with $\A_{1m}$ and $\B_{1m}$ re-estimated using least squares estimation. This is indeed integrated in our estimation procedure in Section \ref{subsec:estimation}.) 

Define
\begin{equation*}
    \wt\Y_t^{(m)} := \Y_t^{(m)}-\E_t^{(m)} - c_t\1_{p_m}\1_{q_m}' - \a_t^{(m)}\1_{q_m}' - \1_{p_m}\b_t^{(m)'}.
\end{equation*}
If we know $\v_1^{(m)}$ as a vector aligning more to the direction of the strongest factors in $\Q_m$, and $\v_2^{(m)}$ as such a vector for $\J_m$ (practically this can be done using similar pre-averaging ideas in \cite{ChenLam2024}, which is also integrated in our estimation procedure), then since for $m\neq n$, by (IC2),
\begin{align}
    &\;\;\;\;
    \bb{E}(\wt\Y_t^{(m)}\v_2^{(m)}\v_2^{(n)'}\wt\Y_t^{(n)'})
    = \Q_m\bb{E}(\G_t\J_m'\v_2^{(m)}\v_2^{(n)'}\J_n'\G_t')\Q_n' \notag \\
    &=
    \Q_m\bb{E}\Big\{ \vec(\G_t\J_m'\v_2^{(m)}) \vec(\G_t\J_n\v_2^{(n)})' \Big\} \Q_n' \notag \\
    &=
    \Q_m\bb{E}\Big\{ (\v_2^{(m)'} \J_m \otimes \I_{k_r}) \vec(\G_t) \vec'(\G_t) ( \J_m'\v_2^{(m)} \otimes \I_{k_r}) \Big\} \Q_n' \notag \\
    &=
    \Q_m (\v_2^{(m)'} \J_m \otimes\I_{k_r} ) \bb{E}(\vec(\G_t))^{\otimes 2} (\J_n' \v_2^{(n)} \otimes \I_{k_r}) \Q_n' ,
    \label{eqn: tildeY_vv_tildeY}
\end{align}
which is of rank $k_r$ through having $\bb{E}(\vec(\G_t))^{\otimes 2}$ of full rank by (IC3), we can identify $\Q_m$ as the first $k_r$ eigenvectors of
\[\sum_{n=1;n\neq m}^M \{\bb{E}(\wt\Y_t^{(m)}\v_2^{(m)}\v_2^{(n)'}\wt\Y_t^{(n)'})\}^{\otimes 2}.\]
Similarly, we can identify $\J_m$ as the first $k_c$ eigenvectors of
\[\sum_{n=1;n\neq m}^M \{\bb{E}(\wt\Y_t^{(m)'}\v_1^{(m)}\v_1^{(n)'}\wt\Y_t^{(n)})\}^{\otimes 2}.\]
We can also identify $\R_m$ as the first $k_r^{(m)}$ eigenvectors of
\[\bb{E}(\wt\Y_t^{(m)}\J_m^{\perp})^{\otimes 2}
= \R_m\bb{E}[\F_t^{(m)}\C_m'\J_m^{\perp}\J_m^{\perp'}\C_m\F_t^{(m)'}]\R_m',\]
which is of full rank since
$\bb{E}(\vec(\F_t^{(m)}))^{\otimes 2}$ and $\J_m^{\perp'}\C_m$ have full rank by (IC3). Similarly, $\C_m$ can be identified as the first $k_c^{(m)}$ eigenvectors of
\[\bb{E}(\wt\Y_t^{(m)'}\Q_m^{\perp})^{\otimes 2} = \C_m\bb{E}[\F_t^{(m)'}\R_m'\Q_m^{\perp}\Q_m^{\perp'}\R_m\F_t^{(m)}]\C_m'.\]
Then $\F_t^{(m)}$ can be identified as
\[\F_t^{(m)} = (\R_m'\Q_m^{\perp}\Q_m^{\perp'}\R_m)^{-1}\R_m'\Q_m^{\perp}\Q_m^{\perp'}\wt\Y_t^{(m)}\C_m.\]
Finally $\G_t$ can be identified as
\[\G_t = \Q_m'(\wt\Y_t^{(m)} - \R_m\F_t^{(m)}\C_m')\J_m.\]
Again, we can consider the average of the above over all $m\in[M]$ as an estimator of $\G_t$, and re-estimate $\Q_m$ and $\J_m$ using least square method, which is indeed integrated into our estimation procedure. $\square$

\textbf{\textit{Proof of Theorem~\ref{thm:main_effect}.}}
We first derive the convergence rate for grand mean and its corresponding global and local components. By left- and right-multiplying $\1_{p_m}'$ and $\1_{q_m}$ on \eqref{eqn:MMEFM}, we have
\begin{align}\label{eqn:left_right_multiply_ones_MMEFM}
    \1_{p_m}'\Y_t^{(m)}\1_{q_m} = p_mq_mc_t^{(m)} + \1_{p_m}'\E_t^{(m)}\1_{q_m}.
\end{align}
Thus, 
\begin{align}
    \bb{E}(\wh{c}_t^{(m)} - c_t^{(m)})^2 &= (p_mq_m)^{-2}\bb{E}(\1_{p_m}'\E_t^{(m)}\1_{q_m})^2 = (p_mq_m)^{-2}\bb{E}\big(\sum_{i=1}^{p_m}\sum_{j=1}^{q_m}E_{t,ij}^{(m)}\big)^2 \nonumber \\
    &= (p_mq_m)^{-2}\sum_{i=1}^{p_m}\sum_{l=1}^{p_m}\sum_{j=1}^{q_m}\sum_{h=1}^{q_m}\bb{E}(E_{t,ij}^{(m)}E_{t,lh}^{(m)}) = O(p_m^{-1}q_m^{-1}), \label{eqn:sum_of_error_rate}
\end{align}
where we apply Lemma~\ref{lemma:correlation_Etm} in the last equality. Hence $|\wh{c}_t^{(m)} - c_t^{(m)}| = O_P\big((p_mq_m)^{-1/2}\big)$. Taking sum on both sides of \eqref{eqn:left_right_multiply_ones_MMEFM} \text{w.r.t.} $m$, we then have $\mu_t = M^{-1}\sum_{m=1}^M(p_mq_m)^{-1}\1_{p_m}'(\Y_t^{(m)} - \E_t^{(m)})\1_{q_m}$. Therefore, since $\wh\mu_t = M^{-1}\sum_{m=1}^{M}\wh{c}_t^{(m)}$, we have
\begin{align*}
    (\wh\mu_t - \mu_t)^2 
    &\lesssim \big(\sum_{m=1}^{M}(p_mq_m)^{-1}\1_{p_m}'\E_t^{(m)}\1_{q_m}\big)^2 
    \lesssim \sum_{m=1}^{M}\big((p_mq_m)^{-1}\1_{p_m}'\E_t^{(m)}\1_{q_m}\big)^2 \\
    &= \sum_{m=1}^M\big(\wh{c}_t^{(m)} - c_t^{(m)}\big)^2 
    = \sum_{m=1}^{M}O_P\big((p_mq_m)^{-1}\big) = O_P\Big(\max_{m\in[M]}(p_mq_m)^{-1}\Big).
\end{align*}
Furthermore, as $\wh\mu_t^{(m)} = \wh{c}_t^{(m)} - \wh\mu_t$ and $\mu_t^{(m)} = c_t^{(m)} - \mu_t$, for each $m\in[M]$, 
\begin{align*}
    \big(\wh\mu_t^{(m)}-\mu_t^{(m)}\big)^2
    &\lesssim \big(\wh c_t^{(m)}-c_t^{(m)}\big)^2
    +\big(\wh\mu_t-\mu_t\big)^2
    =O_P\Big(\max_{n\in[M]}(p_nq_n)^{-1}\Big),
\end{align*}
which is the stated rate.

By \eqref{eqn:main_effects}, the definitions of $\wh\a_t^{(m)}$ and $\wh\b_t^{(m)}$, and (IC1), we have
\begin{align*}
    \norm{\wh\a_t^{(m)} - \a_t^{(m)}}^2 
    = q_m^{-2}\norm{\M_{p_m}\E_t^{(m)}\1_{q_m}}^2 
    = q_m^{-2}\norm{\E_t^{(m)}\1_{q_m} - p_m^{-1}\1_{p_m}\1_{p_m}'\E_t^{(m)}\1_{q_m}}^2.
\end{align*}
For the first part, by Lemma~\ref{lemma:correlation_Etm},
\begin{align*}
    \bb{E}\norm{\E_t^{(m)}\1_{q_m}}^2
    =\sum_{i=1}^{p_m}\sum_{j=1}^{q_m}\sum_{h=1}^{q_m}
    \bb{E}(E_{t,ij}^{(m)}E_{t,ih}^{(m)})
    \le\sum_{i=1}^{p_m}\sum_{j=1}^{q_m}
    \sum_{k=1}^{p_m}\sum_{h=1}^{q_m}
    |\bb{E}(E_{t,ij}^{(m)}E_{t,kh}^{(m)})|
    =O(p_mq_m).
\end{align*}
And for the second part, write
\begin{align*}
    \bb{E}\norm{p_m^{-1}\1_{p_m}\1_{p_m}'\E_t^{(m)}\1_{q_m}}^2 &= p_m^{-2}\bb{E}(\1_{q_m}'\E_t^{(m)'}\1_{p_m}\1_{p_m}'\1_{p_m}\1_{p_m}'\E_t^{(m)}\1_{q_m}) \\
    &= p_m^{-1}\bb{E}(\1_{p_m}'\E_t^{(m)}\1_{q_m})^2 = O(q_m),
\end{align*}
where the last equality follows by \eqref{eqn:sum_of_error_rate}. By triangle inequality, combining the two bounds, we obtain $\bb{E}\norm{\wh\a_t^{(m)}-\a_t^{(m)}}^2=O(p_mq_m^{-1})$. Therefore, by Markov's inequality, we have
\begin{align}
    p_m^{-1}\norm{\wh\a_t^{(m)} - \a_t^{(m)}}^2 &= O_P(q_m^{-1}), \label{eqn:row_main_effect_rate} \\
    q_m^{-1}\norm{\wh\b_t^{(m)} - \b_t^{(m)}}^2 &= O_P(p_m^{-1}), \label{eqn:column_main_effect_rate}
\end{align}
where column effect $\wh\b_t^{(m)}$ follows similarly. This ends the proof. $\square$

For the remaining main effect proofs and lemmas, we use the notation $\gamma_{r,\max}^{(m)}:=\gamma_r\vee\gamma_r^{(m)}$ and $\gamma_{r,\min}^{(m)}:=\gamma_r\wedge\gamma_r^{(m)}$, and define $\gamma_{c,\max}^{(m)}$ and $\gamma_{c,\min}^{(m)}$ analogously. These quantities are related to the notation in (R1) by
\begin{align*}
    p_m^{\gamma_{r,\max}^{(m)}-\gamma_r}=1\vee\nu_{r,m},\qquad
    p_m^{\gamma_r-\gamma_{r,\min}^{(m)}}&=1\vee\frac{1}{\nu_{r,m}},\qquad
    p_m^{\gamma_{r,\max}^{(m)}-\gamma_{r,\min}^{(m)}}=\nu_{r,m}\vee\frac{1}{\nu_{r,m}}, \\
    p_m^{\gamma_{r,\max}^{(m)}-\gamma_r^{(m)}}=\frac{1}{1\wedge\nu_{r,m}}&,\qquad
    \frac{p_m^{1-\gamma_{r,\min}^{(m)}}}{q_m}
    =\frac{p_m^{1-\gamma_r}}{q_m(1\wedge\nu_{r,m})}.
\end{align*}
The column-side identities follow similarly.

\textbf{\textit{Proof of Theorem~\ref{thm:initial_A1m}.}}
For $m\ne n\in[M]$, define 
\begin{align}
    \S_{A_1,mn}
    &:=T^{-1}q_m^{-1}q_n^{-1}\M_{p_m}
    \sum_{t=1}^T\bb{E}\left(\Y_t^{(m)}\1_{q_m}\1_{q_n}'\Y_t^{(n)'}\right)\M_{p_n}, \label{eqn:S_A1} \\
    \wh\S_{A_1,mn}
    &:=T^{-1}q_m^{-1}q_n^{-1}\M_{p_m}
    \sum_{t=1}^T\left(\Y_t^{(m)}\1_{q_m}\1_{q_n}'\Y_t^{(n)'}\right)\M_{p_n}. \label{eqn:S_A1_hat}
\end{align}
The corresponding population matrix and its estimator are
\begin{align*}
    \bOmega_{A_1,m}
    = (M-1)^{-1}\sum_{n=1,n\neq m}^{M}\S_{A_1,mn}^{\otimes 2}, \qquad 
    \wh\bOmega_{A_1,m}
    = (M-1)^{-1}\sum_{n=1,n\neq m}^{M}\wh\S_{A_1,mn}^{\otimes 2}.
\end{align*}
By \eqref{eqn: Omega_population}, we have $\S_{A_1,mn}=T^{-1}\A_{1m}\left(\sum_{t=1}^T\bb{E}(\balpha_t\balpha_t')\right)\A_{1n}'$. Hence, 
\begin{align}
    \bOmega_{A_1,m}
    =T^{-2}(M-1)^{-1}\A_{1m}\left(\sum_{t=1}^T\bb{E}(\balpha_t\balpha_t')\right)
    \left(\sum_{n\ne m}\A_{1n}'\A_{1n}\right)
    \left(\sum_{t=1}^T\bb{E}(\balpha_t\balpha_t')\right)\A_{1m}'.
    \label{eqn:Omega_A1}
\end{align}

Define $\omega_{A_1,m}:=p_m^{\gamma_r}\sum_{n\ne m}p_n^{\gamma_r}$. Since \eqref{eqn:Omega_A1} implies that the column space of $\bOmega_{A_1,m}$ is contained in that of $\A_{1m}$, with $\P_{A_1,m}:=\check{\A}_{1m}\check{\A}_{1m}'$, we have
\begin{align}
    (\I_{p_m}-\P_{A_1,m})\bOmega_{A_1,m}=\0.
    \label{eqn:projected_Omega_A1_zero}
\end{align}
Since $\wh\A_{1m}$ consists of the leading $r_1$ eigenvectors of $\wh\bOmega_{A_1,m}$,
\begin{align}
    \wh\bOmega_{A_1,m}\wh\A_{1m}
    =\wh\A_{1m}\wh\D_{A_1,m},\qquad
    \wh\D_{A_1,m}:=\diag\big(\lambda_1(\wh\bOmega_{A_1,m}),\ldots,
    \lambda_{r_1}(\wh\bOmega_{A_1,m})\big).
    \label{eqn:Omega_A1_A1_D1}
\end{align}
Define $\bDelta_{A_1,mn}:=\wh\S_{A_1,mn}-\S_{A_1,mn}$. Then,
\begin{align}
    \wh\bOmega_{A_1,m}-\bOmega_{A_1,m}
    =(M-1)^{-1}\sum_{n\ne m}
    \left(\bDelta_{A_1,mn}\S_{A_1,mn}'+\S_{A_1,mn}\bDelta_{A_1,mn}'+\bDelta_{A_1,mn}\bDelta_{A_1,mn}'\right).
    \label{eqn:Omega_A1_full_decomposition}
\end{align}
Left-multiplying \eqref{eqn:Omega_A1_full_decomposition} by $\I_{p_m}-\P_{A_1,m}$ and using $(\I_{p_m}-\P_{A_1,m})\S_{A_1,mn}=\0$ gives
\begin{align*}
    &\quad\;(\I_{p_m}-\P_{A_1,m})(\wh\bOmega_{A_1,m}-\bOmega_{A_1,m}) \\
    &=(M-1)^{-1}\sum_{n\ne m}\left\{(\I_{p_m}-\P_{A_1,m})\bDelta_{A_1,mn}\S_{A_1,mn}'+(\I_{p_m}-\P_{A_1,m})\bDelta_{A_1,mn}\bDelta_{A_1,mn}'\right\}.
\end{align*} 
By \eqref{eqn:S_A1_bound} and \eqref{eqn:Delta_A1_oP} in Lemma~\ref{lemma:bound_S_A1_Delta_A1}, $\norm{\S_{A_1,mn}}_F^2+\norm{\bDelta_{A_1,mn}}_F^2=O_P(p_m^{\gamma_r}p_n^{\gamma_r})$. Hence, \eqref{eqn:Delta_A1_projected_bound} gives
\begin{align}
    &\norm{(\I_{p_m}-\P_{A_1,m})(\wh\bOmega_{A_1,m}-\bOmega_{A_1,m})}_F^2 \nonumber\\
    &=O_P\Big(R(T,d_{r,\star})p_m^{\gamma_r^{(m)}+\gamma_r}
    \sum_{n\ne m}p_n^{\gamma_r^{(n)}+\gamma_r}
    +T^{-1}p_m^{\gamma_r+1}q_m^{-1}
    \sum_{n\ne m}p_n^{\gamma_{r,\max}^{(n)}+\gamma_r}\Big) \nonumber\\
    &=O_P\Bigg(\omega_{A_1,m}^2\Big(
    R(T,d_{r,\star})p_m^{\gamma_r^{(m)}-\gamma_r}
    \max_{n\in[M]\setminus\{m\}}p_n^{\gamma_r^{(n)}-\gamma_r}
    +T^{-1}p_m^{1-\gamma_r}q_m^{-1}
    \max_{n\in[M]\setminus\{m\}}
    p_n^{\gamma_{r,\max}^{(n)}-\gamma_r}\Big)\Bigg).
    \label{eqn:Omega_A1_projected_bound}
\end{align}

Define $\H_{A_1,m}:=\check{\A}_{1m}'\wh\A_{1m}$.
Since $\check{\A}_{1m}$ and $\wh\A_{1m}$ both have orthonormal columns, Lemma~\ref{lemma:rotation_orthogonality} gives
\begin{align}
    \wh\A_{1m}-\check{\A}_{1m}\H_{A_1,m}
    =(\I_{p_m}-\P_{A_1,m})\wh\A_{1m}.
    \label{eqn:A1_projected_rotation}
\end{align}
Left-multiplying \eqref{eqn:Omega_A1_A1_D1} by $\I_{p_m}-\P_{A_1,m}$, and using \eqref{eqn:projected_Omega_A1_zero} and \eqref{eqn:A1_projected_rotation}, gives
\begin{align}
    \left(\wh\A_{1m}-\check{\A}_{1m}\H_{A_1,m}\right)\wh\D_{A_1,m}
    =(\I_{p_m}-\P_{A_1,m})(\wh\bOmega_{A_1,m}-\bOmega_{A_1,m})\wh\A_{1m}.
    \label{eqn:A1_minus_A1H_times_D}
\end{align}
Combining \eqref{eqn:Omega_A1_projected_bound}, \eqref{eqn:A1_minus_A1H_times_D}, and Lemma~\ref{lemma:bound_D_A1}, together with $\norm{\wh\A_{1m}}=1$, we obtain
\begin{align*}
    &\quad\;\norm{\wh\A_{1m}-\check{\A}_{1m}\H_{A_1,m}}_F^2 
    \leq \norm{(\I_{p_m}-\P_{A_1,m})(\wh\bOmega_{A_1,m}-\bOmega_{A_1,m})}_F^2
    \norm{\wh\A_{1m}}^2\norm{\wh\D_{A_1,m}^{-1}}_F^2 \\
    &=O_P\Big(R(T,d_{r,\star})p_m^{\gamma_r^{(m)}-\gamma_r}
    \max_{n\in[M]\setminus\{m\}}p_n^{\gamma_r^{(n)}-\gamma_r}
    +T^{-1}p_m^{1-\gamma_r}q_m^{-1}
    \max_{n\in[M]\setminus\{m\}}p_n^{\gamma_{r,\max}^{(n)}-\gamma_r}\Big).
\end{align*}
Using $p_n^{\gamma_{r,\max}^{(n)}-\gamma_r}=1\vee\nu_{r,n}$, this is exactly the row-side rate stated in Theorem~\ref{thm:initial_A1m}. Both terms above are $o(1)$ under (R1), so that $\norm{\wh\A_{1m}-\check{\A}_{1m}\H_{A_1,m}}_F=o_P(1)$. Hence, Lemma~\ref{lemma:rotation_orthogonality} gives $\H_{A_1,m}'\H_{A_1,m}=\I_{r_1}+o_P(1)$, $\H_{A_1,m}\H_{A_1,m}'=\I_{r_1}+o_P(1)$, and $\H_{A_1,m}$ is invertible with probability tending to one. Moreover, defining $\wh\P_{A_1,m}:=\wh\A_{1m}\wh\A_{1m}'$, we have
\begin{align}
    \norm{\wh\P_{A_1,m}-\P_{A_1,m}}_F^2
    =2\norm{\wh\A_{1m}-\check{\A}_{1m}\H_{A_1,m}}_F^2.
    \label{eqn:A1_projector_rate}
\end{align}
The proof for the global column-loading estimator $\wh\B_{1m}$ follows similarly, with $\H_{B_1,m}:=\check\B_{1m}'\wh\B_{1m}$. This ends the proof. $\square$

\textbf{\textit{Proof of Theorem~\ref{thm:A2m_alphatm}.}}
Recall that
\begin{align}
    \S_{A_2,m}:=T^{-1}\sum_{t=1}^T\a_t^{(m)}\a_t^{(m)'}\P_{A_1,m}^{\perp},\qquad
    \wh\S_{A_2,m}:=T^{-1}\sum_{t=1}^T\wh\a_t^{(m)}\wh\a_t^{(m)'}\wh\P_{A_1,m}^{\perp},
    \label{eqn:S_A2}
\end{align}
where $\P_{A_1,m}^{\perp}:=\I_{p_m}-\P_{A_1,m}$ and $\wh\P_{A_1,m}^{\perp}:=\I_{p_m}-\wh\P_{A_1,m}$. Since $\wh\A_{1m}^{\perp}\wh\A_{1m}^{\perp'}=\wh\P_{A_1,m}^{\perp}$, the sample matrix in the estimation procedure satisfies
\begin{align}
    \wh\bOmega_{A_2,m}
    &=\left(T^{-1}\sum_{t=1}^T\wh\a_t^{(m)}\wh\a_t^{(m)'}\right)
    \wh\P_{A_1,m}^{\perp}
    \left(T^{-1}\sum_{t=1}^T\wh\a_t^{(m)}\wh\a_t^{(m)'}\right)
    =\wh\S_{A_2,m}\wh\S_{A_2,m}'.
    \label{eqn:Omega_A2_sample}
\end{align}
Define its population counterpart as
\begin{align}
    \bOmega_{A_2,m}:=\left(T^{-1}\sum_{t=1}^T\a_t^{(m)}\a_t^{(m)'}\right)
    \P_{A_1,m}^{\perp}\left(T^{-1}\sum_{t=1}^T\a_t^{(m)}\a_t^{(m)'}\right)
    =\S_{A_2,m}\S_{A_2,m}'.
    \label{eqn:Omega_A2_population}
\end{align}
By (IC2),
\begin{align*}
    \S_{A_2,m}=\A_{2m}\left(T^{-1}\sum_{t=1}^T\balpha_t^{(m)}\balpha_t^{(m)'}\right)\A_{2m}'\P_{A_1,m}^{\perp}.
\end{align*}
Hence, with $\P_{A_2,m}:=\check{\A}_{2m}\check{\A}_{2m}'$, 
\begin{align}
    (\I_{p_m}-\P_{A_2,m})\S_{A_2,m}=\0,\qquad
    (\I_{p_m}-\P_{A_2,m})\bOmega_{A_2,m}=\0.
    \label{eqn:projected_Omega_A2_zero}
\end{align}
Since $\wh\A_{2m}$ consists of the leading $r_2^{(m)}$ eigenvectors of $\wh\bOmega_{A_2,m}$,
\begin{align}
    \wh\bOmega_{A_2,m}\wh\A_{2m}=\wh\A_{2m}\wh\D_{A_2,m},\qquad
    \wh\D_{A_2,m}:=\diag\big(\lambda_1(\wh\bOmega_{A_2,m}),\ldots,
    \lambda_{r_2^{(m)}}(\wh\bOmega_{A_2,m})\big).
    \label{eqn:Omega_A2_A2_D2}
\end{align}
Define $\H_{A_2,m}:=\check{\A}_{2m}'\wh\A_{2m}$. Since $\check{\A}_{2m}$ and $\wh\A_{2m}$ both have orthonormal columns, Lemma~\ref{lemma:rotation_orthogonality} gives
\begin{align}
    \wh\A_{2m}-\check{\A}_{2m}\H_{A_2,m}
    =(\I_{p_m}-\P_{A_2,m})\wh\A_{2m}.
    \label{eqn:A2_projected_rotation}
\end{align}
Left-multiplying \eqref{eqn:Omega_A2_A2_D2} by $\I_{p_m}-\P_{A_2,m}$, and using \eqref{eqn:projected_Omega_A2_zero} and \eqref{eqn:A2_projected_rotation}, gives
\begin{align}
    \left(\wh\A_{2m}-\check{\A}_{2m}\H_{A_2,m}\right)\wh\D_{A_2,m}
    =(\I_{p_m}-\P_{A_2,m})(\wh\bOmega_{A_2,m}-\bOmega_{A_2,m})\wh\A_{2m}.
    \label{eqn:A2_minus_A2H_times_D}
\end{align}

Let $\bDelta_{A_2,m}:=\wh\S_{A_2,m}-\S_{A_2,m}$. Since
\begin{align*}
    \wh\bOmega_{A_2,m}-\bOmega_{A_2,m}
    =\bDelta_{A_2,m}\S_{A_2,m}'
    +\S_{A_2,m}\bDelta_{A_2,m}'
    +\bDelta_{A_2,m}\bDelta_{A_2,m}',
\end{align*}
using $(\I_{p_m}-\P_{A_2,m})\S_{A_2,m}=\0$ gives
\begin{align}
    \norm{(\I_{p_m}-\P_{A_2,m})
    (\wh\bOmega_{A_2,m}-\bOmega_{A_2,m})}_F^2
    \lesssim
    \norm{(\I_{p_m}-\P_{A_2,m})\bDelta_{A_2,m}}_F^2
    \left(\norm{\S_{A_2,m}}_F^2+\norm{\bDelta_{A_2,m}}_F^2\right).
    \label{eqn:bound_Omega_A2_hat_minus_Omega}
\end{align}
By \eqref{eqn:S_A2_bound} and \eqref{eqn:Delta_A2_oP} in Lemma~\ref{lemma:bound_S_A2_Delta_A2}, $\norm{\S_{A_2,m}}_F^2+\norm{\bDelta_{A_2,m}}_F^2=O_P(p_m^{2\gamma_r^{(m)}})$. Combining \eqref{eqn:bound_Omega_A2_hat_minus_Omega}, \eqref{eqn:Delta_A2_projected_bound}, \eqref{eqn:A2_minus_A2H_times_D}, and Lemma~\ref{lemma:bound_D_A2}, together with $\norm{\wh\A_{2m}}=1$, we obtain
\begin{align*}
    &\quad\;\norm{\wh\A_{2m}-\check{\A}_{2m}\H_{A_2,m}}_F^2
    \leq\norm{(\I_{p_m}-\P_{A_2,m})
    (\wh\bOmega_{A_2,m}-\bOmega_{A_2,m})}_F^2
    \norm{\wh\A_{2m}}^2\norm{\wh\D_{A_2,m}^{-1}}_F^2 \\
    &=O_P\Big(R(T,d_{r,\star})p_m^{\gamma_r-\gamma_r^{(m)}}
    \max_{n\in[M]\setminus\{m\}}p_n^{\gamma_r^{(n)}-\gamma_r}
    +T^{-1}p_m^{1+\gamma_r-2\gamma_r^{(m)}}q_m^{-1}
    \max_{n\in[M]}p_n^{\gamma_{r,\max}^{(n)}-\gamma_r}
    +p_m^{1-2\gamma_r^{(m)}}q_m^{-2}\Big).
\end{align*}
Together with the definition of $\nu_{r,m}$, this proves the row-side loading rate in Theorem~\ref{thm:A2m_alphatm}. All three terms above are $o(1)$ under (R1), so that $\norm{\wh\A_{2m}-\check{\A}_{2m}\H_{A_2,m}}_F=o_P(1)$. Hence, Lemma~\ref{lemma:rotation_orthogonality} gives $\H_{A_2,m}'\H_{A_2,m}=\I_{r_2^{(m)}}+o_P(1)$, $\H_{A_2,m}\H_{A_2,m}'=\I_{r_2^{(m)}}+o_P(1)$, and $\H_{A_2,m}$ is invertible with probability tending to one. Moreover, defining $\wh\P_{A_2,m}:=\wh\A_{2m}\wh\A_{2m}'$, we have
\begin{align}
    \norm{\wh\P_{A_2,m}-\P_{A_2,m}}_F^2
    =2\norm{\wh\A_{2m}-\check{\A}_{2m}\H_{A_2,m}}_F^2.
    \label{eqn:A2_projector_rate}
\end{align}
The proof for the local column-loading estimator $\wh\B_{2m}$ follows similarly, with $\H_{B_2,m}:=\check\B_{2m}'\wh\B_{2m}$.

We next prove the local main effect factor series. For $m\in[M]$, define
\begin{align*}
    \cL_{\alpha,m}
    :=\H_{A_2,m}^{-1}
    (\check\A_{2m}'\P_{A_1,m}^{\perp}\check\A_{2m})^{-1}
    \check\A_{2m}'\P_{A_1,m}^{\perp},\qquad
    \wh\cL_{\alpha,m}
    :=(\wh\A_{2m}'\wh\P_{A_1,m}^{\perp}\wh\A_{2m})^{-1}
    \wh\A_{2m}'\wh\P_{A_1,m}^{\perp}.
\end{align*}
By (S1) and the asymptotic orthogonality of $\H_{A_2,m}$, we have $\norm{\cL_{\alpha,m}}=O_P(1)$. Moreover, $\cL_{\alpha,m}\A_{1m}=\0$, $\cL_{\alpha,m}\check\A_{2m}=\H_{A_2,m}^{-1}$, $\wh\cL_{\alpha,m}\wh\A_{2m}=\I_{r_2^{(m)}}$, and $\wh\cL_{\alpha,m}\wh\P_{A_1,m}=\0$.

We first control $\wh\cL_{\alpha,m}-\cL_{\alpha,m}$. Note that
\begin{align*}
    &\quad\;\wh\A_{2m}'\wh\P_{A_1,m}^{\perp}\wh\A_{2m}
    -\H_{A_2,m}'\check\A_{2m}'\P_{A_1,m}^{\perp}
    \check\A_{2m}\H_{A_2,m} \\
    &=(\wh\A_{2m}-\check\A_{2m}\H_{A_2,m})'\wh\P_{A_1,m}^{\perp}\wh\A_{2m}
    +\H_{A_2,m}'\check\A_{2m}'(\wh\P_{A_1,m}^{\perp}-\P_{A_1,m}^{\perp})\wh\A_{2m} \\
    &\quad+\H_{A_2,m}'\check\A_{2m}'\P_{A_1,m}^{\perp}
    (\wh\A_{2m}-\check\A_{2m}\H_{A_2,m}).
\end{align*}
Since $\check\A_{2m}$ and $\wh\A_{2m}$ have orthonormal columns, $\norm{\H_{A_2,m}}=O_P(1)$, and $\norm{\wh\P_{A_1,m}^{\perp}-\P_{A_1,m}^{\perp}}=\norm{\wh\P_{A_1,m}-\P_{A_1,m}}$, it follows that
\begin{align}
    &\quad\;\norm{\wh\A_{2m}'\wh\P_{A_1,m}^{\perp}\wh\A_{2m}
    -\H_{A_2,m}'\check\A_{2m}'\P_{A_1,m}^{\perp}
    \check\A_{2m}\H_{A_2,m}} \nonumber \\
    &=O_P\Big(\norm{\wh\A_{2m}-\check\A_{2m}\H_{A_2,m}}_F
    +\norm{\wh\P_{A_1,m}-\P_{A_1,m}}\Big).
    \label{eqn:local_gram_perturbation}
\end{align}
By (S1), the eigenvalues of $\check\A_{2m}'\P_{A_1,m}^{\perp}\check\A_{2m}$ are bounded away from zero. Together with the asymptotic orthogonality of $\H_{A_2,m}$, this implies $\lambda_{r_2^{(m)}}\big(\H_{A_2,m}'\check\A_{2m}'\P_{A_1,m}^{\perp}\check\A_{2m}\H_{A_2,m}\big)\gtrsim_P 1$.
Since the right-hand side of \eqref{eqn:local_gram_perturbation} is $o_P(1)$, Weyl's inequality further gives
\begin{align*}
    \norm{(\wh\A_{2m}'\wh\P_{A_1,m}^{\perp}\wh\A_{2m})^{-1}}
    =O_P(1), \qquad
    \norm{(\H_{A_2,m}'\check\A_{2m}'\P_{A_1,m}^{\perp}\check\A_{2m}\H_{A_2,m})^{-1}}
    =O_P(1).
\end{align*}
Therefore, using $\A^{-1}-\B^{-1}=\A^{-1}(\B-\A)\B^{-1}$,
\begin{align}
    &\quad\;\norm{(\wh\A_{2m}'\wh\P_{A_1,m}^{\perp}\wh\A_{2m})^{-1}
    -(\H_{A_2,m}'\check\A_{2m}'\P_{A_1,m}^{\perp}
    \check\A_{2m}\H_{A_2,m})^{-1}} \nonumber \\
    &=O_P\Big(\norm{\wh\A_{2m}-\check\A_{2m}\H_{A_2,m}}_F
    +\norm{\wh\P_{A_1,m}-\P_{A_1,m}}\Big).
    \label{eqn:local_gram_inverse_perturbation}
\end{align}
Moreover,
\begin{align*}
    \wh\A_{2m}'\wh\P_{A_1,m}^{\perp}
    -\H_{A_2,m}'\check\A_{2m}'\P_{A_1,m}^{\perp}
    =(\wh\A_{2m}-\check\A_{2m}\H_{A_2,m})'\wh\P_{A_1,m}^{\perp}
    +\H_{A_2,m}'\check\A_{2m}'(\wh\P_{A_1,m}^{\perp}-\P_{A_1,m}^{\perp}),
\end{align*}
and hence
\begin{align}
    \norm{\wh\A_{2m}'\wh\P_{A_1,m}^{\perp}
    -\H_{A_2,m}'\check\A_{2m}'\P_{A_1,m}^{\perp}}
    =O_P\Big(\norm{\wh\A_{2m}-\check\A_{2m}\H_{A_2,m}}_F
    +\norm{\wh\P_{A_1,m}-\P_{A_1,m}}\Big).
    \label{eqn:local_numerator_perturbation}
\end{align}
Since $\cL_{\alpha,m}=(\H_{A_2,m}'\check\A_{2m}'\P_{A_1,m}^{\perp}\check\A_{2m}\H_{A_2,m})^{-1}\H_{A_2,m}'\check\A_{2m}'\P_{A_1,m}^{\perp}$, combining \eqref{eqn:local_gram_inverse_perturbation} and \eqref{eqn:local_numerator_perturbation} yields
\begin{align*}
    \norm{\wh\cL_{\alpha,m}-\cL_{\alpha,m}}
    =O_P\Big(\norm{\wh\A_{2m}-\check\A_{2m}\H_{A_2,m}}_F
    +\norm{\wh\P_{A_1,m}-\P_{A_1,m}}\Big).
\end{align*}
Thus, by Lemma~\ref{lemma:rotation_orthogonality},
\begin{align}
    \norm{\wh\cL_{\alpha,m}-\cL_{\alpha,m}}^2
    =O_P\Big(\norm{\wh\A_{1m}-\check\A_{1m}\H_{A_1,m}}_F^2
    +\norm{\wh\A_{2m}-\check\A_{2m}\H_{A_2,m}}_F^2\Big).
    \label{eqn:L_alpha_perturbation}
\end{align}
In particular, $\norm{\wh\cL_{\alpha,m}}=O_P(1)$.

By the definition of $\wh\balpha_t^{(m)}$ and $\wh\a_t^{(m)}-\a_t^{(m)}=q_m^{-1}\M_{p_m}\E_t^{(m)}\1_{q_m}$,
\begin{align}
    \wh\balpha_t^{(m)}-\H_{A_2,m}^{-1}\check\balpha_t^{(m)}
    &=q_m^{-1}\cL_{\alpha,m}\M_{p_m}\E_t^{(m)}\1_{q_m}
    +q_m^{-1}(\wh\cL_{\alpha,m}-\cL_{\alpha,m})\M_{p_m}\E_t^{(m)}\1_{q_m} \nonumber \\
    &\quad+\wh\cL_{\alpha,m}(\P_{A_1,m}-\wh\P_{A_1,m})\A_{1m}\balpha_t
    -\wh\cL_{\alpha,m}(\wh\A_{2m}-\check\A_{2m}\H_{A_2,m})\H_{A_2,m}^{-1}\check\balpha_t^{(m)} \nonumber \\
    &=:\cI_{\alpha,1}^{(m)}+\cI_{\alpha,2}^{(m)}+\cI_{\alpha,3}^{(m)}+\cI_{\alpha,4}^{(m)},
    \label{eqn:local_alpha_decomposition}
\end{align}
where the first equality uses $\cL_{\alpha,m}\A_{1m}=\0$, $\wh\cL_{\alpha,m}\wh\P_{A_1,m}=\0$, $\cL_{\alpha,m}\check\A_{2m}=\H_{A_2,m}^{-1}$ and $\wh\cL_{\alpha,m}\wh\A_{2m}=\I_{r_2^{(m)}}$.

For the first term in \eqref{eqn:local_alpha_decomposition}, note that the matrix $(\check\A_{2m}'\P_{A_1,m}^{\perp}\check\A_{2m})^{-1}\check\A_{2m}'\P_{A_1,m}^{\perp}$ has bounded Frobenius norm by (S1). Since $\norm{\H_{A_2,m}^{-1}}=O_P(1)$, Lemma~\ref{lemma:correlation_Etm} gives $\norm{\cI_{\alpha,1}^{(m)}}^2=O_P(q_m^{-1})$. Moreover, Lemma~\ref{lemma:correlation_Etm}, \eqref{eqn:L_alpha_perturbation}, (ME1) and (ME2) yield respectively
\begin{align*}
    &\norm{\cI_{\alpha,2}^{(m)}}^2
    =O_P\left(p_mq_m^{-1}\left(\norm{\wh\A_{1m}-\check\A_{1m}\H_{A_1,m}}_F^2
    +\norm{\wh\A_{2m}-\check\A_{2m}\H_{A_2,m}}_F^2\right)\right),\\
    &\norm{\cI_{\alpha,3}^{(m)}}^2
    =O_P\Big(p_m^{\gamma_r}\norm{\wh\A_{1m}-\check\A_{1m}\H_{A_1,m}}_F^2\Big),\qquad
    \norm{\cI_{\alpha,4}^{(m)}}^2
    =O_P\Big(p_m^{\gamma_r^{(m)}}\norm{\wh\A_{2m}-\check\A_{2m}\H_{A_2,m}}_F^2\Big).
\end{align*}
Consequently,
\begin{align*}
    p_m^{-\gamma_r^{(m)}}
    \norm{\wh\balpha_t^{(m)}-\H_{A_2,m}^{-1}\check\balpha_t^{(m)}}^2
    =O_P\Big(&p_m^{-\gamma_r^{(m)}}q_m^{-1}
    +(p_m^{1-\gamma_r^{(m)}}q_m^{-1}
    +p_m^{\gamma_r-\gamma_r^{(m)}})
    \norm{\wh\A_{1m}-\check\A_{1m}\H_{A_1,m}}_F^2 \\
    &+(p_m^{1-\gamma_r^{(m)}}q_m^{-1}+1)
    \norm{\wh\A_{2m}-\check\A_{2m}\H_{A_2,m}}_F^2\Big).
\end{align*}
Substituting the rates from Theorems~\ref{thm:initial_A1m} and \ref{thm:A2m_alphatm} obtained above, and using $1+p_m^{\gamma_r-\gamma_r^{(m)}}\lesssim p_m^{\gamma_{r,\max}^{(m)}-\gamma_r^{(m)}}$, and noting that $p_m^{1-2\gamma_r^{(m)}}q_m^{-2}=o(p_m^{-\gamma_r^{(m)}}q_m^{-1})$ under (R1), gives
\begin{align}
    p_m^{-\gamma_r^{(m)}}\norm{\wh\balpha_t^{(m)}-\H_{A_2,m}^{-1}\check\balpha_t^{(m)}}^2
    =O_P\Big(&p_m^{-\gamma_r^{(m)}}q_m^{-1}
    +R(T,d_{r,\star})p_m^{\gamma_{r,\max}^{(m)}-\gamma_r^{(m)}}
    \max_{n\in[M]\setminus\{m\}}p_n^{\gamma_r^{(n)}-\gamma_r} \nonumber \\
    &+T^{-1}p_m^{1+\gamma_{r,\max}^{(m)}-2\gamma_r^{(m)}}q_m^{-1}
    \max_{n\in[M]}p_n^{\gamma_{r,\max}^{(n)}-\gamma_r}\Big).
    \label{eqn:local_alpha_explicit_rate}
\end{align}
Since $p_m^{\gamma_{r,\max}^{(m)}-\gamma_r^{(m)}}=(1\wedge\nu_{r,m})^{-1}$ and
$p_n^{\gamma_{r,\max}^{(n)}-\gamma_r}=1\vee\nu_{r,n}$, this is the row-side factor rate stated in Theorem~\ref{thm:A2m_alphatm}. The column-side rate follows by symmetry. This ends the proof. $\square$

\textbf{\textit{Proof of Theorem~\ref{thm:A1m_alphat}.}}
We prove the row-side results; the column-side results follow by symmetry. Let $\A_\alpha:=(\balpha_1,\ldots,\balpha_T)$,
$\bGamma_{\alpha,T}:=T^{-1}\A_\alpha\A_\alpha'$, 
$\check\A_\alpha:=\bGamma_{\alpha,T}^{-1/2}\A_\alpha=(\check\balpha_1,\ldots,\check\balpha_T)$.
Also write $\wh\A_\alpha:=(\wh\balpha_1,\ldots,\wh\balpha_T)$ and define
$\P_\alpha:=T^{-1}\check\A_\alpha'\check\A_\alpha$.
By Lemma~\ref{lemma:prop_piecewise}, (S1), and Weyl's inequality, $\lambda_{r_1}(\bGamma_{\alpha,T})\asymp_P \lambda_1(\bGamma_{\alpha,T})\asymp_P1$. Hence $\norm{\bGamma_{\alpha,T}^{-1/2}}=O_P(1)$ and $\norm{\check\balpha_t}=O_P(1)$ for each $t\in[T]$.

For each $m\in[M]$, define $\Z_{\alpha,m}:=\wh\A_{1m}'\A_{1m}\A_\alpha$ and $\bDelta_{\alpha,m}:=\wh\Z_{\alpha,m}-\Z_{\alpha,m}$, and let $\bdelta_{\alpha,m,t}$ denote the $t$th column of $\bDelta_{\alpha,m}$. Since $\wh\A_{1m}'\A_{1m}=\H_{A_1,m}'(\A_{1m}'\A_{1m})^{1/2}$, Theorem~\ref{thm:initial_A1m} and (ME1) imply that $\wh\A_{1m}'\A_{1m}$ is invertible with probability tending to one and has squared singular values of order $p_m^{\gamma_r}$. Thus, 
\begin{align}
    \P_{\Z_{\alpha,m}}
    =\A_\alpha'(\A_\alpha\A_\alpha')^{-1}\A_\alpha
    =\P_\alpha,\qquad
    \sigma_{r_1}^2(\Z_{\alpha,m})\asymp_P Tp_m^{\gamma_r}.
    \label{eqn:Z_alpham}
\end{align}

Write 
\begin{align*}
    \wh\y_t^{(m)}-\A_{1m}\balpha_t&=q_m^{-1}\M_{p_m}\E_t^{(m)}\1_{q_m}-(\wh\A_{2m}\wh\balpha_t^{(m)}-\A_{2m}\balpha_t^{(m)}), \\
    \bdelta_{\alpha,m,t}&=\wh\A_{1m}'\left(q_m^{-1}\M_{p_m}\E_t^{(m)}\1_{q_m}-(\wh\A_{2m}\wh\balpha_t^{(m)}-\A_{2m}\balpha_t^{(m)})\right).
\end{align*}
Using $\wh\A_{2m}\wh\balpha_t^{(m)}-\A_{2m}\balpha_t^{(m)}=\wh\A_{2m}\big(\wh\balpha_t^{(m)}-\H_{A_2,m}^{-1}\check\balpha_t^{(m)}\big)+(\wh\A_{2m}-\check\A_{2m}\H_{A_2,m})\H_{A_2,m}^{-1}\check\balpha_t^{(m)}$, the loading and series rates in Theorem~\ref{thm:A2m_alphatm}, the asymptotic orthogonality of $\H_{A_2,m}$, and $\norm{\check\balpha_t^{(m)}}^2=O_P(p_m^{\gamma_r^{(m)}})$ give
\begin{align}
    \norm{\wh\A_{2m}\wh\balpha_t^{(m)}-\A_{2m}\balpha_t^{(m)}}^2
    =O_P\Big(&q_m^{-1}
    +R(T,d_{r,\star})p_m^{\gamma_{r,\max}^{(m)}}
    \max_{n\in[M]\setminus\{m\}}p_n^{\gamma_r^{(n)}-\gamma_r}
    \nonumber\\
    &+T^{-1}p_m^{1+\gamma_{r,\max}^{(m)}-\gamma_r^{(m)}}q_m^{-1}
    \max_{n\in[M]}p_n^{\gamma_{r,\max}^{(n)}-\gamma_r}\Big).
    \label{eqn:global_alpha_local_reconstruction}
\end{align}
The same rate holds for $T^{-1}\sum_{t=1}^T\norm{\wh\A_{2m}\wh\balpha_t^{(m)}-\A_{2m}\balpha_t^{(m)}}^2$,
Indeed, the same argument for $\cI_{\alpha,1}^{(m)}$ gives $T^{-1}\sum_t\norm{\cI_{\alpha,1}^{(m)}}^2=O_P(q_m^{-1})$. Applying the remaining bounds in \eqref{eqn:local_alpha_decomposition}, together with
\begin{align*}
    &T^{-1}\sum_t\norm{q_m^{-1}\M_{p_m}\E_t^{(m)}\1_{q_m}}^2=O_P(p_mq_m^{-1}), \\
    T^{-1}\sum_t\norm{&\A_{1m}\balpha_t}^2=O_P(p_m^{\gamma_r}), \qquad
    T^{-1}\sum_t\norm{\check\balpha_t^{(m)}}^2=O_P(p_m^{\gamma_r^{(m)}}),
\end{align*}
gives the stated time-average reconstruction rate.

Next, since 
\begin{align*}
    q_m^{-1}\wh\A_{1m}'\M_{p_m}\E_t^{(m)}\1_{q_m}
    =q_m^{-1}\H_{A_1,m}'\check\A_{1m}'\M_{p_m}\E_t^{(m)}\1_{q_m}
    +q_m^{-1}(\wh\A_{1m}-\check\A_{1m}\H_{A_1,m})'\M_{p_m}\E_t^{(m)}\1_{q_m},
\end{align*}
Lemma~\ref{lemma:correlation_Etm} gives $\norm{q_m^{-1}\check\A_{1m}'\M_{p_m}\E_t^{(m)}\1_{q_m}}^2=O_P(q_m^{-1})$, with the same rate after averaging over $t$. Using Theorem~\ref{thm:initial_A1m} and (R1), this gives
\begin{align*}
    &\quad\;\norm{q_m^{-1}\wh\A_{1m}'\M_{p_m}\E_t^{(m)}\1_{q_m}}^2,\quad
    T^{-1}\sum_{t=1}^T\norm{q_m^{-1}\wh\A_{1m}'\M_{p_m}\E_t^{(m)}\1_{q_m}}^2 \\
    &=O_P\Big(q_m^{-1}
    +R(T,d_{r,\star})p_m^{\gamma_r^{(m)}}\max_{n\in[M]\setminus\{m\}}p_n^{\gamma_r^{(n)}-\gamma_r}
    +T^{-1}p_mq_m^{-1}\max_{n\in[M]\setminus\{m\}}p_n^{\gamma_{r,\max}^{(n)}-\gamma_r}\Big),
\end{align*}
which is bounded by the right-hand side of \eqref{eqn:global_alpha_local_reconstruction}. Since $\norm{\wh\A_{1m}}=1$, we therefore obtain
\begin{align}
    \norm{\bdelta_{\alpha,m,t}}^2,\quad
    T^{-1}\norm{\bDelta_{\alpha,m}}_F^2
    =O_P\Big(&q_m^{-1}
    +R(T,d_{r,\star})p_m^{\gamma_{r,\max}^{(m)}}
    \max_{n\in[M]\setminus\{m\}}p_n^{\gamma_r^{(n)}-\gamma_r}
    \nonumber\\
    &+T^{-1}p_m^{1+\gamma_{r,\max}^{(m)}-\gamma_r^{(m)}}q_m^{-1}
    \max_{n\in[M]}p_n^{\gamma_{r,\max}^{(n)}-\gamma_r}\Big).
    \label{eqn:global_alpha_score_bound}
\end{align}

Define $\wh\P_{\alpha,m}:=\wh\Z_{\alpha,m}'(\wh\Z_{\alpha,m}\wh\Z_{\alpha,m}')^{-1}\wh\Z_{\alpha,m}$. By \eqref{eqn:Z_alpham}, \eqref{eqn:global_alpha_score_bound}, and (R1), $\norm{\bDelta_{\alpha,m}}_F^2=o_P(Tp_m^{\gamma_r})$.
Hence Weyl's inequality gives $\sigma_{r_1}^2(\wh\Z_{\alpha,m})\asymp_P Tp_m^{\gamma_r}$, so that $\wh\Z_{\alpha,m}\wh\Z_{\alpha,m}'$ is invertible with probability tending to one. Since $\P_\alpha^\perp\Z_{\alpha,m}'=\0$,
\begin{align}
    \P_\alpha^\perp\wh\P_{\alpha,m}
    =\P_\alpha^\perp\bDelta_{\alpha,m}'
    (\wh\Z_{\alpha,m}\wh\Z_{\alpha,m}')^{-1}\wh\Z_{\alpha,m}.
    \label{eqn:projected_group_temporal}
\end{align}
Using
$\norm{(\wh\Z_{\alpha,m}\wh\Z_{\alpha,m}')^{-1}\wh\Z_{\alpha,m}} =\sigma_{r_1}^{-1}(\wh\Z_{\alpha,m})$ and $\norm{\wh\P_{\alpha,m}-\P_\alpha}_F^2=2\norm{\P_\alpha^\perp\wh\P_{\alpha,m}}_F^2$, we obtain
\begin{align}
    \norm{\wh\P_{\alpha,m}-\P_\alpha}_F^2
    =O_P\Big(&p_m^{-\gamma_r}q_m^{-1}
    +R(T,d_{r,\star})
    p_m^{\gamma_{r,\max}^{(m)}-\gamma_r}
    \max_{n\in[M]\setminus\{m\}}p_n^{\gamma_r^{(n)}-\gamma_r}
    \nonumber\\
    &+T^{-1}p_m^{1-\gamma_{r,\min}^{(m)}}q_m^{-1}
    \max_{n\in[M]}p_n^{\gamma_{r,\max}^{(n)}-\gamma_r}\Big).
    \label{eqn:global_alpha_group_projector}
\end{align}
All terms on the right-hand side are $o(1)$ under (R1).

For the rowwise bound, let $\U_\alpha:=T^{-1/2}\check\A_\alpha'$, so that $\P_\alpha=\U_\alpha\U_\alpha'$ and $\norm{\e_t'\U_\alpha}^2=T^{-1}\norm{\check\balpha_t}^2=O_P(T^{-1})$. Hence
\begin{align*}
    \norm{\e_t'\P_\alpha^\perp\bDelta_{\alpha,m}'}^2
    \lesssim
    \norm{\bdelta_{\alpha,m,t}}^2
    +T^{-1}\norm{\check\balpha_t}^2\norm{\bDelta_{\alpha,m}}_F^2.
\end{align*}
Combining this with \eqref{eqn:projected_group_temporal} and \eqref{eqn:global_alpha_score_bound} yields
\begin{align}
    T\norm{\e_t'\P_\alpha^\perp
    (\wh\P_{\alpha,m}-\P_\alpha)}^2
    =O_P\Big(&p_m^{-\gamma_r}q_m^{-1}
    +R(T,d_{r,\star})
    p_m^{\gamma_{r,\max}^{(m)}-\gamma_r}
    \max_{n\in[M]\setminus\{m\}}p_n^{\gamma_r^{(n)}-\gamma_r}
    \nonumber\\
    &+T^{-1}p_m^{1-\gamma_{r,\min}^{(m)}}q_m^{-1}
    \max_{n\in[M]}p_n^{\gamma_{r,\max}^{(n)}-\gamma_r}\Big).
    \label{eqn:global_alpha_group_projector_row}
\end{align}

Now let $\wh\K_\alpha:=M^{-1}\sum_{m=1}^M\wh\P_{\alpha,m}$ and $\wh\U_\alpha:=T^{-1/2}\wh\A_\alpha'$, and define $\H_\alpha:=\U_\alpha'\wh\U_\alpha=T^{-1}\check\A_\alpha\wh\A_\alpha'$. By construction,
$\wh\K_\alpha\wh\U_\alpha=\wh\U_\alpha\wh\D_\alpha$, where $\wh\D_\alpha$ contains the leading $r_1$ eigenvalues of
$\wh\K_\alpha$. Since $M$ is fixed, \eqref{eqn:global_alpha_group_projector} gives $\norm{\wh\K_\alpha-\P_\alpha}_F=o_P(1)$. The nonzero eigenvalues of $\P_\alpha$ are all equal to one, so $\lambda_{r_1}(\wh\K_\alpha)\asymp_P1$ and $\norm{\wh\D_\alpha^{-1}}=O_P(1)$.

By Lemma~\ref{lemma:rotation_orthogonality}, $\wh\U_\alpha-\U_\alpha\H_\alpha
=\P_\alpha^\perp\wh\U_\alpha$. Left-multiplying the preceding eigenequation by $\P_\alpha^\perp$ therefore gives
\begin{align}
    (\wh\U_\alpha-\U_\alpha\H_\alpha)\wh\D_\alpha
    =\P_\alpha^\perp(\wh\K_\alpha-\P_\alpha)\wh\U_\alpha.
    \label{eqn:global_alpha_projected_eigen}
\end{align}
Equations \eqref{eqn:global_alpha_group_projector} and \eqref{eqn:global_alpha_projected_eigen} imply $\norm{\wh\U_\alpha-\U_\alpha\H_\alpha}_F=o_P(1)$. Hence Lemma~\ref{lemma:rotation_orthogonality} shows that $\H_\alpha$ is asymptotically orthogonal.

For each $t\in[T]$, left-multiplying \eqref{eqn:global_alpha_projected_eigen} by $\e_t'$ and using \eqref{eqn:global_alpha_group_projector_row} gives
\begin{align*}
    \norm{\wh\balpha_t-\H_\alpha'\check\balpha_t}^2
    =O_P\Bigg(\max_{m\in[M]}\Big\{
    &p_m^{-\gamma_r}q_m^{-1}
    +R(T,d_{r,\star})
    p_m^{\gamma_{r,\max}^{(m)}-\gamma_r}
    \max_{n\in[M]\setminus\{m\}}p_n^{\gamma_r^{(n)}-\gamma_r}\\
    &+T^{-1}p_m^{1-\gamma_{r,\min}^{(m)}}q_m^{-1}
    \max_{n\in[M]}p_n^{\gamma_{r,\max}^{(n)}-\gamma_r}
    \Big\}\Bigg).
\end{align*}
Moreover, $\H_\alpha'\H_\alpha=\I-(\wh\U_\alpha-\U_\alpha\H_\alpha)'(\wh\U_\alpha-\U_\alpha\H_\alpha)$. Since $\H_\alpha^{-1}=(\H_\alpha'\H_\alpha)^{-1}\H_\alpha'$ and $\H_\alpha$ is asymptotically orthogonal, $\norm{\H_\alpha^{-1}-\H_\alpha'}=O_P(\norm{\wh\U_\alpha-\U_\alpha\H_\alpha}_F^2)$. Together with $\norm{\check\balpha_t}=O_P(1)$, replacing $\H_\alpha'$ by $\H_\alpha^{-1}$ contributes only the square of the preceding bound and is therefore negligible under (R1). Therefore, 
\begin{align*}
    \norm{\wh\balpha_t-\H_\alpha^{-1}\check\balpha_t}^2
    =O_P\Bigg(\max_{m\in[M]}\Big\{
    &p_m^{-\gamma_r}q_m^{-1}
    +R(T,d_{r,\star})
    p_m^{\gamma_{r,\max}^{(m)}-\gamma_r}
    \max_{n\in[M]\setminus\{m\}}p_n^{\gamma_r^{(n)}-\gamma_r}\\
    &+T^{-1}p_m^{1-\gamma_{r,\min}^{(m)}}q_m^{-1}
    \max_{n\in[M]}p_n^{\gamma_{r,\max}^{(n)}-\gamma_r}
    \Big\}\Bigg),
\end{align*}
which proves the stated rate for the global row main effect series.

We next consider the refitted global loading $\wt\A_{1m}$. Let $\wh\Y_m:=(\wh\y_1^{(m)},\ldots,\wh\y_T^{(m)})$ and $\Y_m:=\A_{1m}\A_\alpha$. By \eqref{eqn:A1mtilde} and the definitions of $\check{\check\A}_{1m}$ and $\H_\alpha$,
\begin{align}
    \wt\A_{1m}-\check{\check\A}_{1m}\H_\alpha
    =T^{-1}(\wh\Y_m-\Y_m)\wh\A_\alpha'.
    \label{eqn:final_A1_exact}
\end{align}

Define $\bPi_{2\mid1,m}:=\check\A_{2m}(\check\A_{2m}'\P_{A_1,m}^\perp\check\A_{2m})^{-1}\check\A_{2m}'\P_{A_1,m}^\perp$, and define $\wh\bPi_{2\mid1,m}$ analogously using $\wh\A_{2m}$ and $\wh\P_{A_1,m}^\perp$. By (S1) and Theorem~\ref{thm:A2m_alphatm}, $\norm{\wh\bPi_{2\mid1,m}}=O_P(1)$. Since $\wh\A_{2m}\wh\balpha_t^{(m)}=\wh\bPi_{2\mid1,m}\wh\a_t^{(m)}$ and $\A_{2m}\balpha_t^{(m)}=\bPi_{2\mid1,m}\a_t^{(m)}$, we have
\begin{align*}
    \wh\y_t^{(m)}-\A_{1m}\balpha_t
    =q_m^{-1}(\I_{p_m}-\wh\bPi_{2\mid1,m})\M_{p_m}\E_t^{(m)}\1_{q_m}
    -(\wh\bPi_{2\mid1,m}-\bPi_{2\mid1,m})\a_t^{(m)}.
\end{align*}
By (IC2), $T^{-1}\sum_t\a_t^{(m)}\check\balpha_t'=\A_{1m}\bGamma_{\alpha,T}^{1/2}=\check{\check\A}_{1m}$ and $\bPi_{2\mid1,m}\check{\check\A}_{1m}=\0$.
Consequently,
\begin{align*}
    T^{-1}(\wh\Y_m-\Y_m)\check\A_\alpha'
    =(\I_{p_m}-\wh\bPi_{2\mid1,m})
    \Big(T^{-1}\sum_{t=1}^Tq_m^{-1}\M_{p_m}\E_t^{(m)}\1_{q_m}\check\balpha_t'\Big)
    -\wh\bPi_{2\mid1,m}\check{\check\A}_{1m}.
\end{align*}
Since $T^{-1}\sum_tq_m^{-1}\M_{p_m}\E_t^{(m)}\1_{q_m}\check\balpha_t'=q_m^{-1}\M_{p_m}(T^{-1}\sum_t\E_t^{(m)}\1_{q_m}\balpha_t')\bGamma_{\alpha,T}^{-1/2}$, \eqref{eqn:E1u} in Lemma~\ref{lemma:bound_Delta} gives its squared Frobenius norm as $O_P(T^{-1}p_mq_m^{-1})$. Also $\P_{A_1,m}^\perp\check{\check\A}_{1m}=\0$, and hence
\begin{align*}
    \norm{\wh\bPi_{2\mid1,m}\check{\check\A}_{1m}}_F^2
    \lesssim_P\norm{(\wh\P_{A_1,m}-\P_{A_1,m})\check{\check\A}_{1m}}_F^2
    =O_P\Big(p_m^{\gamma_r}\norm{\wh\A_{1m}-\check\A_{1m}\H_{A_1,m}}_F^2\Big).
\end{align*}
Substituting Theorem~\ref{thm:initial_A1m} gives
\begin{align}
    \norm{T^{-1}(\wh\Y_m-\Y_m)\check\A_\alpha'}_F^2
    =O_P\Big(&R(T,d_{r,\star})p_m^{\gamma_r^{(m)}}
    \max_{n\in[M]\setminus\{m\}}p_n^{\gamma_r^{(n)}-\gamma_r} \nonumber \\
    &+T^{-1}p_mq_m^{-1}\max_{n\in[M]\setminus\{m\}}p_n^{\gamma_{r,\max}^{(n)}-\gamma_r}\Big).
    \label{eqn:final_A1_cross_moment}
\end{align}

Furthermore, \eqref{eqn:global_alpha_local_reconstruction} and $T^{-1}\sum_t
\norm{q_m^{-1}\M_{p_m}\E_t^{(m)}\1_{q_m}}^2=O_P(p_mq_m^{-1})$ give
\begin{align}
    T^{-1}\norm{\wh\Y_m-\Y_m}_F^2
    =O_P\Big(p_mq_m^{-1}
    +R(T,d_{r,\star})p_m^{\gamma_{r,\max}^{(m)}}
    \max_{n\in[M]\setminus\{m\}}p_n^{\gamma_r^{(n)}-\gamma_r}\Big),
    \label{eqn:final_A1_residual_energy}
\end{align}
where the remaining term in \eqref{eqn:global_alpha_local_reconstruction} is absorbed into $p_mq_m^{-1}$ under (R1).

Finally, since $\P_\alpha\wh\A_\alpha'=\check\A_\alpha'\H_\alpha$, \eqref{eqn:final_A1_exact} gives
\begin{align*}
    \wt\A_{1m}-\check{\check\A}_{1m}\H_\alpha
    =\big(T^{-1}(\wh\Y_m-\Y_m)\check\A_\alpha'\big)\H_\alpha
    +T^{-1}(\wh\Y_m-\Y_m)\P_\alpha^\perp\wh\A_\alpha'.
\end{align*}
The first term is controlled by \eqref{eqn:final_A1_cross_moment}. For the second term, since $\P_\alpha^\perp\wh\A_\alpha'=\sqrt{T}\P_\alpha^\perp\wh\U_\alpha$,
\begin{align*}
    \norm{T^{-1}(\wh\Y_m-\Y_m)\P_\alpha^\perp\wh\A_\alpha'}_F^2
    \le\big\{T^{-1}\norm{\wh\Y_m-\Y_m}_F^2\big\}
    \norm{\P_\alpha^\perp\wh\U_\alpha}_F^2.
\end{align*}
By \eqref{eqn:global_alpha_group_projector} and \eqref{eqn:global_alpha_projected_eigen}, $\norm{\P_\alpha^\perp\wh\U_\alpha}_F^2$ satisfies the same bound obtained above for the global row main effect series. Combining this with \eqref{eqn:final_A1_cross_moment} and \eqref{eqn:final_A1_residual_energy}, and dividing the resulting bound by $p_m^{\gamma_r}$, gives 
\begin{align*}
    p_m^{-\gamma_r}
    \norm{\wt\A_{1m}-\check{\check\A}_{1m}\H_\alpha}_F^2
    =O_P\Big(\eta_{A_1,m}
    +\Big\{\frac{p_m^{1-\gamma_r}}{q_m}
    +R(T,d_{r,\star})(1\vee\nu_{r,m})
    \max_{n\in[M]\setminus\{m\}}\nu_{r,n}\Big\}\eta_\alpha\Big),
\end{align*}
which proves Theorem~\ref{thm:A1m_alphat}. The column-side result follows by symmetry.

Expanding $\eta_{A_1,m}$ and $\eta_\alpha$ gives the equivalent explicit rate:
\begin{align*}
    p_m^{-\gamma_r}\norm{\wt\A_{1m}&-\check{\check\A}_{1m}\H_\alpha}_F^2
    =O_P\Big(R(T,d_{r,\star})p_m^{\gamma_r^{(m)}-\gamma_r}
    \max_{n\in[M]\setminus\{m\}}p_n^{\gamma_r^{(n)}-\gamma_r} \\
    &+T^{-1}p_m^{1-\gamma_r}q_m^{-1}\max_{n\in[M]\setminus\{m\}}p_n^{\gamma_{r,\max}^{(n)}-\gamma_r} 
    +\Big\{p_m^{1-\gamma_r}q_m^{-1}
    +R(T,d_{r,\star})p_m^{\gamma_{r,\max}^{(m)}-\gamma_r}
    \max_{n\in[M]\setminus\{m\}}p_n^{\gamma_r^{(n)}-\gamma_r}\Big\} \\
    &\cdot\max_{n\in[M]}\Big\{
    p_n^{-\gamma_r}q_n^{-1} 
    +R(T,d_{r,\star})p_n^{\gamma_{r,\max}^{(n)}-\gamma_r}
    \max_{s\in[M]\setminus\{n\}}p_s^{\gamma_r^{(s)}-\gamma_r}
    +T^{-1}p_n^{1-\gamma_{r,\min}^{(n)}}q_n^{-1}
    \max_{s\in[M]}p_s^{\gamma_{r,\max}^{(s)}-\gamma_r}
    \Big\}\Big), \\
    q_m^{-\gamma_c}\norm{\wt\B_{1m}&-\check{\check\B}_{1m}\H_\beta}_F^2
    =O_P\Big(R(T,d_{c,\star})q_m^{\gamma_c^{(m)}-\gamma_c}
    \max_{n\in[M]\setminus\{m\}}q_n^{\gamma_c^{(n)}-\gamma_c} \\ 
    &+T^{-1}q_m^{1-\gamma_c}p_m^{-1}
    \max_{n\in[M]\setminus\{m\}}q_n^{\gamma_{c,\max}^{(n)}-\gamma_c}
    +\Big\{q_m^{1-\gamma_c}p_m^{-1}
    +R(T,d_{c,\star})q_m^{\gamma_{c,\max}^{(m)}-\gamma_c}
    \max_{n\in[M]\setminus\{m\}}q_n^{\gamma_c^{(n)}-\gamma_c}\Big\} \\
    &\cdot\max_{n\in[M]}\Big\{
    q_n^{-\gamma_c}p_n^{-1}
    +R(T,d_{c,\star})q_n^{\gamma_{c,\max}^{(n)}-\gamma_c}
    \max_{s\in[M]\setminus\{n\}}q_s^{\gamma_c^{(s)}-\gamma_c}
    +T^{-1}q_n^{1-\gamma_{c,\min}^{(n)}}p_n^{-1}
    \max_{s\in[M]}q_s^{\gamma_{c,\max}^{(s)}-\gamma_c}
    \Big\}\Big),
\end{align*}
where the column-side results follow by symmetry. This completes the proof. $\square$

\textbf{\textit{Proof of Corollary~\ref{corollary:simplified_main_effect_rates}.}}
The result follows by substituting $\gamma_r=\gamma_c=\gamma_r^{(m)}=\gamma_c^{(m)}=1$ and $R(T,d_{r,\star})=R(T,d_{c,\star})=T^{-1}$ into Theorems~\ref{thm:initial_A1m}--\ref{thm:A1m_alphat}. $\square$

\newpage

\textbf{\textit{Proof of Theorem~\ref{thm:common_comp_global_local_init}.}}
For the local factor loading estimators, it suffices to show the results for $\wh\C_m$, and analogous arguments follow for $\wh\R_m$.
To this end, we adapt the decomposition and arguments used in the proof of Lemma~\ref{lemma: common_comp_global_local_init}, with the roles of rows and columns interchanged appropriately. 
Recall from \eqref{eqn: check_Yt_m_simplify} that $\check\Y_t^{(m)} = 
\Q_m\G_t\J_m' + \R_m\F_t^{(m)}\C_m' + \M_{p_m}\E_t^{(m)}\M_{q_m}$.
Hence
\begin{align}
    \check\Y_t^{(m)\prime} \wh\Q_m^{\perp}
    &=
    \J_m\G_t'\Q_m'\wh\Q_m^{\perp}
    + \C_m\F_t^{(m)\prime}\R_m'\wh\Q_m^{\perp}
    + \M_{q_m} \E_t^{(m)\prime} \M_{p_m} \wh\Q_m^{\perp}
    \notag \\
    &=:
    \cA_t + \cB_t + \cC_t .
    \label{eqn: local_C_decomp_components}
\end{align}
With \eqref{eqn: local_C_decomp_components}, we may expand $\X_{C,m}$ as
\begin{align}
    \X_{C,m} &= \frac{1}{T} \sum_{t=1}^{T} \big( \check\Y_t^{(m)\prime} \wh\Q_m^{\perp} \big) \big( \check\Y_t^{(m)\prime} \wh\Q_m^{\perp} \big)'  \notag\\
    &=
    \frac{1}{T}\sum_{t=1}^{T} \cB_t \cB_t'
    + \frac{1}{T}\sum_{t=1}^{T} \cB_t \cC_t' 
    + \frac{1}{T}\sum_{t=1}^{T} \cB_t \cA_t' 
    + \frac{1}{T}\sum_{t=1}^{T} \cC_t \cB_t' 
    + \frac{1}{T}\sum_{t=1}^{T} \cC_t \cC_t' \notag\\
    &\;\quad
    + \frac{1}{T}\sum_{t=1}^{T} \cC_t \cA_t' 
    + \frac{1}{T}\sum_{t=1}^{T} \cA_t \cB_t' 
    + \frac{1}{T}\sum_{t=1}^{T} \cA_t \cC_t' 
    + \frac{1}{T} \sum_{t=1}^{T} \cA_t \cA_t'  
    = :
    \sum_{i=1}^9 \cJ_i . \label{eqn:XCm-decomposition}
\end{align}

For convenience, define the matrices $\P_{Q,m}^{\perp} := \Q_m^{\perp} \Q_m^{\perp \prime}$ and 
$\wh{\P}_{Q,m}^{\perp} := \wh{\Q}_m^{\perp} \wh{\Q}_m^{\perp \prime}$.
Then we can write $\cB_t\cB_t' = \C_m\F_t^{(m) \prime}\R_m' \wh{\P}_{Q,m}^{\perp} \R_m\F_t^{(m)}\C_m'$.
Further define
\begin{align}
    \cB_t^0 := \C_m \F_t^{(m) \prime} \R_m' \Q_m^{\perp},
    \quad
    \cJ_1^0 := \frac{1}{T}\sum_{t=1}^{T} \cB_t^0 \cB_t^{0 \prime}.
    \label{eqn: local_C_J1_pop}
\end{align}
Then by Lemmas~\ref{lemma:rotation_orthogonality},
\ref{lemma: common_comp_global_local_init}, 
and following Lemma~S5 in \cite{Cen2026_multiorder},
we have
\begin{align}
    &\;\quad
    \Big\| \wh{\P}_{Q,m}^{\perp}  - \P_{Q,m}^{\perp} \Big\|_F^2
    = 
    \Big\| \wh{\P}_{Q,m} - \P_{Q,m} \Big\|_F^2 \notag \\
    &=
    O_P\left\{
    R(T,d_{\text{fac},\star}) \;
    p_m^{2 (\delta_{q,1}^{(m)} -\delta_{q,k_r}^{(m)})}
    \left( \frac{  \max_{n\in[M] \setminus \{m\}} s_{J,n}^2 p_n^{\delta_{q,1}^{(n)}}}{\max_{n\in[M] \setminus\{m\}} s_{J,n}^2 p_n^{\delta_{q,k_r}^{(n)}}} \right)^2
    \right\} .
    \label{eqn: local_C_projection_perturbation}
\end{align}

Consider $\cJ_1$.
Note that $\cJ_1 - \cJ_1^0 = T^{-1} \sum_{t=1}^{T} \C_m \F_t^{(m)\prime} \R_m' \big( \wh{\P}_{Q,m}^{\perp} - \P_{Q,m}^{\perp} \big) \R_m \F_t^{(m)} \C_m'$,
we have
\begin{align}
    \big\| \cJ_1 - \cJ_1^0 \big\|_F^2
    &\leq
    \frac{1}{T} \sum_{t=1}^{T} \Big\| \C_m \F_t^{(m)\prime} \R_m' \big( \wh{\P}_{Q,m}^{\perp}  - \P_{Q,m}^{\perp} \big)  \R_m \F_t^{(m)} \C_m' \Big\|_F^2 \notag \\
    &\lesssim
    \big\| \C_m \big\|^4 
    \big\| \R_m \big\|^4
    \Big\| \wh{\P}_{Q,m}^{\perp} - \P_{Q,m}^{\perp} \Big\|_F^2
    \cdot \frac{1}{T} \sum_{t=1}^{T} \big\| \F_t^{(m)} \big\|_F^4 ,
    \label{eqn: local_C_J1_perturbation}
\end{align}
where the last inequality used the sub-multiplicativity and the fact that the Frobenius norm of a product is bounded by products of spectral and Frobenius norms. 
By Assumption~(F1), $\F_t^{(m)}$ has uniformly bounded fourth moments, so $T^{-1}\sum_{t=1}^{T}\|\F_t^{(m)}\|_F^4=O_P(1)$. 
Combining \eqref{eqn: local_C_J1_perturbation},
\eqref{eqn: local_C_projection_perturbation},
and Assumption~(L1), we thus have
\begin{align}
    \big\| \cJ_1 - \cJ_1^0 \big\|_F^2
    \lesssim_P
    q_m^{2 \delta_{c,1}^{(m)}}
    p_m^{2 \delta_{r,1}^{(m)}}
    \left\{
    R(T,d_{\text{fac},\star}) \;
    p_m^{2 (\delta_{q,1}^{(m)} -\delta_{q,k_r}^{(m)})}
    \left( \frac{  \max_{n\in[M] \setminus \{m\}} s_{J,n}^2 p_n^{\delta_{q,1}^{(n)}}}{\max_{n\in[M] \setminus\{m\}} s_{J,n}^2 p_n^{\delta_{q,k_r}^{(n)}}} \right)^2
    \right\} .
    \label{eqn: local_C_J1_perturbation_rate}
\end{align}

Next, by Assumptions~(L1) and (S1), the nonzero eigenvalues of $\cJ_1^0$ satisfy that, for any $i\in [k_c^{(m)}]$,
\begin{align}
    \lambda_i\big( \cJ_1^0 \big)
    \asymp
    q_m^{\delta_{c,i}^{(m)}}
    p_m^{\delta_{r,1}^{(m)}} ,
    \label{eqn: local_C_J1_eigenvalues}
\end{align}
which follows analogous arguments as for \eqref{eqn: Omega_Q_m_krth_eigenvalue_lower} in the proof of Lemma~\ref{lemma: common_comp_global_local_init}.

It remains to bound $\cJ_i$, $i=2,\dots, 9$.
Note that these terms all involve at least one of $\cA_t$, $\cC_t$, or the projection perturbation (cf.\ \eqref{eqn: local_C_projection_perturbation}). 
We bound their squared Frobenius norms separately.
To this end, for $\cJ_2$ and $\cJ_4$, 
by Assumptions~(E2), (F1), and the same calculation as in \eqref{eqn: cIII_Q_mn}, we have
\begin{align}
    \big\| \cJ_2 \big\|_F^2, \;
    \big\| \cJ_4 \big\|_F^2
    &=
    \Bigg\| \frac{1}{T} \sum_{t=1}^{T} \cB_t \cC_t' \Bigg\|_F^2
    \lesssim_P
    \frac{1}{T} q_m^{\delta_{c,1}^{(m)}} p_m^{\delta_{r,1}^{(m)}} q_m p_m
    =
    \frac{1}{T}  q_m^{1 + \delta_{c,1}^{(m)}} p_m^{1 + \delta_{r,1}^{(m)}} .
    \label{eqn: local_C_J2}
\end{align}

For $\cJ_5$, using Lemma~\ref{lemma:bound_Delta} and the proof of \eqref{eqn:Em11Em}, it holds that
\begin{align}
    \big\| \cJ_5 \big\|_F^2
    \lesssim_P
    q_m^{2 - 2 \delta_{c,1}^{(m)}}
    + \frac{1}{T} q_m^2 p_m .
    \label{eqn: local_C_J5}
\end{align}

For $\cJ_6$ and $\cJ_8$, it suffices to consider the former and the latter follows in similar arguments.
Recall that $\cA_t = \J_m\G_t'\Q_m' \wh\Q_m^{\perp}$ from \eqref{eqn: local_C_decomp_components}.
Note that
\begin{align}
    \big\| \Q_m' \wh{\Q}_m^{\perp} \big\|_F
    &=
    \big\| \Q_m'
    (\wh{\Q}_m^{\perp} \wh{\Q}_m^{\perp \prime} - \Q_m^{\perp} \Q_m^{\perp \prime})
    \wh{\Q}_m^{\perp}
    \big\| \notag \\
    &\leq
    \big\| \Q_m \big\| \cdot
    \big\| \wh{\P}_{Q,m}^{\perp} - \P_{Q,m}^{\perp} \big\|_F
    \cdot \big\| \wh{\Q}_m^{\perp} \big\| 
    =
    \big\| \Q_m \big\| 
    \cdot \big\| \wh{\P}_{Q,m}^{\perp} - \P_{Q,m}^{\perp} \big\|_F ,
    \label{eqn: local_C_Q_perp_bound}
\end{align}
Also, with $f(\X) = \|\X\|_F^2$ being convex, we have $\big\| T^{-1} \sum_{t=1}^{T} \X_t \big\|_F^2 = f\big( T^{-1}\sum_{t=1}^{T} \X_t \big) \leq T^{-1} \sum_{t=1}^{T} f(\X_t) = T^{-1} \sum_{t=1}^{T} \|\X_t\|_F^2$.
Altogether, by Assumption~(F1), and the sub-multiplicativity of the Frobenius norm, we arrive at
\begin{align}
    \big\| \cJ_6 \big\|_F^2 &=
    \Bigg\| \frac{1}{T} \sum_{t=1}^{T} \cC_t \cA_t' \Bigg\|_F^2
    \leq
    \frac{1}{T} \sum_{t=1}^{T}
    \big\| \cC_t \big\|_F^2
    \big\| \cA_t \big\|_F^2 \notag \\
    &\leq
    \frac{1}{T} \sum_{t=1}^{T}
    \big\| \cC_t \big\|_F^2 \cdot
    \big\| \J_m \big\|^2 \cdot
    \big\| \G_t \big\|_F^2 \cdot
    \big\| \Q_m \big\|_F^2 \cdot
    \big\| \wh{\P}_{Q,m}^{\perp} - \P_{Q,m}^{\perp} \big\|_F^2  \notag \\
    &\lesssim_P
    q_m^{1 + \delta_{j,1}^{(m)}}
    p_m^{1 + \delta_{q,1}^{(m)}}
    \left\{
    R(T,d_{\text{fac},\star}) \;
    p_m^{2 (\delta_{q,1}^{(m)} -\delta_{q,k_r}^{(m)})}
    \left( \frac{  \max_{n\in[M] \setminus \{m\}} s_{J,n}^2 p_n^{\delta_{q,1}^{(n)}}}{\max_{n\in[M] \setminus\{m\}} s_{J,n}^2 p_n^{\delta_{q,k_r}^{(n)}}} \right)^2
    \right\} ,
    \label{eqn: local_C_J6}
\end{align}
where the second inequality used
\eqref{eqn: local_C_projection_perturbation},
\eqref{eqn: local_C_Q_perp_bound},
and $\| \cC_t \|_F^2 = O_P(p_m q_m)$ which follows from similar arguments for \eqref{eqn: unit_vector_error_bound}.
Next, $\cJ_3$ and $\cJ_7$ can be bounded analogously by the above arguments, with $\cC_t$ replaced by $\cB_t$ where $\| \cB_t \|_F^2 = O_P(q_m^{\delta_{c,1}^{(m)}} p_m^{\delta_{r,1}^{(m)}})$, thus giving
\begin{align}
    \big\| \cJ_3 \big\|_F^2, \;
    \big\| \cJ_7 \big\|_F^2
    &\lesssim_P
    q_m^{\delta_{c,1}^{(m)} + \delta_{j,1}^{(m)}}
    p_m^{\delta_{r,1}^{(m)} + \delta_{q,1}^{(m)}}
    \left\{
    R(T,d_{\text{fac},\star}) \;
    p_m^{2 (\delta_{q,1}^{(m)} -\delta_{q,k_r}^{(m)})}
    \left( \frac{  \max_{n\in[M] \setminus \{m\}} s_{J,n}^2 p_n^{\delta_{q,1}^{(n)}}}{\max_{n\in[M] \setminus\{m\}} s_{J,n}^2 p_n^{\delta_{q,k_r}^{(n)}}} \right)^2
    \right\} .
    \label{eqn: local_C_J3}
\end{align}

For $\cJ_9$, using Assumption~(F1),
the bound of $\| \cA_t \|_F^2$ as in \eqref{eqn: local_C_Q_perp_bound} and \eqref{eqn: local_C_J6}, we have
\begin{align}
    \big\| \cJ_9 \big\|_F^2
    &\leq
    \frac{1}{T} \sum_{t=1}^T \big\| \cA_t \big\|_F^4
    &\lesssim_P
    q_m^{2 \delta_{j,1}^{(m)}}
    p_m^{2 \delta_{q,1}^{(m)}}
    \left\{
    R(T,d_{\text{fac},\star}) \;
    p_m^{2 (\delta_{q,1}^{(m)} -\delta_{q,k_r}^{(m)})}
    \left( \frac{  \max_{n\in[M] \setminus \{m\}} s_{J,n}^2 p_n^{\delta_{q,1}^{(n)}}}{\max_{n\in[M] \setminus\{m\}} s_{J,n}^2 p_n^{\delta_{q,k_r}^{(n)}}} \right)^2
    \right\}^2 .
    \label{eqn: local_C_J9}
\end{align}

Finally, recall \eqref{eqn: local_C_J1_pop}.
Then combining \eqref{eqn: local_C_J1_perturbation_rate},
\eqref{eqn: local_C_J2},
\eqref{eqn: local_C_J5},
\eqref{eqn: local_C_J6},
\eqref{eqn: local_C_J3},
and \eqref{eqn: local_C_J9}
give the rate on $\big\| \X_{C,m} - \cJ_1^0 \big\|_F^2$,
which, together with \eqref{eqn: local_C_J1_eigenvalues} and applying the Davis--Kahan theorem \citep[Theorem~2]{yu2015useful}, complete the proof for $\big\| \wh\C_m - \check{\C}_m \H_{C,m} \big\|_F^2$.

It remains to show the results for the local factor estimators.
To this end, define
\begin{align*}
    \cL_{R,m}
    &:=
    \big( \check{\R}_m' \P_{Q,m}^{\perp} \check{\R}_m \big)^{-1} \check{\R}_m' \P_{Q,m}^{\perp}
    ,  \quad
    \wh{\cL}_{R,m}
    := \big( \wh{\R}_m' \wh{\P}_{Q,m}^{\perp} \wh{\R}_m \big)^{-1} \wh{\R}_m' \wh{\P}_{Q,m}^{\perp} .
\end{align*}
By Assumptions~(L1) and (S1), the matrix $\check{\R}_m' \P_{Q,m}^{\perp} \check{\R}_m$ is positive definite with eigenvalues bounded away from zero, and it holds that
\begin{align}
    \big\| \cL_{R,m} \big\|, \;
    \big\| \wh{\cL}_{R,m} \big\| =
    O_P\Big(
    p_m^{-\delta_{r,1}^{(m)}/2} \Big),
    \quad
    \cL_{R,m} \check{\R}_m = \I,
    \quad
    \cL_{R,m} \Q_m = \0 .
    \label{eqn: local_F_L_properties}
\end{align}
Then by the definition of $\wh\F_t^{(m)}$ in Step~3 of Section~\ref{subsec:estimation} and \eqref{eqn: check_Yt_m_simplify}, we may decompose
\begin{align}
    \wh\F_t^{(m)} &= \wh{\cL}_{R,m} \check{\Y}_t^{(m)} \wh{\C}_m \notag \\
    &= 
    \wh{\cL}_{R,m} \Q_m \G_t \J_m' \wh{\C}_m
    + \wh{\cL}_{R,m} \check{\R}_m \check{\F}_t^{(m)} \check{\C}_m' \wh{\C}_m
    + \wh{\cL}_{R,m} \M_{p_m} \E_t^{(m)} \M_{q_m} \wh{\C}_m 
    =:
    \cK_1 + \cK_2 + \cK_3 .
    \label{eqn: local_F_decomp}
\end{align}

For $\cK_1$, by \eqref{eqn: local_F_L_properties}, \eqref{eqn: local_C_Q_perp_bound}, and Assumption~(F1), we have
\begin{align}
    \big\| \cK_1 \big\|_F^2
    &\leq
    \big\| \wh{\cL}_{R,m} \big\|^2 \big\| \Q_m'\wh{\P}_{Q,m}^{\perp} \big\|_F^2 \big\| \G_t \big\|_F^2 \big\| \J_m \big\|^2 \big\| \wh{\C}_m \big\|^2 \notag \\
    &\lesssim_P
    p_m^{\delta_{q,1}^{(m)} -\delta_{r,1}^{(m)}}
    q_m^{\delta_{j,1}^{(m)}}
    \eta_{Q,m}
    \big\| \G_t \big\|_F^2 
    \lesssim_P
    p_m^{\delta_{q,1}^{(m)} -\delta_{r,1}^{(m)}}
    q_m^{\delta_{j,1}^{(m)}}
    \eta_{Q,m} ,
    \label{eqn: local_F_K1_rate}
\end{align}
with $\eta_{Q,m}$ defined in the statement of Theorem~\ref{thm:common_comp_global_local_init}.

Consider $\cK_2$ next.
Note that we can write
\begin{align}
    \cK_2 &=
    \big( \H_{R,m}^{-1} + \wh{\cL}_{R,m} \check{\R}_m - \H_{R,m}^{-1} \big)
    \check{\F}_t^{(m)}
    \big( \H_{C,m} + \check{\C}_m' \wh{\C}_m - \H_{C,m} \big) \notag \\
    &=
    \H_{R,m}^{-1} \check{\F}_t^{(m)} \H_{C,m}
    + (\wh{\cL}_{R,m} \check{\R}_m - \H_{R,m}^{-1}) \check{\F}_t^{(m)} \H_{C,m} \notag \\
    &\;\quad
    + \H_{R,m}^{-1} \check{\F}_t^{(m)} (\check{\C}_m' \wh{\C}_m - \H_{C,m})
    + (\wh{\cL}_{R,m} \check{\R}_m - \H_{R,m}^{-1}) \check{\F}_t^{(m)} (\check{\C}_m' \wh{\C}_m - \H_{C,m}) .
    \label{eqn: local_F_K2_expansion}
\end{align}
First, by the rates of convergence of the local loading matrices, we have
\begin{align}
    &\;\quad
    \big\| \wh{\cL}_{R,m} \check{\R}_m - \H_{R,m}^{-1} \big\|_F^2 
    =
    \big\| \big( \wh{\R}_m' \wh{\P}_{Q,m}^{\perp} \wh{\R}_m \big)^{-1} \wh{\R}_m' \wh{\P}_{Q,m}^{\perp} \check{\R}_m
    - \H_{R,m}^{-1} \big\|_F^2 \notag \\
    &=
    \big\| \big( \wh{\R}_m' \wh{\P}_{Q,m}^{\perp} \wh{\R}_m \big)^{-1} \wh{\R}_m' \wh{\P}_{Q,m}^{\perp} \check{\R}_m \H_{R,m}
    - \I \big\|_F^2 \notag \\
    &=
    \big\| \big( \wh{\R}_m' \wh{\P}_{Q,m}^{\perp} \wh{\R}_m \big)^{-1} \wh{\R}_m' \wh{\P}_{Q,m}^{\perp} (\check{\R}_m \H_{R,m} -\wh{\R}_m)
    + \big( \wh{\R}_m' \wh{\P}_{Q,m}^{\perp} \wh{\R}_m \big)^{-1} \wh{\R}_m' \wh{\P}_{Q,m}^{\perp} \wh{\R}_m
    - \I \big\|_F^2 \notag \\
    &\leq
    \big\| \big( \wh{\R}_m' \wh{\P}_{Q,m}^{\perp} \wh{\R}_m \big)^{-1} \wh{\R}_m' \wh{\P}_{Q,m}^{\perp} \big\|^2
    \cdot \big\| \check{\R}_m \H_{R,m} -\wh{\R}_m \big\|_F^2 
    \lesssim_P
    \Delta_{R,m} ,
    \label{eqn: local_F_E_rate_row}
\end{align}
where $\Delta_{R,m}$ is defined in the statement of Theorem~\ref{thm:common_comp_global_local_init}.
Similarly, we have
\begin{align}
    &\;\quad
    \big\| \check{\C}_m' \wh{\C}_m - \H_{C,m} \big\|_F^2 
    = 
    \big\| \check{\C}_m' (\wh{\C}_m - \check{\C}_m \H_{C,m}) + \check{\C}_m'\check{\C}_m \H_{C,m} - \H_{C,m} \big\|_F^2 \notag \\
    &=
    \big\| \check{\C}_m' (\wh{\C}_m - \check{\C}_m \H_{C,m}) + (\C_m'\C_m)^{-1/2} \C_m'\C_m (\C_m'\C_m)^{-1/2} \H_{C,m} - \H_{C,m} \big\|_F^2 
    \lesssim_P
    \Delta_{C,m} ,
    \label{eqn: local_F_E_rate_col} .
\end{align}
Also, recall the definition $\check{\F}_t^{(m)} = \big( \R_m'\R_m \big)^{1/2} \F_t^{(m)} \big( \C_m'\C_m \big)^{1/2}$, and it is straightforward from Assumptions~(F1) and (L1) that $\big\| \check{\F}_t^{(m)} \big\|_F^2 = O_P\big( p_m^{\delta_{r,1}^{(m)}} q_m^{\delta_{c,1}^{(m)}} \big)$,
which, combined with 
\eqref{eqn: local_F_K2_expansion},
\eqref{eqn: local_F_E_rate_row},
\eqref{eqn: local_F_E_rate_col},
give
\begin{align}
    \Big\| \cK_2 - \H_{R,m}^{-1} \check{\F}_t^{(m)} \H_{C,m} \Big\|_F^2
    &\lesssim_P
    p_m^{\delta_{r,1}^{(m)}}
    q_m^{\delta_{c,1}^{(m)}}
    \big( \Delta_{R,m} +
    \Delta_{C,m} \big) .
    \label{eqn: local_F_K2_rate}
\end{align}

Finally, for $\cK_3$, by \eqref{eqn: local_F_L_properties} and the same argument as for \eqref{eqn: unit_vector_error_bound},
\begin{align}
    &\;\quad
    \big\| \cK_3 \big\|_F^2 =
    \Big\| \wh{\cL}_{R,m}\M_{p_m} \E_t^{(m)} \M_{q_m} \wh{\C}_m \Big\|_F^2  \notag \\
    &\leq
    \big\| \wh{\cL}_{R,m} \big\|^2
    \cdot \Big\| \M_{p_m} \E_t^{(m)} \M_{q_m} 
    \big\{ \check{\C}_m \H_{C,m} +
    (\wh{\C}_m - \check{\C}_m \H_{C,m})
    \big\} 
    \Big\|_F^2 \notag \\
    &\lesssim
    \big\| \wh{\cL}_{R,m} \big\|^2
    \cdot \Big\| \M_{p_m} \E_t^{(m)} \M_{q_m} \check{\C}_m \H_{C,m} \Big\|_F^2 
    + \big\| \wh{\cL}_{R,m} \big\|^2
    \cdot \Big\| \M_{p_m} \E_t^{(m)}
    \M_{q_m} (\wh{\C}_m - \check{\C}_m \H_{C,m}) \Big\|_F^2 \notag \\
    &\lesssim_P
    p_m^{1 -\delta_{r,1}^{(m)}} + p_m^{1 -\delta_{r,1}^{(m)}} q_m \Delta_{C,m} .
    \label{eqn: local_F_K3}
\end{align}

By combining 
\eqref{eqn: local_F_decomp},
\eqref{eqn: local_F_K1_rate},
\eqref{eqn: local_F_K2_rate}, 
and \eqref{eqn: local_F_K3},
we conclude that
\begin{align*}
    \frac{1}{
    p_m^{\delta_{r,1}^{(m)}} 
    q_m^{\delta_{c,1}^{(m)}}
    }
    \Big\| \wh\F_t^{(m)} - \H_{R,m}^{-1} \check{\F}_t^{(m)} \H_{C,m} \Big\|_F^2 
    &\lesssim_P
    \Delta_{R,m}
    +
    \Delta_{C,m}
    +
    p_m^{1 - 2\delta_{r,1}^{(m)}}
    q_m^{1 - \delta_{c,1}^{(m)}}
    \Delta_{C,m}
    +
    p_m^{1 - 2\delta_{r,1}^{(m)}}
    q_m^{ - \delta_{c,1}^{(m)}} ,
\end{align*}
where the rate in \eqref{eqn: local_F_K1_rate}, after scaled by $p_m^{\delta_{r,1}^{(m)}} q_m^{\delta_{c,1}^{(m)}}$, is omitted as it is dominated by $\Delta_{C,m}$, particularly:
\[
q_m^{-2 \delta_{c,k_c^{(m)}}^{(m)}}
    p_m^{-2 \delta_{r,1}^{(m)}} 
    \big(q_m^{2 \delta_{c,1}^{(m)}} p_m^{2 \delta_{r,1}^{(m)}} 
    + q_m^{1 + \delta_{j,1}^{(m)}} p_m^{1 + \delta_{q,1}^{(m)}} \big) \eta_{Q,m} .
\]
Noting that all these arguments hold if we alternatively work on $\wh\F_t^{(m) \prime}$, with the resulting rates of convergence symmetric (i.e., with row and column dimensions swapped),
and taking the minimum of these two completes the proof of this theorem.
$\square$

\textbf{\textit{Proof of Theorem~\ref{thm:common_comp_global}.}}
Recall $\check\G_{t,m}$ from Lemma~\ref{lemma: aligned_projected_cores} and we can write
\begin{align*}
    &\;\quad
    \check\G_{t,m} 
    = \check\Q_m' \Q_m \G_t\J_m' \check\J_m  \\
    &=  
    (\Q_m'\Q_m)^{-1/2} \Q_m'\Q_m \G_t \J_m'\J_m (\J_m'\J_m)^{-1/2}
    = 
    (\Q_m'\Q_m)^{1/2} \G_t  (\J_m'\J_m)^{1/2} ,
\end{align*}
so that for the reference group $m_0$, we have $\check\G_{t,m_0} = \check\G_t$, 
where $\check\G_t = \big( \Q_{m_0}'\Q_{m_0} \big)^{1/2} \G_t \big( \J_{m_0}'\J_{m_0} \big)^{1/2}$ as defined in the statement of Theorem~\ref{thm:common_comp_global}.

For each group $m\in[M]$ and $t\in[T]$, define the population quantities
\begin{align*}
    & \bXi_{Q,m}^* := 
     \H_{Q,m_0}'
     \big( \Q_{m_0}'\Q_{m_0} \big)^{1/2} \big( \Q_m'\Q_m \big)^{-1/2} \H_{Q,m} , \\
    & \bXi_{J,m}^* := 
    \H_{J,m_0}'
    \big( \J_{m_0}'\J_{m_0} \big)^{1/2} \big( \J_m'\J_m \big)^{-1/2} \H_{J,m} , \\
    & \G_{t,m}^* :=
    \H_{Q,m}' \big( \Q_m'\Q_m \big)^{1/2} \G_t \big( \J_m'\J_m \big)^{1/2} \H_{J,m} .
\end{align*}
First, note that for all $t\in[T]$,
\begin{align}
    \vec\big( \G_{t,m_0}^* \big) =
    \big( \bXi_{J,m}^* \otimes \bXi_{Q,m}^* \big) \vec\big( \G_{t,m}^* \big) 
    +
    o_P\Big(
    p_m^{\delta_{q,1}^{(m)}/2}
    q_m^{\delta_{j,1}^{(m)}/2}
    \Big) . 
    \label{eqn: Gt_m0_pop}
\end{align}
As $\H_{Q,m}$ and $\H_{J,m}$ are asymptotically orthogonal by Lemmas~\ref{lemma: common_comp_global_local_init} and \ref{lemma: aligned_projected_cores_rot_rate}, 
it also holds that
\begin{align*}
    \bXi_{Q,m}^* \H_{Q,m}' \big( \Q_m'\Q_m \big)^{1/2}
    &=
    \H_{Q,m_0}'
    \big( \Q_{m_0}'\Q_{m_0} \big)^{1/2}
    \big( \Q_m'\Q_m \big)^{-1/2}
    \H_{Q,m} \H_{Q,m}'
    \big( \Q_m'\Q_m \big)^{1/2}
    \\
    &=
    \H_{Q,m_0}'
    \big( \Q_{m_0}'\Q_{m_0} \big)^{1/2}
    + O_P\Big( \eta_{Q,m}^{1/2} p_{m_0}^{\delta_{q,1}^{(m_0)}/2} \Big),
    \\
    \bXi_{J,m}^* \H_{J,m}' \big( \J_m'\J_m \big)^{1/2}
    &=
    \H_{J,m_0}'
    \big( \J_{m_0}'\J_{m_0} \big)^{1/2}
    \big( \J_m'\J_m \big)^{-1/2}
    \H_{J,m} \H_{J,m}'
    \big( \J_m'\J_m \big)^{1/2} \\
    &=
    \H_{J,m_0}'
    \big( \J_{m_0}'\J_{m_0} \big)^{1/2}
    +
    O_P\Big( 
    \eta_{J,m}^{1/2}
    q_{m_0}^{\delta_{j,1}^{(m_0)}/2} \Big).
\end{align*}
so that together with $\lambda_{\min}\big( \Q_{m_0}'\Q_{m_0} \big) \lambda_{\min}\big( \J_{m_0}'\J_{m_0} \big) = O\big\{ \lambda_{\min}\big( \Q_m'\Q_m \big) \lambda_{\min}\big( \J_m'\J_m \big) \big\}$ from the statement of Theorem~\ref{thm:common_comp_global},
we arrive at
\begin{align}
    &\;\quad
    \bXi_{Q,m}^* \G_{t,m}^* \bXi_{J,m}^{*\prime}
    =
    \bXi_{Q,m}^* \H_{Q,m}' \big( \Q_m'\Q_m \big)^{1/2} \G_t \big( \J_m'\J_m \big)^{1/2}  \H_{J,m} \bXi_{J,m}^{*\prime} \notag \\
    &=
    \H_{Q,m_0}'
    \big( \Q_{m_0}'\Q_{m_0} \big)^{1/2} 
    \G_t
    \big( \J_{m_0}'\J_{m_0} \big)^{1/2} 
    \H_{J,m_0}
    + O_P\big( 
    \eta_{Q,m}^{1/2}
    \eta_{J,m}^{1/2}
    p_{m}^{\delta_{q,1}^{(m)} /2} q_{m}^{\delta_{j,1}^{(m)} /2} \big) \notag \\
    &=
    \G_{t,m_0}^* 
    + O_P\big( 
    \eta_{Q,m}^{1/2}
    \eta_{J,m}^{1/2}
    p_{m}^{\delta_{q,1}^{(m)} /2} q_{m}^{\delta_{j,1}^{(m)} /2} \big) 
    =
    \H_{Q,m_0}' \check\G_t  \H_{J,m_0}
    + O_P\big( 
    \eta_{Q,m}^{1/2}
    \eta_{J,m}^{1/2}
    p_{m}^{\delta_{q,1}^{(m)} /2} q_{m}^{\delta_{j,1}^{(m)} /2} \big) .
    \label{eqn: XiQ_G_XiJ_pop}
\end{align}
Also by Lemma~\ref{lemma: aligned_projected_cores}, we have
\begin{align}
    \wh\G_{t,m} = 
    \G_{t,m}^* + O_P(\Delta_{G,m}^{1/2}) 
    =
    \H_{Q,m}' \big( \Q_m'\Q_m \big)^{1/2} \G_t \big( \J_m'\J_m \big)^{1/2} \H_{J,m} + O_P( \Delta_{G,m}^{1/2} ) ,
    \label{eqn: wh_Gt_m}
\end{align}
where $O_P(\cdot)$ is in terms of Frobenius norm,
and $\Delta_{G,m}$ is defined in the statement of Theorem~\ref{thm:common_comp_global} such that
\[
\Delta_{G,m} = 
\min(p_m,q_m)
    + p_m^{\delta_{r,1}^{(m)}}
    q_m^{\delta_{c,1}^{(m)}}
    \big( \Delta_{R,m} + \Delta_{C,m} + \Delta_{F,m} \big)
    + \big( \eta_{Q,m} + \eta_{J,m} \big) p_m q_m .
\]
By the construction of $\wh\Z_{G,m}$ in Step~3 of Section~\ref{subsec:estimation} and with $\wh\G_{t,m}$ defined in Lemma~\ref{lemma: aligned_projected_cores},
note that
\begin{align*}
    \wh\Z_{G,m} &= 
    \left( \vec(\wh\Q_m'\check{\Y}_{G,1}^{(m)}\wh\J_m), \dots, \vec(\wh\Q_m'\check{\Y}_{G,T}^{(m)}\wh\J_m) \right)
    = 
    \left( \vec(\wh\G_{1,m}), \dots, \vec(\wh\G_{T,m}) \right) .
\end{align*}
Then we may define the population version of $\wh\Z_{G,m}$ as $\Z_{G,m}^* := \left( \vec(\G_{1,m}^*), \dots, \vec(\G_{T,m}^*) \right)$,
so that by \eqref{eqn: wh_Gt_m}, we have
\begin{align}
    \big\| \wh\Z_{G,m} - \Z_{G,m}^* \big\|_F^2
    \lesssim_P
    T \cdot \Delta_{G,m} .
    \label{eqn: Z_Gm_est_rate}
\end{align}

Next, recall the matrix $\wh\bXi_{m} = \wh\Z_{G,m_0} \wh\Z_{G,m}' (\wh\Z_{G,m} \wh\Z_{G,m}')^{-1}$ for the two-sided alignment in Step~3 of Section~\ref{subsec:estimation}.
For any group $m\neq m_0$, the population version of $\wh\bXi_m$ is $\bXi_m^* := \Z_{G,m_0}^* \Z_{G,m}^{*\prime} \big( \Z_{G,m}^* \Z_{G,m}^{*\prime} \big)^{-1}$.
With \eqref{eqn: Gt_m0_pop}, we can write
\begin{align*}
    \Z_{G,m_0}^* &= \left( \vec(\G_{1,m_0}^*), \dots, \vec(\G_{T,m_0}^*) \right) 
    =
    \big( \bXi_{J,m}^* \otimes \bXi_{Q,m}^* \big) \Z_{G,m}^* 
    + o_P\Big(
    \big\|
    \Z_{G,m}^*
    \big\|_F
    \Big) ,
\end{align*}
which implies
$\bXi_m^* = \Z_{G,m_0}^* \Z_{G,m}^{*\prime} \big( \Z_{G,m}^* \Z_{G,m}^{*\prime} \big)^{-1} 
= \bXi_{J,m}^* \otimes \bXi_{Q,m}^* +o_P(1)$
given Assumption~(S1).
Meanwhile, also by Assumption~(S1), we have
\begin{align}
    \Big\| \big( \Z_{G,m}^* \Z_{G,m}^{*\prime} \big)^{-1} \Big\|
    = O_P\left( \frac{1}{T S_m} \right) ,
    \label{eqn: global_Gram_inv}
\end{align}
where 
$S_m$ is defined in the main text.
Hence we may expand
\begin{align}
    \wh\bXi_m &=
    \big\{ \Z_{G,m_0}^* +
    (\wh\Z_{G,m_0} - \Z_{G,m_0}^*) \big\}
    \big\{ \Z_{G,m}^* +
    (\wh\Z_{G,m} - \Z_{G,m}^*)
    \big\}' \notag \\
    &\;\quad
    \cdot
    \Big[ \big\{ \Z_{G,m}^* +
    (\wh\Z_{G,m} - \Z_{G,m}^*)
    \big\}
    \big\{ \Z_{G,m}^* +
    (\wh\Z_{G,m} - \Z_{G,m}^*)
    \big\}' \Big]^{-1} \notag \\
    &=:
    \Z_{G,m_0}^* \Z_{G,m}^{*\prime} \big( \Z_{G,m}^* \Z_{G,m}^{*\prime} \big)^{-1}
    +  \cR_{\Xi,m} ,
    \label{eqn: global_hat_Xi_expansion}
\end{align}
where $\cR_{\Xi,m}$ denotes all remaining terms in the decomposition and would involve at least one $(\wh\Z_{G,m} - \Z_{G,m}^*)$ or $(\wh\Z_{G,m_0} - \Z_{G,m_0}^*)$.
As detailed, we bound all terms in $\cR_{\Xi,m}$.
To this end, by \eqref{eqn: Z_Gm_est_rate} and \eqref{eqn: global_Gram_inv}, we first have
\begin{align}
    &\;\quad
    \Big\| (\wh\Z_{G,m_0} - \Z_{G,m_0}^*) \Z_{G,m}^{*\prime}
    \big( \Z_{G,m}^* \Z_{G,m}^{*\prime} \big)^{-1} \Big\|_F^2 
    \lesssim_P
    \big\| \wh\Z_{G,m_0} - \Z_{G,m_0}^* \big\|_F^2
    \cdot \big\| \Z_{G,m}^{*\prime} \big\|_F^2
    \cdot \Big\| \big( \Z_{G,m}^* \Z_{G,m}^{*\prime} \big)^{-1} \Big\|^2 \notag \\
    &=
    O_P\big( T \cdot \Delta_{G,m_0} \big)
    \cdot O_P\big( T S_m \big)
    \cdot O_P\left( (T S_m)^{-2} \right)
    =
    O_P\left( \frac{\Delta_{G,m_0}}{S_m}
    \right) ,
    \label{eqn: global_Xi_error_1}
\end{align}
where $\big\| \Z_{G,m}^{*\prime} \big\|_F^2 = O_P(T S_m)$ used Assumption~(F1).
Similarly,
\begin{align}
    &\;\quad
    \Big\| \Z_{G,m_0}^*
    (\wh\Z_{G,m} - \Z_{G,m}^*)'
    \big( \Z_{G,m}^* \Z_{G,m}^{*\prime} \big)^{-1}
    \Big\|_F^2 
    \lesssim_P
    \big\| \Z_{G,m_0}^* \big\|^2
    \cdot \big\| \wh\Z_{G,m} - \Z_{G,m}^* \big\|_F^2
    \cdot \Big\| \big( \Z_{G,m}^* \Z_{G,m}^{*\prime} \big)^{-1} \Big\|^2 \notag \\
    &= 
    O_P\big( T S_{m_0} \big)
    \cdot O_P\big( T\cdot \Delta_{G,m} \big)
    \cdot O_P\left( (T S_m)^{-2} \right)
    = 
    O_P\left( \frac{\Delta_{G,m} S_{m_0} }{S_{m}^2}
    \right) .
    \label{eqn: global_Xi_error_2}
\end{align}
To bound the cross term in $\cR_{\Xi,m}$ involving both
$\wh\Z_{G,m_0} - \Z_{G,m_0}^*$ and
$\wh\Z_{G,m} - \Z_{G,m}^*$,
let $\E_{G,m} := \wh\Z_{G,m} - \Z_{G,m}^*$.
By the condition $\Delta_{G,m} = o\big\{ \lambda_{\min}\big( \Q_m'\Q_m \big) \lambda_{\min}\big( \J_m'\J_m \big) \big\}$ in the statement of Theorem~\ref{thm:common_comp_global},
we have $\Delta_{G,m} = o(S_m)$.
Then the cross term satisfies
\begin{align}
    &\;\quad
    \Big\| \E_{G,m_0} \E_{G,m}'
    \Big[ \big( \Z_{G,m}^* + \E_{G,m} \big) 
    \big( \Z_{G,m}^* + \E_{G,m}
    \big)' \Big]^{-1} \Big\|_F^2 \notag \\
    &\lesssim_P
    \big\| \E_{G,m_0} \big\|_F^2
    \big\| \E_{G,m} \big\|_F^2
    \Big\| \big[ \big( \Z_{G,m}^* + \E_{G,m} \big)
    \big( \Z_{G,m}^* + \E_{G,m} \big)' \Big]^{-1} \Big\|^2 \notag \\
    &\lesssim_P
    \big( T\Delta_{G,m_0} \big) 
    \big( T\Delta_{G,m} \big)
    \big( T^{-2} S_m^{-2} \big) 
    =
    \frac{\Delta_{G,m_0} \Delta_{G,m}}{ S_m^2 }
    ,
    \label{eqn: global_cross_Xi}
\end{align}
where we used $\Big\| \big[ \big( \Z_{G,m}^* + \E_{G,m} \big) \big( \Z_{G,m}^* + \E_{G,m} \big)' \Big]^{-1} \Big\|^2 \lesssim_P T^{-2} S_m^{-2}$, which follows from applying Weyl's inequality on $\lambda_{\min}\big( A_m \big) = O_P(T S_m)$ (as in \eqref{eqn: ZZ_diff_inv_gram_pert} below), combined with \eqref{eqn: global_Bm_bound} and $\Delta_{G,m} = o(S_m)$.

Finally, we bound
\begin{align}
    \Z_{G,m_0}^* \Z_{G,m}^{*\prime}
    \Big\{ \big( \Z_{G,m}^* \Z_{G,m}^{*\prime} \big)^{-1}
    - \big( \Z_{G,m}^* \Z_{G,m}^{*\prime}
    + B_m \big)^{-1} \Big\}
    =
    \Z_{G,m_0}^* \Z_{G,m}^{*\prime} \Big\{ A_m^{-1} - \big( A_m + B_m \big)^{-1} \Big\} ,
    \label{eqn: ZZ_diff_inv_gram_pert}
\end{align}
where $A_m := \Z_{G,m}^* \Z_{G,m}^{*\prime}$
and $B_m := \E_{G,m} \Z_{G,m}^{*\prime} + \Z_{G,m}^* \E_{G,m}' + \E_{G,m} \E_{G,m}'$.
Note that $\lambda_{\min}\big( A_m \big) = O_P(T S_m)$ by Assumption~(S1) and the definition of $S_m$.
Also, by the condition $\Delta_{G,m} = o(S_m)$ from the statement of Theorem~\ref{thm:common_comp_global} and \eqref{eqn: Z_Gm_est_rate}, we have
\begin{align}
    \big\| B_m \big\|
    &\lesssim_P
    \big\| \E_{G,m} \big\|_F
    \big\| \Z_{G,m}^* \big\| +
    \big\| \E_{G,m} \big\|_F^2
    =
    O_P\Big(
    \sqrt{T \Delta_{G,m}}
    \cdot \sqrt{T S_m} \Big)
    + O_P\big( T \Delta_{G,m} \big) 
    =
    O_P\Big( T \sqrt{ \Delta_{G,m} S_m} \Big),
    \label{eqn: global_Bm_bound}
\end{align}
where we used $\big\| \Z_{G,m}^* \big\|^2 \asymp_P T S_m$ under Assumptions~(S1) and (F1).

Using the matrix identity $\A^{-1} - (\A+\B)^{-1} = \A^{-1} \B (\A + \B)^{-1}$ for any matrices $\A, \B$,
we obtain
\begin{align*}
    &\;\quad
    \Big\| A_m^{-1}
    - \big( A_m + B_m \big)^{-1} \Big\| 
    \leq
    \big\| A_m^{-1} \big\|
    \big\| B_m \big\|
    \Big\| \big( A_m + B_m \big)^{-1} \Big\| \notag \\
    &=
    O_P\Big( 
    T^{-1} S_m^{-1} T \sqrt{ \Delta_{G,m} S_m} T^{-1} S_m^{-1}
    \Big) 
    =
    O_P\Big( 
    T^{-1} S_m^{-3/2} \Delta_{G,m}^{1/2}
    \Big) .
\end{align*}
Hence the corresponding term in $\cR_{\Xi,m}$ from \eqref{eqn: ZZ_diff_inv_gram_pert} is bounded by
\begin{align}
    &\;\quad
    \Big\| \Z_{G,m_0}^* \Z_{G,m}^{*\prime}
    \Big[ A_m^{-1} - \big( A_m + B_m \big)^{-1} \Big] \Big\|_F^2 
    \lesssim
    \big\| \Z_{G,m_0}^* \big\|^2
    \big\| \Z_{G,m}^* \big\|^2
    \Big\| A_m^{-1} - \big( A_m + B_m \big)^{-1} \Big\|^2 \notag \\
    &=
    O_P\Big(
    T S_{m_0}
    T S_m
    T^{-2}
    S_m^{-3}
    \Delta_{G,m}
    \Big)
    =
    O_P\left(
    \frac{ S_{m_0}}{S_m^2}
    \Delta_{G,m}
    \right)
    = 
    O_P\left(
    \frac{\Delta_{G,m}}{S_m}
    \right) ,
    \label{eqn: global_inv_gram_term}
\end{align}
where the last equality used $\lambda_{\min}\big( \Q_{m_0}'\Q_{m_0} \big) \lambda_{\min}\big( \J_{m_0}'\J_{m_0} \big)
= O\big\{ \lambda_{\min}\big( \Q_m'\Q_m \big) \lambda_{\min}\big( \J_m'\J_m \big) \big\}$ so that $S_{m_0} = O(S_m)$ from the statement of Theorem~\ref{thm:common_comp_global}.

Combining \eqref{eqn: global_Xi_error_1}, 
\eqref{eqn: global_Xi_error_2}, 
\eqref{eqn: global_cross_Xi}, 
and \eqref{eqn: global_inv_gram_term},
we conclude $\|\cR_{\Xi,m}\|_F^2 = O_P(\Delta_{G,m_0}/S_m + \Delta_{G,m} / S_m)$, and thus
\begin{align}
    \big\| \wh\bXi_m - \bXi_m^* \big\|_F^2
    \lesssim_P
    \max_{m'\in[M]} \frac{\Delta_{G,m'}}{S_{m'}} .
    \label{eqn: global_hat_Xi_rate}
\end{align}

Next, recall from Step~3 of Section~\ref{subsec:estimation} the definition of the notation $\cR(\W)$ for a block matrix $\W = (\W_{ij})_{i\in[k_c], j\in[k_c]}$, with each $\W_{ij}$ of size $k_r\times k_r$, such that
\begin{align*}
    \cR(\W) = (\vec(\W_{11}),\ldots,\vec(\W_{k_c1}),\ldots,\vec(\W_{1k_c}),\ldots,\vec(\W_{k_ck_c})).
\end{align*}
Note that $\cR(\cdot)$ is a fixed permutation of the entries of the matrix input and hence preserves the Frobenius norm.
Thus, $\big\| \cR(\wh\bXi_m) - \cR(\bXi_m^*) \big\|_F = \big\| \wh\bXi_m - \bXi_m^* \big\|_F$, and hence by \eqref{eqn: global_hat_Xi_rate},
\begin{align}
    \big\| \cR(\wh\bXi_m) - \cR(\bXi_m^*) \big\|_F^2
    =
    O_P\Bigg(
    \max_{m'\in[M]} \frac{\Delta_{G,m'}}{S_{m'}}
    \Bigg) .
    \label{eqn: global_R_Xi_perturbation}
\end{align}

By construction in Step~3 of Section~\ref{subsec:estimation},
the matrix $\wh\sigma_{\Xi,m} \wh\u_{\Xi,m} \wh\v_{\Xi,m}'$ is the rank-1 singular value decomposition of $\cR(\wh\bXi_m)$.
Then let $\sigma_{\Xi,m}^* \u_{\Xi,m}^* \v_{\Xi,m}^{*\prime}$ be the rank-1 singular value decomposition of $\cR(\bXi_m^*)$.
From Lemma~\ref{lemma: alignment_equality}, we have
\[
\cR\big( \wh\bXi_{J,m} \otimes \wh\bXi_{Q,m} \big)
= 
\wh\sigma_{\Xi,m} \wh\u_{\Xi,m} \wh\v_{\Xi,m}' .
\]
Recall previously that $\bXi_m^* = \bXi_{J,m}^* \otimes \bXi_{Q,m}^*$, and by analogous arguments in the proof of Lemma~\ref{lemma: alignment_equality}, we also have
\[
\sigma_{\Xi,m}^* \u_{\Xi,m}^* \v_{\Xi,m}^{*\prime} 
= \cR\big( \bXi_{J,m}^* \otimes \bXi_{Q,m}^* \big)
= \vec\big( \bXi_{Q,m}^* \big) \vec\big( \bXi_{J,m}^* \big)' ,
\]
which implies
\begin{align}
    \sigma_{\Xi,m}^* = \big\| \bXi_{Q,m}^* \big\|_F \cdot \big\| \bXi_{J,m}^* \big\|_F
    =O_P( 1 ) ,
    \label{eqn: sigma_Xi_m_pop}
\end{align}
where we used the definitions of $\bXi_{Q,m}^*$ and $\bXi_{J,m}^*$ at the beginning of this proof, together with the rate condition $\lambda_{\min}\big( \Q_{m_0}'\Q_{m_0} \big) \lambda_{\min}\big( \J_{m_0}'\J_{m_0} \big) = O\big\{ \lambda_{\min}\big( \Q_m'\Q_m \big) \lambda_{\min}\big( \J_m'\J_m \big) \big\}$ in the statement of Theorem~\ref{thm:common_comp_global}.
Altogether, using the fact that $\cR(\cdot)$ preserves the Frobenius norm (as in \eqref{eqn: global_R_Xi_perturbation}), we have
\begin{align}
    &\;\quad
    \Big\| \wh\bXi_{J,m} \otimes \wh\bXi_{Q,m}
    - \bXi_{J,m}^* \otimes \bXi_{Q,m}^* \Big\|_F^2
    =
    \Big\| \cR(\wh\bXi_{J,m} \otimes \wh\bXi_{Q,m})
    - \cR(\bXi_{J,m}^* \otimes \bXi_{Q,m}^*) \Big\|_F^2 \notag \\
    &=
    \Big\| \wh\sigma_{\Xi,m} \wh\u_{\Xi,m} \wh\v_{\Xi,m}'
    - \sigma_{\Xi,m}^* \u_{\Xi,m}^* \v_{\Xi,m}^{*\prime} \Big\|_F^2  \notag \\
    &\lesssim
    \big| \wh\sigma_{\Xi,m} - \sigma_{\Xi,m}^* \big|^2
    \cdot \big\| \wh\u_{\Xi,m} \wh\v_{\Xi,m}' \big\|_F^2
    + (\sigma_{\Xi,m}^*)^2 \big\| \wh\u_{\Xi,m} \wh\v_{\Xi,m}'
    - \u_{\Xi,m}^* \v_{\Xi,m}^{*\prime} \big\|_F^2 \notag \\
    &\lesssim_P
    \big\| \cR(\wh\bXi_m) - \cR(\bXi_m^*) \big\|_F^2 
    =
    \big\| \wh\bXi_m - \bXi_m^* \big\|_F^2 
    =
    O_P\Bigg(
    \max_{m'\in[M]} \frac{\Delta_{G,m'}}{S_{m'}}
    \Bigg) ,
    \label{eqn: global_product_alignment}
\end{align}
where the last line used Weyl's inequality and the Davis--Kahan theorem \citep[Theorem~2]{yu2015useful},
in addition to \eqref{eqn: global_R_Xi_perturbation} and \eqref{eqn: sigma_Xi_m_pop}.

By the construction of the global factor estimator $\wh\G_t$ in Step~3 of Section~\ref{subsec:estimation}, we have
\begin{align}
    &\;\quad
    \Big\| \wh\G_t
    - \H_{Q,m_0}' \check\G_t \H_{J,m_0} \Big\|_F^2
    \lesssim
    \max_{m\in[M]}
    \Big\| \wh\bXi_{Q,m} \wh\G_{t,m} \wh\bXi_{J,m}'
    - \H_{Q,m_0}' \check\G_t \H_{J,m_0}
    \Big\|_F^2.
    \label{eqn: global_final_start}
\end{align}
For each fixed $m$, we decompose
\begin{align}
    \wh\bXi_{Q,m} \wh\G_{t,m} \wh\bXi_{J,m}'
    - \H_{Q,m_0}' \check\G_t \H_{J,m_0} 
    &=
    \wh\bXi_{Q,m} \big( \wh\G_{t,m} - \G_{t,m}^* \big) \wh\bXi_{J,m}' 
    + (\wh\bXi_{Q,m} \G_{t,m}^* \wh\bXi_{J,m}'
    - \H_{Q,m_0}' \check\G_t \H_{J,m_0}) \notag \\
    &=:
    \cI_{\Xi,G,1} + \cI_{\Xi,G,2} .
    \label{eqn: global_final_decomp}
\end{align}
For $\cI_{\Xi,G,1}$, by \eqref{eqn: sigma_Xi_m_pop} and \eqref{eqn: wh_Gt_m}, we have
$\big\| \wh\bXi_{Q,m} \big\|, \big\| \wh\bXi_{J,m} \big\| = O_P(1)$, and
$\big\| \wh\G_{t,m} - \G_{t,m}^* \big\|_F^2 = O_P\big( \Delta_{G,m} \big)$, so that
\begin{align}
    \big\| \cI_{\Xi,G,1} \big\|_F^2
    \lesssim_P 
    \Delta_{G,m} .
    \label{eqn: global_final_I1}
\end{align}

For $\cI_{\Xi,G,2}$, we can write
\begin{align}
    &\;\quad
    \vec(\cI_{\Xi,G,2}) =
    \big( \wh\bXi_{J,m} \otimes \wh\bXi_{Q,m} \big)
    \vec\big( \G_{t,m}^* \big)
    - \big( \H_{J,m_0}' \otimes \H_{Q,m_0}' \big)
    \vec\big( \check\G_t \big) \notag \\
    &=
    \Big[ \big( \wh\bXi_{J,m} \otimes \wh\bXi_{Q,m} \big)
    - \big( \bXi_{J,m}^* \otimes \bXi_{Q,m}^* \big) \Big]
    \vec\big( \G_{t,m}^* \big) \notag \\
    &\;\quad
    + \Big[ \big(
    \bXi_{J,m}^* \otimes \bXi_{Q,m}^*
    \big)
    \vec\big( \G_{t,m}^* \big)
    - \big( \H_{J,m_0}' \otimes \H_{Q,m_0}' \big)
    \vec\big( \check\G_t \big) \Big]
    =:
    \cI_{\Xi,G,21}
    + \cI_{\Xi,G,22} .
    \label{eqn: global_final_I2_decomp}
\end{align}
In particular, for $\cI_{\Xi,G,21}$,
by \eqref{eqn: global_product_alignment} and $\big\| \G_{t,m}^* \big\|_F^2 = O_P(S_m)$,
we have
\begin{align}
    \big\| \cI_{\Xi,G,21} \big\|_F^2
    &\lesssim_P
    \Big\| \wh\bXi_{J,m} \otimes \wh\bXi_{Q,m} - \bXi_{J,m}^* \otimes \bXi_{Q,m}^* \Big\|_F^2
    \cdot \big\| \G_{t,m}^* \big\|_F^2 
    \lesssim_P
    \max_{m'\in[M]}
    \frac{\Delta_{G,m'}}{S_{m'}}
    S_m .
    \label{eqn: global_final_I21}
\end{align}
Also, by \eqref{eqn: XiQ_G_XiJ_pop},
\begin{align}
    \big\| \cI_{\Xi,G,22} \big\|_F^2 &=
    O_P\Big(
    \eta_{Q,m}
    \eta_{J,m}
    p_m^{\delta_{q,1}^{(m)}}
    q_m^{\delta_{j,1}^{(m)}}
    \Big) .
    \label{eqn: global_final_I22}
\end{align}

Combining \eqref{eqn: global_final_start},
\eqref{eqn: global_final_decomp},
\eqref{eqn: global_final_I2_decomp},
\eqref{eqn: global_final_I1}, 
\eqref{eqn: global_final_I21}, 
and \eqref{eqn: global_final_I22}, 
we obtain
\[
\Big\| \wh\G_t - \H_{Q,m_0}' \check\G_t \H_{J,m_0} \Big\|_F^2
\lesssim_P
\max_{m\in[M]} \Delta_{G,m}
+ \max_{m'\in[M]} \frac{\Delta_{G,m'}}{S_{m'}} \max_{m\in[M]} S_m 
+ \max_{m\in[M]} \eta_{Q,m} \eta_{J,m} p_m^{\delta_{q,1}^{(m)}} q_m^{\delta_{j,1}^{(m)}}   ,
\]
which thus concludes the proof for $\big\| \wh\G_t - \H_{Q,m_0}' \check\G_t \H_{J,m_0} \big\|_F^2$, by noting that the last term above is dominated by the first term.

It remains to derive the rate for the re-estimated global loading matrices $\wt\Q_m$ and $\wt\J_m$ in Step~3 of Section~\ref{subsec:estimation}. 
First, as $\H_{Q,m_0}$ and $\H_{J,m_0}$ are asymptotically orthogonal by Lemma~\ref{lemma: aligned_projected_cores_rot_rate}, the population version of the re-estimated Kronecker product matrix is
\begin{align}
    \K_m^0 :=
    \Big\{ \J_m \big( \J_{m_0}'\J_{m_0} \big)^{-1/2} \H_{J,m_0} \Big\}
    \otimes \Big\{ \Q_m \big( \Q_{m_0}'\Q_{m_0} \big)^{-1/2} \H_{Q,m_0} \Big\} .
    \label{eqn: reloading_kron_target}
\end{align}
For each $m\in[M]$, define the matrices
\begin{align*}
    \Y_m &:=
    \Big( \vec\big( \check\Y_{G,1}^{(m)} \big),
    \dots,
    \vec\big( \check\Y_{G,T}^{(m)} \big) \Big), 
    \quad
    \G^* :=
    \Big( \vec\big( \H_{Q,m_0}' \check\G_1 \H_{J,m_0} \big),
    \dots,
    \vec\big( \H_{Q,m_0}' \check\G_T \H_{J,m_0} \big) \Big) , \\
    \wh\G &:=
    \Big( \vec\big( \wh\G_1 \big),
    \dots,
    \vec\big( \wh\G_T \big) \Big) ,
    \quad
    \U_m := \Y_m - \K_m^0 \G^* ,
\end{align*}
so that $\U_m$ satisfies
\begin{align}
    \U_m &= \Y_m - \K_m^0 \G^*
    =
    \big( \Y_m - \Y_m^{0}\big)
    + \big( \Y_m^{0} - \K_m^0 \G^* \big) ,
    \label{eqn: reloading_U_decomp}
\end{align}
where $\Y_m^{0}$ denotes the matrix whose columns are $\vec(\Q_m \G_t \J_m')$, $t\in[T]$.
Consider the first term in \eqref{eqn: reloading_U_decomp}, which has columns $\vec\big( \check\Y_{G,t}^{(m)} - \Q_m \G_t \J_m' \big)$.
By our already constructed result in Theorem~\ref{thm:common_comp_global} and with the notation $\Delta_G$ in the theorem's statement,
we have $\big\|\wh\G - \G^* \big\|_F^2 = O_P(T \Delta_G)$.
For each $m\in[M]$, by Theorem~\ref{thm:common_comp_global_local_init} and Lemma~\ref{lemma:bound_Delta},
we have
\begin{align}
    \frac{1}{T} \big\| \Y_m - \Y_m^{0} \big\|_F^2
    &=
    \frac{1}{T} \sum_{t=1}^{T}
    \big\| \big( \R_m \F_t^{(m)} \C_m'
    - \wh\R_m \wh\F_t^{(m)} \wh\C_m' \big)
    + \M_{p_m} \E_t^{(m)} \M_{q_m} \big\|_F^2 \notag \\
    &\lesssim_P
    p_m^{\delta_{r,1}^{(m)}}
    q_m^{\delta_{c,1}^{(m)}}
    \big( \Delta_{R,m} + \Delta_{C,m} + \Delta_{F,m} \big)
    + p_m q_m .
    \label{eqn: reloading_Utm_rate1}
\end{align}
For the second term in \eqref{eqn: reloading_U_decomp}, for each $t\in[T]$,
\begin{align}
    &\;\quad
    \Big\| \big( \Q_m \big( \Q_{m_0}' \Q_{m_0} \big)^{-1/2} \H_{Q,m_0} \big) 
    \H_{Q,m_0}' \check\G_t \H_{J,m_0}
    \big( \J_m \big( \J_{m_0}' \J_{m_0} \big)^{-1/2} \H_{J,m_0} \big)'
    - \Q_m \G_t \J_m' \Big\|_F^2 \notag \\
    &\lesssim_P
    \big( \eta_{Q,m_0} + \eta_{J,m_0} \big)
    \big\| \Q_m \big\|^2
    \big\| \J_m \big\|^2
    \big\| \big( \Q_{m_0}' \Q_{m_0} \big)^{-1/2} \big\|^2
    \big\|  \big( \J_{m_0}' \J_{m_0} \big)^{-1/2} \big\|^2
    \big\| \check\G_t \big\|_F^2 \notag \\
    &=
    O_P\Bigg( \big( \eta_{Q,m_0} + \eta_{J,m_0}
    \big)
    \frac{p_m^{\delta_{q,1}^{(m)}}
    q_m^{\delta_{j,1}^{(m)}}}{p_{m_0}^{\delta_{q,k_r}^{(m_0)}}
    q_{m_0}^{\delta_{j,k_c}^{(m_0)}}}
    S_{m_0}
    \Bigg),
    \label{eqn: reloading_Utm_rate2}
\end{align}
where the last line used Assumption~(S1), the definition of $S_{m_0}$ in Theorem~\ref{thm:common_comp_global}, and $\|\check\G_t\|_F^2 = O_P(S_{m_0})$.
Combining \eqref{eqn: reloading_U_decomp}, \eqref{eqn: reloading_Utm_rate1}, and \eqref{eqn: reloading_Utm_rate2},
we have
\begin{align}
    \frac{1}{T} \big\| \U_m \big\|_F^2
    \lesssim_P
    p_m^{\delta_{r,1}^{(m)}}
    q_m^{\delta_{c,1}^{(m)}}
    \big( \Delta_{R,m} + \Delta_{C,m} + \Delta_{F,m} \big)
    + p_m q_m
    + \big( \eta_{Q,m_0} + \eta_{J,m_0}
    \big)
    \frac{p_m^{\delta_{q,1}^{(m)}}
    q_m^{\delta_{j,1}^{(m)}}}{p_{m_0}^{\delta_{q,k_r}^{(m_0)}}
    q_{m_0}^{\delta_{j,k_c}^{(m_0)}}}
    S_{m_0} .
    \label{eqn: reloading_Um_rate}
\end{align}

We next bound $\big\| T^{-1}\U_m \wh\G' \big\|_F^2$.
By definition, the $(i,j)$ entry of $T^{-1}\U_m \wh\G'$ is
\begin{align}
    \big( T^{-1}\U_m \wh\G' \big)_{ij}
    = \frac{1}{T}\sum_{t=1}^{T} \big(\U_m\big)_{it} \big(\wh\G\big)_{jt},
    \label{eqn: reloading_UTG_entry}
\end{align}
where $\big(\U_m\big)_{it}$ denotes the $(i,t)$ entry of $\U_m$ and $\big(\wh\G\big)_{jt}$ denotes the $(j,t)$ entry of $\wh\G$.
Recall the decomposition $\U_m = (\Y_m - \Y_m^0) + (\Y_m^0 - \K_m^0\G^*)$, where $\Y_m^0$ is the matrix whose columns are $\vec(\Q_m\G_t\J_m')$, $t\in[T]$.
Thus we have
\begin{align*}
    \big(\U_m\big)_{it} &=
    \Big[ \vec\big( \check\Y_{G,t}^{(m)} - \Q_m \G_t \J_m' \big) \Big]_i
    + \Big[ \vec(\Q_m\G_t\J_m') - \K_m^0 \G^*_{\cdot t} \Big]_i,
\end{align*}
where $\G^*_{\cdot t}$ denotes the $t$-th column of $\G^*$.
By the Cauchy--Schwarz inequality,
\begin{align}
    \big( T^{-1}\U_m \wh\G' \big)_{ij}^2
    &\leq 
    2 \Bigg[ \frac{1}{T}\sum_{t=1}^{T} \Big[ \vec\big( \check\Y_{G,t}^{(m)} - \Q_m \G_t \J_m' \big) \Big]_i \big(\wh\G\big)_{jt} \Bigg]^2
    \notag \\
    &\;\quad 
    + 2 \Bigg[ \frac{1}{T}\sum_{t=1}^{T} \Big[ \vec(\Q_m\G_t\J_m') - \K_m^0 \G^*_{\cdot t} \Big]_i \big(\wh\G\big)_{jt} \Bigg]^2.
    \label{eqn: reloading_UTG_decomp}
\end{align}
For the second term on the right-hand side of \eqref{eqn: reloading_UTG_decomp},
by \eqref{eqn: reloading_Utm_rate2} and $\|\wh\G\|_F^2 = O_P(T S_{m_0})$, we have
\begin{align}
    &\;\quad
    \Bigg( \frac{1}{T}\sum_{t=1}^{T}
    \Big[ \vec(\Q_m\G_t\J_m') - \K_m^0 \G^*_{\cdot t} \Big]_i \big(\wh\G\big)_{jt}
    \Bigg)^2 \notag \\
    &\lesssim_P
    \frac{1}{T} \frac{1}{p_m q_m}
    \big( \eta_{Q,m_0} + \eta_{J,m_0} \big)
    \frac{p_m^{\delta_{q,1}^{(m)}} q_m^{\delta_{j,1}^{(m)}}}{S_{m_0}}
    S_{m_0}
    =
    \frac{1}{T p_m q_m}
    \big( \eta_{Q,m_0} + \eta_{J,m_0} \big)
    p_m^{\delta_{q,1}^{(m)}} q_m^{\delta_{j,1}^{(m)}} .
    \label{eqn: reloading_UTG_rot}
\end{align}

For the first term on the right-hand side of \eqref{eqn: reloading_UTG_decomp}, we further decompose
\begin{align*}
    \Big[ \vec\big( \check\Y_{G,t}^{(m)} - \Q_m \G_t \J_m' \big) \Big]_i
    =
    \Big[ \vec\big( \M_{p_m}\E_t^{(m)}\M_{q_m} \big) \Big]_i
    + \Big[ \vec\big( \R_m\F_t^{(m)}\C_m' - \wh\R_m\wh\F_t^{(m)}\wh\C_m' \big) \Big]_i .
\end{align*}
By Assumption~(E2), the noise process $\{\M_{p_m} \E_t^{(m)} \M_{q_m}\}$ is independent of the factor process $\{\G_t\}$.
By Lemma~\ref{lemma:correlation_Etm}, the entries of $\{\M_{p_m} \E_t^{(m)} \M_{q_m}\}$ have summable autocovariance bounded by $C\rho(\cdot)$.
By Assumptions~(S1), (F1), and (L1), the entries of $\wh\G$ have second moment of order $S_{m_0}$.
Altogether, we obtain
\begin{align}
    \Bigg(
    \frac{1}{T}\sum_{t=1}^{T}
    \Big[ \vec\big( \M_{p_m}\E_t^{(m)}\M_{q_m} \big) \Big]_i
    \big(\wh\G\big)_{jt} \Bigg)^2
    &\lesssim_P
    \frac{S_{m_0}}{T} .
    \label{eqn: reloading_UTG_noise_entry}
\end{align}
For $\R_m\F_t^{(m)}\C_m' - \wh\R_m\wh\F_t^{(m)}\wh\C_m'$, by Assumption~(IC2), the local factor $\F_t^{(m)}$ is uncorrelated with the global factor $\G_t$. Hence by Theorem~\ref{thm:common_comp_global_local_init}, we have
$\big\| \R_m\F_t^{(m)}\C_m' - \wh\R_m\wh\F_t^{(m)}\wh\C_m' \big\|_F^2 \lesssim_P p_m^{\delta_{r,1}^{(m)}} q_m^{\delta_{c,1}^{(m)}} \big( \Delta_{R,m} + \Delta_{C,m} + \Delta_{F,m} \big)$,
so that
\begin{align}
    \Bigg(\frac{1}{T}\sum_{t=1}^{T}
    \Big[ \vec\big( \R_m\F_t^{(m)}\C_m' - \wh\R_m\wh\F_t^{(m)}\wh\C_m' \big) \Big]_i
    \big(\wh\G\big)_{jt} \Bigg)^2
    &\lesssim_P
    \frac{1}{T p_m q_m}
    p_m^{\delta_{r,1}^{(m)}} q_m^{\delta_{c,1}^{(m)}}
    \big( \Delta_{R,m} + \Delta_{C,m} + \Delta_{F,m} \big)
    S_{m_0} .
    \label{eqn: reloading_UTG_local_entry}
\end{align}
Combining \eqref{eqn: reloading_UTG_decomp}, 
\eqref{eqn: reloading_UTG_rot},
\eqref{eqn: reloading_UTG_noise_entry}, and \eqref{eqn: reloading_UTG_local_entry},
summing over $i\in[p_m q_m]$ and $j\in[k_r k_c]$, we conclude
\begin{align}
    \Bigg\| \frac{1}{T} \U_m \wh\G' \Bigg\|_F^2
    &\lesssim_P
    \frac{p_m q_m
    + p_m^{\delta_{r,1}^{(m)}} q_m^{\delta_{c,1}^{(m)}}
    \big( \Delta_{R,m} + \Delta_{C,m} + \Delta_{F,m} \big)}{T}
    S_{m_0} \notag \\
    &\;\quad
    + \frac{1}{T} \big( \eta_{Q,m_0} + \eta_{J,m_0} \big)
    p_m^{\delta_{q,1}^{(m)}} q_m^{\delta_{j,1}^{(m)}} 
    \lesssim
    \frac{p_m q_m
    + p_m^{\delta_{r,1}^{(m)}} q_m^{\delta_{c,1}^{(m)}}
    \big( \Delta_{R,m} + \Delta_{C,m} + \Delta_{F,m} \big)}{T}
    S_{m_0} .
    \label{eqn: reloading_UTG_final}
\end{align}

With our notations above, by the construction in Step~3 of Section~\ref{subsec:estimation}, we can write
\begin{align}
    \wh{\J_m\otimes\Q_m} &=
    \Y_m \wh\G' \big( \wh\G \wh\G' \big)^{-1}
    =
    \big( \K_m^0 \G^* + \U_m \big)
    \wh\G' \big( \wh\G \wh\G' \big)^{-1} \notag \\
    &=
    \K_m^0 
    + \K_m^0 \big( \G^* \wh\G' - \wh\G \wh\G' \big) \big( \wh\G \wh\G' \big)^{-1} 
    + \U_m \wh\G' \big( \wh\G \wh\G' \big)^{-1} .
    \label{eqn: reloading_decomp}
\end{align}

We bound the two terms in \eqref{eqn: reloading_decomp}. 
First, by Assumption~(S1), the smallest eigenvalue of $T^{-1} \G^* \G^{*\prime}$ is of order $S_{m_0}$. Then we obtain
\begin{align}
    \Big\| \big( \wh\G \wh\G' \big)^{-1} \Big\|
    = O_P\Big( T^{-1} S_{m_0}^{-1} \Big).
    \label{eqn: reloading_inv}
\end{align}
Also, note that $\big\| \G^* \big\|_F^2 = O_P\big( T S_{m_0} \big)$ and $\big\| \wh\G \big\|_F^2  = O_P\big( T S_{m_0} \big)$.
Then for the first term in \eqref{eqn: reloading_decomp}, we have
\begin{align}
    &\;\quad
    \Bigg\| \K_m^0 \big( \G^* \wh\G'
    - \wh\G \wh\G' \big)
    \big( \wh\G \wh\G' \big)^{-1} \Bigg\|_F^2 
    \lesssim
    \big\| \K_m^0 \big\|^2
    \big\| \G^* - \wh\G \big\|_F^2
    \big\| \wh\G \big\|^2
    \Big\| \big( \wh\G \wh\G' \big)^{-1} \Big\|^2 \notag \\
    &=
    O_P\Bigg(
    \big\| \K_m^0 \big\|^2
    \cdot T \Delta_G
    \cdot T S_{m_0}
    \cdot T^{-2} S_{m_0}^{-2}
    \Bigg)
    =
    O_P\Bigg(
    \big\| \K_m^0 \big\|^2
    \frac{ \Delta_G}{S_{m_0} }
    \Bigg) .
    \label{eqn: reloading_term1}
\end{align}
For the second term in \eqref{eqn: reloading_decomp}, 
by \eqref{eqn: reloading_UTG_final} and \eqref{eqn: reloading_inv}
we have
\begin{align}
    &\;\quad
    \Bigg\| \U_m \wh\G' \big( \wh\G \wh\G' \big)^{-1} \Bigg\|_F^2 
    =
    \Bigg\| \Big(\frac{1}{T} \U_m \wh\G' \Big) \Big( \frac{1}{T} \wh\G \wh\G' \Big)^{-1} \Bigg\|_F^2 
    \lesssim_P
    \Big\| \frac{1}{T} \U_m \wh\G' \Big\|_F^2
    \cdot \Big\| \Big( \frac{1}{T} \wh\G \wh\G' \Big)^{-1} \Big\|^2 \notag \\
    &=
    O_P\Bigg(
    \frac{S_{m_0}}{T}
    \big\{ p_m q_m
    + p_m^{\delta_{r,1}^{(m)}} q_m^{\delta_{c,1}^{(m)}}
    \big( \Delta_{R,m} + \Delta_{C,m} + \Delta_{F,m} \big)\big\}
    \cdot S_{m_0}^{-2}
    \Bigg) \notag \\
    &=
    O_P\Bigg(
    \frac{ p_m q_m
    + p_m^{\delta_{r,1}^{(m)}} q_m^{\delta_{c,1}^{(m)}}
    \big( \Delta_{R,m} + \Delta_{C,m} + \Delta_{F,m} \big)}{T S_{m_0}}
    \Bigg) .
    \label{eqn: reloading_term2}
\end{align}

Combining \eqref{eqn: reloading_decomp}, 
\eqref{eqn: reloading_term1}, 
and \eqref{eqn: reloading_term2}, 
we arrive at
\begin{align}
    \Bigg\| \wh{\J_m\otimes\Q_m}
    - \K_m^0 \Bigg\|_F^2 
    &\lesssim_P
    \big\| \K_m^0 \big\|^2
    \frac{ \Delta_G }{ S_{m_0} }
    +
    \frac{ p_m q_m
    + p_m^{\delta_{r,1}^{(m)}} q_m^{\delta_{c,1}^{(m)}}
    \big( \Delta_{R,m} + \Delta_{C,m} + \Delta_{F,m} \big)}{T S_{m_0}}  
    = \Delta_{K,m}  ,
    \label{eqn: reloading_final_rate}
\end{align}
where the last equality used the definition of $\Delta_{K,m}$ in the statement of Theorem~\ref{thm:common_comp_global}, the fact that $\|\K_m^0\|_F^2 = O_P\big( p_m^{\delta_{q,1}^{(m)}}  q_m^{\delta_{j,1}^{(m)}} /S_{m_0} \big)$ which is direct from \eqref{eqn: reloading_kron_target},
and the condition from the statement of Theorem~\ref{thm:common_comp_global} that
$\lambda_{\min}\big( \Q_{m_0}'\Q_{m_0} \big) \lambda_{\min}\big( \J_{m_0}'\J_{m_0} \big)
= O\big\{ \lambda_{\min}\big( \Q_m'\Q_m \big) \lambda_{\min}\big( \J_m'\J_m \big) \big\}$
and thus 
\[
\frac{p_m^{\delta_{r,1}^{(m)}} q_m^{\delta_{c,1}^{(m)}}
\big( \Delta_{R,m} + \Delta_{C,m} + \Delta_{F,m} \big)}{T S_{m_0}}
\lesssim
\frac{\Delta_{G,m}}{T S_{m_0}}
\lesssim
\frac{\Delta_{G,m}}{S_{m_0}} \frac{1}{T} \frac{p_m^{\delta_{q,1}^{(m)}}  q_m^{\delta_{j,1}^{(m)}}}{S_{m_0}}
= o\left( \big\| \K_m^0 \big\|^2 \frac{ \Delta_G }{ S_{m_0} } \right) .
\]

The remaining arguments are similar to the previous ones in this proof on $\big\| \wh\G_t - \H_{Q,m_0}' \check\G_t \H_{J,m_0} \big\|_F^2$.
In particular, by the construction in Step~3 of Section~\ref{subsec:estimation}, the matrix $\cR(\wh{\J_m \otimes \Q_m})$ has a rank-1 singular value decomposition
\begin{align}
    \cR\big( \wh{\J_m\otimes\Q_m} \big)
    &=
    \wh\sigma_{\Xi,m}
    \wh\u_{\Xi,m}
    \wh\v_{\Xi,m}'
    +
    \wh\Xi_m^{(\text{rem})},
    \label{eqn: reloading_SVD_hat}
\end{align}
where the remainder $\wh\Xi_m^{(\text{rem})}$ includes the remaining singular components.
Similarly, write the population version of the above as the rank-1 singular value decomposition and the remaining singular components (which are zero due to the following), such that
\begin{align}
    \cR\big( \K_m^0 \big)
    &=
    \sigma_{*,m}
    \u_{*,m}
    \v_{*,m}'
    = \vec(\Q_m^*) \vec(\J_m^*)' ,
    \label{eqn: reloading_SVD_pop}
\end{align}
where $\sigma_{*,m}$ denotes the largest singular value of $\cR(\K_m^0)$.
By Assumptions~(S1), (L1), and the asymptotic orthogonality of $\H_{Q,m_0}$ and $\H_{J,m_0}$ from Lemma~\ref{lemma: aligned_projected_cores_rot_rate},
the leading singular value of $\cR(\K_m^0)$ satisfies $\sigma_{*,m} \asymp_P 1$.
Then by Weyl's inequality, \eqref{eqn: reloading_final_rate},
and recall that $\cR(\cdot)$ preserves the Frobenius norm,
\begin{align}
    \big| \wh\sigma_{\Xi,m}
    - \sigma_{*,m} \big|
    &\lesssim_P
    \big\| \cR(\wh{\J_m \otimes \Q_m}) - \cR(\K_m^0) \big\|
    \lesssim_P
    \Delta_{K,m}^{1/2} ,
    \label{eqn: reloading_sigma_perturbation}
\end{align}
which then implies $\wh\sigma_{\Xi,m} \asymp_P 1$.
By the Davis--Kahan theorem \citep[Theorem~2]{yu2015useful}, we have
\begin{align*}
    \big\| \wh\u_{\Xi,m} - \u_{*,m} \big\|^2
    &\lesssim_P
    \frac{\big\| \cR(\wh{\J_m \otimes \Q_m}) - \cR(\K_m^0) \big\|_F^2}{(\sigma_{*,m} - \sigma_{*,m}^{(2)})^2 } 
    , \quad
    \big\| \wh\v_{\Xi,m} - \v_{*,m} \big\|^2
    \lesssim_P
    \frac{\big\|\cR(\wh{\J_m \otimes \Q_m}) - \cR(\K_m^0) \big\|_F^2}{(\sigma_{*,m} - \sigma_{*,m}^{(2)})^2} ,
\end{align*}
where $\sigma_{*,m}^{(2)}$ denotes the second largest singular value of $\cR(\K_m^0)$ such that $\sigma_{*,m}^{(2)} =0$ due to \eqref{eqn: reloading_SVD_pop}, and thus $|\sigma_{*,m} - \sigma_{*,m}^{(2)}| \asymp_P 1$.
Then with \eqref{eqn: reloading_sigma_perturbation}, we may conclude
\begin{align}
    \big\| \wh\u_{\Xi,m} - \u_{*,m} \big\|^2
    &\lesssim_P
    \Delta_{K,m} 
    , \quad
    \big\| \wh\v_{\Xi,m} - \v_{*,m} \big\|^2
    \lesssim_P
    \Delta_{K,m} .
    \label{eqn: reloading_singular_vectors_rate}
\end{align}

Recall from Step~3 of Section~\ref{subsec:estimation} that $\vec(\wt\Q_m) = \wh\sigma_{\Xi,m}^{1/2} \wh\u_{\Xi,m}$ and $\vec(\wt\J_m) = \wh\sigma_{\Xi,m}^{1/2} \wh\v_{\Xi,m}$.
Similarly, define their population versions $\vec(\Q_m^*) := \sigma_{*,m}^{1/2} \u_{*,m}$ and $\vec(\J_m^*) = \sigma_{*,m}^{1/2} \v_{*,m}$.
Combining \eqref{eqn: reloading_sigma_perturbation} and \eqref{eqn: reloading_singular_vectors_rate}, we obtain
\begin{align}
    &\;\quad
    \big\| \vec(\wt\Q_m) - \vec(\Q_m^*) \big\|^2
    =
    \big\| \wh\sigma_{\Xi,m}^{1/2} \wh\u_{\Xi,m}
    - \sigma_{*,m}^{1/2} \u_{*,m} \big\|^2 \notag \\
    &\lesssim
    \big| \wh\sigma_{\Xi,m}^{1/2} - \sigma_{*,m}^{1/2} \big|^2
    \big\| \wh\u_{\Xi,m} \big\|^2
    + \sigma_{*,m} \big\| \wh\u_{\Xi,m} - \u_{*,m} \big\|^2 
    =
    O_P\big( \Delta_{K,m} \big),
    \label{eqn: reloading_Q_rate}
\end{align}
and analogously,
\begin{align}
    \big\| \vec(\wt\J_m) - \vec(\J_m^*) \big\|^2
    =
    O_P\big( \Delta_{K,m} \big).
    \label{eqn: reloading_J_rate}
\end{align}
Thus finally, the re-estimated global loading matrices $\wt\Q_m$ and $\wt\J_m$ satisfy
\begin{align}
    \big\| \wt\Q_m - \Q_m^* \big\|_F^2
    &=
    O_P\big( \Delta_{K,m} \big),
    \quad
    \big\| \wt\J_m - \J_m^* \big\|_F^2 
    = 
    O_P\big( \Delta_{K,m} \big) ,
    \label{eqn: reloading_final_individual}
\end{align}
where
\[
\Q_m^* := \Q_m \big( \Q_{m_0}' \Q_{m_0} \big)^{-1/2} \H_{Q,m_0}
, \quad
\J_m^* := \J_m \big( \J_{m_0}' \J_{m_0} \big)^{-1/2} \H_{J,m_0} .
\]
This completes the proof of the theorem.
$\square$

\textbf{\textit{Proof of Theorem~\ref{thm:global_factor_detection}.}} For each $m\in[M]$ and any $c_m\in[k_r]$ and $d_m\in[k_c]$, we have 
\begin{align}
    \wh\Q_{m,\cdot c_m}'\widecheck\Y_t^{(m)}\wh\J_{m,\cdot d_m} &= (\wh\J_{m,\cdot d_m}'\J_m \otimes \wh\Q_{m,\cdot c_m}'\Q_m)\vec(\G_t) + (\wh\J_{m,\cdot d_m}'\C_m \otimes \wh\Q_{m,\cdot c_m}'\R_m)\vec(\F_t^{(m)})\notag\\ 
    &\;\;\;\;+ \wh\Q_{m,\cdot c_m}'\M_{p_m}\E_t^{(m)}\M_{q_m}\wh\J_{m,\cdot d_m}. \label{eqn:Thm9-1}
\end{align}
From the proof of Theorem \ref{thm:common_comp_global_local_init}, for some invertible constant order matrices $\H_q$ and $\H_j$, we have
\begin{align}
    \norm{\wh\Q_m - \Q_m(\Q_m'\Q_m)^{-1/2}\H_{q,m}} = O_P(\eta_{Q,m}) = o_P(1), \;\;\norm{\wh\J_m - \J_m(\J_m'\J_m)^{-1/2}\H_{j,m}} = O_P(\eta_{J,m}) = o_P(1). \label{eqn:Thm9-eta_QJ}
\end{align}
Hence, defining $\e_{p,k}$ as the real unit vector of length $p$ with 1 at its $k$th position, we have
\begin{align}
    \max_{c_m \in[k_r]}\norm{\wh\Q_{m,\cdot c_m}'\Q_m} &= \max_{c_m\in[k_r]} \norm{\e_{k_r, c_m}'\wh\Q_m'\Q_m} \asymp \norm{\wh\Q_m'\Q_m}\notag\\ 
    &\geq \norm{(\Q_m'\Q_m)^{1/2}\H_{q,m}} - \norm{\Q_m'(\wh\Q_m - \Q_m(\Q_m'\Q_m)^{-1/2}\H_{q,m})} \asymp_P p_m^{\delta_{q,1}^{(m)}/2}. \label{eqn:Thm9-2}
\end{align}
Similarly,
\begin{equation}\label{eqn:Thm9-3}
    \max_{d_m \in[k_c]}\norm{\wh\J_{m,\cdot d_m}'\J_m} \asymp_P q_m^{\delta_{j,1}^{(m)}/2}. 
\end{equation}
Consider $m\in\cA_G$. Then (\ref{eqn:Thm9-1}) implies that
\begin{align*}
    &T^{-1}\sum_{t=1}^T\wh\Q_{m,\cdot c_m}'\widecheck\Y_t^{(m)}\wh\J_{m,\cdot d_m}\wh\J_{n,\cdot d_n}'\widecheck\Y_t^{(n)'}\wh\Q_{m,\cdot c_n} = \sum_{j=0}^{9}I_j, \text{ where}\\
    I_0 &:=  (\wh\J_{m,\cdot d_m}'\J_m \otimes \wh\Q_{m,\cdot c_m}'\Q_m)T^{-1}\sum_{t=1}^T\bb{E}(\vec(\G_t))^{\otimes 2}(\wh\J_{n,\cdot d_n}'\J_n \otimes \wh\Q_{n,\cdot c_n}'\Q_n)',\\
    I_1 &:=  (\wh\J_{m,\cdot d_m}'\J_m \otimes \wh\Q_{m,\cdot c_m}'\Q_m)T^{-1}\sum_{t=1}^T(\vec(\G_t)^{\otimes 2} - \bb{E}(\vec(\G_t))^{\otimes 2})(\wh\J_{n,\cdot d_n}'\J_n \otimes \wh\Q_{n,\cdot c_n}'\Q_n)',\\
    I_2 &:= (\wh\J_{m,\cdot d_m}'\J_m \otimes \wh\Q_{m,\cdot c_m}'\Q_m)T^{-1}\sum_{t=1}^T\vec(\G_t)\vec(\F_t^{(n)})'(\wh\J_{n,\cdot d_n}'\C_n \otimes \wh\Q_{n,\cdot c_n}'\R_n)',\\
    I_3 &:= (\wh\J_{m,\cdot d_m}'\J_m \otimes \wh\Q_{m,\cdot c_m}'\Q_m)T^{-1}\sum_{t=1}^T\vec(\G_t)\wh\J_{n,\cdot d_n}'\M_{q_n}\E_t^{(n)'}\M_{p_n}\wh\Q_{n,\cdot c_n},\\
    I_4 &:= (\wh\J_{m,\cdot d_m}'\C_m \otimes \wh\Q_{m,\cdot c_m}'\R_m)T^{-1}\sum_{t=1}^T\vec(\F_t^{(m)})\vec(\G_t)'(\wh\J_{n,\cdot d_n}'\J_n \otimes \wh\Q_{n,\cdot c_n}'\Q_n)',\\
    I_5 &:= (\wh\J_{m,\cdot d_m}'\C_m \otimes \wh\Q_{m,\cdot c_m}'\R_m)T^{-1}\sum_{t=1}^T\vec(\F_t^{(m)})\vec(\F_t^{(n)})'(\wh\J_{n,\cdot d_n}'\C_n \otimes \wh\Q_{n,\cdot c_n}'\R_n)',\\
    I_6 &:= (\wh\J_{m,\cdot d_m}'\C_m \otimes \wh\Q_{m,\cdot c_m}'\R_m)T^{-1}\sum_{t=1}^T\vec(\F_t^{(m)})\wh\J_{n,\cdot d_n}'\M_{q_n}\E_t^{(n)'}\M_{p_n}\wh\Q_{n,\cdot c_n},\\
    I_7 &:= \wh\Q_{m,\cdot c_m}'\M_{p_m}T^{-1}\sum_{t=1}^T\E_t^{(m)}\M_{q_m}\wh\J_{m,\cdot d_m}\vec(\G_t)'(\wh\J_{n,\cdot d_n}'\J_n \otimes \wh\Q_{n,\cdot c_n}'\Q_n)',\\
    I_8 &:= \wh\Q_{m,\cdot c_m}'\M_{p_m}T^{-1}\sum_{t=1}^T\E_t^{(m)}\M_{q_m}\wh\J_{m,\cdot d_m}\vec(\F_t^{(n)})'(\wh\J_{n,\cdot d_n}'\C_n \otimes \wh\Q_{n,\cdot c_n}'\R_n)',\\
    I_9 &:= \wh\Q_{m,\cdot c_m}'\M_{p_m}T^{-1}\sum_{t=1}^T\E_t^{(m)}\M_{q_m}\wh\J_{m,\cdot d_m}\wh\J_{n,\cdot d_n}'\M_{q_n}\E_t^{(n)'}\M_{p_n}\wh\Q_{n,\cdot c_n},
\end{align*}
with $I_0 = I_1=I_4=I_7=0$ if $n\notin \cA_G$. By Assumption (S1) on $T^{-1}\sum_{t=1}^T\bb{E}\vec(\G_t)^{\otimes 2}$, and that the set of vectors
$\{\e_{k_c,d_n}'\H_{j,n}'(\J_n'\J_n)^{1/2}\otimes \e_{k_r,c_n}'\H_{q,n}'(\Q_n'\Q_n)^{1/2}\}_{c_n\in[k_r],d_n\in[k_c]}$ is linearly independent by Assumption (L1),
for each $n\neq m$ with $n\in\cA_G$, there must exist $c_n\in[k_r], d_n\in[k_c]$ (through using (\ref{eqn:Thm9-eta_QJ}), so that we can replace $\wh\Q_{n,\cdot c_n}$ by $\Q_n(\Q_n'\Q_n)^{-1/2}\H_{q,n}\e_{k_r,c_n}$  and similarly $\wh\J_{n,\cdot d_n}$ by $\J_n(\J_n'\J_n)^{-1/2}\H_{j,n}\e_{k_c,d_n}$) such that 
\begin{align}
    \max_{c_m\in[k_r], d_m\in[k_c]}I_0^2 &\asymp_P \max_{c_m\in[k_r], d_m\in[k_c]} \norm{\e_{k_c,d_m}'\H_{j,m}'(\J_m'\J_m)^{1/2}\otimes \e_{k_r,c_m}'\H_{q,m}'(\Q_m'\Q_m)^{1/2}}^2\notag\\ &\cdot\norm{\e_{k_c,d_n}'\H_{j,n}'(\J_n'\J_n)^{1/2}\otimes \e_{k_r,c_n}'\H_{q,n}'(\Q_n'\Q_n)^{1/2}}^2\notag\\
    &\gtrsim_P p_m^{\delta_{q,1}^{(m)}}q_m^{\delta_{j,1}^{(m)}}\cdot p_n^{\delta_{q,k_r}^{(n)}}q_n^{\delta_{j,k_c}^{(n)}}, \label{eqn:Thm9-I0}
\end{align}
where the last line used (\ref{eqn:Thm9-2}), (\ref{eqn:Thm9-3}) and Assumption (L1). 
By Lemma \ref{lemma:prop_piecewise}-2 and \ref{lemma:prop_piecewise}-3, we also have
\begin{align*}
    \max_{c_m\in[k_r], d_m\in[k_c]}I_1^2 &\lesssim_P  p_m^{\delta_{q,1}^{(m)}}q_m^{\delta_{j,1}^{(m)}}\cdot p_n^{\delta_{q,1}^{(n)}}q_n^{\delta_{j,1}^{(n)}}R(T,d_{\text{fac},\star}),\;\; \max_{c_m\in[k_r], d_m\in[k_c]}I_2^2 \lesssim_P  p_m^{\delta_{q,1}^{(m)}}q_m^{\delta_{j,1}^{(m)}}\cdot p_n^{\delta_{r,1}^{(n)}}q_n^{\delta_{c,1}^{(n)}}R(T,d_{\text{fac},\star}),\\
    \max_{c_m\in[k_r], d_m\in[k_c]}I_4^2 &\lesssim_P p_m^{\delta_{r,1}^{(m)}}q_m^{\delta_{c,1}^{(m)}}\cdot p_n^{\delta_{q,1}^{(n)}}q_n^{\delta_{j,1}^{(n)}}R(T,d_{\text{fac},\star}), \;\;
    \max_{c_m\in[k_r], d_m\in[k_c]}I_5^2
    \lesssim_P p_m^{\delta_{r,1}^{(m)}}q_m^{\delta_{c,1}^{(m)}}\cdot p_n^{\delta_{r,1}^{(n)}}q_n^{\delta_{c,1}^{(n)}}R(T,d_{\text{fac},\star}).
\end{align*}
Using similar proof for Lemma \ref{lemma:bound_Delta}, and the fact that 
\[\max_{i\in[p_m],j\in[k_r]}|(\wh\Q_{m} - \Q_m(\Q_m'\Q_m)^{-1/2}\H_{q,m})_{ij}| = O_P(\norm{\wh\Q_{m} - \Q_m(\Q_m'\Q_m)^{-1/2}\H_{q,m}}) = o_P(\eta_{Q,m}) = o_P(1),\]
and similarly 
\[\max_{i\in[q_m],j\in[k_c]}|(\wh\J_{m} - \J_m(\J_m'\J_m)^{-1/2}\H_{j,m})_{ij}| = O_P(\norm{\wh\J_{m} - \J_m(\J_m'\J_m)^{-1/2}\H_{j,m}}) = o_P(\eta_{J,m}) = o_P(1),\]
we can prove the following in parallel to (\ref{eqn:E1u}), (\ref{eqn:Em11En}) and (\ref{eqn:Em11Em})(details omitted):
\begin{align}
    \norm{T^{-1}\sum_{t=1}^T\E_t^{(m)}\wh\v_m\u_t'}_F^2 &= O_P(T^{-1}p_m),\label{eqn:Thm9-d33}\\
    \norm{T^{-1}\sum_{t=1}^T\E_t^{(m)}\wh\v_m\wh\v_n'\E_t^{(n)'}}_F^2 &= O_P(T^{-1}p_mp_n),\label{eqn:Thm9-d34}\\
    \norm{T^{-1}\sum_{t=1}^T\E_t^{(m)}\wh\v_m\wh\v_m'\E_t^{(m)'}}_F^2 &= O_P(p_m + T^{-1}p_m^2),\label{eqn:Thm9-d35}
\end{align}
where $\{\u_t\}$ is independent of $\{\E_t^{(m)}\}$ for each $m\in[M]$, and $\wh{\v}_m,\wh\v_n$ are unit vectors potentially dependent on $\{\E_t^{(m)}\}$ for some $m\in[M]$, but are consistent to some non-random unit vectors $\v_m$ and $\v_n$ respectively. 
With (\ref{eqn:Thm9-d33}) and (\ref{eqn:Thm9-d34}), we can show that
\begin{align*}
    \max_{c_m\in[k_r],d_m\in[k_c]} I_3^2 &\asymp_P p_m^{\delta_{q,1}^{(m)}}q_m^{\delta_{j,1}^{(m)}}\cdot T^{-1}p_n, \;\;  \max_{c_m\in[k_r],d_m\in[k_c]} I_6^2 \asymp_P p_m^{\delta_{r,1}^{(m)}}q_m^{\delta_{c,1}^{(m)}}\cdot T^{-1}p_n,\\
    \max_{c_m\in[k_r],d_m\in[k_c]} I_7^2 &\asymp_P T^{-1}p_m\cdot p_n^{\delta_{q,1}^{(n)}}q_n^{\delta_{j,1}^{(n)}}, \;\; \max_{c_m\in[k_r],d_m\in[k_c]} I_8^2 \asymp_P T^{-1}p_m\cdot p_n^{\delta_{r,1}^{(n)}}q_n^{\delta_{c,1}^{(n)}},\\
    \max_{c_m\in[k_r],d_m\in[k_c]} I_9^2 &\asymp_P T^{-1}p_mp_n.
\end{align*}
Using the rate assumption in the statement of the theorem, together with Assumption (R2), (\ref{eqn:Thm9-I0}) and the  rates for $I_1^2$ to $I_9^2$ result in
\begin{align*}
    &\sum_{n\neq m}\max_{c_m,c_m\in[k_r]}\max_{d_m,d_n\in[k_c]}\big\{T^{-1}\sum_{t=1}^T\wh\Q_{m,\cdot c_m}'\widecheck\Y_t^{(m)}\wh\J_{m,\cdot d_m}\wh\J_{n,\cdot d_n}'\widecheck\Y_t^{(n)'}\wh\Q_{n,\cdot c_n}\big\}^2 \gtrsim_P p_m^{\delta_{q,1}^{(m)}}q_m^{\delta_{j,1}^{(m)}}\max_{n\neq m,n\in\cA_G}p_n^{\delta_{q,k_r}^{(n)}}q_n^{\delta_{j,k_c}^{(n)}}.
\end{align*}
On the other hand, if $m=k\notin\cA_G$, then $I_0=I_1=I_2=I_3=0$, and the rates for $I_4^2$ to $I_9^2$ lead to 
\begin{align*}
    &\sum_{n\neq k}\max_{c_k,c_n\in[k_r]}\max_{d_k,d_n\in[k_c]}\big\{T^{-1}\sum_{t=1}^T\wh\Q_{k,\cdot c_k}'\widecheck\Y_t^{(k)}\wh\J_{k,\cdot d_k}\wh\J_{n,\cdot d_n}'\widecheck\Y_t^{(n)'}\wh\Q_{n,\cdot c_n}\big\}^2\\ &\lesssim_P p_k^{\delta_{r,1}^{(k)}}q_k^{\delta_{c,1}^{(k)}}\max(\max_{n\in\cA_G}p_n^{\delta_{q,1}^{(n)}}q_n^{\delta_{j,1}^{(n)}}, \max_{n\neq k, n\notin \cA_G} p_n^{\delta_{r,1}^{(n)}}q_n^{\delta_{c,1}^{(n)}})R(T,d_{\text{fac},\star}).
\end{align*}
Now we rewrite $\Y_t^{(m)}$ as
\[\widecheck\Y_t^{(m)} = \U_m\H_t^{(m)}\V_m' + \M_{p_m}\E_t^{(m)}\M_{q_m},\]
where $\U_m := (\Q_m,\R_m)$, $\V_m := (\J_m,\C_m)$ and $\H_t^{(m)} := \diag(\G_t,\F_t^{(m)})$. We then have $\text{rank}(\U_m) \leq k_r+k_r^{(m)}$ and $\text{rank}(\V_m) \leq k_c + k_c^{(m)}$. Repeating our estimation methodology for just the local common components for individual groups using data $\{\widecheck\Y_t^{(m)}\}$, we can use the same proof as Theorem \ref{thm:common_comp_global_local_init} to arrive at $\norm{\wh\U_m - \U_m(\U_m'\U_m)^{-1/2}\H_{u,m}}_F^2 = o_P(1)$ and $\norm{\wh\V_m - \V_m(\V_m'\V_m)^{-1/2}\H_{v,m}}_F^2 = o_P(1)$, where $\H_{u,m}$ and $\H_{v,m}$ are two constant order invertible matrices. Then using (\ref{eqn:Thm9-d35}) and Assumption (R2) and (S1), similar to the treatment before, we have for $m\in\cA_G$,
\begin{align*}
    &\max_{c_m\in[k_r+k_r^{(m)}], d_m\in[k_c+k_c^{(m)}]} \big(T^{-1}\sum_{t=1}^T(\wh\U_{m,\cdot c_m}'\Q_m\G_t\J_m'\wh\V_{m,\cdot d_m} + \wh\U_{m,\cdot c_m}'\R_m\F_t^{(m)}\C_m'\wh\V_{m,\cdot d_m}\\ &\hspace{1.3in}+ \wh\U_{m,\cdot c_m}'\M_{p_m}\E_t^{(m)}\M_{q_m}\wh\V_{m,\cdot d_m})^{2}\big)^2\\
    &\gtrsim_P p_m^{2\delta_{q,1}^{(m)}}q_m^{2\delta_{j,1}^{(m)}} - p_m^{2\delta_{r,1}^{(m)}}q_m^{2\delta_{c,1}^{(m)}} - p_m - T^{-1}p_m^2 \asymp_P p_m^{2\delta_{q,1}^{(m)}}q_m^{2\delta_{j,1}^{(m)}}.
\end{align*}
At the same time, if $m\notin\cA_G$, 
\begin{align*}
    &\max_{c_m\in[k_r+k_r^{(m)}], d_m\in[k_c+k_c^{(m)}]} \big(T^{-1}\sum_{t=1}^T(\wh\U_{m,\cdot c_m}'\R_m\F_t^{(m)}\C_m'\wh\V_{m,\cdot d_m} + \wh\U_{m,\cdot c_m}'\M_{p_m}\E_t^{(m)}\M_{q_m}\wh\V_{m,\cdot d_m})^{2}\big)^2\\
    &\gtrsim_P  p_m^{2\delta_{r,1}^{(m)}}q_m^{2\delta_{c,1}^{(m)}} - p_m - T^{-1}p_m^2 \asymp_P p_m^{2\delta_{r,1}^{(m)}}q_m^{2\delta_{c,1}^{(m)}}.
\end{align*}
This completes the proof of the theorem. $\square$

\textbf{\textit{Proof of Theorem~\ref{thm:rank_estimation}.}} 
Consider $\wh{r}_1$. We omit the details for $\wh{\ell}_1$ since its proof is parallel to that for $\wh{r}_1$. It is not difficult to see that, through the finiteness of $M$,
\begin{equation}\label{eqn:Thm10-1}
\max_{n\in[M]}\max_{n\neq m}\sigma_{r_1}^2\Big\{T^{-1}\sum_{t=1}^T\wh{\a}_t^{(m)}\wh{\a}_t^{(n)'}\Big\} \lesssim \max_{m\in[M]}\lambda_{i}(\wh\bOmega_{\A_1,m}) \lesssim \max_{m\in[M]}\max_{n\neq m}\sigma_{1}^2\Big\{T^{-1}\sum_{t=1}^T\wh{\a}_t^{(m)}\wh{\a}_t^{(n)'}\Big\}, \;\; i\in[r_1],
\end{equation}
where $\sigma_i(\A)$ denotes the $i$th largest singular value of the matrix $\A$. At the same time, for $i\in[r_1]$,
\begin{equation}\label{eqn:Thm10-2}
\sigma_{i}\Big\{\A_{1m}\big(T^{-1}\sum_{t=1}^T\balpha_t^{\otimes 2}\big)\A_{1n}'\Big\} - \sum_{j=2}^5I_j\leq \sigma_{i}\Big\{T^{-1}\sum_{t=1}^T\wh{\a}_t^{(m)}\wh{\a}_t^{(n)'}\Big\} \leq
\sigma_{1}\Big\{\A_{1m}\big(T^{-1}\sum_{t=1}^T\balpha_t^{\otimes 2}\big)\A_{1n}'\Big\} + \sum_{j=2}^5I_j,
\end{equation}
where
\begin{align*}
    I_2 &:= \sigma_1\Big\{\A_{2m}\big(T^{-1}\sum_{t=1}^T\balpha_t^{(m)}\balpha_t^{(n)'}\big)\A_{2n}'\Big\}, \;\; I_3 := \sigma_1\Big\{ T^{-1}\sum_{t=1}^T q_m^{-1}\M_{p_m}\E_t^{(m)}\1_{q_m}\big(\A_{1n}\balpha_t + \A_{2n}\balpha_t^{(n)}\big)' \Big\},\\
    I_4 &:= \sigma_1\Big\{ T^{-1}\sum_{t=1}^T \big(\A_{1m}\balpha_t + \A_{2m}\balpha_t^{(m)}\big)q_n^{-1}\1_{q_n}' \E_t^{(n)'}\M_{p_n} \Big\},\\
    I_5 &:= \sigma_1\big\{T^{-1}q_m^{-1}q_n^{-1}\sum_{t=1}^T\M_{p_m}\E_t^{(m)}\1_{q_m}\1_{q_n}'\E_t^{(n)'}\M_{p_n}\big\}.
\end{align*}
Firstly, for $i\in[r_1]$, using (\ref{eqn:sample_covariance_G}) and Assumption (S1),
\begin{align*}
\sigma_i\Big\{\A_{1m}\big(T^{-1}\sum_{t=1}^T\balpha_t^{\otimes 2}\big)\A_{1n}'\Big\} &\geq \lambda_i^{1/2}(\A_{1m}'\A_{1m})\lambda_{r_1}^{1/2}(\A_{1n}'\A_{1n})\lambda_{r_1}(T^{-1}\sum_{t=1}^T\balpha_t^{\otimes 2})\\&\gtrsim_P p_m^{\gamma_r/2}p_n^{\gamma_r/2}(1 + R(T,d_{r,\star})^{1/2}).
\end{align*}
Using (ME1), and  (\ref{eqn:sample_cross_moment_G}) on $I_2$, (\ref{eqn:E1u}) on $I_3$ and $I_4$ and (\ref{eqn:Em11En}) on $I_5$, 
\begin{align*}
    I_2 &:= O_P(p_m^{\gamma_r^{(m)}/2}p_n^{\gamma_r^{(n)}/2}R(T,d_{r,\star})^{1/2}) = o_P(p_m^{\gamma_r/2}p_n^{\gamma_r/2}),\\
    I_3 &:= O_P(T^{-1/2}p_m^{1/2}q_m^{-1/2}\cdot (p_n^{\gamma_r/2} + p_n^{\gamma_r^{(n)}/2})) = o_P(p_m^{\gamma_r/2}p_n^{\gamma_r/2}),\\
    I_4 &:= O_P(T^{-1/2}p_n^{1/2}q_n^{-1/2}\cdot(p_m^{\gamma_r/2} + p_m^{\gamma_r^{(m)}/2})) = o_P(p_m^{\gamma_r/2}p_n^{\gamma_r/2}),\\
    I_5 &:= O_P(T^{-1/2}p_m^{1/2}p_n^{1/2}q_m^{-1/2}q_n^{-1/2}) = o_P(p_m^{\gamma_r/2}p_n^{\gamma_r/2}),
\end{align*}
where we use (R1) as well for concluding that all the rates above are $o_P(p_m^{\gamma_r/2}p_n^{\gamma_r/2})$.
Hence (\ref{eqn:Thm10-2}) implies that
for $i\in[r_1]$,
\[
\max_{m,n\in[M], n\neq m}\sigma_i^2\Big\{T^{-1}\sum_{t=1}^T\wh\a_t^{(m)}\wh\a_t^{(n)'}\Big\} \asymp_P \max_{m,n\in[M],n\neq m}p_m^{\gamma_r}p_n^{\gamma_r},
\]
so that by (\ref{eqn:Thm10-1}), for $i\in[r_1]$,
\begin{equation}\label{eqn:Thm10-3}
\max_{m\in[M]}\lambda_i(\wh\bOmega_{\A_1,m}) \asymp_P \max_{m,n\in[M],n\neq m}p_m^{\gamma_r}p_n^{\gamma_r}.
\end{equation}
On the other hand, we can write
\begin{align*}
    \wh\bOmega_{\A_1,m} &=: (M-1)^{-1}\sum_{n\neq m} \Big(\A_{1m}\big(T^{-1}\sum_{t=1}^T\balpha_t^{\otimes 2}\big)\A_{1n}' + R_{m,n}\Big)^{\otimes 2}\\
    &= (M-1)^{-1}\sum_{n\neq m}\Big(\A_{1m}\big(T^{-1}\sum_{t=1}^T\balpha_t^{\otimes 2}\big)\A_{1n}'\A_{1n}\big(T^{-1}\sum_{t=1}^T\balpha_t^{\otimes 2}\big)\A_{1m}' + \A_{1m}\big(T^{-1}\sum_{t=1}^T\balpha_t^{\otimes 2}\big)\A_{1n}'R_{m,n}'\\ 
    &+ R_{m,n}\A_{1n}\big(T^{-1}\sum_{t=1}^T\balpha_t^{\otimes 2}\big)\A_{1m}' + R_{m,n}R_{m,n}'\Big), 
\end{align*}
so that for $i\in\{r_1+1,\ldots,p_m\}$,
\begin{align}
    \lambda_i(\wh\bOmega_{\A_1,m}) &\leq \lambda_1\Big\{(M-1)^{-1}\sum_{n\neq m}R_{m,n}R_{m,n}'\Big\} \leq \max_{n\neq m}\sigma_1^2(R_{m,n})\notag\\
    &\lesssim \max_{n\neq m}\sum_{j=2}^5I_j^2 = O_P(\max_{n\neq m}p_m^{\gamma_r^{(m)}}p_n^{\gamma_r^{(n)}}R(T,d_{r,\star})), \label{eqn:Thm10-4}
\end{align}
where the last equality used Assumption (R1). Hence combining (\ref{eqn:Thm10-3}) and (\ref{eqn:Thm10-4}), we have
\begin{align*}
    \frac{\max_{m\in[M]}\lambda_{i+1}(\wh\bOmega_{\A_1,m})}{\max_{m\in[M]}\lambda_i(\wh\bOmega_{\A_1,m})} \left\{
  \begin{array}{ll}
    \asymp_P 1, & \hbox{$i\in[r_1-1]$;} \\
    \lesssim_P \frac{\max_{m\in[M],n\neq m}p_m^{\gamma_r^{(m)}}p_n^{\gamma_r^{(n)}}R(T,d_{r,\star})}{\max_{m\in[M],n\neq m}p_m^{\gamma_r}p_n^{\gamma_r}}  = o_P(1), & \hbox{$i=r_1$,}
  \end{array}
\right. 
\end{align*}
where the equation for the case $i=r_1$ used Assumption (R1). This implies $P(\wh{r}_1 \geq r_1) \rightarrow 1$. It is also clear that using (\ref{eqn:Thm10-3}) and (\ref{eqn:Thm10-4}), the ratio
\begin{align*}
    \frac{\max_{m\in[M]}\lambda_{i+1}(\wh\bOmega_{\A_1,m}) + \max_{m\neq n\in[M]}Cp_m^{\gamma_r^{(m)}}p_n^{\gamma_r^{(n)}}R(T,d_{r,\star})}{\max_{m\in[M]}\lambda_i(\wh\bOmega_{\A_1,m}) + \max_{m\neq n\in[M]}Cp_m^{\gamma_r^{(m)}}p_n^{\gamma_r^{(n)}}R(T,d_{r,\star})} \left\{
  \begin{array}{ll}
    \asymp_P 1, & \hbox{$i\in[r_1-1]$;} \\
     = o_P(1), & \hbox{$i=r_1$,} \\
    \asymp_P 1, & \hbox{$i=r_1+1,\ldots,p_m$,}
  \end{array}
\right.  
\end{align*}
so that $P(\wh{r}_1 = r_1) \rightarrow 1$, which aligns with the statement of adding $\max_{m\neq n\in[M]}CT^{-1}p_mp_n$ to the ratio when all factors are strong and all processes have short memory.

For $\wh{r}_2^{(m)}$ (We omit $\wh{\ell}_2^{(m)}$ since the proof is parallel), given $i\in[r_2^{(m)}]$, consider 
\begin{align}
   &J_1 - \sum_{i=2}^6J_i  \lesssim \sigma_i\Big(T^{-1}\sum_{t=1}^T(\wh\a_t^{(m)})^{\otimes 2}\wh\A_{1m}^{\perp}\Big) \lesssim J_1 + \sum_{i=2}^6J_i, \; \text{ where } \label{eqn:Thm10-5}\\
       J_1 &:= \sigma_i\Big(\A_{2m}\big(T^{-1}\sum_{t=1}^T(\balpha_t^{(m)})^{\otimes 2}\big)\A_{2m}'\A_{1m}^{\perp}\Big), \notag\\
       J_2 &:= \sigma_1\Big(\A_{1m}\big(T^{-1}\sum_{t=1}^T\balpha_t^{\otimes 2}\big)(\A_{1m}'\A_{1m})^{1/2}(\H_{\A_1,m}^{-1})'(\check\A_{1m}\H_{A_1,m} - \wh\A_{1m})'\wh\A_{1m}^{\perp}\Big),\notag\\
       J_3 &:= \sigma_i\Big(\A_{2m}\big(T^{-1}\sum_{t=1}^T(\balpha_t^{(m)})^{\otimes 2}\big)\A_{2m}'(\wh\A_{1m}^{\perp} - \A_{1m}^{\perp})\Big), \notag\\
       J_4 &:= \sigma_1\Big(T^{-1}\sum_{t=1}^Tq_m^{-1}\M_{p_m}\E_t^{(m)}\1_{q_m}(\A_{1m}\balpha_t + \A_{2m}\balpha_t^{(m)})'\wh\A_{1m}^{\perp}\Big),\notag\\
       J_5 &:= \sigma_1\Big(T^{-1}\sum_{t=1}^T (\A_{1m}\balpha_t + \A_{2m}\balpha_t^{(m)})\1_{q_m}'\E_t^{(m)'}\M_{p_m}q_m^{-1}\wh\A_{1m}^{\perp}\Big), \notag\\
       J_6 &:= \sigma_1\Big(T^{-1}q_m^{-2}\sum_{t=1}^T\M_{p_m}\E_t^{(m)}\1_{q_m}\1_{q_m}'\E_t^{(m)'}\M_{p_m}\wh\A_{1m}^{\perp}\Big). \notag
\end{align}
For $i\in[r_2^{(m)}]$, using (\ref{eqn:sample_covariance_G}) and Assumption (S1) and (L1),
\begin{align*}
    J_1 &= \sigma_i\Big(\big(T^{-1}\sum_{t=1}^T(\balpha_t^{(m)})^{\otimes 2}\big)(\A_{2m}'\A_{1m}^{\perp})^{\otimes 2}\Big) \geq \sigma_i^2(\A_{2m}'\A_{1m}^{\perp})\sigma_{r_2^{(m)}}\Big(T^{-1}\sum_{t=1}^T(\balpha_t^{(m)})^{\otimes 2}\Big) 
    \asymp_P p_m^{\gamma_r^{(m)}},
\end{align*}
Also,
\begin{align*}
    J_2 &= O_P(p_m^{\gamma_r}\eta_{A_{1m}}^{1/2}) = o_P(p_m^{\gamma_r^{(m)}}),\\
    J_3 &= O_P(p_m^{\gamma_r^{(m)}}\norm{\wh\A_{1m}^{\perp} - \A_{1m}^{\perp}}) = O_P(p_m^{\gamma_r^{(m)}}\norm{\wh{\A}_{1m} - \check\A_{1m}\H_{A_1,m}}) = o_P(p_m^{\gamma_r^{(m)}}),\\
    J_4, J_5 &= O_P(T^{-1/2}p_m^{1/2}q_m^{-1/2}\cdot(p_m^{\gamma_r/2} + p_m^{\gamma_r^{(m)}/2})) = o_P(p_m^{\gamma_r^{(m)}}),\\
    J_6 &= O_P(p_m^{1/2}q_m^{-1} + T^{-1/2}p_mq_m^{-1}) = o_P(p_m^{\gamma_r^{(m)}}),
\end{align*}
where $\eta_{A_{1m}}$ is the rate defined in Theorem \ref{thm:initial_A1m}, and the equation for $J_2$ used the assumption in the statement of the theorem. We use (\ref{eqn:E1u}) for $J_4$ and $J_5$ and that $\eta_{A_{1m}} = o(\nu_{r,m}^2)$ and (R1) to conclude that they are $o_P(p_m^{\gamma_r^{(m)}})$, while $J_6$ uses (\ref{eqn:Em11Em}).

On the other hand, for $i\in\{r_2^{(m)+1},\ldots,p_m\}$,
\begin{align}
    \sigma_i\Big(T^{-1}\sum_{t=1}^T(\wh\a_t^{(m)})^{\otimes 2}\wh\A_{1m}^{\perp}\Big) \lesssim_P\sum_{i=2}^6 J_i = o_P(p_m^{\gamma_r^{(m)}}). \label{eqn:Thm10-6}
\end{align}
The rates for $J_1$ to $J_6$, combined with (\ref{eqn:Thm10-5}) and (\ref{eqn:Thm10-6}), allow us to conclude that
\begin{align*}
    \frac{\lambda_{j+1}(\wh\bOmega_{A_2,m})}{\lambda_j(\wh\bOmega_{A_2,m})} \left\{
  \begin{array}{ll}
    \asymp_P 1, & \hbox{$j\in[r_2^{(m)}-1]$;} \\
    = o_P(1), & \hbox{$j=r_2^{(m)}$,}
  \end{array}
\right.
\end{align*}
so that $P(\wh{r}_2^{(m)} \geq r_2^{(m)}) \rightarrow 1$. It is also clear that when all factors are strong and all processes have short memory, $\eta_{A_{1m}} = T^{-1}$, so that the rate in (\ref{eqn:Thm10-6}) is $O_P(T^{-1/2}p_m)$, and 
the ratio
\begin{align*}
    \frac{\lambda_{j+1}(\wh\bOmega_{A_2,m}) + T^{-1}p_m^2}{\lambda_j(\wh\bOmega_{A_2,m}) + T^{-1}p_m^2} \left\{
  \begin{array}{ll}
    \asymp_P 1, & \hbox{$j\in[r_2^{(m)}-1]$;} \\
    = o_P(1), & \hbox{$j=r_2^{(m)}$;}\\
    \asymp_P 1 & \hbox{$j=r_2^{(m)}+1,\ldots,p_m$,}
  \end{array}
\right.
\end{align*}
so that $P(\wh{r}_2^{(m)} = r_2^{(m)}) \rightarrow 1$.

Now for $\wh{k}_r$ (we omit the details for $\wh{k}_c$ since it involves parallel arguments), consider for $i\in[k_r]$,
\begin{align}
  \max_{m\in[M]}\max_{n\neq m}\sigma_i^2\Big\{T^{-1}\sum_{t=1}^T\check\Y_t^{(m)}\v_b^{(m)}\v_b^{(n)'}\check\Y_t^{(n)'}\Big\} &\lesssim \max_{m\in[M]}\lambda_i(\X_{Q,m}) \notag\\&\lesssim \max_{m\in[M]}\max_{n\neq m}\sigma_1^2\Big\{T^{-1}\sum_{t=1}^T\check\Y_t^{(m)}\v_b^{(m)}\v_b^{(n)'}\check\Y_t^{(n)'}\Big\}.\label{eqn:Thm10-7}
\end{align}
At the same time, for $i\in[k_r]$, 
\begin{align*}
    K_{1,i} - \sum_{j=2}^9K_j &\leq \sigma_i\Big\{T^{-1}\sum_{t=1}^T\check\Y_t^{(m)}\v_b^{(m)}\v_b^{(n)'}\check\Y_t^{(n)'}\Big\} \leq K_{1,i} + \sum_{j=2}^9K_j, \; \text{ where}\\
    K_{1,i} &:= \sigma_i\Big( (\v_b^{(m)'}\J_m \otimes \Q_m)T^{-1}\sum_{t=1}^T\vec(\G_t)^{\otimes 2}(\v_b^{(n)'}\J_n \otimes \Q_n)'\Big),\\
    K_2 &:= \sigma_1\Big( (\v_b^{(m)'}\J_m \otimes \Q_m)T^{-1}\sum_{t=1}^T\vec(\G_t)\vec(\F_t^{(n)})'(\v_b^{(n)'}\C_n \otimes \R_n)'\Big),\\
    K_3 &:= \sigma_1\Big( (\v_b^{(m)'}\J_m \otimes \Q_m)T^{-1}\sum_{t=1}^T\vec(\G_t)\v_b^{(n)'}\M_{q_n}\E_t^{(n)'}\M_{p_n}\Big),\\
    K_4 &:= \sigma_1\Big( (\v_b^{(m)'}\C_m \otimes \R_m)T^{-1}\sum_{t=1}^T\vec(\F_t^{(m)})\vec(\G_t)'(\v_b^{(n)'}\J_n \otimes \Q_n)'\Big),\\
    K_5 &:= \sigma_1\Big( (\v_b^{(m)'}\C_m \otimes \R_m)T^{-1}\sum_{t=1}^T\vec(\F_t^{(m)})\vec(\F_t^{(n)})'(\v_b^{(n)'}\C_n \otimes \R_n)'\Big),\\
    K_6 &:=\sigma_1\Big( (\v_b^{(m)'}\C_m \otimes \R_m)T^{-1}\sum_{t=1}^T\vec(\F_t^{(m)})\v_b^{(n)'}\M_{q_n}\E_t^{(n)'}\M_{p_n}\Big),\\
    K_7 &:= \sigma_1\Big( \M_{p_m}\cdot T^{-1}\sum_{t=1}^T\E_t^{(m)}\M_{q_m}\v_b^{(m)}\vec(\G_t)'(\v_b^{(n)'}\J_n \otimes \Q_n)' \Big),\\
    K_8 &:= \sigma_1\Big( \M_{p_m}\cdot T^{-1}\sum_{t=1}^T\E_t^{(m)}\M_{q_m}\v_b^{(m)}\vec(\F_t^{(m)})'(\v_b^{(n)'}\C_n \otimes \R_n)' \Big),\\
    K_9 &:= \sigma_1\Big( \M_{p_m}\cdot T^{-1}\sum_{t=1}^T\E_t^{(m)}\M_{q_m}\v_b^{(m)}\v_b^{(n)'}\M_{q_n}\E_t^{(n)'}\cdot \M_{p_n}\Big).
\end{align*}
To bound the above terms, firstly, we have $\|\v_b^{(m)'}\J_m\| \asymp q_m^{\delta_{j,1}^{(m)}/2}$ and $\|\v_a^{(m)'}\Q_m\| \asymp p_m^{\delta_{q,1}/2}$ by construction, and hence using (\ref{eqn:sample_covariance_G}) and Assumption (S1), for $i\in[k_r]$,
\begin{align*}
    K_{1,i} &\gtrsim_P \lambda_1(\J_n'\v_b^{(n)}\v_b^{(m)}\J_m')\lambda_i(\Q_n'\Q_m)\lambda_{k_r+k_c}\Big(T^{-1}\sum_{t=1}^T\vec(\G_t)^{\otimes 2}\Big)\\
    &\gtrsim_P q_m^{\delta_{j,1}^{(m)}/2}q_n^{\delta_{j,1}^{(n)}/2}p_m^{\delta_{q,i}^{(m)}/2}p_n^{\delta_{q,k_r}^{(m)}/2}.
\end{align*}
Also, using (\ref{eqn:sample_cross_moment_G}), (\ref{eqn:Thm9-d33}) and (\ref{eqn:Thm9-d34}),
\begin{align*}
    K_2 &= O_P(R(T,d_{\text{fac},\star})^{1/2}q_m^{\delta_{j,1}^{(m)}/2}q_n^{\delta_{c,1}^{(n)}/2}p_m^{\delta_{q,1}^{(m)}/2}p_n^{\delta_{r,1}^{(n)}/2}), &\; K_3 &= O_P(T^{-1/2}p_n^{1/2}q_n^{1/2}\cdot q_m^{\delta_{j,1}^{(m)}/2}p_m^{\delta_{q,1}^{(m)}/2}),\\
    K_4 &= O_P(R(T,d_{\text{fac},\star})^{1/2}q_m^{\delta_{c,1}^{(m)}/2}q_n^{\delta_{j,1}^{(n)}/2}p_m^{\delta_{r,1}^{(m)}/2}p_n^{\delta_{q,1}^{(n)}/2}), & K_5 &= O_P(R(T,d_{\text{fac},\star})^{1/2}q_m^{\delta_{c,1}^{(m)}/2}q_n^{\delta_{c,1}^{(n)}/2}p_m^{\delta_{r,1}^{(m)}/2}p_n^{\delta_{r,1}^{(n)}/2}),\\
    K_6 &= O_P(T^{-1/2}p_n^{1/2}q_n^{1/2}\cdot q_m^{\delta_{c,1}^{(m)}/2}p_m^{\delta_{r,1}^{(m)}/2}), & K_7 &= O_P(T^{-1/2}p_m^{1/2}q_m^{1/2}\cdot q_n^{\delta_{j,1}^{(n)}/2}p_n^{\delta_{q,1}^{(n)}/2}),\\
    K_8 &= O_P(T^{-1/2}p_m^{1/2}q_m^{1/2}\cdot q_n^{\delta_{c,1}^{(n)}/2}p_n^{\delta_{r,1}^{(n)}/2}), & K_9 &= O_P(T^{-1/2}p_m^{1/2}p_n^{1/2}q_m^{1/2}q_n^{1/2}).
\end{align*}
By the assumption in the statement of the theorem, we also have for $i\in[k_r]$,
\[\max_{m,n\in[M],n\neq m,j=2,\ldots,9}K_j^2 = o_P(\max_{m,n\in[M],n\neq m}q_m^{\delta_{j,1}^{(m)}}q_n^{\delta_{j,1}^{(n)}}p_m^{\delta_{q,i}^{(m)}}p_n^{\delta_{q,k_r}^{(m)}}) \lesssim_P \max_{m,n\in[M],n\neq m}K_{1,i}^2\asymp_P \max_{m\in[M]}\lambda_i(\X_{Q,m}),\]
after using (\ref{eqn:Thm10-7}).
On the other hand, for $i\in\{k_r+1,\ldots,p_m\}$,
\begin{align*}
    \max_{m\in[M]}\lambda_i(\X_{Q,m}) &\lesssim_P \max_{m,n\in[M],n\neq m,j=2,\ldots,9}K_j^2\\
    &= O_P\Big(\max_{m,n\in[M],n\neq m}\big(T^{-1}p_mp_nq_mq_n + R(T,d_{\text{fac},\star})p_m^{\delta_{q,1}^{(m)}}q_m^{\delta_{j,1}^{(m)}} p_n^{\delta_{q,1}^{(n)}}q_n^{\delta_{j,1}^{(n)}}\big)\Big).
\end{align*}
With these intermediate results, we can conclude that for $i\in[k_r-1]$,
\begin{align}
    &\frac{\max_{m\in[M]}\lambda_{i+1}(\X_{Q,m})}{\max_{m\in[M]}\lambda_i(\X_{Q,m})} 
    \gtrsim_P \frac{\max_{m,n\in[M],n\neq m}q_m^{\delta_{j,1}^{(m)}}q_n^{\delta_{j,1}^{(n)}}p_m^{\delta_{q,i+1}^{(m)}}p_n^{\delta_{q,k_r}^{(m)}}}{\max_{m,n\in[M],n\neq m}q_m^{\delta_{j,1}^{(m)}}q_n^{\delta_{j,1}^{(n)}}p_m^{\delta_{q,1}^{(m)}}p_n^{\delta_{q,1}^{(m)}}}\notag\\
    &\succ_P \frac{\max_{m,n\in[M],n\neq m}\big(T^{-1}p_mp_nq_mq_n + R(T,d_{\text{fac},\star})p_m^{\delta_{q,1}^{(m)}}q_m^{\delta_{j,1}^{(m)}} p_n^{\delta_{q,1}^{(n)}}q_n^{\delta_{j,1}^{(n)}}\big)}{\max_{m,n\in[M],n\neq m}q_m^{\delta_{j,1}^{(m)}}q_n^{\delta_{j,1}^{(n)}}p_m^{\delta_{q,k_r}^{(m)}}p_n^{\delta_{q,k_r}^{(m)}}} \;\; (=o_P(1))\notag\\
    &\gtrsim_P \frac{\max_{m\in[M]}\lambda_{k_r+1}(\X_{Q,m})}{\max_{m\in[M]}\lambda_{k_r}(\X_{Q,m})}, \label{eqn:Thm10-8}
\end{align}
where the second line is by the rate assumption in the statement of the theorem. The above implies \linebreak $P(\wh{k}_r \geq k_r) \rightarrow 1$. When all factors are strong and all processes have short memory, then for $i\in\{k_r+1,\ldots,p_m\}$, 
\[\max_{m\in[M]}\lambda_i(\X_{Q,m}) = O_P(\max_{m,n\in[M],n\neq m}T^{-1}p_mp_nq_mq_n),\]
so we can easily see that 
\[\frac{\max_{m\in[M]}\lambda_{i+1}(\X_{Q,m}) + \max_{m,n\in[M],n\neq m}CT^{-1}p_mp_nq_mq_n}{\max_{m\in[M]}\lambda_i(\X_{Q,m}) + \max_{m,n\in[M],n\neq m}CT^{-1}p_mp_nq_mq_n} \asymp_P 1 \]
for $i\in\{k_r+1,\ldots,p_m\}$. This, together with (\ref{eqn:Thm10-8}), proves $P(\wh{k}_r = k_r) \rightarrow 1$.

For $\wh{k}_c^{(m)}$ (the proof for $\wh{k}_r^{(m)}$ runs parallel and is omitted), consider the decomposition of $\X_{C,m}$ (\ref{eqn:XCm-decomposition}), where we can see from the rate in (\ref{eqn: local_C_J1_eigenvalues}) that for $i\in[k_c^{(m)}]$, 
\[\lambda_i(\X_{C,m}) \asymp_P q_m^{\delta_{c,i}^{(m)}}p_m^{\delta_{r,1}^{(m)}},\]
while for $i\in\{k_c^{(m)}+1,\ldots,q_m\}$, the rates $\norm{\cJ_5}_F^2$ and $\norm{\cJ_6}_F^2$ from (\ref{eqn: local_C_J5}) and (\ref{eqn: local_C_J6}) respectively dominated, and hence we have
\[\lambda_i(\X_{C,m}) = O_P(q_m^{1-\delta_{c,1}^{(m)}} + T^{-1/2}q_mp_m^{1/2} + q_m^{(1+\delta_{j,1}^{(m)})/2}p_m^{(1+\delta_{q,1}^{(m)})/2}\eta_{Q,m}^{1/2}).\]
With these two intermediate results, we can conclude that for $i\in[k_c^{(m)}-1]$,
\begin{align}
    \frac{\lambda_{i+1}(\X_{C,m})}{\lambda_i(\X_{C,m})} \gtrsim_P \frac{q_m^{\delta_{c,i+1}^{(m)}}p_m^{\delta_{r,1}^{(m)}}}{q_m^{\delta_{c,1}^{(m)}}p_m^{\delta_{r,1}^{(m)}}} &\succ_P \frac{q_m^{1-\delta_{c,1}^{(m)}} + T^{-1/2}q_mp_m^{1/2} + q_m^{(1+\delta_{j,1}^{(m)})/2}p_m^{(1+\delta_{q,1}^{(m)})/2}\eta_{Q,m}^{1/2}}{q_m^{\delta_{c,k_c^{(m)}}^{(m)}}p_m^{\delta_{r,1}^{(m)}}}  \;\;(=o_P(1))\notag\\
    &\gtrsim_P \frac{\lambda_{k_c^{(m)}+1}(\X_{C,m})}{\lambda_{k_c^{(m)}}(\X_{C,m})}, \label{eqn:Thm10-9}
\end{align}
where the second line is by the rate assumption in the statement of the theorem. The above implies $P(\hat{k}_c^{(m)}\geq k_c^{(m)}) \rightarrow 1$. When all factors are strong and all processes have short memory, $\eta_{Q,m} = T^{-1}$, and hence for $i\in\{k_c^{(m)}+1,\ldots,q_m\}$,
\[\lambda_i(\X_{C,m}) = O_P(T^{-1/2}p_mq_m).\]
Hence, we can see that for $i\in\{k_c^{(m)}+1,\ldots,q_m\}$,
\[\frac{\lambda_{i+1}(\X_{C,m}) + CT^{-1/2}p_mq_m}{\lambda_i(\X_{C,m}) + CT^{-1/2}p_mq_m} \asymp_P 1.\]
This, together with (\ref{eqn:Thm10-9}), proves $P(\hat{k}_c^{(m)} = k_c^{(m)}) \rightarrow 1$. $\square$

\clearpage

\section{Auxiliary Results and Proofs}\label{app:proof_of_lemmas}

\begin{proposition}{(Contraction properties of orthogonal projectors)}\label{prop:projector_prop}
    Let $\A\in\bb{R}^{a\times b}$ be any matrix, and let $\P\in\bb{R}^{a\times a}$ and $\Q\in\bb{R}^{b\times b}$ be orthogonal projectors. Then,
    \begin{align}
        \norm{\P\A\Q} \leq \norm{\A}, \qquad
        \norm{\P\A\Q}_F \leq \norm{\A}_F.
        \label{eqn:projector_contraction}
    \end{align}
    In particular, for any positive integer $n$, define $\M_n:=\I_n-n^{-1}\1_n\1_n'$. Then, $\M_n$ is the orthogonal projector onto $\1_n^\perp$, and
    \begin{align}
        \M_n'=\M_n,\qquad
        \M_n^2=\M_n,\qquad
        \M_n\1_n=\0,\qquad
        \norm{\M_n}\leq 1.
        \label{eqn:Mn_properties}
    \end{align}
    Moreover, 
    \begin{align}
        \norm{\M_n}_{\max} \leq1, \qquad
        \norm{\M_n}_1=\norm{\M_n}_{\infty}=2(1-n^{-1})\leq 2.
        \label{eqn:Mn_l1_lmax_norms}
    \end{align}
\end{proposition}
\textbf{\textit{Proof of Proposition~\ref{prop:projector_prop}.}}
Since $\P$ and $\Q$ are orthogonal projectors, $\norm{\P}\leq1$ and $\norm{\Q}\leq1$. Hence,
\begin{align*}
    \norm{\P\A\Q}
    &\leq \norm{\P}\norm{\A}\norm{\Q}
    \leq \norm{\A},\\
    \norm{\P\A\Q}_F
    &\leq \norm{\P}\norm{\A}_F\norm{\Q}
    \leq \norm{\A}_F.
\end{align*}

By direct calculation, $\M_n'=\M_n$, $\M_n^2=\M_n$, and $\M_n\1_n=\0$. Thus, $\M_n$ is the orthogonal projector onto $\1_n^\perp$. Since the eigenvalues of an orthogonal projector belong to $\{0,1\}$, we have $\norm{\M_n}\leq1$. Finally, its diagonal entries are $1-1/n$ and its off-diagonal entries are $-1/n$. Therefore,
\begin{align*}
    \norm{\M_n}_{\max}\leq 1,\qquad
    \norm{\M_n}_1=\norm{\M_n}_{\infty}
    =(1-n^{-1})+(n-1)n^{-1}
    =2(1-n^{-1})\leq2.
\end{align*}
This completes the proof. $\square$

\begin{lemma}{(Covariance bounds for $\E_t^{(m)}$)}\label{lemma:correlation_Etm}
    Let Assumptions~(E1) and (E2) hold. Then there exist a nonnegative sequence $\{\rho(h):h\in\bb{Z}\}$ and a constant $C<\infty$, such that $\sum_{h\in\bb{Z}}\rho(h)<\infty$, $\sup_{h\in\bb{Z}}\rho(h)\le 1$, and, uniformly over $h\in\bb{Z}$ and all $t$ such that $t,t+h\in[T]$, 
    \begin{align}
        \max_{m\in[M]}\sup_{i\in[p_m]}\sup_{j\in[q_m]}\sum_{k=1}^{p_m}\sum_{l=1}^{q_m}
        \Big| \bb{E}\big(E_{t+h,ij}^{(m)}E_{t,kl}^{(m)}\big) \Big|
        \le C\rho(h).
        \label{eqn:error_cov_summable}
    \end{align}
    In particular, uniformly over all $t$ such that $t,t+h\in[T]$,
    \begin{align}
        \max_{m\in[M]}\sup_{h\in\bb{Z}}\sup_{i\in[p_m]}\sup_{j\in[q_m]}\sum_{k=1}^{p_m}\sum_{l=1}^{q_m}
        \Big| \bb{E}\big(E_{t+h,ij}^{(m)}E_{t,kl}^{(m)}\big) \Big|
        \le C.
        \label{eqn:error_cov_uniform}
    \end{align}
    Moreover, uniformly over $h\in\bb{Z}$, all $t$ such that $t,t+h\in[T]$, the covariance of the error products satisfies
    \begin{align}
        \max_{m,n\in[M]}\sup_{i\in[p_m]}\sup_{j\in[q_m]}\sup_{k\in[p_n]}\sup_{l\in[q_n]}
        \sum_{a=1}^{q_m}\sum_{b=1}^{q_n}
        \left|\cov\left(
        E_{t+h,ij}^{(m)}E_{t+h,kl}^{(n)},
        E_{t,ia}^{(m)}E_{t,kb}^{(n)}
        \right)\right|
        \le C\rho(h).
        \label{eqn:error_product_cov}
    \end{align}
\end{lemma}
\textbf{\textit{Proof of Lemma~\ref{lemma:correlation_Etm}.}}
For $h\in\bb{Z}$, let $\gamma_e(h):=\sum_{w\geq 0}|a_{e,w}a_{e,w+|h|}|$, and for each $m\in[M]$, define analogously $\gamma_e^{(m)}(h):=\sum_{w\geq 0}|a_{e,w}^{(m)}a_{e,w+|h|}^{(m)}|$ and $\gamma_\epsilon(h)^{(m)}:=\sum_{w\geq 0}|a_{\epsilon,w}^{(m)}a_{\epsilon,w+|h|}^{(m)}|$. By (E2),
\begin{align*}
    \sum_{h\in\bb{Z}}\gamma_e(h)
    &= \sum_{w\geq0}|a_{e,w}|^2
    +2\sum_{h=1}^\infty\sum_{w\geq0}|a_{e,w}||a_{e,w+h}| \\
    &= \sum_{w\geq0}|a_{e,w}|^2
    +2\sum_{0\leq w<r<\infty}|a_{e,w}||a_{e,r}| 
    = \bigg(\sum_{w\geq0}|a_{e,w}|\bigg)^2
    \leq c^2<\infty,
\end{align*}
and analogously, for every $m\in[M]$, $\sum_{h\in\bb{Z}}\gamma_e^{(m)}(h)\leq c^2$ and $\sum_{h\in\bb{Z}}\gamma^{(m)}_\epsilon(h)\leq c^2$, where $c$ is the constant in (E2). Moreover, by the Cauchy--Schwarz inequality and (E2),
\begin{align}
    \gamma_e(h)
    \leq \big(\sum_{w\geq 0} a_{e,w}^2\big)^{1/2}
    \big(\sum_{w\geq 0} a_{e,w+|h|}^2\big)^{1/2}
    \leq 1,
    \label{eqn:lag_kernel_bound}
\end{align}
so that $\sup_h \gamma_e(h)\leq 1$. The same argument gives $\sup_h\gamma_e^{(m)}(h)\leq 1$ and $\sup_h\gamma^{(m)}_\epsilon(h)\leq 1$. Since $M$ is fixed, define
\begin{align*}
    \rho(h):=\frac{1}{3}\left\{\gamma_e(h)+\max_{m\in[M]}\gamma_e^{(m)}(h)+\max_{m\in[M]}\gamma^{(m)}_\epsilon(h)\right\}.
\end{align*}
Then $\rho(h)\geq0$, $\sup_{h\in\bb{Z}}\rho(h)\leq1$, and
\begin{align*}
    \sum_{h\in\bb{Z}}\rho(h)
    \leq \frac{1}{3}\left\{
    \sum_{h\in\bb{Z}}\gamma_e(h)
    +\sum_{m=1}^M\sum_{h\in\bb{Z}}\gamma_e^{(m)}(h)
    +\sum_{m=1}^M\sum_{h\in\bb{Z}}\gamma^{(m)}_\epsilon(h)
    \right\}
    \leq \frac{1+2M}{3}c^2<\infty.
\end{align*}

We next prove \eqref{eqn:error_cov_summable}. Fix $m\in[M]$. By (E1), write
\begin{align}
    E_{t,ij}^{(m)} 
    = (\Q_{e,m}\G_{e,t}\J_{e,m}')_{ij} + (\R_{e,m}\F_{e,t}^{(m)}\C_{e,m}')_{ij} + (\bSigma_{\epsilon,m}\circ \bepsilon_t^{(m)})_{ij} =: U_{t,ij}^{(m)} + V_{t,ij}^{(m)} + W_{t,ij}^{(m)}.
    \label{eqn:Et_decomposition}
\end{align}
Since the three component processes are mutually independent and have mean zero under (E1) and (E2), we have
\begin{align}\label{eqn:EtEt}
    \bb{E}(E_{t+h,ij}^{(m)} E_{t,kl}^{(m)}) 
    = \bb{E}(U_{t+h,ij}^{(m)}U_{t,kl}^{(m)}) + \bb{E}(V_{t+h,ij}^{(m)}V_{t,kl}^{(m)}) + \bb{E}(W_{t+h,ij}^{(m)}W_{t,kl}^{(m)}).
\end{align}
We bound the three terms on the r.h.s.\ separately. First, 
\begin{align*}
    U_{t,ij}^{(m)} = \sum_{a=1}^{k_{e,r}}\sum_{b=1}^{k_{e,c}}(\Q_{e,m})_{ia}(\G_{e,t})_{ab}(\J_{e,m})_{jb}.
\end{align*}
Hence 
\begin{align*}
    \bb{E}(U_{t+h,ij}^{(m)}U_{t,kl}^{(m)}) = \sum_{a=1}^{k_{e,r}}\sum_{b=1}^{k_{e,c}}\sum_{c=1}^{k_{e,r}}\sum_{d=1}^{k_{e,c}}
    (\Q_{e,m})_{ia}(\Q_{e,m})_{kc}(\J_{e,m})_{jb}(\J_{e,m})_{ld}\bb{E}\{(\G_{e,t+h})_{ab}(\G_{e,t})_{cd}\}.
\end{align*}
By (E2), we have $\bb{E}\{(\X_{e,r})_{ab}(\X_{e,s})_{cd}\}= \bb{1}\{r=s\}\bb{1}\{a=c,b=d\}$, so that
\begin{align*}
    \bb{E}\{(\G_{e,t+h})_{ab}(\G_{e,t})_{cd}\}
    &= \sum_{u\geq0}\sum_{v\geq0}
    a_{e,u}a_{e,v}
    \bb{E}\{(\X_{e,t+h-u})_{ab}(\X_{e,t-v})_{cd}\} \\
    &= \bb{1}\{a=c,b=d\}
    \sum_{u\geq0}\sum_{v\geq0}
    a_{e,u}a_{e,v}\bb{1}\{t+h-u=t-v\} \\
    &= \bb{1}\{a=c,b=d\}\sum_{w\geq0}a_{e,w+|h|}a_{e,w},
\end{align*}
where the last equality follows by setting $w=v$ when $h\geq 0$ and
$w=u$ when $h<0$.
Hence,
\begin{align*}
    \big|\bb{E}\{(\G_{e,t+h})_{ab}(\G_{e,t})_{cd}\}\big|
    \leq \gamma_e(h)\bb{1}\{a=c,b=d\}.
\end{align*}
Therefore, 
\begin{align*}
    \big|\bb{E}(U_{t+h,ij}^{(m)}U_{t,kl}^{(m)})\big|
    \leq \gamma_e(h)\sum_{a=1}^{k_{e,r}}\sum_{b=1}^{k_{e,c}}
    |(\Q_{e,m})_{ia}||(\Q_{e,m})_{ka}||(\J_{e,m})_{jb}||(\J_{e,m})_{lb}|.
\end{align*}
Taking absolute values and summing over $k$ and $l$, we obtain
\begin{align*}
    &\quad\;\sum_{k=1}^{p_m}\sum_{l=1}^{q_m}\big|\bb{E}(U_{t+h,ij}^{(m)}U_{t,kl}^{(m)})\big| \\ 
    &\leq \gamma_e(h)\sum_{a=1}^{k_{e,r}}\sum_{b=1}^{k_{e,c}} |(\Q_{e,m})_{ia}||(\J_{e,m})_{jb}|\Big(\sum_{k=1}^{p_m}|(\Q_{e,m})_{ka}|\Big)\Big(\sum_{l=1}^{q_m}|(\J_{e,m})_{lb}|\Big) \\
    &\leq \gamma_e(h)\Big(\sum_{a=1}^{k_{e,r}} |(\Q_{e,m})_{ia}|\Big)\Big(\max_{a}\sum_{k=1}^{p_m}|(\Q_{e,m})_{ka}|\Big)\Big(\sum_{b=1}^{k_{e,c}} |(\J_{e,m})_{jb}|\Big)\Big(\max_{b}\sum_{l=1}^{q_m} |(\J_{e,m})_{lb}|\Big) \\
    &\leq \gamma_e(h)\big(k_{e,r}\|\Q_{e,m}\|_1\big)\|\Q_{e,m}\|_1\big(k_{e,c}\|\J_{e,m}\|_1\big)\|\J_{e,m}\|_1 
    = O(\gamma_e(h)),
\end{align*}
where we use $\sum_{k}|(\Q_{e,m})_{ka}| \leq \norm{\Q_{e,m}}_1$ and $\sum_{l}|(\J_{e,m})_{lb}| \leq \norm{\J_{e,m}}_1$ in the first inequality, $\sum_a|(\Q_{e,m})_{ia}|\leq k_{e,r}\norm{\Q_{e,m}}_1$ and $\sum_b|(\J_{e,m})_{jb}|\leq k_{e,c}\norm{\J_{e,m}}_1$ in the second inequality, and Assumption~(E1) in the last equality. By the same argument, 
\begin{align*}
    \sum_{k=1}^{p_m}\sum_{l=1}^{q_m}|\bb{E}(V_{t+h,ij}^{(m)}V_{t,kl}^{(m)})|=O(\gamma_e^{(m)}(h)).
\end{align*}
Finally, (E2) gives
\begin{align*}
    |\bb{E}(\epsilon_{t+h,ij}^{(m)}\epsilon_{t,kl}^{(m)})| \leq \gamma^{(m)}_\epsilon(h)\bb{1}\{(i,j)=(k,l)\}.
\end{align*}
Therefore, 
\begin{align*}
    \sum_{k=1}^{p_m}\sum_{l=1}^{q_m}|\bb{E}(W_{t+h,ij}^{(m)}W_{t,kl}^{(m)})| 
    &\leq \sum_{k=1}^{p_m}\sum_{l=1}^{q_m}|(\bSigma_{\epsilon,m})_{ij}(\bSigma_{\epsilon,m})_{kl}|\gamma^{(m)}_\epsilon(h)\bb{1}\{(i,j)=(k,l)\} \\
    &= \gamma^{(m)}_\epsilon(h)(\bSigma_{\epsilon,m})_{ij}^2 = O(\gamma^{(m)}_\epsilon(h)),
\end{align*}
by the uniform boundedness of $\bSigma_{\epsilon,m}$ in (E1). Combining the above bounds, there exists an absolute constant $C$ such that
\begin{align*}
    \sum_{k=1}^{p_m}\sum_{l=1}^{q_m}
    |\bb{E}(U_{t+h,ij}^{(m)}U_{t,kl}^{(m)})|
    &\leq \frac{C}{3}\gamma_e(h), \\
    \sum_{k=1}^{p_m}\sum_{l=1}^{q_m}
    |\bb{E}(V_{t+h,ij}^{(m)}V_{t,kl}^{(m)})|
    &\leq \frac{C}{3}\gamma_e^{(m)}(h), \\
    \sum_{k=1}^{p_m}\sum_{l=1}^{q_m}
    |\bb{E}(W_{t+h,ij}^{(m)}W_{t,kl}^{(m)})|
    &\leq \frac{C}{3}\gamma^{(m)}_\epsilon(h),
\end{align*}
uniformly over $m\in[M]$, $h\in\bb{Z}$, $i\in[p_m]$, and $j\in[q_m]$. By \eqref{eqn:EtEt} and the triangle inequality,
\begin{align*}
    \sum_{k=1}^{p_m}\sum_{l=1}^{q_m}
    |\bb{E}(E_{t+h,ij}^{(m)}E_{t,kl}^{(m)})|
    &\leq 
    \frac{C}{3}\left(\gamma_e(h)+\gamma_e^{(m)}(h)+\gamma^{(m)}_\epsilon(h)\right) \\
    &\leq 
    \frac{C}{3}\left(\gamma_e(h)+\max_{m\in[M]}\gamma_e^{(m)}(h)+\max_{m\in[M]}\gamma^{(m)}_\epsilon(h)\right) 
    = C\rho(h) ,
\end{align*}
which concludes \eqref{eqn:error_cov_summable}. Since $\sup_{h\in\bb{Z}}\rho(h)\leq1$, \eqref{eqn:error_cov_uniform} follows immediately. 

We next show \eqref{eqn:error_product_cov}. For $h\in\bb{Z}$, define $\kappa_e(h):=\sum_{w\ge0}a_{e,w}^2a_{e,w+|h|}^2$, and for each $m\in[M]$, define $\kappa_e^{(m)}(h):=\sum_{w\ge0}\left(a_{e,w}^{(m)}\right)^2\left(a_{e,w+|h|}^{(m)}\right)^2$ and $\kappa_\epsilon^{(m)}(h):=\sum_{w\ge0}\left(a_{\epsilon,w}^{(m)}\right)^2\left(a_{\epsilon,w+|h|}^{(m)}\right)^2$. Since 
\begin{align*}
    \kappa_e(h)\le\gamma_e(h)^2\le\gamma_e(h)\le 3\rho(h),
\end{align*}
and analogously $\kappa_e^{(m)}(h)\le 3\rho(h)$ and $\kappa_\epsilon^{(m)}(h)\le 3\rho(h)$ for every $m\in[M]$, we have 
\begin{align*}
    \kappa_e(h)+\max_{m\in[M]}\kappa_e^{(m)}(h)+\max_{m\in[M]}\kappa_\epsilon^{(m)}(h)
    \le 9\rho(h).
\end{align*}
Moreover, for any two sequences $\gamma_1$ and $\gamma_2$ chosen from
$\{\gamma_e, \gamma_e^{(m)}, \gamma_\epsilon^{(m)}:m\in[M]\}$, we have
\begin{align*}
    \gamma_1(h)\gamma_2(h)\le 9\rho(h)^2\le 9\rho(h),
\end{align*}
where the second inequality used $\rho(h)\le 1$.

Recall the decomposition in \eqref{eqn:Et_decomposition}. Expanding each of the four error entries in the covariance yields finitely many terms. We first consider
\begin{align*}
    S_U(h):=\sum_{a=1}^{q_m}\sum_{b=1}^{q_n}
    \left|\cov\left(U_{t+h,ij}^{(m)}U_{t+h,kl}^{(n)},U_{t,ia}^{(m)}U_{t,kb}^{(n)}\right)\right|.
\end{align*}
By the definition of $U_{t,ij}^{(m)}$,
\begin{align*}
    U_{t,ij}^{(m)}
    =\sum_{c=1}^{k_{e,r}}\sum_{d=1}^{k_{e,c}}
    (\Q_{e,m})_{ic}(\J_{e,m})_{jd}(\G_{e,t})_{cd}.
\end{align*}
It follows that
\begin{align*}
    &\quad\;\cov\left(
    U_{t+h,ij}^{(m)}U_{t+h,kl}^{(n)},
    U_{t,ia}^{(m)}U_{t,kb}^{(n)}
    \right) \\
    &=\sum_{c_1=1}^{k_{e,r}}\sum_{d_1=1}^{k_{e,c}}
    \sum_{c_2=1}^{k_{e,r}}\sum_{d_2=1}^{k_{e,c}}
    \sum_{c_3=1}^{k_{e,r}}\sum_{d_3=1}^{k_{e,c}}
    \sum_{c_4=1}^{k_{e,r}}\sum_{d_4=1}^{k_{e,c}}
    (\Q_{e,m})_{ic_1}(\J_{e,m})_{jd_1}
    (\Q_{e,n})_{kc_2}(\J_{e,n})_{ld_2} \\
    &\qquad \cdot
    (\Q_{e,m})_{ic_3}(\J_{e,m})_{ad_3}
    (\Q_{e,n})_{kc_4}(\J_{e,n})_{bd_4} \\
    &\qquad \cdot
    \cov\left(
    (\G_{e,t+h})_{c_1d_1}(\G_{e,t+h})_{c_2d_2},
    (\G_{e,t})_{c_3d_3}(\G_{e,t})_{c_4d_4}
    \right).
\end{align*}
By (E2), we have
\begin{align*}
    &\left|
    \cov\left(
    (\G_{e,t+h})_{c_1d_1}(\G_{e,t+h})_{c_2d_2},
    (\G_{e,t})_{c_3d_3}(\G_{e,t})_{c_4d_4}
    \right)
    \right| \\
    &\leq C\gamma_e(h)^2\bb{1}\{(c_1,d_1)=(c_3,d_3),\,(c_2,d_2)=(c_4,d_4)\} \\
    &\quad + C\gamma_e(h)^2\bb{1}\{(c_1,d_1)=(c_4,d_4),\,(c_2,d_2)=(c_3,d_3)\} \\
    &\quad + C\kappa_e(h)\bb{1}\{(c_1,d_1)=(c_2,d_2)=(c_3,d_3)=(c_4,d_4)\}.
\end{align*}
Substituting the above into the preceding expansion and using the triangle inequality, we obtain
\begin{align*}
    S_U(h)
    &\leq C\gamma_e(h)^2
    \sum_{c_1=1}^{k_{e,r}}\sum_{d_1=1}^{k_{e,c}}
    \sum_{c_2=1}^{k_{e,r}}\sum_{d_2=1}^{k_{e,c}}
    |(\Q_{e,m})_{ic_1}(\J_{e,m})_{jd_1}
    (\Q_{e,n})_{kc_2}(\J_{e,n})_{ld_2}| \\
    &\qquad \cdot
    \sum_{a=1}^{q_m}|(\Q_{e,m})_{ic_1}(\J_{e,m})_{ad_1}|
    \sum_{b=1}^{q_n}|(\Q_{e,n})_{kc_2}(\J_{e,n})_{bd_2}| \\
    &\quad + C\gamma_e(h)^2
    \sum_{c_1=1}^{k_{e,r}}\sum_{d_1=1}^{k_{e,c}}
    \sum_{c_2=1}^{k_{e,r}}\sum_{d_2=1}^{k_{e,c}}
    |(\Q_{e,m})_{ic_1}(\J_{e,m})_{jd_1}
    (\Q_{e,n})_{kc_2}(\J_{e,n})_{ld_2}| \\
    &\qquad \cdot
    \sum_{a=1}^{q_m}|(\Q_{e,m})_{ic_2}(\J_{e,m})_{ad_2}|
    \sum_{b=1}^{q_n}|(\Q_{e,n})_{kc_1}(\J_{e,n})_{bd_1}| \\
    &\quad + C\kappa_e(h)
    \sum_{c=1}^{k_{e,r}}\sum_{d=1}^{k_{e,c}}
    |(\Q_{e,m})_{ic}(\J_{e,m})_{jd}
    (\Q_{e,n})_{kc}(\J_{e,n})_{ld}| \\
    &\qquad \cdot
    \sum_{a=1}^{q_m}|(\Q_{e,m})_{ic}(\J_{e,m})_{ad}|
    \sum_{b=1}^{q_n}|(\Q_{e,n})_{kc}(\J_{e,n})_{bd}|.
\end{align*}
Since $k_{e,r}$ and $k_{e,c}$ are fixed, and since $\|\Q_{e,m}\|_1=\|\J_{e,m}\|_1=O(1)$ and $\|\Q_{e,n}\|_1=\|\J_{e,n}\|_1=O(1)$ for all $m,n\in[M]$, we have, uniformly over $i\in[p_m]$, $j\in[q_m]$, $k\in[p_n]$, and $l\in[q_n]$,
\begin{align}
    S_U(h)\leq C\gamma_e(h)^2+C\kappa_e(h)\leq C\rho(h).
    \label{eqn:four_global_error_bound}
\end{align}
Next, define
\begin{align*}
    S_V(h):=\sum_{a=1}^{q_m}\sum_{b=1}^{q_n}
    \left|\cov\left(
    V_{t+h,ij}^{(m)}V_{t+h,kl}^{(n)},
    V_{t,ia}^{(m)}V_{t,kb}^{(n)}
    \right)\right|.
\end{align*}
When $m=n$, the above arguments apply to the local error process. When $m\neq n$, the cross-group independence of the local error innovation processes in (E2) yields, uniformly over $i\in[p_m]$, $j\in[q_m]$, $k\in[p_n]$, and $l\in[q_n]$,
\begin{align}
    S_V(h)\le C\gamma_e^{(m)}(h)\gamma_e^{(n)}(h)
    +C\bb{1}\{m=n\}\kappa_e^{(m)}(h)
    \le C\rho(h).
    \label{eqn:four_local_error_bound}
\end{align}
Similarly, define
\begin{align*}
    S_W(h):=\sum_{a=1}^{q_m}\sum_{b=1}^{q_n}
    \left|\cov\left(
    W_{t+h,ij}^{(m)}W_{t+h,kl}^{(n)},
    W_{t,ia}^{(m)}W_{t,kb}^{(n)}
    \right)\right|.
\end{align*}
Using the independence and uniformly bounded fourth moments of the idiosyncratic innovations in (E2), together with the uniform boundedness of the entries of $\bSigma_{\epsilon,m}$ in (E1), we obtain
\begin{align}
    S_W(h)\le C\gamma_\epsilon^{(m)}(h)\gamma_\epsilon^{(n)}(h)
    +C\bb{1}\{m=n\}\kappa_\epsilon^{(m)}(h)
    \le C\rho(h).
    \label{eqn:four_idiosyncratic_error_bound}
\end{align}

It remains to bound the terms involving at least two components among $U_{t,ij}^{(m)}$, $V_{t,ij}^{(m)}$, and $W_{t,ij}^{(m)}$. These remaining terms fall into the following two cases. 

\paragraph{(i) At least one driving innovation process appears only once in the covariance.} 
Such terms have zero covariance. For example,
\begin{align*}
    &\quad\;\cov\left(
    U_{t+h,ij}^{(m)}U_{t+h,kl}^{(n)},
    U_{t,ia}^{(m)}V_{t,kb}^{(n)}
    \right) \\
    &=\bb{E}\left(U_{t+h,ij}^{(m)}U_{t+h,kl}^{(n)}U_{t,ia}^{(m)}V_{t,kb}^{(n)}\right)
    -\bb{E}\left(U_{t+h,ij}^{(m)}U_{t+h,kl}^{(n)}\right)\bb{E}\left(U_{t,ia}^{(m)}V_{t,kb}^{(n)}\right) \\
    &=\bb{E}\left(U_{t+h,ij}^{(m)}U_{t+h,kl}^{(n)}U_{t,ia}^{(m)}\right)
    \bb{E}\left(V_{t,kb}^{(n)}\right)
    -\bb{E}\left(U_{t+h,ij}^{(m)}U_{t+h,kl}^{(n)}\right)
    \bb{E}\left(U_{t,ia}^{(m)}\right)\bb{E}\left(V_{t,kb}^{(n)}\right) 
    =0 .
\end{align*}
The same argument applies whenever one of the driving innovation processes occurs only once. For example,
\begin{align*}
    \cov\left(
    U_{t+h,ij}^{(m)}V_{t+h,kl}^{(n)},
    U_{t,ia}^{(m)}W_{t,kb}^{(n)}
    \right)
    =0.
\end{align*}

\paragraph{(ii) Two independent driving innovation processes each appear twice.} 
There are two possible arrangements. 

\begin{itemize}
    \item [(a)]
\textit{The two occurrences of each process lie on the same side of the covariance.} 
In this case, the two products are independent, so their covariance is zero. For example,
\begin{align*}
    \cov\left(
    U_{t+h,ij}^{(m)}U_{t+h,kl}^{(n)},
    V_{t,ia}^{(m)}V_{t,kb}^{(n)}
    \right)
    =0.
\end{align*}

    \item [(b)]
\textit{Each process contributes once on each side of the covariance.}
We first consider the arrangements in which each innovation process contributes one factor at time $t+h$ and one factor at time $t$. As a representative case, write
\begin{align*}
    S_{UV}(h)&:=\sum_{a=1}^{q_m}\sum_{b=1}^{q_n}
    \left|\cov\left(
    U_{t+h,ij}^{(m)}V_{t+h,kl}^{(n)},
    U_{t,ia}^{(m)}V_{t,kb}^{(n)}
    \right)\right| \\
    &=\sum_{a=1}^{q_m}\sum_{b=1}^{q_n}
    \left|
    \bb{E}\left(U_{t+h,ij}^{(m)}U_{t,ia}^{(m)}\right)
    \bb{E}\left(V_{t+h,kl}^{(n)}V_{t,kb}^{(n)}\right)
    \right| \\
    &\le
    \left\{\sum_{a=1}^{q_m}\left|\bb{E}\left(U_{t+h,ij}^{(m)}U_{t,ia}^{(m)}\right)\right|\right\}
    \left\{\sum_{b=1}^{q_n}\left|\bb{E}\left(V_{t+h,kl}^{(n)}V_{t,kb}^{(n)}\right)
    \right|\right\} \\
    &\le C\gamma_e(h)\gamma_e^{(n)}(h)
    \le C\rho(h).
\end{align*}
We finally consider the arrangement in which the two component types exchange their group assignments across the two sides of the covariance. For example,
\begin{align*}
    \cov\left(
    U_{t+h,ij}^{(m)}V_{t+h,kl}^{(n)},
    V_{t,ia}^{(m)}U_{t,kb}^{(n)}
    \right).
\end{align*}
If $m\neq n$, then the covariance is zero by case (i). If $m=n$, uniformly over $m,n\in[M]$, independence of the global and local innovation processes gives
\begin{align*}
    &\quad\;\sum_{a=1}^{q_m}\sum_{b=1}^{q_n}
    \left|\cov\left(
    U_{t+h,ij}^{(m)}V_{t+h,kl}^{(n)},
    V_{t,ia}^{(m)}U_{t,kb}^{(n)}
    \right)\right| \\
    &=\sum_{a=1}^{q_m}\sum_{b=1}^{q_n}
    \left|\bb{E}\left(U_{t+h,ij}^{(m)}U_{t,kb}^{(m)}\right)
    \bb{E}\left(V_{t+h,kl}^{(m)}V_{t,ia}^{(m)}\right)\right| \\
    &\le C\gamma_e(h)\gamma_e^{(m)}(h)
    \leq C\rho(h).
\end{align*}

\end{itemize}

Cases (i) and (ii) cover all terms involving at least two of $U_{t,ij}^{(m)}$, $V_{t,ij}^{(m)}$, and $W_{t,ij}^{(m)}$, where these terms are not necessarily corresponding to the same group $m$. Combining these bounds with \eqref{eqn:four_global_error_bound}, \eqref{eqn:four_local_error_bound} and \eqref{eqn:four_idiosyncratic_error_bound}, we obtain, uniformly over
$h\in\bb{Z}$ and all $t$ such that $t,t+h\in[T]$:
\begin{align*}
    \max_{m,n\in[M]}\sup_{i\in[p_m]}\sup_{j\in[q_m]}\sup_{k\in[p_n]}\sup_{l\in[q_n]}\sum_{a=1}^{q_m}\sum_{b=1}^{q_n}
    \left|
    \cov\left(
    E_{t+h,ij}^{(m)}E_{t+h,kl}^{(n)},
    E_{t,ia}^{(m)}E_{t,kb}^{(n)}
    \right)
    \right|
    \le C\rho(h).
\end{align*}
This proves
\eqref{eqn:error_product_cov} and completes the proof. $\square$

\begin{lemma}{(Basic properties of $\cG_T$ processes)}\label{lemma:prop_piecewise}
    Let $\{\u_t\}_{t\in[T]}\in \cG_T(\{d_{u,1},\ldots,d_{u,m_u}\}|\{\x_{u,i},i\in\mathbb{Z}\})$ be a $p_u$-dimensional process, where $p_u$ and $m_u$ are bounded. Define $d_u:=\max_{1\le j\le m_u} d_{u,j}$. Then: %
    \begin{itemize}
        \item [1.] There exists a constant $C<\infty$ such that the second moments
        are uniformly bounded as
        \begin{align}
            \sup_{t\in[T]}\bb{E}\|\u_t\|^2\le C.
            \label{eqn:uniform_second_moment_G}
        \end{align}
        \item [2.] For any $h\in\{0,\dots,T-1\}$, the centered sample autocovariance satisfies
        \begin{align}
            \left\|T^{-1}\sum_{t=1}^{T-h}\big(\u_t\u_{t+h}'-\bb{E}(\u_t\u_{t+h}')\big)\right\|_F^2 = O_P(R(T,d_u)).
            \label{eqn:sample_covariance_G}
        \end{align}
        \item [3.] Let $\{\v_t\}_{t\in[T]}\in \cG_T(\{d_{v,1},\ldots,d_{v,m_v}\}|\{\x_{v,i},i\in\mathbb{Z}\})$ be another $p_v$-dimensional process, independent of $\{\u_t\}_{t\in[T]}$, where $p_v$ and $m_v$ are bounded. Define $d_v:=\max_{1\le j\le m_v} d_{v,j}$. For $d_{uv}:=(d_u+d_v)/2$, the sample cross-moment satisfies
        \begin{align}
            \left\|T^{-1}\sum_{t=1}^T \u_t\v_t'\right\|_F^2 = O_P(R(T,d_{uv})).
            \label{eqn:sample_cross_moment_G}
        \end{align}
    \end{itemize}
\end{lemma}
\textbf{\textit{Proof of Lemma~\ref{lemma:prop_piecewise}.}}
We first show \eqref{eqn:uniform_second_moment_G}. Fix $t\in[T]$, and suppose that $t$ belongs to segment $a\in[m_u]$ of $\{\u_t\}$. By the definition of $\cG_T(\cdot)$, 
\begin{align*}
    \u_t=\sum_{w\geq 0}\psi_{u,a}(w)\x_{u,t-w}, 
\end{align*}
where $\psi_{u,a}(0)=1$, and for $w\geq 1$,
\begin{align*}
    \psi_{u,a}(w)=w^{d_{u,a}-1}l_{u,a}(w),\qquad d_{u,a}<1/2.
\end{align*}
Since $l_{u,a}(w)$ converges to a finite constant and $m_u$ is bounded, 
\begin{align*}
    |\psi_{u,a}(w)|\lesssim w^{d_{u,a}-1},\qquad w\geq 1.
\end{align*}
Since the entries of the innovation sequence $\{\x_{u,t}\}$ are independent across coordinates and over time and have zero means and uniformly bounded second moments,
\begin{align*}
    \bb{E}\|\u_t\|^2
    &= \sum_{w\ge 0}\psi_{u,a}(w)^2\bb{E}\norm{\x_{u,t-w}}^2 \\
    &\lesssim p_u\sum_{w\ge0}\psi_{u,a}(w)^2
    \lesssim p_u\left(1+\sum_{w\ge1}w^{2d_{u,a}-2}\right)
    \lesssim 1,
\end{align*}
because $d_{u,a}<1/2$, while $p_u$ and $m_u$ are bounded. Taking the supremum over $t\in[T]$ gives \eqref{eqn:uniform_second_moment_G}.

We next show \eqref{eqn:sample_covariance_G}. Fix $h\in\{0,\dots,T-1\}$, and define
\begin{align*}
    \wt\bGamma_u(h):=T^{-1}\sum_{t=1}^{T-h}\big(\u_t\u_{t+h}'-\bb{E}(\u_t\u_{t+h}')\big).
\end{align*}
Since $p_u$ is bounded, it is enough to bound the variance of each entry. The proof proceeds in three steps. We first control the covariance between observations associated with arbitrary segment pairs. We then bound the covariance between the corresponding products. Finally, we group the pairs of time indices according to their lag.

First, fix an entry $(i,j)$. For $t,s\in[T-h]$, let
\begin{align*}
    k=J_u(t),\qquad r=J_u(t+h),\qquad
    k'=J_u(s),\qquad r'=J_u(s+h),
\end{align*}
where $J_u(t)\in[m_u]$ denotes the segment containing time $t$, and write
$\tau=t-s$. Then
\begin{align}
    \var\left(\wt\Gamma_{u,ij}(h)\right)
    =T^{-2}\sum_{t=1}^{T-h}\sum_{s=1}^{T-h}
    \cov\left(u_{t,i}u_{t+h,j},u_{s,i}u_{s+h,j}\right).
    \label{eqn:sample_covariance_variance_expansion}
\end{align}
Because all coordinates use the same filter within each segment and the innovation coordinates are identically distributed, the covariance between two observations from the same coordinate index depends only on their segment labels and time lag. For $k,k'\in[m_u]$ and $\tau\geq 0$, define
\begin{align}
    \gamma_{k,k'}(\tau)
    :=\sigma_u^2\sum_{w\geq 0}
    \psi_{u,k}(w+\tau)\psi_{u,k'}(w),
    \label{eqn:segment_pair_covariance}
\end{align}
where $\sigma_u^2:=\bb{E}(x_{u,0,1}^2)<\infty$, and, for $\tau<0$, set $\gamma_{k,k'}(\tau):=\gamma_{k',k}(-\tau)$. By the temporal independence of the innovations, whenever $J_u(t)=k$, $J_u(s)=k'$, and $t-s=\tau$, it holds that $\cov(u_{t,i},u_{s,i})=\gamma_{k,k'}(\tau)$ for every $i\in[p_u]$. For each segment $a\in[m_u]$, since each $l_{u,a}(w)$ converges to a finite constant, we have
\begin{align}
    |\psi_{u,a}(w)|
    \lesssim w^{d_{u,a}-1},
    \qquad \quad w\geq1.
    \label{eqn:segment_filter_bound}
\end{align}
Moreover, by the Cauchy--Schwarz inequality and the square summability of the filters,
\begin{align*}
    \max_{k,k'\in[m_u]}|\gamma_{k,k'}(0)|
    \lesssim \max_{k,k'\in[m_u]}
    \left\{\sum_{w\geq0}\psi_{u,k}(w)^2\right\}^{1/2}
    \left\{\sum_{w\geq0}\psi_{u,k'}(w)^2\right\}^{1/2}
    \lesssim 1.
\end{align*}

Suppose first that $d_u>0$. For $\tau\geq1$, since $d_{u,a}\leq d_u$ for every $a\in[m_u]$,
\begin{align*}
    |\gamma_{k,k'}(\tau)|
    &\lesssim \tau^{d_u-1}+\sum_{w\geq1}(w+\tau)^{d_u-1}w^{d_u-1} \\
    &\lesssim \tau^{d_u-1}+\tau^{d_u-1}\sum_{w=1}^{\tau}w^{d_u-1}+\sum_{w>\tau}w^{2d_u-2}.
\end{align*}
Since $0<d_u<1/2$, integral comparison gives
\begin{align*}
    \sum_{w=1}^{\tau}w^{d_u-1}
    \leq 1+\int_1^\tau x^{d_u-1}\mathrm{d}x
    =1+\frac{\tau^{d_u}-1}{d_u}
    \lesssim \tau^{d_u},
\end{align*}
and
\begin{align*}
    \sum_{w>\tau}w^{2d_u-2}
    \leq \int_\tau^\infty x^{2d_u-2}\mathrm{d}x
    \lesssim \tau^{2d_u-1}.
\end{align*}
Combining these bounds yields $|\gamma_{k,k'}(\tau)|\lesssim \tau^{2d_u-1}$ for $\tau\geq 1$. Hence, 
\begin{align}
    \max_{k,k'\in[m_u]}|\gamma_{k,k'}(\tau)|^2
    \lesssim (1+|\tau|)^{4d_u-2},
    \qquad \tau\in\bb{Z}.
    \label{eqn:segment_covariance_long_pointwise}
\end{align}
Therefore, by the standard partial-sum bounds for $p$-series,
\begin{align}
    \sum_{\tau=-2(T-1)}^{2(T-1)}
    \max_{k,k'\in[m_u]}|\gamma_{k,k'}(\tau)|^2
    \lesssim 1+\sum_{\tau=1}^{2(T-1)}\tau^{4d_u-2}
    \lesssim
    \begin{cases}
        1, & 0<d_u<1/4,\\
        \log T, & d_u=1/4,\\
        T^{4d_u-1}, & 1/4<d_u<1/2.
    \end{cases}
    \label{eqn:segment_covariance_long_memory_bound}
\end{align}

If $d_u=0$, then \eqref{eqn:segment_filter_bound}, together with $\psi_{u,a}(0)=1$, gives
\begin{align*}
    |\psi_{u,a}(w)|
    \lesssim (1+w)^{-1},
    \qquad a\in[m_u],\quad w\geq0.
\end{align*}
Hence, for $\tau\geq 1$, splitting the sum in \eqref{eqn:segment_pair_covariance} at $w=\tau$ gives
\begin{align*}
    |\gamma_{k,k'}(\tau)|
    &\lesssim
    \sum_{w=0}^{\tau}\frac{1}{(1+w)(1+w+\tau)}
    +\sum_{w>\tau}\frac{1}{(1+w)(1+w+\tau)} \\
    &\lesssim
    (1+\tau)^{-1}\sum_{w=0}^{\tau}(1+w)^{-1}+\sum_{w>\tau}(1+w)^{-2} \\
    &\lesssim (1+\tau)^{-1}\log(2+\tau).
\end{align*}
Together with the bound at $\tau=0$ and the symmetry for negative lags, we have
\begin{align}
    \max_{k,k'\in[m_u]}|\gamma_{k,k'}(\tau)|
    \lesssim (1+|\tau|)^{-1}\log(2+|\tau|),
    \qquad \tau\in\bb{Z},
    \label{eqn:segment_covariance_edge_pointwise}
\end{align}
so that
\begin{align}
    \sum_{\tau\in\bb{Z}}
    \max_{k,k'\in[m_u]}|\gamma_{k,k'}(\tau)|^2
    <\infty.
    \label{eqn:segment_covariance_edge_point_bound}
\end{align}

If $d_u<0$, all segment filters are absolutely summable. Extending each filter by zero on negative indices, we have
\begin{align*}
    \sum_{\tau\in\bb{Z}}
    |\gamma_{k,k'}(\tau)|
    \lesssim \sum_{\tau\in\bb{Z}}\sum_{w\in\bb{Z}}
    |\psi_{u,k}(w+\tau)||\psi_{u,k'}(w)|
    = \left\{\sum_{v\in\bb{Z}}|\psi_{u,k}(v)|\right\}
    \left\{\sum_{w\in\bb{Z}}|\psi_{u,k'}(w)|\right\}
    <\infty.
\end{align*}
Hence, it follows that 
\begin{align}
    \sum_{\tau\in\bb{Z}}\max_{k,k'\in[m_u]}|\gamma_{k,k'}(\tau)|<\infty, \qquad
    \sum_{\tau\in\bb{Z}}\max_{k,k'\in[m_u]}|\gamma_{k,k'}(\tau)|^2<\infty.
    \label{eqn:segment_covariance_absolute_sum}
\end{align}
Combining \eqref{eqn:segment_covariance_long_memory_bound}, \eqref{eqn:segment_covariance_edge_point_bound}, and \eqref{eqn:segment_covariance_absolute_sum}, we have
\begin{align}
    B_u(T):=\sum_{\tau=-2(T-1)}^{2(T-1)}
    \max_{k,k'\in[m_u]}|\gamma_{k,k'}(\tau)|^2
    \lesssim
    \begin{cases}
        1, & d_u<1/4,\\
        \log T, & d_u=1/4,\\
        T^{4d_u-1}, & d_u>1/4,
    \end{cases}
    \label{eqn:segment_covariance_square_sum}
\end{align}
In particular, $T^{-1}B_u(T)=O(R(T,d_u))$.

We next bound the covariance between the products. Recall that $k=J_u(t)$, $r=J_u(t+h)$, $k'=J_u(s)$, $r'=J_u(s+h)$, and $\tau=t-s$. If $i\neq j$, independence across coordinates and the zero-mean property give
\begin{align*}
    \cov(u_{t,i}u_{t+h,j},u_{s,i}u_{s+h,j})
    =\gamma_{k,k'}(\tau)\gamma_{r,r'}(\tau).
\end{align*}
If $i=j$, the fourth-moment expansion gives
\begin{align}
    \cov(u_{t,i}u_{t+h,i},u_{s,i}u_{s+h,i})
    &=\gamma_{k,k'}(\tau)\gamma_{r,r'}(\tau)
    +\gamma_{k,r'}(\tau-h)\gamma_{r,k'}(\tau+h)
    +\cK_{k,r,k',r'}(\tau,h),
    \label{eqn:cov_uuuu}
\end{align}
where $\cK_{k,r,k',r'}(\tau,h)$ denotes the fourth-cumulant contribution, viz.\ after extending each filter by zero on negative indices,
\begin{align*}
    \cK_{k,r,k',r'}(\tau,h)
    :=\kappa_{4,u}
    \sum_{\ell\in\bb{Z}}
    \psi_{u,k}(\ell)
    \psi_{u,r}(\ell+h)
    \psi_{u,k'}(\ell-\tau)
    \psi_{u,r'}(\ell+h-\tau),
\end{align*}
where $\kappa_{4,u}:=\bb{E}(x_{u,0,1}^4)-3\{\bb{E}(x_{u,0,1}^2)\}^2$ is finite by the bounded fourth moment assumption. Define
\begin{align*}
    a_q:=\psi_{u,k}(q)\psi_{u,r}(q+h),
    \qquad
    b_q:=\psi_{u,k'}(q)\psi_{u,r'}(q+h).
\end{align*}
Then, $|\cK_{k,r,k',r'}(\tau,h)| \lesssim \sum_{q\in\bb{Z}}|a_q|\,|b_{q-\tau}|$. Summing over $\tau$ and applying the Cauchy--Schwarz inequality gives
\begin{align*}
    &\;\quad
    \sum_{\tau\in\bb{Z}}|\cK_{k,r,k',r'}(\tau,h)|
    \lesssim \left(\sum_{q\in\bb{Z}}|a_q|\right)
    \left(\sum_{q\in\bb{Z}}|b_q|\right) \\
    &\lesssim
    \left\{\sum_{q\in\bb{Z}}|\psi_{u,k}(q)|^2\right\}^{1/2}
    \left\{\sum_{q\in\bb{Z}}|\psi_{u,r}(q+h)|^2\right\}^{1/2} 
    \left\{\sum_{q\in\bb{Z}}|\psi_{u,k'}(q)|^2\right\}^{1/2}
    \left\{\sum_{q\in\bb{Z}}|\psi_{u,r'}(q+h)|^2\right\}^{1/2} 
    \lesssim 1.
\end{align*}
Therefore, we have
\begin{align}
    \sum_{\tau\in\bb{Z}}\max_{k,r,k',r'\in[m_u]}
    |\cK_{k,r,k',r'}(\tau,h)|
    \lesssim 1.
    \label{eqn:cumulant_sum_G}
\end{align}
For the two covariance-product terms in \eqref{eqn:cov_uuuu}, we have
\begin{align}
    \sum_{\tau=-(T-1)}^{T-1}\max_{k,r,k',r'\in[m_u]}
    |\gamma_{k,k'}(\tau)\gamma_{r,r'}(\tau)|
    \leq \sum_{\tau=-(T-1)}^{T-1}
    \left\{\max_{a,b\in[m_u]}|\gamma_{a,b}(\tau)|\right\}^2
    \leq B_u(T),
    \label{eqn:first_covariance_product_G}
\end{align}
and, by the Cauchy--Schwarz inequality,
\begin{align}
    &\quad\;\sum_{\tau=-(T-1)}^{T-1}
    \max_{k,r,k',r'\in[m_u]}
    |\gamma_{k,r'}(\tau-h)\gamma_{r,k'}(\tau+h)| \nonumber \\
    &\leq
    \left\{\sum_{\tau=-(T-1)}^{T-1}\max_{k,r'\in[m_u]}|\gamma_{k,r'}(\tau-h)|^2\right\}^{1/2}
    \left\{\sum_{\tau=-(T-1)}^{T-1}\max_{r,k'\in[m_u]}|\gamma_{r,k'}(\tau+h)|^2\right\}^{1/2} %
    \leq B_u(T),
    \label{eqn:shifted_covariance_product_G}
\end{align}
where the last inequality follows because both shifted ranges $\{\tau-h:|\tau|\leq T-1\}$ and $\{\tau+h:|\tau|\leq T-1\}$ are contained in $\{-2(T-1),\dots,2(T-1)\}$.
For fixed $i,j$ and $h$, define
\begin{align*}
    D_{u,ij,h}(\tau)
    &:=\max_{\substack{t,s\in[T-h]\\t-s=\tau}}
    \left|\cov\left(u_{t,i}u_{t+h,j},u_{s,i}u_{s+h,j}\right)\right|,
\end{align*}
where the maximum over an empty set is taken to be zero by convention. Combining \eqref{eqn:cumulant_sum_G}, \eqref{eqn:first_covariance_product_G}, and
\eqref{eqn:shifted_covariance_product_G}, we obtain
\begin{align}
    \sum_{\tau=-(T-1)}^{T-1}
    D_{u,ij,h}(\tau)
    \lesssim B_u(T)+1.
    \label{eqn:product_covariance_sum_G}
\end{align}
For each $\tau$, the number of pairs $(t,s)\in[T-h]^2$ satisfying $t-s=\tau$ is $\max\{T-h-|\tau|,0\}$. Therefore, by \eqref{eqn:sample_covariance_variance_expansion},
\begin{align*}
    \var\left(\wt\Gamma_{u,ij}(h)\right)
    &\leq T^{-2}\sum_{\tau=-(T-1)}^{T-1}
    \max\{T-h-|\tau|,0\}D_{u,ij,h}(\tau) \\
    &\leq T^{-1}\sum_{\tau=-(T-1)}^{T-1}D_{u,ij,h}(\tau) \\
    &\lesssim T^{-1}(B_u(T)+1).
\end{align*}
Since $T^{-1}B_u(T)=O(R(T,d_u))$ and $T^{-1}=O(R(T,d_u))$, it follows that
\begin{align*}
    \var\left(\wt\Gamma_{u,ij}(h)\right)
    =O(R(T,d_u)).
\end{align*}
Since each entry of $\wt\bGamma_u(h)$ is centered and $p_u$ is bounded,
\begin{align*}
    \bb{E}\norm{\wt\bGamma_u(h)}_F^2
    =\sum_{i=1}^{p_u}\sum_{j=1}^{p_u}\bb{E}\left(\wt\Gamma_{u,ij}(h)^2\right)
    =\sum_{i=1}^{p_u}\sum_{j=1}^{p_u}\var\left(\wt\Gamma_{u,ij}(h)\right)
    =O(R(T,d_u)).
\end{align*}
Markov's inequality then gives
\begin{align*}
    \norm{\wt\bGamma_u(h)}_F^2=O_P(R(T,d_u)),
\end{align*}
which proves \eqref{eqn:sample_covariance_G}.

To prove \eqref{eqn:sample_cross_moment_G}, define
\begin{align*}
    \wt\bSigma_{uv}:=T^{-1}\sum_{t=1}^T\u_t\v_t'.
\end{align*}
Fix an entry $(i,j)\in[p_u]\times[p_v]$. Since the two processes $\{\u_t\}$ and $\{\v_t\}$ are independent and have mean zero, we have $\bb{E}(u_{t,i}v_{t,j})=0$ for every $t\in[T]$. Hence,
\begin{align}
    \var\left(\wt\Sigma_{uv,ij}\right)
    =T^{-2}\sum_{t=1}^T\sum_{s=1}^T
    \cov(u_{t,i}v_{t,j},u_{s,i}v_{s,j})
    =T^{-2}\sum_{t=1}^T\sum_{s=1}^T
    \cov(u_{t,i},u_{s,i})\cov(v_{t,j},v_{s,j}),
    \label{eqn:sample_cross_moment_variance}
\end{align}
where the second equality follows from the independence of $\{\u_t\}$ and $\{\v_t\}$.

Let $J_u(t)$ and $J_v(t)$ denote the segment indices of $\u_t$ and $\v_t$, respectively. For $a,a'\in[m_u]$ and $\tau\geq0$, define
\begin{align*}
    \gamma_{u,a,a'}(\tau)
    &:=\sigma_u^2\sum_{w\geq0}\psi_{u,a}(w+\tau)\psi_{u,a'}(w),
\end{align*}
where $\sigma_u^2:=\bb{E}(x_{u,0,1}^2)<\infty$, and set $\gamma_{u,a,a'}(\tau):=\gamma_{u,a',a}(-\tau)$ when $\tau<0$. Define $\gamma_{v,b,b'}(\tau)$ analogously for $b,b'\in[m_v]$. By the temporal independence of the innovations, whenever $J_u(t)=a$, $J_u(s)=a'$, and $t-s=\tau$, we have $\cov(u_{t,i},u_{s,i})=\gamma_{u,a,a'}(\tau)$, and the analogous equality holds for $\{\v_t\}$. In particular,
\begin{align*}
    \max_{\substack{a,a'\in[m_u]\\ b,b'\in[m_v]}}
    \left|\gamma_{u,a,a'}(\tau)\gamma_{v,b,b'}(\tau)\right|
    =\left\{\max_{a,a'\in[m_u]}|\gamma_{u,a,a'}(\tau)|\right\}
    \left\{\max_{b,b'\in[m_v]}|\gamma_{v,b,b'}(\tau)|\right\}.
\end{align*}

We first consider the case $d_u>0$ and $d_v>0$. Taking square roots in \eqref{eqn:segment_covariance_long_pointwise}, applied separately to $\{\u_t\}$ and $\{\v_t\}$, gives
\begin{align}
    \max_{\substack{a,a'\in[m_u]\\ b,b'\in[m_v]}}
    \left|\gamma_{u,a,a'}(\tau)\gamma_{v,b,b'}(\tau)\right|
    \lesssim (1+|\tau|)^{2d_u-1}(1+|\tau|)^{2d_v-1}
    = (1+|\tau|)^{4d_{uv}-2},
    \label{eqn:cross_covariance_decay}
\end{align}
where $d_{uv}:=(d_u+d_v)/2$. 

Next, consider the case where $d_u\leq 0$ or $d_v\leq 0$. We divide it into two disjoint subcases. First, suppose that $\min\{d_u,d_v\}<0$. Without loss of generality, let $d_u<0$. The argument in \eqref{eqn:segment_covariance_absolute_sum} gives 
\begin{align*}
    \sum_{\tau\in\bb{Z}}\max_{a,a'\in[m_u]}|\gamma_{u,a,a'}(\tau)|<\infty.
\end{align*}
Moreover, by the Cauchy--Schwarz inequality and \eqref{eqn:uniform_second_moment_G},
\begin{align*}
    \sup_{\tau\in\bb{Z}}\max_{b,b'\in[m_v]}|\gamma_{v,b,b'}(\tau)|
    \leq \sigma_v^2\max_{b,b'\in[m_v]}
    \left\{\sum_{w\geq 0}\psi_{v,b}(w)^2\right\}^{1/2}
    \left\{\sum_{w\geq 0}\psi_{v,b'}(w)^2\right\}^{1/2}
    \lesssim 1,
\end{align*}
where $\sigma_v^2:=\bb{E}(x_{v,0,1}^2)$. It follows that
\begin{align*}
    \sum_{\tau\in\bb{Z}}\max_{\substack{a,a'\in[m_u]\\ b,b'\in[m_v]}}
    \left|\gamma_{u,a,a'}(\tau)\gamma_{v,b,b'}(\tau)\right|
    <\infty.
\end{align*}
It remains to consider the case $\min\{d_u,d_v\}=0$. Without loss of generality, let $d_u=0$ and $d_v\geq 0$. The bound in \eqref{eqn:segment_covariance_edge_pointwise} gives
\begin{align*}
    \max_{a,a'\in[m_u]}|\gamma_{u,a,a'}(\tau)|
    \lesssim (1+|\tau|)^{-1}\log(2+|\tau|).
\end{align*}
If $d_v>0$, then
\begin{align*}
    \max_{\substack{a,a'\in[m_u]\\ b,b'\in[m_v]}}
    \left|\gamma_{u,a,a'}(\tau)\gamma_{v,b,b'}(\tau)\right|
    \lesssim (1+|\tau|)^{2d_v-2}\log(2+|\tau|),
\end{align*}
which is summable because $2d_v-2<-1$. If $d_v=0$, the same bound in \eqref{eqn:segment_covariance_edge_pointwise} gives
\begin{align*}
    \max_{\substack{a,a'\in[m_u]\\ b,b'\in[m_v]}}
    \left|\gamma_{u,a,a'}(\tau)\gamma_{v,b,b'}(\tau)\right|
    \lesssim (1+|\tau|)^{-2}\log^2(2+|\tau|),
\end{align*}
which is also summable. Hence, whenever $d_u\leq 0$ or $d_v\leq 0$, we have
\begin{align}
    \sum_{\tau\in\bb{Z}}\max_{\substack{a,a'\in[m_u]\\ b,b'\in[m_v]}}
    \left|\gamma_{u,a,a'}(\tau)\gamma_{v,b,b'}(\tau)\right|
    <\infty.
    \label{eqn:cross_covariance_short_sum}
\end{align}
In this case, since $d_u\leq 0$ or $d_v\leq 0$, while both $d_u,d_v<1/2$, we have $d_{uv}<1/4$.

Define
\begin{align*}
    B_{uv}(T)
    &:=\sum_{\tau=-(T-1)}^{T-1}\max_{\substack{a,a'\in[m_u]\\ b,b'\in[m_v]}}
    \left|\gamma_{u,a,a'}(\tau)\gamma_{v,b,b'}(\tau)\right|.
\end{align*}
Combining the two cases \eqref{eqn:cross_covariance_decay} and \eqref{eqn:cross_covariance_short_sum}, we obtain
\begin{align*}
    B_{uv}(T)\lesssim
    \begin{cases}
        1, & d_{uv}<1/4,\\
        \log T, & d_{uv}=1/4,\\
        T^{4d_{uv}-1}, & d_{uv}>1/4.
    \end{cases}
\end{align*}
Consequently, $T^{-1}B_{uv}(T)=O(R(T,d_{uv}))$. For each $\tau\in\{-(T-1),\ldots,T-1\}$, the number of pairs $(t,s)\in[T]^2$ satisfying $t-s=\tau$ is $T-|\tau|$. Therefore, grouping the pairs according to their lag gives
\begin{align*}
    \var\left(\wt\Sigma_{uv,ij}\right)
    &\le T^{-2}\sum_{\tau=-(T-1)}^{T-1}(T-|\tau|)
    \max_{\substack{a,a'\in[m_u]\\ b,b'\in[m_v]}}
    \left|\gamma_{u,a,a'}(\tau)\gamma_{v,b,b'}(\tau)\right| \\
    &\le T^{-1}B_{uv}(T)=O(R(T,d_{uv})).
\end{align*}
Since $p_u$ and $p_v$ are bounded and $\bb{E}(\wt\Sigma_{uv,ij})=0$,
\begin{align*}
    \bb{E}\norm{\wt\bSigma_{uv}}_F^2
    =\sum_{i=1}^{p_u}\sum_{j=1}^{p_v}\var\left(\wt\Sigma_{uv,ij}\right)
    =O(R(T,d_{uv})).
\end{align*}
Markov's inequality then gives
\begin{align*}
    \norm{\wt\bSigma_{uv}}_F^2=O_P(R(T,d_{uv})),
\end{align*}
which proves \eqref{eqn:sample_cross_moment_G} and completes the proof. $\square$

\begin{lemma}(Auxiliary bounds for error interactions)
\label{lemma:bound_Delta}
    Let Assumptions~(E1) and (E2) hold. Let $\{\u_t\}_{t\in[T]}\in\cG_T(\{d_{u,1},\ldots,d_{u,m_u}\}|\{\x_{u,i},i\in\bb{Z}\})$ be a $p_u$-dimensional piecewise-stationary time series, where $p_u$ and $m_u$ are bounded, and assume that $\{\u_t\}_{t\in[T]}$ is independent of $\{\E_t^{(m)}:t\in[T],\,m\in[M]\}$. Let $d_u:=\max_{1\le j\le m_u}d_{u,j}$. For each $m,n\in[M]$ with $m\neq n$, the following bounds hold:
    \begin{align}
        \big\| T^{-1}\sum_{t=1}^T \E_{t}^{(m)}\1_{q_m}\u_t' \big\|_F^2 
        &= O_P(T^{-1}p_mq_m), \label{eqn:E1u} \\
        \big\| T^{-1}\sum_{t=1}^T \E_t^{(m)}\1_{q_m}\1_{q_n}'\E_t^{(n)'} \big\|_F^2  
        &= O_P(T^{-1}p_mp_nq_mq_n), \label{eqn:Em11En} \\
        \big\| T^{-1}\sum_{t=1}^T \E_t^{(m)}\1_{q_m}\1_{q_m}'\E_t^{(m)'} \big\|_F^2
        &= O_P(p_mq_m^2+T^{-1}p_m^2q_m^2). \label{eqn:Em11Em}
	\end{align}
    Moreover, for any deterministic vector $\w=(w_1,\dots,w_T)'$ and for each $m\in[M]$,
    \begin{align}
        \Big\|T^{-1}\sum_{t=1}^T w_t\M_{p_m}\E_t^{(m)}\1_{q_m}\Big\|^2
        =O_P\big(T^{-2}p_mq_m\|\w\|^2\big). \label{eqn:weighted_row_error}
    \end{align}
\end{lemma}
\textbf{\textit{Proof of Lemma~\ref{lemma:bound_Delta}.}}
We first prove \eqref{eqn:E1u}. 
\begin{align*}
    \bb{E} \bigg\| T^{-1}\sum_{t=1}^T \E_{t}^{(m)}\1_{q_m}\u_t' \bigg\|_F^2 
    &= T^{-2}\sum_{t=1}^{T}\sum_{s=1}^T 
    \bb{E}\left(\tr(\E_{t}^{(m)}\1_{q_m}\u_t'\u_s\1_{q_m}'\E_{s}^{(m)'})\right) \\
    &= T^{-2}\sum_{t=1}^{T}\sum_{s=1}^T 
    \bb{E}(\1_{q_m}'\E_{s}^{(m)'}\E_{t}^{(m)}\1_{q_m})
    \bb{E}(\u_t'\u_s).
\end{align*}
The second equality follows because the whole process $\{\u_t\}_{t\in[T]}$ is independent of $\{\E_t^{(m)}\}_{t\in[T]}$. Define the admissible index set $I_T(h):=\{t\in[T]: t+h\in[T]\}$. Re-indexing by lag $h:= s-t$ gives
\begin{align*}
    \bb{E} \bigg\| T^{-1}\sum_{t=1}^T \E_{t}^{(m)}\1_{q_m}\u_t' \bigg\|_F^2 
    &= T^{-2} \sum_{h=-(T-1)}^{T-1} \sum_{t\in I_T(h)}
    \bb{E}\big(\1_{q_m}'\E_{t+h}^{(m)'}\E_{t}^{(m)}\1_{q_m}\big)
    \bb{E}(\u_{t}'\u_{t+h}).
\end{align*}
We bound the two expectations on the r.h.s.\ separately. First, expand
\begin{align*}
    \big|\bb{E}(\1_{q_m}'\E_{t+h}^{(m)'}\E_{t}^{(m)}\1_{q_m})\big|
    = \Big|\bb{E}\big(\sum_{i=1}^{p_m}\sum_{j=1}^{q_m}\sum_{l=1}^{q_m}E_{t+h,ij}^{(m)}E_{t,il}^{(m)}\big)\Big|
    \le \sum_{i=1}^{p_m}\sum_{j=1}^{q_m}\sum_{l=1}^{q_m}\big|\bb{E}(E_{t+h,ij}^{(m)}E_{t,il}^{(m)})\big|.
\end{align*}
By Lemma~\ref{lemma:correlation_Etm}, for each fixed $(i,j)$,
\begin{align*}
    \sum_{l=1}^{q_m}\big| \bb{E}(E_{t+h,ij}^{(m)}E_{t,il}^{(m)}) \big|
    \le \sum_{k=1}^{p_m}\sum_{l=1}^{q_m}\big| \bb{E}(E_{t+h,ij}^{(m)}E_{t,kl}^{(m)}) \big|
    \le C\rho(h),
\end{align*}
uniformly over $t,h,m,i,j$. Summing over $i$ and $j$, we obtain
\begin{align}
    \big|\bb{E}(\1_{q_m}'\E_{t+h}^{(m)'}\E_{t}^{(m)}\1_{q_m})\big|
    \le Cp_mq_m\rho(h).
    \label{eqn:E_inner_rho_bound}
\end{align}
Moreover, by \eqref{eqn:uniform_second_moment_G} in Lemma~\ref{lemma:prop_piecewise} and the Cauchy--Schwarz inequality,
\begin{align}
    \sup_{t\in I_T(h)}|\bb{E}(\u_t'\u_{t+h})|
    \le \sup_{t,s\in[T]}\left(\bb{E}\|\u_t\|^2\right)^{1/2}\left(\bb{E}\|\u_s\|^2\right)^{1/2}
    \le C.
    \label{eqn:u_cov_uniform_bound}
\end{align}
Combining \eqref{eqn:E_inner_rho_bound} and \eqref{eqn:u_cov_uniform_bound},
\begin{align*}
    \bb{E} \bigg\| T^{-1}\sum_{t=1}^T \E_{t}^{(m)}\1_{q_m}\u_t' \bigg\|_F^2 
    &\le CT^{-2}p_mq_m\sum_{h=-(T-1)}^{T-1}\sum_{t\in I_T(h)}\rho(h) \\
    &\le CT^{-1}p_mq_m\sum_{h\in\bb{Z}}\rho(h) \\
    &=O(T^{-1}p_mq_m),
\end{align*}
and Markov's inequality proves \eqref{eqn:E1u}.

We next prove \eqref{eqn:Em11En}. First, for each group $m\in[M]$, define $\bzeta_t^{(m)}:=q_m^{-1}\E_t^{(m)}\1_{q_m}$. Then
\begin{align*}
    \bb{E}\bigg\|T^{-1}\sum_{t=1}^T 
    \E_t^{(m)}\1_{q_m}\1_{q_n}'\E_t^{(n)'}\bigg\|_F^2 
    = q_m^2q_n^2\sum_{i=1}^{p_m}\sum_{k=1}^{p_n}
    \bb{E}\left(T^{-1}\sum_{t=1}^T\zeta_{t,i}^{(m)}\zeta_{t,k}^{(n)}\right)^2 .
\end{align*}
We first show that, for $m\ne n$ and $t,s\in[T]$,
\begin{align}
    \bb{E}\left(\zeta_{t,i}^{(m)}\zeta_{s,k}^{(n)}\right)=0.
    \label{eqn:zeta_cross_zero}
\end{align}
Write $\zeta_{t,i}^{(m)}=\zeta_{G,t,i}^{(m)}+\zeta_{F,t,i}^{(m)}+\zeta_{\epsilon,t,i}^{(m)}$, where the three terms are the corresponding row-averaged components induced respectively by
$\Q_{e,m}\G_{e,t}\J_{e,m}'$, $\R_{e,m}\F_{e,t}^{(m)}\C_{e,m}'$, and
$\bSigma_{\epsilon,m}\circ\bepsilon_t^{(m)}$ in (E1). By the mean-zero assumptions and the mutual independence of the global error, local error, and idiosyncratic error innovation processes in (E1) and (E2), all cross terms vanish, and hence
\begin{align*}
    \bb{E}\left(\zeta_{t,i}^{(m)}\zeta_{s,k}^{(n)}\right)
    =\bb{E}\left(\zeta_{G,t,i}^{(m)}\zeta_{G,s,k}^{(n)}\right)
    +\bb{E}\left(\zeta_{F,t,i}^{(m)}\zeta_{F,s,k}^{(n)}\right)
    +\bb{E}\left(\zeta_{\epsilon,t,i}^{(m)}\zeta_{\epsilon,s,k}^{(n)}\right).
\end{align*}
The last two terms are zero by the cross-group independence of the group-specific local error processes and the idiosyncratic error processes in (E2). It remains to consider the global-global term. By (E2),
\begin{align*}
    \bb{E}\left((\G_{e,t})_{ab}(\G_{e,s})_{cd}\right)
    =\bb{1}\{a=c,b=d\}
    \sum_{u,v\ge0}a_{e,u}a_{e,v}\bb{1}\{t-u=s-v\}.
\end{align*}
Therefore,
\begin{align*}
    \bb{E}\{\zeta_{G,t,i}^{(m)}\zeta_{G,s,k}^{(n)}\}
    &=(q_mq_n)^{-1}
    \sum_{j=1}^{q_m}\sum_{l=1}^{q_n}
    \bb{E}\left((\Q_{e,m}\G_{e,t}\J_{e,m}')_{ij}
    (\Q_{e,n}\G_{e,s}\J_{e,n}')_{kl}\right) \\
    &=(q_mq_n)^{-1}
    \left(\sum_{u,v\ge0}a_{e,u}a_{e,v}\bb{1}\{t-u=s-v\}\right)
    (\Q_{e,m}\Q_{e,n}')_{ik}
    \big(\1_{q_m}'\J_{e,m}\J_{e,n}'\1_{q_n}\big) \\
    &=0,
\end{align*}
where the last equality follows from the cross-group orthogonality $\Q_{e,m}\Q_{e,n}'=\0$ or $\J_{e,m}\J_{e,n}'=\0$ in (E1). This proves \eqref{eqn:zeta_cross_zero}.

By \eqref{eqn:zeta_cross_zero}, for each fixed $i\in[p_m]$ and $k\in[p_n]$,
\begin{align*}
    &\quad\;\bb{E}\left(
    T^{-1}\sum_{t=1}^T
    \zeta_{t,i}^{(m)}\zeta_{t,k}^{(n)}\right)^2
    =T^{-2}\sum_{h=-(T-1)}^{T-1}\sum_{t\in I_T(h)}
    \cov\left(
    \zeta_{t+h,i}^{(m)}\zeta_{t+h,k}^{(n)},
    \zeta_{t,i}^{(m)}\zeta_{t,k}^{(n)}\right)  \\
    &\le T^{-2}q_m^{-2}q_n^{-2}
    \sum_{h=-(T-1)}^{T-1}\sum_{t\in I_T(h)}
    \sum_{j=1}^{q_m}\sum_{l=1}^{q_n}
    \sum_{a=1}^{q_m}\sum_{b=1}^{q_n}
    \left|\cov\left(
    E_{t+h,ij}^{(m)}E_{t+h,kl}^{(n)},
    E_{t,ia}^{(m)}E_{t,kb}^{(n)}
    \right)\right|.
\end{align*}
By \eqref{eqn:error_product_cov} in Lemma~\ref{lemma:correlation_Etm},
\begin{align*}
    \sum_{a=1}^{q_m}\sum_{b=1}^{q_n}
    \left|\cov\left(
    E_{t+h,ij}^{(m)}E_{t+h,kl}^{(n)},
    E_{t,ia}^{(m)}E_{t,kb}^{(n)}
    \right)\right|
    \le C\rho(h),
\end{align*}
uniformly over all admissible indices. Therefore,
\begin{align*}
    \bb{E}\left(
    T^{-1}\sum_{t=1}^T
    \zeta_{t,i}^{(m)}\zeta_{t,k}^{(n)}\right)^2
    &\le CT^{-2}q_m^{-2}q_n^{-2}
    \sum_{h=-(T-1)}^{T-1}\sum_{t\in I_T(h)}
    \sum_{j=1}^{q_m}\sum_{l=1}^{q_n}\rho(h) \\
    &\le CT^{-1}q_m^{-1}q_n^{-1}
    \sum_{h\in\bb{Z}}\rho(h) \\
    &=O(T^{-1}q_m^{-1}q_n^{-1}).
\end{align*}
Consequently,
\begin{align*}
    \bb{E}\bigg\|T^{-1}\sum_{t=1}^T 
    \E_t^{(m)}\1_{q_m}\1_{q_n}'\E_t^{(n)'}\bigg\|_F^2 
    &\le Cq_m^2q_n^2
    \sum_{i=1}^{p_m}\sum_{k=1}^{p_n}T^{-1}q_m^{-1}q_n^{-1} \\
    &=O(T^{-1}p_mp_nq_mq_n).
\end{align*}
Markov's inequality proves \eqref{eqn:Em11En}.

We next prove \eqref{eqn:Em11Em}. Using the same notation $\bzeta_t^{(m)}=q_m^{-1}\E_t^{(m)}\1_{q_m}$, we have
\begin{align*}
    T^{-1}\sum_{t=1}^T
    \E_t^{(m)}\1_{q_m}\1_{q_m}'\E_t^{(m)'}
    =T^{-1}q_m^2\sum_{t=1}^T\bzeta_t^{(m)}\bzeta_t^{(m)'}.
\end{align*}
Hence,
\begin{align*}
    &\quad\;\bb{E}\Big\|T^{-1}\sum_{t=1}^T
    \E_t^{(m)}\1_{q_m}\1_{q_m}'\E_t^{(m)'}\Big\|_F^2  \\
    &= q_m^4\bb{E}\Big\|T^{-1}\sum_{t=1}^T\bzeta_t^{(m)}\bzeta_t^{(m)'}\Big\|_F^2  \\
    &\le 2q_m^4
    \Big\|T^{-1}\sum_{t=1}^T
    \bb{E}\left(\bzeta_t^{(m)}\bzeta_t^{(m)'}\right)\Big\|_F^2
    +2q_m^4\bb{E}\Big\|T^{-1}\sum_{t=1}^T
    \Big[\bzeta_t^{(m)}\bzeta_t^{(m)'}-\bb{E}\left(\bzeta_t^{(m)}\bzeta_t^{(m)'}\right)\Big]\Big\|_F^2 .
\end{align*}
We first bound the deterministic part. Let $\bSigma_{\zeta,t}^{(m)}:=\bb{E}\left(\bzeta_t^{(m)}\bzeta_t^{(m)'}\right)$. For each fixed $t\in[T]$ and $i\in[p_m]$, Lemma~\ref{lemma:correlation_Etm} with $h=0$ gives
\begin{align*}
    \sum_{k=1}^{p_m}
    \big|(\bSigma_{\zeta,t}^{(m)})_{ik}\big|
    &=q_m^{-2}\sum_{k=1}^{p_m}
    \left|\sum_{j=1}^{q_m}\sum_{l=1}^{q_m}
    \bb{E}(E_{t,ij}^{(m)}E_{t,kl}^{(m)})\right| \\
    &\le q_m^{-2}
    \sum_{j=1}^{q_m}
    \sum_{k=1}^{p_m}\sum_{l=1}^{q_m}
    \big|\bb{E}(E_{t,ij}^{(m)}E_{t,kl}^{(m)})\big| \\
    &\le Cq_m^{-1}.
\end{align*}
Therefore,
\begin{align*}
    \big\|\bSigma_{\zeta,t}^{(m)}\big\|_F^2
    =\sum_{i=1}^{p_m}\sum_{k=1}^{p_m}\big|(\bSigma_{\zeta,t}^{(m)})_{ik}\big|^2
    \le\sum_{i=1}^{p_m}\left\{\sum_{k=1}^{p_m}
    \big|(\bSigma_{\zeta,t}^{(m)})_{ik}\big|\right\}^2
    =O(p_mq_m^{-2}),
\end{align*}
uniformly over $t\in[T]$. By Jensen's inequality,
\begin{align*}
    \Big\|T^{-1}\sum_{t=1}^T
    \bb{E}\left(\bzeta_t^{(m)}\bzeta_t^{(m)'}\right)\Big\|_F^2
    \le T^{-1}\sum_{t=1}^T
    \big\|\bSigma_{\zeta,t}^{(m)}\big\|_F^2
    =O(p_mq_m^{-2}).
\end{align*}

We next bound the centered stochastic part. Write
\begin{align*}
    &\quad\;\bb{E}\Big\|T^{-1}\sum_{t=1}^T
    \Big[\bzeta_t^{(m)}\bzeta_t^{(m)'}
    -\bb{E}\left(\bzeta_t^{(m)}\bzeta_t^{(m)'}\right)
    \Big]\Big\|_F^2  \\
    &=T^{-2}\sum_{i=1}^{p_m}\sum_{k=1}^{p_m}
    \sum_{h=-(T-1)}^{T-1}\sum_{t\in I_T(h)}
    \cov\left(
    \zeta_{t+h,i}^{(m)}\zeta_{t+h,k}^{(m)},
    \zeta_{t,i}^{(m)}\zeta_{t,k}^{(m)}
    \right).
\end{align*}
Taking absolute values and using $\bzeta_t^{(m)}=q_m^{-1}\E_t^{(m)}\1_{q_m}$, we obtain
\begin{align*}
    &\quad\;\bb{E}\Big\|T^{-1}\sum_{t=1}^T
    \Big[\bzeta_t^{(m)}\bzeta_t^{(m)'}
    -\bb{E}\left(\bzeta_t^{(m)}\bzeta_t^{(m)'}\right)
    \Big]\Big\|_F^2  \\
    &\le T^{-2}q_m^{-4}
    \sum_{i=1}^{p_m}\sum_{k=1}^{p_m}
    \sum_{h=-(T-1)}^{T-1}\sum_{t\in I_T(h)}
    \sum_{j=1}^{q_m}\sum_{l=1}^{q_m}
    \sum_{a=1}^{q_m}\sum_{b=1}^{q_m}
    \left|\cov\left(
    E_{t+h,ij}^{(m)}E_{t+h,kl}^{(m)},
    E_{t,ia}^{(m)}E_{t,kb}^{(m)}
    \right)\right|.
\end{align*}
By \eqref{eqn:error_product_cov} in Lemma~\ref{lemma:correlation_Etm}, with the two group indices both equal to $m$,
\begin{align*}
    \sum_{a=1}^{q_m}\sum_{b=1}^{q_m}
    \left|\cov\left(
    E_{t+h,ij}^{(m)}E_{t+h,kl}^{(m)},
    E_{t,ia}^{(m)}E_{t,kb}^{(m)}
    \right)\right|
    \le C\rho(h),
\end{align*}
uniformly over all admissible indices. Hence,
\begin{align*}
    &\quad\;\bb{E}\Big\|T^{-1}\sum_{t=1}^T
    \Big[\bzeta_t^{(m)}\bzeta_t^{(m)'}
    -\bb{E}\left(\bzeta_t^{(m)}\bzeta_t^{(m)'}\right)
    \Big]\Big\|_F^2  \\
    &\le CT^{-2}q_m^{-4}
    \sum_{i=1}^{p_m}\sum_{k=1}^{p_m}
    \sum_{h=-(T-1)}^{T-1}\sum_{t\in I_T(h)}
    \sum_{j=1}^{q_m}\sum_{l=1}^{q_m}
    \rho(h) \\
    &\le CT^{-1}p_m^2q_m^{-2}
    \sum_{h\in\bb Z}\rho(h) \\
    &=O(T^{-1}p_m^2q_m^{-2}).
\end{align*}
Combining the deterministic and centered stochastic parts, we obtain
\begin{align*}
    \bb{E}\Big\|T^{-1}\sum_{t=1}^T
    \E_t^{(m)}\1_{q_m}\1_{q_m}'\E_t^{(m)'}\Big\|_F^2
    =O(p_mq_m^2+T^{-1}p_m^2q_m^2).
\end{align*}
Markov's inequality proves \eqref{eqn:Em11Em}.

To prove \eqref{eqn:weighted_row_error}, retain the notation $\bzeta_t^{(m)}=q_m^{-1}\E_t^{(m)}\1_{q_m}$. Then, 
\begin{align*}
    \bb{E}\Big\|T^{-1}\sum_{t=1}^T
    w_t\M_{p_m}\E_t^{(m)}\1_{q_m}\Big\|^2
    &=T^{-2}q_m^2\sum_{t=1}^T\sum_{s=1}^T w_tw_s
    \bb{E}\left(\bzeta_t^{(m)'}\M_{p_m}\bzeta_s^{(m)}\right) \\
    &\le T^{-2}q_m^2\sum_{t=1}^T\sum_{s=1}^T
    |w_tw_s|\left|\bb{E}\left(\bzeta_t^{(m)'}\M_{p_m}\bzeta_s^{(m)}\right)\right|.
\end{align*}
For any $h\in\bb Z$ and $t\in I_T(h)$, write
\begin{align*}
    \left|\bb{E}\left(\bzeta_{t+h}^{(m)'}\M_{p_m}\bzeta_t^{(m)}\right)\right|
    &=q_m^{-2}\left|\sum_{i=1}^{p_m}\sum_{k=1}^{p_m}
    (\M_{p_m})_{ik}\sum_{j=1}^{q_m}\sum_{l=1}^{q_m}
    \bb{E}\left(E_{t+h,ij}^{(m)}E_{t,kl}^{(m)}\right)\right| \\
    &\le q_m^{-2}\sum_{i=1}^{p_m}\sum_{j=1}^{q_m}
    \sum_{k=1}^{p_m}\sum_{l=1}^{q_m}
    \left|\bb{E}\left(E_{t+h,ij}^{(m)}E_{t,kl}^{(m)}\right)\right| \\
    &\le Cq_m^{-2}\sum_{i=1}^{p_m}\sum_{j=1}^{q_m}\rho(h) \\
    &=Cp_mq_m^{-1}\rho(h),
\end{align*}
where the first inequality uses $|(\M_{p_m})_{ik}|\le 1$, and the last inequality follows from \eqref{eqn:error_cov_summable} in Lemma~\ref{lemma:correlation_Etm}. Therefore,
\begin{align*}
    \bb{E}\Big\|T^{-1}\sum_{t=1}^T
    w_t\M_{p_m}\E_t^{(m)}\1_{q_m}\Big\|^2
    \le CT^{-2}p_mq_m
    \sum_{h=-(T-1)}^{T-1}\sum_{t\in I_T(h)}|w_tw_{t+h}|\rho(h).
\end{align*}
Recall the $p=1,q=2,r=2$ case of the discrete Young inequality: if $\sum_{h\in\bb Z}|a_h|<\infty$ and $\sum_{t\in\bb Z}b_t^2<\infty$, then
\begin{align*}
    \sum_{s\in\bb Z}
    \left(\sum_{t\in\bb Z}|a_{s-t}||b_t|\right)^2
    \le\left(\sum_{h\in\bb Z}|a_h|\right)^2
    \sum_{t\in\bb Z}b_t^2 .
\end{align*}
Extending $w_t$ by zero outside $[T]$, and taking $a_h=\rho(h)$ and $b_t=w_t$ in the preceding inequality, we have
\begin{align*}
    &\quad\;\sum_{h=-(T-1)}^{T-1}\sum_{t\in I_T(h)}
    |w_tw_{t+h}|\rho(h)
    =\sum_{s\in\bb Z}|w_s|\sum_{t\in\bb Z}\rho(s-t)|w_t| \\
    &\le\left(\sum_{s\in\bb Z}w_s^2\right)^{1/2}
    \left\{\sum_{s\in\bb Z}
    \left(\sum_{t\in\bb Z}\rho(s-t)|w_t|\right)^2
    \right\}^{1/2} 
    \le\left(\sum_{h\in\bb Z}\rho(h)\right)\sum_{t\in\bb Z}w_t^2 
    \le C\|\w\|^2,
\end{align*}
where the second inequality follows from the discrete Young inequality, and the last inequality follows from $\sum_{h\in\bb Z}\rho(h)<\infty$ in Lemma~\ref{lemma:correlation_Etm}. Consequently,
\begin{align*}
    \bb{E}\Big\|T^{-1}\sum_{t=1}^T
    w_t\M_{p_m}\E_t^{(m)}\1_{q_m}\Big\|^2
    \le CT^{-2}p_mq_m\|\w\|^2.
\end{align*}
Markov's inequality proves \eqref{eqn:weighted_row_error}. This completes the proof. $\square$

\begin{lemma}(Asymptotic orthogonality of projected rotation)
\label{lemma:rotation_orthogonality}
    Let $\U,\wh\U\in\mathbb{R}^{d\times r}$ satisfy $\U'\U=\I_r$ and $\wh\U'\wh\U=\I_r$. Define $\P_U:=\U\U'$, $\wh\P_U:=\wh\U\wh\U'$, and $\H_U:=\U'\wh\U$. Then, 
    \begin{align}
        \wh\U-\U\H_U=(\I_d-\P_U)\wh\U.
    \end{align}
    Moreover, if $\norm{\wh\U-\U\H_U}_F=o_P(1)$, then
    \begin{align}
        \H_U'\H_U=\I_r+o_P(1),
        \qquad \H_U\H_U'=\I_r+o_P(1),
    \end{align}
    and $\H_U$ is invertible with probability tending to one. In addition,
    \begin{align}
        \norm{\wh\P_U-\P_U}_F^2
        = 2\norm{\wh\U-\U\H_U}_F^2.
    \end{align}
\end{lemma}

\textbf{\textit{Proof of Lemma~\ref{lemma:rotation_orthogonality}.}}
By the definition of $\H_U$, $\U\H_U=\U\U'\wh\U=\P_U\wh\U$. Therefore,
\begin{align*}
    \wh\U-\U\H_U
    =\wh\U-\P_U\wh\U
    =(\I_d-\P_U)\wh\U.
\end{align*}
Let $\E_U:=\wh\U-\U\H_U$. Since $\U'\E_U=0$, we have
\begin{align*}
    \I_r
    =\wh\U'\wh\U
    =(\U\H_U+\E_U)'(\U\H_U+\E_U)
    =\H_U'\H_U+\E_U'\E_U.
\end{align*}
Hence
\begin{align*}
    \norm{\H_U'\H_U-\I_r}
    =\norm{\E_U}^2
    \le\norm{\E_U}_F^2
    =o_P(1).
\end{align*}
Since $\H_U$ is square, $\H_U'\H_U$ and $\H_U\H_U'$ have the same eigenvalues, given by the squared singular values of $\H_U$. Hence
\begin{align*}
    \norm{\H_U\H_U'-\I_r}
    =\max_{1\le j\le r}|\sigma_j^2(\H_U)-1|
    =\norm{\H_U'\H_U-\I_r}
    =o_P(1).
\end{align*}
Also, since $\lambda_{\min}(\H_U'\H_U)=1+o_P(1)$, $\H_U$ is invertible with probability tending to one.

Finally, write
\begin{align*}
    \norm{\wh\P_U-\P_U}_F^2
    &=\tr(\wh\P_U)+\tr(\P_U)
    -2\tr(\wh\P_U\P_U)  \\
    &=2r-2\tr(\wh\U'\U\U'\wh\U)  \\
    &=2r-2\tr(\H_U'\H_U).
\end{align*}
On the other hand, from 
$\I_r=\H_U'\H_U+\E_U'\E_U$, we have
\begin{align*}
    \norm{\E_U}_F^2
    =\tr(\E_U'\E_U)
    =r-\tr(\H_U'\H_U).
\end{align*}
Therefore,
\begin{align*}
    \norm{\wh\P_U-\P_U}_F^2
    =2\norm{\E_U}_F^2
    =2\norm{\wh\U-\U\H_U}_F^2.
\end{align*}
This completes the proof.

\begin{lemma}(Bounds for $\S_{A_1,mn}$ and $\bDelta_{A_1,mn}$)\label{lemma:bound_S_A1_Delta_A1}
    For any $m,n\in[M]$ with $m\neq n$, let $\S_{A_1,mn}$ and $\wh\S_{A_1,mn}$ be defined in \eqref{eqn:S_A1} and \eqref{eqn:S_A1_hat}, respectively, and define $\bDelta_{A_1,mn}:=\wh\S_{A_1,mn}-\S_{A_1,mn}$. Under the assumptions of Theorem~\ref{thm:initial_A1m},
    \begin{align}
        \norm{\S_{A_1,mn}}_F^2
        &= O(p_m^{\gamma_r}p_n^{\gamma_r}), \label{eqn:S_A1_bound} \\
        \norm{\bDelta_{A_1,mn}}_F^2
        &= o_P(p_m^{\gamma_r}p_n^{\gamma_r}). \label{eqn:Delta_A1_oP}
    \end{align}
    Furthermore, defining $\P_{A_1,m}:=\check{\A}_{1m}\check{\A}_{1m}'$, we have
    \begin{align}
        \norm{(\I_{p_m}-\P_{A_1,m})\bDelta_{A_1,mn}}_F^2
        = O_P\Big(R(T,d_{r,\star})p_m^{\gamma_r^{(m)}}p_n^{\gamma_r^{(n)}}
        +T^{-1}p_mp_n^{\gamma_{r,\max}^{(n)}}q_m^{-1}\Big).
        \label{eqn:Delta_A1_projected_bound}
    \end{align}
\end{lemma}
\textbf{\textit{Proof of Lemma~\ref{lemma:bound_S_A1_Delta_A1}.}}
Recall from \eqref{eqn:S_A1} and \eqref{eqn:S_A1_hat} that
\begin{align*}
    \S_{A_1,mn}
    &:=T^{-1}q_m^{-1}q_n^{-1}\M_{p_m}
    \sum_{t=1}^T\bb{E}\left(\Y_t^{(m)}\1_{q_m}\1_{q_n}'\Y_t^{(n)'}\right)\M_{p_n}, \\
    \wh\S_{A_1,mn}
    &:=T^{-1}q_m^{-1}q_n^{-1}\M_{p_m}
    \sum_{t=1}^T\left(\Y_t^{(m)}\1_{q_m}\1_{q_n}'\Y_t^{(n)'}\right)\M_{p_n}.
\end{align*}

We first prove \eqref{eqn:S_A1_bound}. By \eqref{eqn:uniform_second_moment_G} in
Lemma~\ref{lemma:prop_piecewise},
\begin{align*}
    \left\|T^{-1}\sum_{t=1}^T\bb{E}(\balpha_t\balpha_t')\right\|_F
    \leq T^{-1}\sum_{t=1}^T\bb{E}\|\balpha_t\balpha_t'\|_F
    = T^{-1}\sum_{t=1}^T\bb{E}\|\balpha_t\|^2
    = O(1).
\end{align*}
By \eqref{eqn: Omega_population}, we have $\S_{A_1,mn}=\A_{1m}\left(T^{-1}\sum_{t=1}^T\bb{E}(\balpha_t\balpha_t')\right)\A_{1n}'$. Hence, by (ME1), 
\begin{align*}
    \norm{\S_{A_1,mn}}_F^2
    \leq \norm{\A_{1m}}_F^2
    \left\|T^{-1}\sum_{t=1}^T\bb{E}(\balpha_t\balpha_t')\right\|_F^2
    \norm{\A_{1n}}_F^2
    =O\bigl(p_m^{\gamma_r}p_n^{\gamma_r}\bigr),
\end{align*}
which proves \eqref{eqn:S_A1_bound}.

We next prove \eqref{eqn:Delta_A1_oP} and \eqref{eqn:Delta_A1_projected_bound}. By (IC1), for each $m\in[M]$, 
\begin{align*}
    q_m^{-1}\M_{p_m}\Y_t^{(m)}\1_{q_m}
    =\a_t^{(m)}+q_m^{-1}\M_{p_m}\E_t^{(m)}\1_{q_m}.
\end{align*}
Substituting these identities into the definitions of $\S_{A_1,mn}$ and $\wh\S_{A_1,mn}$ gives 
\begin{align*}
    \bDelta_{A_1,mn} = \wh\S_{A_1,mn} - \S_{A_1,mn} 
    = \cI_{A_1,1}+\cI_{A_1,2}+\cI_{A_1,3}+\cI_{A_1,4}, 
\end{align*}
where
\begin{alignat*}{2}
    \cI_{A_1,1} &:= T^{-1}\sum_{t=1}^T \a_t^{(m)} \a_t^{(n)'} - \S_{A_1,mn}, \qquad
    &&\cI_{A_1,2} := (Tq_n)^{-1} \sum_{t=1}^T \a_t^{(m)}\1_{q_n}'\E_t^{(n)'}\M_{p_n}, \\
    \cI_{A_1,3} &:= (Tq_m)^{-1} \sum_{t=1}^T \M_{p_m}\E_t^{(m)}\1_{q_m}\a_t^{(n)'}, \qquad
    &&\cI_{A_1,4} := (T q_m q_n)^{-1} \sum_{t=1}^T \M_{p_m}\E_t^{(m)}\1_{q_m}\1_{q_n}'\E_t^{(n)'}\M_{p_n}.
\end{alignat*}
We bound the four components separately. Throughout the remainder of the proof, we repeatedly use the contraction properties \eqref{eqn:projector_contraction} in Proposition~\ref{prop:projector_prop} to remove orthogonal projectors from upper bounds without further mention. By \eqref{eqn: Omega_population}, since $\S_{A_1,mn}=T^{-1}\sum_{t=1}^T\bb{E}\left(\a_t^{(m)}\a_t^{(n)'}\right)$, we have
\begin{align*}
    \cI_{A_1,1}
    =T^{-1}\sum_{t=1}^T
    \left\{\a_t^{(m)}\a_t^{(n)'}-\bb{E}\left(\a_t^{(m)}\a_t^{(n)'}\right)\right\}.
\end{align*}
Using $\a_t^{(m)}=\A_{1m}\balpha_t+\A_{2m}\balpha_t^{(m)}$ and $\a_t^{(n)}=\A_{1n}\balpha_t+\A_{2n}\balpha_t^{(n)}$, together with the sample orthogonality and cross-group uncorrelatedness conditions in (IC2), we obtain
\begin{align*}
    \cI_{A_1,1}=
    \A_{1m}\left\{T^{-1}\sum_{t=1}^T
    \left(\balpha_t\balpha_t'-\bb{E}(\balpha_t\balpha_t')\right)\right\}\A_{1n}'
    +\A_{2m}\left(T^{-1}\sum_{t=1}^T\balpha_t^{(m)}\balpha_t^{(n)'}\right)\A_{2n}'.
\end{align*}
It follows from \eqref{eqn:sample_covariance_G} and \eqref{eqn:sample_cross_moment_G} in Lemma~\ref{lemma:prop_piecewise}, together with (ME1) and (ME2), that
\begin{align}
    \norm{\cI_{A_1,1}}_F^2
    =O_P\left(R(T,d_{r,\star})\left(p_m^{\gamma_r}p_n^{\gamma_r}
    +p_m^{\gamma_r^{(m)}}p_n^{\gamma_r^{(n)}}\right)\right).
    \label{eqn:I_A1_1_full}
\end{align}
Moreover, since $(\I_{p_m}-\P_{A_1,m})\A_{1m}=\0$,
\begin{align*}
    (\I_{p_m}-\P_{A_1,m})\cI_{A_1,1}
    =(\I_{p_m}-\P_{A_1,m})\A_{2m}
    \left(T^{-1}\sum_{t=1}^T\balpha_t^{(m)}\balpha_t^{(n)'}\right)\A_{2n}'.
\end{align*}
Thus,
\begin{align}
    \norm{(\I_{p_m}-\P_{A_1,m})\cI_{A_1,1}}_F^2
    =O_P\left(R(T,d_{r,\star})p_m^{\gamma_r^{(m)}}p_n^{\gamma_r^{(n)}}\right).
    \label{eqn:I_A1_1_projected}
\end{align}

For $\cI_{A_1,2}$, write
\begin{align*}
    \cI_{A_1,2}
    =\A_{1m}\left\{(Tq_n)^{-1}\sum_{t=1}^T
    \balpha_t\1_{q_n}'\E_t^{(n)'}\M_{p_n}\right\}
    +\A_{2m}\left\{(Tq_n)^{-1}\sum_{t=1}^T\balpha_t^{(m)}\1_{q_n}'\E_t^{(n)'}\M_{p_n}\right\}.
\end{align*}
Applying \eqref{eqn:E1u} in Lemma~\ref{lemma:bound_Delta} to $\{\balpha_t\}$ and $\{\balpha_t^{(m)}\}$, together with (ME1),
\begin{align}
    \norm{\cI_{A_1,2}}_F^2
    =O_P\left(T^{-1}\left(p_m^{\gamma_r}
    +p_m^{\gamma_r^{(m)}}\right)p_nq_n^{-1}\right).
    \label{eqn:I_A1_2_full}
\end{align}
Since left-multiplication by $(\I_{p_m}-\P_{A_1,m})$ eliminates the term involving $\A_{1m}$,
\begin{align*}
    (\I_{p_m}-\P_{A_1,m})\cI_{A_1,2}
    =(\I_{p_m}-\P_{A_1,m})\A_{2m}
    \left\{(Tq_n)^{-1}\sum_{t=1}^T\balpha_t^{(m)}\1_{q_n}'\E_t^{(n)'}\M_{p_n}\right\}.
\end{align*}
Consequently,
\begin{align}
    \norm{(\I_{p_m}-\P_{A_1,m})\cI_{A_1,2}}_F^2
    =O_P\left(T^{-1}p_m^{\gamma_r^{(m)}}p_nq_n^{-1}\right).
    \label{eqn:I_A1_2_projected}
\end{align}

For $\cI_{A_1,3}$, we have
\begin{align*}
    \cI_{A_1,3}
    =q_m^{-1}\left(T^{-1}\sum_{t=1}^T\M_{p_m}\E_t^{(m)}\1_{q_m}\balpha_t'\right)\A_{1n}'
    +q_m^{-1}\left(T^{-1}\sum_{t=1}^T\M_{p_m}\E_t^{(m)}\1_{q_m}\balpha_t^{(n)'}\right)\A_{2n}'.
\end{align*}
Applying \eqref{eqn:E1u} in Lemma~\ref{lemma:bound_Delta} and (ME1) gives
\begin{align}
    \norm{\cI_{A_1,3}}_F^2
    =O_P\left(T^{-1}p_mq_m^{-1}\left(p_n^{\gamma_r}+p_n^{\gamma_r^{(n)}}\right)\right)
    =O_P\left(T^{-1}p_mp_n^{\gamma_{r,\max}^{(n)}}q_m^{-1}\right).
    \label{eqn:I_A1_3_full}
\end{align}
Consequently,
\begin{align}
    \norm{(\I_{p_m}-\P_{A_1,m})\cI_{A_1,3}}_F^2
    \leq \norm{\cI_{A_1,3}}_F^2
    = O_P\Big(T^{-1}p_mp_n^{\gamma_{r,\max}^{(n)}}q_m^{-1}\Big).
    \label{eqn:I_A1_3_projected}
\end{align}

For $\cI_{A_1,4}$, by \eqref{eqn:Em11En} in Lemma~\ref{lemma:bound_Delta},
\begin{align}
    \norm{\cI_{A_1,4}}_F^2
    &=O_P\Big(T^{-1}p_mp_nq_m^{-1}q_n^{-1}\Big),
    \label{eqn:I_A1_4_full} \\
    \norm{(\I_{p_m}-\P_{A_1,m})\cI_{A_1,4}}_F^2
    &=O_P\Big(T^{-1}p_mp_nq_m^{-1}q_n^{-1}\Big).
    \label{eqn:I_A1_4_projected}
\end{align}
Using $\norm{\sum_{j=1}^4\cI_{A_1,j}}_F^2\lesssim\sum_{j=1}^4\norm{\cI_{A_1,j}}_F^2$, together with \eqref{eqn:I_A1_1_full}, \eqref{eqn:I_A1_2_full}, \eqref{eqn:I_A1_3_full}, and \eqref{eqn:I_A1_4_full}, we obtain
\begin{align}
    \norm{\bDelta_{A_1,mn}}_F^2
    =O_P\Bigg(&R(T,d_{r,\star})\left(p_m^{\gamma_r}p_n^{\gamma_r}
    +p_m^{\gamma_r^{(m)}}p_n^{\gamma_r^{(n)}}\right)
    +T^{-1}p_m^{\gamma_{r,\max}^{(m)}}p_nq_n^{-1} \nonumber \\
    &+T^{-1}p_mp_n^{\gamma_{r,\max}^{(n)}}q_m^{-1}
    +T^{-1}p_mp_nq_m^{-1}q_n^{-1}\Bigg).
    \label{eqn:Delta_A1_bound}
\end{align}

To deduce \eqref{eqn:Delta_A1_oP}, first note that $T^{-1}=O(R(T,d_{r,\star}))$. Moreover, by (R1), $p_s^{1-\gamma_r}q_s^{-1}$ and $p_s^{1-\gamma_r^{(s)}}q_s^{-1}$ are both $o(1)$ uniformly in $s\in[M]$. Since $p_m^{\gamma_{r,\max}^{(m)}} \leq p_m^{\gamma_r}+p_m^{\gamma_r^{(m)}}$, we have
\begin{align*}
    T^{-1}p_m^{\gamma_{r,\max}^{(m)}}p_nq_n^{-1}
    &\lesssim R(T,d_{r,\star})
    \Big\{p_m^{\gamma_r}p_n^{\gamma_r}p_n^{1-\gamma_r}q_n^{-1}
    +p_m^{\gamma_r^{(m)}}p_n^{\gamma_r^{(n)}}p_n^{1-\gamma_r^{(n)}}q_n^{-1}\Big\} \\
    &=o\left(R(T,d_{r,\star})
    \big\{p_m^{\gamma_r}p_n^{\gamma_r}
    +p_m^{\gamma_r^{(m)}}p_n^{\gamma_r^{(n)}}\big\}
    \right).
\end{align*}
The term $T^{-1}p_mp_n^{\gamma_{r,\max}^{(n)}}q_m^{-1}$ follows symmetrically, while the term $T^{-1}p_mp_nq_m^{-1}q_n^{-1}$ is of still smaller order since $p_m^{1-\gamma_{r,\max}^{(m)}}q_m^{-1}=o(1)$. Hence, \eqref{eqn:Delta_A1_bound} reduces to
\begin{align*}
    \norm{\bDelta_{A_1,mn}}_F^2
    =O_P\left(R(T,d_{r,\star})
    \big\{p_m^{\gamma_r}p_n^{\gamma_r}
    +p_m^{\gamma_r^{(m)}}p_n^{\gamma_r^{(n)}}\big\}\right).
\end{align*}
After division by $p_m^{\gamma_r}p_n^{\gamma_r}$, the r.h.s.\ is 
$O_P\left(R(T,d_{r,\star})\{1+\nu_{r,m}\nu_{r,n}\}\right)$. For $m\ne n$,
\begin{align*}
    1+\nu_{r,m}\nu_{r,n}\leq 2(\nu_{r,m}\vee\nu_{r,m}^{-1})(1\vee\nu_{r,n}),
\end{align*}
so (R1) gives $R(T,d_{r,\star})\{1+\nu_{r,m}\nu_{r,n}\}=o(1)$ uniformly over $m\ne n$. This proves \eqref{eqn:Delta_A1_oP}.

Finally, by $T^{-1}=O(R(T,d_{r,\star}))$ and (R1), we have $T^{-1}p_m^{\gamma_r^{(m)}}p_nq_n^{-1}=o\left(R(T,d_{r,\star})p_m^{\gamma_r^{(m)}}p_n^{\gamma_r^{(n)}}\right)$, while $T^{-1}p_mp_nq_m^{-1}q_n^{-1}=o\left(T^{-1}p_mp_n^{\gamma_{r,\max}^{(n)}}q_m^{-1}\right)$. Combining \eqref{eqn:I_A1_1_projected}, \eqref{eqn:I_A1_2_projected}, \eqref{eqn:I_A1_3_projected}, and \eqref{eqn:I_A1_4_projected}, we obtain
\begin{align*}
    \norm{(\I_{p_m}-\P_{A_1,m})\bDelta_{A_1,mn}}_F^2
    =O_P\Big(
    R(T,d_{r,\star})p_m^{\gamma_r^{(m)}}p_n^{\gamma_r^{(n)}}
    +T^{-1}p_mp_n^{\gamma_{r,\max}^{(n)}}q_m^{-1}
    \Big),
\end{align*}
which proves \eqref{eqn:Delta_A1_projected_bound}. This completes the proof. $\square$

\begin{lemma}(Bounding $\wh\D_{A_1,m}^{-1}$)\label{lemma:bound_D_A1}
Under the assumptions of Theorem~\ref{thm:initial_A1m}, define $\omega_{A_1,m}:=p_m^{\gamma_r}\sum_{n\ne m}p_n^{\gamma_r}$, and let $\wh\D_{A_1,m}$ be defined in \eqref{eqn:Omega_A1_A1_D1}. Then, for each $m\in[M]$,
\begin{align}
    \lambda_{r_1}(\bOmega_{A_1,m})\asymp \omega_{A_1,m},\qquad
    \lambda_{r_1}(\wh\bOmega_{A_1,m})\asymp_P \omega_{A_1,m},\qquad
    \norm{\wh\D_{A_1,m}^{-1}}_F^2=O_P(\omega_{A_1,m}^{-2}).
    \label{eqn:D_A1_inv}
\end{align}
\end{lemma}
\textbf{\textit{Proof of Lemma~\ref{lemma:bound_D_A1}.}}
By (ME1), we have $\lambda_{r_1}(\A_{1m}'\A_{1m})\asymp\lambda_1(\A_{1m}'\A_{1m})\asymp p_m^{\gamma_r}$. Moreover, since $\A_{1n}'\A_{1n}\succeq0$ for every $n\ne m$,
\begin{align*}
    \lambda_{r_1}\left(\sum_{n\ne m}\A_{1n}'\A_{1n}\right)
    &\geq\sum_{n\ne m}\lambda_{r_1}(\A_{1n}'\A_{1n})
    \asymp\sum_{n\ne m}p_n^{\gamma_r},\\
    \lambda_1\left(\sum_{n\ne m}\A_{1n}'\A_{1n}\right)
    &\leq\sum_{n\ne m}\lambda_1(\A_{1n}'\A_{1n})
    \asymp\sum_{n\ne m}p_n^{\gamma_r}.
\end{align*}
By (S1), $\lambda_{r_1}\left(T^{-1}\sum_{t=1}^T\bb{E}(\balpha_t\balpha_t')\right)\asymp\lambda_1\left(T^{-1}\sum_{t=1}^T\bb{E}(\balpha_t\balpha_t')\right)\asymp1$. Together with the preceding bounds and (ME1), we have
\begin{align*}
    \lambda_{r_1}(\bOmega_{A_1,m})\asymp\lambda_1(\bOmega_{A_1,m})
    \asymp p_m^{\gamma_r}\sum_{n\ne m}p_n^{\gamma_r}
    =\omega_{A_1,m}.
\end{align*}
By \eqref{eqn:Omega_A1_full_decomposition}, Lemma~\ref{lemma:bound_S_A1_Delta_A1},
\begin{align*}
    \norm{\wh\bOmega_{A_1,m}-\bOmega_{A_1,m}}_F
    &\lesssim\sum_{n\ne m}\left\{\norm{\bDelta_{A_1,mn}}_F\norm{\S_{A_1,mn}}_F+\norm{\bDelta_{A_1,mn}}_F^2\right\}\\
    &=o_P\left(p_m^{\gamma_r}\sum_{n\ne m}p_n^{\gamma_r}\right)
    =o_P(\omega_{A_1,m}).
\end{align*}
Therefore, Weyl's inequality gives
\begin{align*}
    \big|\lambda_{r_1}(\wh\bOmega_{A_1,m})-\lambda_{r_1}(\bOmega_{A_1,m})\big|
    \leq\norm{\wh\bOmega_{A_1,m}-\bOmega_{A_1,m}}
    \leq\norm{\wh\bOmega_{A_1,m}-\bOmega_{A_1,m}}_F
    =o_P(\omega_{A_1,m}),
\end{align*}
and hence $\lambda_{r_1}(\wh\bOmega_{A_1,m})\asymp_P\omega_{A_1,m}$. Thus, $\wh\D_{A_1,m}$ is invertible with probability tending to one and, by \eqref{eqn:Omega_A1_A1_D1},
\begin{align*}
    \norm{\wh\D_{A_1,m}^{-1}}_F^2
    =\sum_{i=1}^{r_1}\lambda_i^{-2}(\wh\bOmega_{A_1,m})
    \lesssim\lambda_{r_1}^{-2}(\wh\bOmega_{A_1,m})
    =O_P(\omega_{A_1,m}^{-2}),
\end{align*}
which completes the proof. $\square$

\begin{lemma}(Bounds for $\S_{A_2,m}$ and $\bDelta_{A_2,m}$)\label{lemma:bound_S_A2_Delta_A2}
    For any $m\in[M]$, let $\S_{A_2,m}$ and $\wh\S_{A_2,m}$ be defined in \eqref{eqn:S_A2}, and define $\bDelta_{A_2,m}:=\wh\S_{A_2,m}-\S_{A_2,m}$. Under the assumptions of Theorem~\ref{thm:A2m_alphatm}, 
    \begin{align}
        \norm{\S_{A_2,m}}_F^2
        &=O_P(p_m^{2\gamma_r^{(m)}}), 
        \label{eqn:S_A2_bound} \\
        \norm{\bDelta_{A_2,m}}_F^2
        &=o_P(p_m^{2\gamma_r^{(m)}}).
        \label{eqn:Delta_A2_oP}
    \end{align}
    Furthermore, defining $\P_{A_2,m}:=\check{\A}_{2m}\check{\A}_{2m}'$, we have
    \begin{align}
        \norm{(\I_{p_m}-\P_{A_2,m})\bDelta_{A_2,m}}_F^2
        =O_P\Big(&R(T,d_{r,\star})p_m^{\gamma_r+\gamma_r^{(m)}}
        \max_{n\in[M]\setminus\{m\}}p_n^{\gamma_r^{(n)}-\gamma_r} \nonumber \\
        &+T^{-1}p_m^{\gamma_r+1}q_m^{-1}
        \max_{n\in[M]}p_n^{\gamma_{r,\max}^{(n)}-\gamma_r}
        +p_mq_m^{-2}\Big).
        \label{eqn:Delta_A2_projected_bound}
    \end{align}
\end{lemma}
\textbf{\textit{Proof of Lemma~\ref{lemma:bound_S_A2_Delta_A2}.}}
Recall from \eqref{eqn:S_A2} that
\begin{align*}
    \S_{A_2,m}
    :=T^{-1}\sum_{t=1}^T
    \a_t^{(m)}\a_t^{(m)'}\P_{A_1,m}^{\perp}, \qquad
    \wh\S_{A_2,m}
    :=T^{-1}\sum_{t=1}^T
    \wh\a_t^{(m)}\wh\a_t^{(m)'}\wh\P_{A_1,m}^{\perp},
\end{align*}
where $\P_{A_1,m}^{\perp}=\I_{p_m}-\P_{A_1,m}$ and $\wh\P_{A_1,m}^{\perp}=\I_{p_m}-\wh\P_{A_1,m}$. Consequently,
\begin{align}
    \bDelta_{A_2,m}
    =T^{-1}\sum_{t=1}^T\a_t^{(m)}\a_t^{(m)'}
    \left(\wh\P_{A_1,m}^{\perp}-\P_{A_1,m}^{\perp}\right)
    +T^{-1}\sum_{t=1}^T
    \left(\wh\a_t^{(m)}\wh\a_t^{(m)'}-\a_t^{(m)}\a_t^{(m)'}\right)
    \wh\P_{A_1,m}^{\perp}.
    \label{eqn:Delta_A2_decomposition}
\end{align}

We first prove \eqref{eqn:S_A2_bound}. By \eqref{eqn:uniform_second_moment_G} and \eqref{eqn:sample_covariance_G} in Lemma~\ref{lemma:prop_piecewise}, and (IC2), it follows that
\begin{align}
    T^{-1}\sum_{t=1}^T
    \a_t^{(m)}\a_t^{(m)'}
    =\A_{1m}\left(T^{-1}\sum_{t=1}^T\balpha_t\balpha_t'\right)\A_{1m}'
    +\A_{2m}\left(T^{-1}\sum_{t=1}^T\balpha_t^{(m)}\balpha_t^{(m)'}\right)\A_{2m}'.
    \label{eqn:sample_cov_a_m}
\end{align}
Since $\A_{1m}'\P_{A_1,m}^{\perp}=\0$,
\begin{align}
    \S_{A_2,m}
    =\A_{2m}\left(T^{-1}\sum_{t=1}^T\balpha_t^{(m)}\balpha_t^{(m)'}\right)\A_{2m}'\P_{A_1,m}^{\perp}.
    \label{eqn:S_A2_rep}
\end{align}
Therefore, by (ME1), Proposition~\ref{prop:projector_prop}, and Lemma~\ref{lemma:prop_piecewise},
\begin{align*}
    \norm{\S_{A_2,m}}_F^2
    \leq\norm{\A_{2m}}_F^4
    \norm{T^{-1}\sum_{t=1}^T\balpha_t^{(m)}\balpha_t^{(m)'}}_F^2
    =O_P(p_m^{2\gamma_r^{(m)}}),
\end{align*}
which proves \eqref{eqn:S_A2_bound}.

We next prove \eqref{eqn:Delta_A2_oP} and \eqref{eqn:Delta_A2_projected_bound}. By (IC1), $\wh\a_t^{(m)}-\a_t^{(m)}=q_m^{-1}\M_{p_m}\E_t^{(m)}\1_{q_m}$. Using \eqref{eqn:Delta_A2_decomposition}, write
\begin{align*}
    \bDelta_{A_2,m}
    =\cI_{A_2,1}+\cI_{A_2,2}+\cI_{A_2,3}+\cI_{A_2,4},
\end{align*}
where
\begin{alignat*}{2}
    \cI_{A_2,1}
    &:=T^{-1}\sum_{t=1}^T\a_t^{(m)}\a_t^{(m)'}
    \left(\wh\P_{A_1,m}^{\perp}-\P_{A_1,m}^{\perp}\right), \;\;
    &&\cI_{A_2,2}
    :=(Tq_m)^{-1}\sum_{t=1}^T\a_t^{(m)}\1_{q_m}'\E_t^{(m)'}\M_{p_m}\wh\P_{A_1,m}^{\perp}, \\
    \cI_{A_2,3}
    &:=(Tq_m)^{-1}\sum_{t=1}^T\M_{p_m}\E_t^{(m)}\1_{q_m}\a_t^{(m)'}\wh\P_{A_1,m}^{\perp}, \;\;
    &&\cI_{A_2,4}
    :=(Tq_m^2)^{-1}\sum_{t=1}^T\M_{p_m}\E_t^{(m)}\1_{q_m}\1_{q_m}'\E_t^{(m)'}\M_{p_m}\wh\P_{A_1,m}^{\perp}.
\end{alignat*}
We bound these four terms separately.

By Theorem~\ref{thm:initial_A1m} and \eqref{eqn:A1_projector_rate},
\begin{align}
    \norm{\wh\P_{A_1,m}-\P_{A_1,m}}_F^2
    =O_P\Big(&R(T,d_{r,\star})p_m^{\gamma_r^{(m)}-\gamma_r}
    \max_{n\in[M]\setminus\{m\}}p_n^{\gamma_r^{(n)}-\gamma_r} \nonumber \\
    &+T^{-1}p_m^{1-\gamma_r}q_m^{-1}
    \max_{n\in[M]\setminus\{m\}}p_n^{\gamma_{r,\max}^{(n)}-\gamma_r}\Big).
    \label{eqn:A1_projector_explicit_rate}
\end{align}

For $\cI_{A_2,1}$, since $\wh\P_{A_1,m}^{\perp}-\P_{A_1,m}^{\perp}=\P_{A_1,m}-\wh\P_{A_1,m}$, together with \eqref{eqn:sample_cov_a_m} and (ME1), we have
\begin{align}
    \norm{\cI_{A_2,1}}_F^2
    =O_P\left((p_m^{2\gamma_r}+p_m^{2\gamma_r^{(m)}})\norm{\wh\P_{A_1,m}-\P_{A_1,m}}_F^2\right).
    \label{eqn:I_A2_1_full}
\end{align}
Furthermore, since
$(\I_{p_m}-\P_{A_2,m})\A_{2m}=\0$,
\begin{align*}
    (\I_{p_m}-\P_{A_2,m})\cI_{A_2,1}
    =(\I_{p_m}-\P_{A_2,m})\A_{1m}\left(T^{-1}\sum_{t=1}^T\balpha_t\balpha_t'\right)\A_{1m}'(\P_{A_1,m}-\wh\P_{A_1,m}).
\end{align*}
Consequently,
\begin{align}
    \norm{(\I_{p_m}-\P_{A_2,m})\cI_{A_2,1}}_F^2
    =O_P\left(p_m^{2\gamma_r}\norm{\wh\P_{A_1,m}-\P_{A_1,m}}_F^2\right).
    \label{eqn:I_A2_1_projected}
\end{align}

For $\cI_{A_2,2}$, write
\begin{align*}
    \cI_{A_2,2}
    &=\A_{1m}\left\{(Tq_m)^{-1}\sum_{t=1}^T\balpha_t\1_{q_m}'\E_t^{(m)'}\M_{p_m}\right\}\wh\P_{A_1,m}^{\perp}\\
    &\quad\;+\A_{2m}\left\{(Tq_m)^{-1}\sum_{t=1}^T\balpha_t^{(m)}\1_{q_m}'\E_t^{(m)'}\M_{p_m}\right\}\wh\P_{A_1,m}^{\perp}.
\end{align*}
Applying \eqref{eqn:E1u} in Lemma~\ref{lemma:bound_Delta}, together with (ME1) and Proposition~\ref{prop:projector_prop}, gives
\begin{align}
    \norm{\cI_{A_2,2}}_F^2
    =O_P\left(T^{-1}p_mq_m^{-1}(p_m^{\gamma_r}+p_m^{\gamma_r^{(m)}})\right).
    \label{eqn:I_A2_2_full}
\end{align}
Since left-multiplication by $\I_{p_m}-\P_{A_2,m}$ eliminates the term involving
$\A_{2m}$,
\begin{align}
    \norm{(\I_{p_m}-\P_{A_2,m})\cI_{A_2,2}}_F^2
    =O_P\left(T^{-1}p_m^{\gamma_r+1}q_m^{-1}\right).
    \label{eqn:I_A2_2_projected}
\end{align}

For $\cI_{A_2,3}$, we have
\begin{align*}
    \cI_{A_2,3}
    &=q_m^{-1}\left\{T^{-1}\sum_{t=1}^T\M_{p_m}\E_t^{(m)}\1_{q_m}\balpha_t'\right\}
    \A_{1m}'\wh\P_{A_1,m}^{\perp}\\
    &\quad\;+q_m^{-1}\left\{T^{-1}\sum_{t=1}^T\M_{p_m}\E_t^{(m)}\1_{q_m}\balpha_t^{(m)'}\right\}
    \A_{2m}'\wh\P_{A_1,m}^{\perp}.
\end{align*}
By \eqref{eqn:E1u} in Lemma~\ref{lemma:bound_Delta}, (ME1), and Proposition~\ref{prop:projector_prop},
\begin{align}
    \norm{\cI_{A_2,3}}_F^2
    =O_P\left(T^{-1}p_mq_m^{-1}(p_m^{\gamma_r}+p_m^{\gamma_r^{(m)}})\right)
    =O_P\left(T^{-1}p_m^{\gamma_{r,\max}^{(m)}+1}q_m^{-1}\right).
    \label{eqn:I_A2_3_full}
\end{align}
The same bound remains valid after left-multiplication by $\I_{p_m}-\P_{A_2,m}$, and hence
\begin{align}
    \norm{(\I_{p_m}-\P_{A_2,m})\cI_{A_2,3}}_F^2
    =O_P\left(T^{-1}p_m^{\gamma_{r,\max}^{(m)}+1}q_m^{-1}\right).
    \label{eqn:I_A2_3_projected}
\end{align}

For $\cI_{A_2,4}$, by \eqref{eqn:Em11Em} in Lemma~\ref{lemma:bound_Delta} and Proposition~\ref{prop:projector_prop},
\begin{align}
    \norm{\cI_{A_2,4}}_F^2
    &=O_P\left(p_mq_m^{-2}+T^{-1}p_m^2q_m^{-2}\right),
    \label{eqn:I_A2_4_full}\\
    \norm{(\I_{p_m}-\P_{A_2,m})\cI_{A_2,4}}_F^2
    &=O_P\left(p_mq_m^{-2}+T^{-1}p_m^2q_m^{-2}\right).
    \label{eqn:I_A2_4_projected}
\end{align}
Using $\norm{\sum_{j=1}^4\cI_{A_2,j}}_F^2\lesssim\sum_{j=1}^4\norm{\cI_{A_2,j}}_F^2$, together with \eqref{eqn:I_A2_1_full}, \eqref{eqn:I_A2_2_full}, \eqref{eqn:I_A2_3_full}, and \eqref{eqn:I_A2_4_full}, we obtain
\begin{align}
    \norm{\bDelta_{A_2,m}}_F^2
    =O_P\Big(
    &(p_m^{2\gamma_r}+p_m^{2\gamma_r^{(m)}})
    \norm{\wh\P_{A_1,m}-\P_{A_1,m}}_F^2
    \nonumber\\
    &+T^{-1}p_m^{\gamma_{r,\max}^{(m)}+1}q_m^{-1}
    +p_mq_m^{-2}+T^{-1}p_m^2q_m^{-2}\Big).
    \label{eqn:Delta_A2_bound}
\end{align}

To deduce \eqref{eqn:Delta_A2_oP}, Theorem~\ref{thm:initial_A1m} and \eqref{eqn:A1_projector_rate} give
\begin{align*}
    \quad\;\big(1+\nu_{r,m}^{-2}\big)
    \norm{\wh\P_{A_1,m}-\P_{A_1,m}}_F^2
    =O_P\Bigg(&R(T,d_{r,\star})(\nu_{r,m}+\nu_{r,m}^{-1})
    \max_{n\in[M]\setminus\{m\}}\nu_{r,n} \\
    &+\frac{1}{T}\left\{\frac{p_m^{1-\gamma_r}}{q_m}
    +\frac{p_m^{1-\gamma_r^{(m)}}}{q_m}\nu_{r,m}^{-1}\right\}
    \left(1\vee\max_{n\in[M]\setminus\{m\}}\nu_{r,n}\right)\Bigg)
    =o_P(1),
\end{align*}
where the last equality follows from (R1) and $T^{-1}=O(R(T,d_{r,\star}))$. Moreover, 
\begin{align*}
    T^{-1}p_m^{\gamma_{r,\max}^{(m)}+1-2\gamma_r^{(m)}}q_m^{-1}
    =T^{-1}p_m^{1-\gamma_r^{(m)}}q_m^{-1}(1\vee\nu_{r,m}^{-1})=o(1),
\end{align*}
while the remaining two terms in \eqref{eqn:Delta_A2_bound}, after division by $p_m^{2\gamma_r^{(m)}}$, are also $o(1)$ by (R1). Hence \eqref{eqn:Delta_A2_oP} follows.

Finally, combining \eqref{eqn:I_A2_1_projected}, \eqref{eqn:I_A2_2_projected}, \eqref{eqn:I_A2_3_projected}, and \eqref{eqn:I_A2_4_projected} with \eqref{eqn:A1_projector_explicit_rate}, we obtain
\begin{align*}
    \norm{(\I_{p_m}-\P_{A_2,m})\bDelta_{A_2,m}}_F^2
    =O_P\Big(&R(T,d_{r,\star})p_m^{\gamma_r+\gamma_r^{(m)}}
    \max_{n\in[M]\setminus\{m\}}p_n^{\gamma_r^{(n)}-\gamma_r}\\
    &+T^{-1}p_m^{\gamma_r+1}q_m^{-1}
    \max_{n\in[M]}p_n^{\gamma_{r,\max}^{(n)}-\gamma_r}
    +p_mq_m^{-2}\Big),
\end{align*}
where $T^{-1}p_m^2q_m^{-2}$ is absorbed into the second term under
(R1). This proves \eqref{eqn:Delta_A2_projected_bound}, and completes the proof. $\square$

\begin{lemma}(Bounding $\wh\D_{A_2,m}^{-1}$)\label{lemma:bound_D_A2}
    Under the assumptions of Theorem~\ref{thm:A2m_alphatm}, define $\omega_{A_2,m}:=p_m^{2\gamma_r^{(m)}}$, and let $\wh\D_{A_2,m}$ be defined in \eqref{eqn:Omega_A2_A2_D2}. Then, for each $m\in[M]$,
    \begin{align}
        \lambda_{r_2^{(m)}}(\bOmega_{A_2,m})\asymp_P \omega_{A_2,m},\qquad
        \lambda_{r_2^{(m)}}(\wh\bOmega_{A_2,m})\asymp_P \omega_{A_2,m},\qquad
        \norm{\wh\D_{A_2,m}^{-1}}_F^2=O_P(\omega_{A_2,m}^{-2}).
        \label{eqn:D_A2_inv}
    \end{align}
\end{lemma}
\textbf{\textit{Proof of Lemma~\ref{lemma:bound_D_A2}.}}
By Lemma~\ref{lemma:prop_piecewise}, (S1), and Weyl's inequality,
\begin{align*}
    \lambda_{r_2^{(m)}}\left(T^{-1}\sum_{t=1}^T\balpha_t^{(m)}\balpha_t^{(m)'}\right)\asymp_P\lambda_1\left(T^{-1}\sum_{t=1}^T\balpha_t^{(m)}\balpha_t^{(m)'}\right)\asymp_P 1.
\end{align*}
Since $\P_{A_1,m}^{\perp}=\A_{1m}^{\perp}\A_{1m}^{\perp'}$,
\begin{align*}
    &\quad\;(\A_{2m}'\A_{2m})^{-1/2}\A_{2m}'\P_{A_1,m}^{\perp}\A_{2m}(\A_{2m}'\A_{2m})^{-1/2} \\
    &=\left\{\A_{1m}^{\perp'}\A_{2m}(\A_{2m}'\A_{2m})^{-1/2}\right\}'
    \left\{\A_{1m}^{\perp'}\A_{2m}(\A_{2m}'\A_{2m})^{-1/2}\right\}.
\end{align*}
By (S1), (ME1), and Proposition~\ref{prop:projector_prop}, this gives
\begin{align*}
    \lambda_{r_2^{(m)}}(\A_{2m}'\P_{A_1,m}^{\perp}\A_{2m})
    \asymp\lambda_1(\A_{2m}'\P_{A_1,m}^{\perp}\A_{2m})
    \asymp p_m^{\gamma_r^{(m)}}.
\end{align*}
Therefore, (ME1) gives
\begin{align*}
    \lambda_{r_2^{(m)}}(\bOmega_{A_2,m})
    \asymp_P\lambda_1(\bOmega_{A_2,m})
    \asymp_P p_m^{2\gamma_r^{(m)}}
    =\omega_{A_2,m}.
\end{align*}
Let $\bDelta_{A_2,m}:=\wh\S_{A_2,m}-\S_{A_2,m}$. Since $\wh\bOmega_{A_2,m}=\wh\S_{A_2,m}\wh\S_{A_2,m}'$ and $\bOmega_{A_2,m}=\S_{A_2,m}\S_{A_2,m}'$,
\begin{align*}
    \wh\bOmega_{A_2,m}-\bOmega_{A_2,m}
    =\bDelta_{A_2,m}\S_{A_2,m}'+\S_{A_2,m}\bDelta_{A_2,m}'+\bDelta_{A_2,m}\bDelta_{A_2,m}'.
\end{align*}
Lemma~\ref{lemma:bound_S_A2_Delta_A2} gives $\norm{\S_{A_2,m}}_F=O_P(p_m^{\gamma_r^{(m)}})$ and $\norm{\bDelta_{A_2,m}}_F=o_P(p_m^{\gamma_r^{(m)}})$. Therefore,
\begin{align*}
    \norm{\wh\bOmega_{A_2,m}-\bOmega_{A_2,m}}_F
    \lesssim\norm{\bDelta_{A_2,m}}_F\norm{\S_{A_2,m}}_F+\norm{\bDelta_{A_2,m}}_F^2
    =o_P(p_m^{2\gamma_r^{(m)}})
    =o_P(\omega_{A_2,m}).
\end{align*}
By Weyl's inequality,
\begin{align*}
    \big|\lambda_{r_2^{(m)}}(\wh\bOmega_{A_2,m})-\lambda_{r_2^{(m)}}(\bOmega_{A_2,m})\big|\leq\norm{\wh\bOmega_{A_2,m}-\bOmega_{A_2,m}}_F
    =o_P(\omega_{A_2,m}),
\end{align*}
so that $\lambda_{r_2^{(m)}}(\wh\bOmega_{A_2,m})\asymp_P\omega_{A_2,m}$. Thus, $\wh\D_{A_2,m}$ is invertible with probability tending to one and, by \eqref{eqn:Omega_A2_A2_D2},
\begin{align*}
    \norm{\wh\D_{A_2,m}^{-1}}_F^2
    =\sum_{i=1}^{r_2^{(m)}}\lambda_i^{-2}(\wh\bOmega_{A_2,m})
    \leq r_2^{(m)}\lambda_{r_2^{(m)}}^{-2}(\wh\bOmega_{A_2,m})
    =O_P(\omega_{A_2,m}^{-2}),
\end{align*}
which completes the proof. $\square$

\newpage

\begin{lemma}\label{lemma: common_comp_global_local_init}(Properties of $\bOmega_{Q,m}$)
For any $m,n\in[M]$ with $n\neq m$, define
\begin{align*}
    & \bOmega_{Q,m} := \frac{1}{M-1} \sum_{n=1;n\neq m}^M \S_{Q,mn}^{\otimes 2},
    \;\;\;
    \S_{Q,mn} := \frac{1}{T} \sum_{t=1}^T \bb{E}\left( \Q_m\G_t\J_m' \v_b^{(m)} \v_b^{(n)'} \J_n\G_t'\Q_n' \right) \\
    & \wh\S_{Q,mn} := \frac{1}{T} \sum_{t=1}^T\check\Y_t^{(m)} \v_b^{(m)} \v_b^{(n)'} \check\Y_t^{(n)'} ,
    \;\;\;
    \bDelta_{Q,mn} := \wh\S_{Q,mn} - \S_{Q,mn} .
\end{align*}
Then under Assumptions~(IC1), (E1), (E2), (F1), and (L1), we have
\begin{align*}
    \| \bDelta_{Q,mn} \|_F^2
    &\lesssim_P
    p_m^{\delta_{q,1}^{(m)}} p_n^{\delta_{q,1}^{(n)}} s_{J,m}^2 s_{J,n}^2 R(T,d_{\text{fac},\star})
    + p_m^{\delta_{r,1}^{(m)}} p_n^{\delta_{q,1}^{(n)}} s_{J,n}^2  q_m^{\delta_{c,1}^{(m)}} R(T,d_{\text{fac},\star}) \\
    &\;\quad
    + p_m^{\delta_{q,1}^{(m)}} p_n^{\delta_{r,1}^{(n)}} s_{J,m}^2 q_n^{\delta_{c,1}^{(n)}} R(T,d_{\text{fac},\star}) 
    + p_m^{\delta_{r,1}^{(m)}} p_n^{\delta_{r,1}^{(n)}} q_m^{\delta_{c,1}^{(m)}} q_n^{\delta_{c,1}^{(n)}}  R(T,d_{\text{fac},\star}) 
    + T^{-1} p_m p_n^{\delta_{q,1}^{(n)}} s_{J,n}^2 \\
    &\;\quad
    + T^{-1} p_m p_n^{\delta_{r,1}^{(n)}} q_n^{\delta_{c,1}^{(n)}} 
    + T^{-1} p_m^{\delta_{q,1}^{(m)}} s_{J,m}^2 p_n
    + T^{-1} p_m^{\delta_{r,1}^{(m)}} q_m^{\delta_{c,1}^{(m)}} p_n 
    + T^{-1} p_m p_n .
\end{align*}
Further define $\check{\Q}_m := \Q_m (\Q_m'\Q_m)^{-1/2}$.
Then under Assumptions~(L2) and (R2), there exists an asymptotically orthogonal matrix $\H_{Q,m}$ with $\H_{Q,m} \H_{Q,m}'=\I+o_P(1)$,
such that
\[
\big\| \wh\Q_m - \check{\Q}_m \H_{Q,m} \big\|^2 = O_P\left\{
R(T,d_{\text{fac},\star}) \;
p_m^{2 (\delta_{q,1}^{(m)} -\delta_{q,k_r}^{(m)})}
\left( \frac{  \max_{n\in[M] \setminus \{m\}} s_{J,n}^2 p_n^{\delta_{q,1}^{(n)}}}{\max_{n\in[M] \setminus\{m\}} s_{J,n}^2 p_n^{\delta_{q,k_r}^{(n)}}} \right)^2
\right\} .
\]
Analogous results hold for $\wh\J_m$, with asymptotically orthogonal matrix $\H_{J,m}$.
\end{lemma}
\textbf{\textit{Proof of Lemma~\ref{lemma: common_comp_global_local_init}.}}
By Assumption~(F1), we note that $T^{-1} \sum_{t=1}^T \bb{E}[\z_{G,t}^{\otimes 2}]$ and $T^{-1} \sum_{t=1}^T \bb{E}[(\z_{F,t}^{(m)})^{\otimes 2}]$ are diagonal due to the independence between cross-sectional entries. Without loss of generality, we further normalize these long-run covariance matrices to be the identity matrix. Namely, we will assume throughout the proof that
\begin{equation}
\label{eqn: long_run_cov_vec_fac}
\frac{1}{T} \sum_{t=1}^T \bb{E}[\z_{G,t}^{\otimes 2}] = \I_{k_rk_c} , \;\;\;
\frac{1}{T} \sum_{t=1}^T \bb{E}[(\z_{F,t}^{(m)})^{\otimes 2}] = \I_{k_r^{(m)} k_c^{(m)}} .
\end{equation}
This is equivalent to absorbing the entry-wise variance into $\bSigma_{Q,m}$, $\bSigma_{J,m}$, $\bSigma_{R,m}$ and $\bSigma_{C,m}$ (from Assumption~(L1)) which are arbitrary positive-definite matrices with eigenvalues uniformly bounded away from 0 and infinity.
Then for any group $m\in[M]$, by definitions we have
\begin{align}
    &\;\quad
    \check\Y_t^{(m)} = \Y_t^{(m)} - \wh{c}_t^{(m)}\1
    _{p_m}\1_{q_m}' - \wh{\a}_t^{(m)}\1_{q_m}' - \1_{p_m}\wh\b_t^{(m)'} \notag \\
    &=
    \Y_t^{(m)} - (p_mq_m)^{-1} \1_{p_m} \1_{p_m}' \Y_t^{(m)}\1_{q_m}\1_{q_m}' - q_m^{-1} \M_{p_m}\Y_t^{(m)}\1_{q_m} \1_{q_m}' - p_m^{-1} \1_{p_m} \1_{p_m}' \Y_t^{(m)} \M_{q_m} \notag \\
    &=
    \M_{p_m} \Y_t^{(m)} \M_{q_m} + p_m^{-1} \1_{p_m} \1_{p_m}' \Y_t^{(m)} \M_{q_m} + q_m^{-1} \Y_t^{(m)} \1_{q_m} \1_{q_m}' 
    - (p_mq_m)^{-1} \1_{p_m} \1_{p_m}' \Y_t^{(m)}\1_{q_m}\1_{q_m}'  \notag \\
    &\;\quad
    - q_m^{-1} \M_{p_m} \Y_t^{(m)} \1_{q_m} \1_{q_m}' - p_m^{-1} \1_{p_m} \1_{p_m}' \Y_t^{(m)} \M_{q_m}
    = \M_{p_m} \Y_t^{(m)} \M_{q_m} . \notag
\end{align}
Together with our model \eqref{eqn:MMEFM}, Assumption~(IC1), and the fact that $\M_a \1_a =\0$ for any positive integer $a$, we further arrive at
\begin{align}
    \check\Y_t^{(m)} &= \M_{p_m} \Y_t^{(m)} \M_{q_m} =
    \M_{p_m} \Q_m\G_t\J_m' \M_{q_m}
    + \M_{p_m} \R_m\F_t^{(m)}\C_m'  \M_{q_m}
    + \M_{p_m} \E_t^{(m)} \M_{q_m}  \notag \\
    &=
    \Q_m\G_t\J_m'
    + \R_m\F_t^{(m)}\C_m' 
    + \M_{p_m} \E_t^{(m)} \M_{q_m} .
    \label{eqn: check_Yt_m_simplify}
\end{align}

Then it holds that
\begin{align}
    &\;\quad
    \wh\S_{Q,mn} \notag \\
    &= 
    \frac{1}{T} \sum_{t=1}^T \Big( \Q_m\G_t\J_m' \v_b^{(m)} + \R_m\F_t^{(m)}\C_m' \v_b^{(m)} + \M_{p_m} \E_t^{(m)} \M_{q_m} \v_b^{(m)} \Big) \notag \\
    &\;\quad
    \cdot \Big( \v_b^{(n)'} \J_n\G_t'\Q_n'  + \v_b^{(n)'} \C_n\F_t^{(n)'} \R_n' + \v_b^{(n)'} \M_{q_n} \E_t^{(n)'} \M_{p_n} \Big) \notag \\
    &=
    \frac{1}{T} \sum_{t=1}^T \Q_m\G_t\J_m' \v_b^{(m)} \v_b^{(n)'} \J_n\G_t'\Q_n'
    + \frac{1}{T} \sum_{t=1}^T \R_m\F_t^{(m)}\C_m' \v_b^{(m)} \v_b^{(n)'} \J_n\G_t'\Q_n' \notag \\
    &\;\quad
    + \frac{1}{T} \sum_{t=1}^T \M_{p_m} \E_t^{(m)} \M_{q_m} \v_b^{(m)} \v_b^{(n)'} \J_n\G_t'\Q_n'
    + \frac{1}{T} \sum_{t=1}^T \Q_m\G_t\J_m' \v_b^{(m)} \v_b^{(n)'} \C_n\F_t^{(n)'} \R_n'  \notag \\
    &\;\quad
    + \frac{1}{T} \sum_{t=1}^T \R_m\F_t^{(m)}\C_m' \v_b^{(m)} \v_b^{(n)'} \C_n\F_t^{(n)'} \R_n' 
    + \frac{1}{T} \sum_{t=1}^T \M_{p_m} \E_t^{(m)} \M_{q_m} \v_b^{(m)} \v_b^{(n)'} \C_n\F_t^{(n)'} \R_n' \notag \\
    &\;\quad 
    + \frac{1}{T} \sum_{t=1}^T \Q_m\G_t\J_m' \v_b^{(m)} \v_b^{(n)'} \M_{q_n} \E_t^{(n)'} \M_{p_n}  
    + \frac{1}{T} \sum_{t=1}^T \R_m\F_t^{(m)}\C_m' \v_b^{(m)} \v_b^{(n)'} \M_{q_n} \E_t^{(n)'} \M_{p_n} \notag \\
    &\;\quad
    + \frac{1}{T} \sum_{t=1}^T \M_{p_m} \E_t^{(m)} \M_{q_m} \v_b^{(m)} \v_b^{(n)'} \M_{q_n} \E_t^{(n)'} \M_{p_n} \notag \\
    &=:
    \cI_{Q,mn} + \cII_{Q,mn} + \cIII_{Q,mn} + \cIV_{Q,mn} + \cV_{Q,mn} + \cVI_{Q,mn} + \cVII_{Q,mn} + \cVIII_{Q,mn} + \cIX_{Q,mn} .
    \label{eqn: decomp_hat_S_Q_mn}
\end{align}

Consider first $\cI_{Q,mn} - \S_{Q,mn}$.
Define $\z_{G,t} := \vec(\G_t)$ for any $t\in[T]$.
As the dimensions of $\G_t$ are bounded, Lemma~\ref{lemma:prop_piecewise} implies that
\begin{equation}
\label{eqn: GG_var}
\left\| \frac{1}{T} \sum_{t=1}^T \left\{ \z_{G,t} \z_{G,t}' - \bb{E}\big[ \z_{G,t} \z_{G,t}' \big] \right\} \right\|_F^2
= O_P(R(T,d_{\text{fac},\star})) ,
\end{equation}
where $R(\cdot, \cdot)$ is defined in \eqref{eqn:R(T,d)} and $d_{\text{fac},\star}$ is the largest long-memory exponent among all processes in $\{\G_t\}$ and $\{\F_t^{(m)}\}$ for $m\in[M]$.
Then from Assumption~(L1), by the triangle inequality we have
\begin{align}
    &\;\quad
    \Big\| \cI_{Q,mn} - \S_{Q,mn} \Big\|_F^2 = 
    \left\| \Q_m \frac{1}{T} \sum_{t=1}^T \left(  \G_t\J_m' \v_b^{(m)} \v_b^{(n)'} \J_n\G_t' - \bb{E}\left[ \G_t\J_m' \v_b^{(m)} \v_b^{(n)'} \J_n\G_t' \right] \right) \Q_n' \right\|_F^2 \notag \\
    &\lesssim
    \frac{1}{T^2} \cdot p_m^{\delta_{q,1}^{(m)}} \cdot p_n^{\delta_{q,1}^{(n)}} \cdot \left\| \sum_{t=1}^T \left(  \G_t\J_m' \v_b^{(m)} \v_b^{(n)'} \J_n\G_t' - \bb{E}\left[ \G_t\J_m' \v_b^{(m)} \v_b^{(n)'} \J_n\G_t' \right] \right) \right\|_F^2 \notag \\
    &=
    \frac{1}{T^2} \cdot p_m^{\delta_{q,1}^{(m)}} \cdot p_n^{\delta_{q,1}^{(n)}} \cdot \left\| \sum_{t=1}^T \left\{ (\G_t \otimes \G_t) \vec(\J_m' \v_b^{(m)} \v_b^{(n)'} \J_n) - \bb{E}\left[ (\G_t \otimes \G_t) \vec(\J_m' \v_b^{(m)} \v_b^{(n)'} \J_n) \right] \right\} \right\|_F^2 \notag \\
    &\leq
    \frac{1}{T^2} \cdot p_m^{\delta_{q,1}^{(m)}} \cdot p_n^{\delta_{q,1}^{(n)}} \cdot \left\| \sum_{t=1}^T \left\{ \G_t \otimes \G_t - \bb{E}[\G_t \otimes \G_t] \right\} \right\|_F^2
    \cdot \left\| \J_m' \v_b^{(m)} \v_b^{(n)'} \J_n \right\|_F^2 \notag \\
    &\asymp
    p_m^{\delta_{q,1}^{(m)}} \cdot p_n^{\delta_{q,1}^{(n)}} \cdot \left\| \frac{1}{T} \sum_{t=1}^T \left\{ \z_{G,t} \z_{G,t}' - \bb{E}\big[ \z_{G,t} \z_{G,t}' \big] \right\} \right\|_F^2
    \cdot \left\| \J_m' \v_b^{(m)} \v_b^{(n)'} \J_n \right\|_F^2 \notag \\
    &\lesssim_P
    p_m^{\delta_{q,1}^{(m)}} \cdot p_n^{\delta_{q,1}^{(n)}} \cdot
    R(T,d_{\text{fac},\star}) \cdot \left\| \J_m' \v_b^{(m)} \v_b^{(n)'} \J_n \right\|_F^2 \notag \\
    &\lesssim_P
    p_m^{\delta_{q,1}^{(m)}} \cdot p_n^{\delta_{q,1}^{(n)}} \cdot
    s_{J,m}^2 \cdot s_{J,n}^2 \cdot
    R(T,d_{\text{fac},\star}) ,
    \label{eqn: cI_Q_mn}
\end{align}
where the first equivalence follows from the fact that the Frobenius norm of $\G_t \otimes \G_t$ equals that of $\z_{G,t} \z_{G,t}'$ under a fixed permutation of their entries, the second last inequality used \eqref{eqn: GG_var}, 
and the last used the definitions of $s_{J,m}$ and $s_{J,n}$.

By Assumption~(L1), analogous arguments in \eqref{eqn: cI_Q_mn},
and the fact $\v_b^{(m)}$ and $\v_b^{(n)}$ are unit vectors by construction,
we may also obtain
\begin{align}
    \|\cII_{Q,mn}\|_F^2 &
    \lesssim_P
    p_m^{\delta_{r,1}^{(m)}} \cdot p_n^{\delta_{q,1}^{(n)}} \cdot
    s_{J,n}^2 \cdot q_m^{\delta_{c,1}^{(m)}} \cdot
    R(T,d_{\text{fac},\star}) , \label{eqn: cII_Q_mn} \\
    \|\cIV_{Q,mn}\|_F^2 &
    \lesssim_P
    p_m^{\delta_{q,1}^{(m)}} \cdot p_n^{\delta_{r,1}^{(n)}} \cdot
    s_{J,m}^2 \cdot q_n^{\delta_{c,1}^{(n)}} \cdot
    R(T,d_{\text{fac},\star}) , \label{eqn: cIV_Q_mn} \\
    \|\cV_{Q,mn}\|_F^2 &
    \lesssim_P
    p_m^{\delta_{r,1}^{(m)}} \cdot p_n^{\delta_{r,1}^{(n)}} \cdot
    q_m^{\delta_{c,1}^{(m)}} \cdot q_n^{\delta_{c,1}^{(n)}} \cdot
    R(T,d_{\text{fac},\star}) . \label{eqn: cV_Q_mn} 
\end{align}

We next bound the noise-involved terms in \eqref{eqn: decomp_hat_S_Q_mn}, i.e., $\cIII_{Q,mn}$, $\cVI_{Q,mn}$, $\cVII_{Q,mn}$, $\cVIII_{Q,mn}$, and $\cIX_{Q,mn}$.
For each $m\in[M]$, define $a_{m} := \M_{q_m} \v_b^{(m)}$ and $b_{m} := \J_m' \v_b^{(m)}$.
By construction, $\|a_{m}\| \leq 1$ and $\|b_{m}\| = s_{J,m}$. 
We first present a simple but useful bound. For any deterministic vector $a\in\mathbb{R}^{q_m}$ with $\|a\|\leq 1$, Lemma~\ref{lemma:correlation_Etm} implies that, uniformly in $t$,
\begin{align}
    \bb{E}\big\| \M_{p_m} \E_t^{(m)} a \big\|^2 
    = O(p_m) ,
    \label{eqn: unit_vector_error_bound}
\end{align}
which used that for each row $i\in[p_m]$, the covariance matrix of the row vector $(\E_t^{(m)})_{i\cdot}$ has row-wise $\ell_1$-norms bounded by a constant under Lemma~\ref{lemma:correlation_Etm},
and as the covariance matrix is symmetric and positive semidefinite, its spectral norm is bounded by the same constant, and thus the above bound follows by summing over $i\in[p_m]$.
Also, by Assumptions~(F1) and (L1), we have
\begin{align}
    \bb{E}\norm{\Q_n \G_t b_n}^{2}
    \lesssim
    p_n^{\delta_{q,1}^{(n)}} s_{J,n}^2 , \qquad
    \bb{E}\norm{\R_n \F_t^{(n)} \C_n' \v_b^{(n)}}^2
    \lesssim
    p_n^{\delta_{r,1}^{(n)}} q_n^{\delta_{c,1}^{(n)}} .
    \label{eqn: factor_vector_second_moments}
\end{align}

Consider $\cIII_{Q,mn}$. For neatness, define $u_{n,t} := \Q_n \G_t b_n $ and $e_{m,t} := \M_{p_m} \E_t^{(m)} a_m$. Then we can write $\cIII_{Q,mn} = T^{-1} \sum_{t=1}^{T} e_{m,t} u_{n,t}'$.
By Assumption~(E2), the processes $\{e_{m,t}\}_{t\in[T]}$ and $\{u_{n,t}\}_{t\in[T]}$ are independent. Hence we have
\begin{align}
    \bb{E}\norm{\cIII_{Q,mn}}_F^2
    &=
    \frac{1}{T^2}
    \sum_{t=1}^{T}
    \sum_{s=1}^{T}
    \bb{E}\Big( u_{n,t}' u_{n,s} \Big)
    \bb{E}\Big( e_{m,t}' e_{m,s} \Big) .
    \label{eqn: cIII_expansion}
\end{align}
We bound the two expectations on the right-hand side separately. First, by the Cauchy--Schwarz inequality and \eqref{eqn: factor_vector_second_moments}, we have
\begin{align}
    \Big| \bb{E}\Big( u_{n,t}' u_{n,s} \Big) \Big|
    &\leq
    \Big( \bb{E}\norm{u_{n,t}}^{2} \Big)^{1/2}
    \Big( \bb{E}\norm{u_{n,s}}^{2} \Big)^{1/2} 
    \lesssim
    p_n^{\delta_{q,1}^{(n)}} s_{J,n}^2 .
    \label{eqn: cIII_factor_expectation_bound}
\end{align}
Second, by the lagged version of \eqref{eqn: unit_vector_error_bound}, for any $t,s \in[T]$,
\begin{align}
    \Big| \bb{E}\Big( e_{m,t}' e_{m,s} \Big) \Big|
    \lesssim
    p_m \rho(|t-s|) ,
    \label{eqn: cIII_error_expectation_bound}
\end{align}
where $\rho(\cdot)$ is the summable sequence from Lemma~\ref{lemma:correlation_Etm}.
Substituting \eqref{eqn: cIII_factor_expectation_bound} and \eqref{eqn: cIII_error_expectation_bound} into \eqref{eqn: cIII_expansion} gives
\begin{align}
    \bb{E}\norm{\cIII_{Q,mn}}_F^2
    &\lesssim
    T^{-2} p_m p_n^{\delta_{q,1}^{(n)}} s_{J,n}^{2}
    \sum_{t=1}^{T} \sum_{s=1}^{T} \rho(|t-s|) \notag \\
    &\lesssim
    T^{-2} \cdot p_m p_n^{\delta_{q,1}^{(n)}} s_{J,n}^{2} \cdot T
    =
    T^{-1} p_m p_n^{\delta_{q,1}^{(n)}} s_{J,n}^{2} ,
    \label{eqn: cIII_Q_mn}
\end{align}
where the second line follows from evaluating the double sum, viz.\ setting $h = t-s$ gives
\begin{align*}
    \sum_{t=1}^{T} \sum_{s=1}^{T} \rho(|t-s|)
    &=
    \sum_{h= -(T-1)}^{T-1} (T-|h|) \rho(|h|)
    \leq
    T \sum_{h= -\infty}^{\infty} \rho(|h|)
    = O(T) ,
\end{align*}
where the final equality follows from the summability of $\rho(\cdot)$ as in Lemma~\ref{lemma:correlation_Etm}. 
We mark that \eqref{eqn: cIII_Q_mn} carries $T^{-1}$ rather than $R(T,d_{\text{fac},\star})$ as in e.g.\ \eqref{eqn: cI_Q_mn}, since here we consider the cross-moment between the factor and noise processes, whereas previously, say \eqref{eqn: cI_Q_mn}, the sample covariance of factor processes is considered.

The same arguments can be applied to $\cVI_{Q,mn}$, $\cVII_{Q,mn}$, $\cVIII_{Q,mn}$ and $\cIX_{Q,mn}$, yielding
\begin{align}
    \norm{\cVI_{Q,mn}}_F^2
    &=
    O_P\Big( T^{-1} p_m p_n^{\delta_{r,1}^{(n)}} q_n^{\delta_{c,1}^{(n)}} \Big) ,
    \label{eqn: cVI_Q_mn} \\
    \norm{\cVII_{Q,mn}}_F^2
    &=
    O_P\Big( T^{-1} p_m^{\delta_{q,1}^{(m)}} s_{J,m}^2 p_n \Big) ,
    \label{eqn: cVII_Q_mn} \\
    \norm{\cVIII_{Q,mn}}_F^2
    &=
    O_P\Big( T^{-1} p_m^{\delta_{r,1}^{(m)}} q_m^{\delta_{c,1}^{(m)}} p_n \Big) , 
    \label{eqn: cVIII_Q_mn} \\
    \norm{\cIX_{Q,mn}}_F^2
    &=
    O_P\Big( T^{-1} p_m p_n \Big) ,
    \label{eqn: cIX_Q_mn}
\end{align}
where in each case, the bound follows by separating the independent vector processes, applying \eqref{eqn: unit_vector_error_bound} to the error part, \eqref{eqn: factor_vector_second_moments} to the factor part, and using the short-memory property of the error process to sum over $t,s$.
Lastly, combining \eqref{eqn: decomp_hat_S_Q_mn}, 
\eqref{eqn: cI_Q_mn},
\eqref{eqn: cII_Q_mn},
\eqref{eqn: cIII_Q_mn},
\eqref{eqn: cIV_Q_mn},
\eqref{eqn: cV_Q_mn},
\eqref{eqn: cVI_Q_mn},
\eqref{eqn: cVII_Q_mn},
\eqref{eqn: cVIII_Q_mn},
and \eqref{eqn: cIX_Q_mn},
together with the definitions of $\bDelta_{Q,mn}$, we may conclude that
\begin{align*}
    &\;\quad
    \|\bDelta_{Q,mn} \|_F^2 
    =  \| \wh\S_{Q,mn} - \S_{Q,mn} \|_F^2 \\
    &\lesssim
    \Big\| \cI_{Q,mn} - \S_{Q,mn} \Big\|_F^2
    + \norm{\cII_{Q,mn}}_F^2
    + \norm{\cIII_{Q,mn}}_F^2
    + \norm{\cIV_{Q,mn}}_F^2
    + \norm{\cV_{Q,mn}}_F^2 \\
    &\;\quad
    + \norm{\cVI_{Q,mn}}_F^2
    + \norm{\cVII_{Q,mn}}_F^2
    + \norm{\cVIII_{Q,mn}}_F^2
    + \norm{\cIX_{Q,mn}}_F^2 \\
    &\lesssim_P
    p_m^{\delta_{q,1}^{(m)}} p_n^{\delta_{q,1}^{(n)}} s_{J,m}^2 s_{J,n}^2 R(T,d_{\text{fac},\star})
    + p_m^{\delta_{r,1}^{(m)}} p_n^{\delta_{q,1}^{(n)}} s_{J,n}^2  q_m^{\delta_{c,1}^{(m)}} R(T,d_{\text{fac},\star}) 
    + p_m^{\delta_{q,1}^{(m)}} p_n^{\delta_{r,1}^{(n)}} s_{J,m}^2 q_n^{\delta_{c,1}^{(n)}} R(T,d_{\text{fac},\star}) \\
    &\;\quad
    + p_m^{\delta_{r,1}^{(m)}} p_n^{\delta_{r,1}^{(n)}} q_m^{\delta_{c,1}^{(m)}} q_n^{\delta_{c,1}^{(n)}}  R(T,d_{\text{fac},\star}) 
    + T^{-1} p_m p_n^{\delta_{q,1}^{(n)}} s_{J,n}^2
    + T^{-1} p_m p_n^{\delta_{r,1}^{(n)}} q_n^{\delta_{c,1}^{(n)}} \\
    &\;\quad
    + T^{-1} p_m^{\delta_{q,1}^{(m)}} s_{J,m}^2 p_n
    + T^{-1} p_m^{\delta_{r,1}^{(m)}} q_m^{\delta_{c,1}^{(m)}} p_n 
    + T^{-1} p_m p_n .
\end{align*}
This completes the proof on $\| \bDelta_{Q,mn} \|_F^2$.

Next, we focus on the $k_r$-th eigenvalue of $\bOmega_{Q,m}$.
First, by Assumption~(F1), for any $m,n\in[M]$, we may simplify and write
\begin{align}
    \S_{Q,mn} &= \Q_m \bb{E}\left( \G_t\J_m' \v_b^{(m)} \v_b^{(n)'} \J_n\G_t' \right) \Q_n' \notag \\
    &=
    \tr\left( \J_m' \v_b^{(m)} \v_b^{(n)'} \J_n \right) \Q_m \Q_n'
    =
    \left( \v_b^{(n)'} \J_n \J_m' \v_b^{(m)} \right) \Q_m \Q_n' 
    =:
    w_{mn} \Q_m \Q_n' ,
    \label{eqn: S_Q_mn_simplify}
\end{align}
where by Assumption~(L2), the selected projection directions satisfy
\begin{align}
    |w_{mn}| \asymp s_{J,m}s_{J,n}
    \label{eqn: w_mn_order}
\end{align}
for all $m,n\in[M]$ with $n\neq m$. 
Hence we have $\S_{Q,mn}$ of rank $k_r$, and
\begin{equation}
\label{eqn: S_Q_mn_bound}
\|\S_{Q,mn}\| \asymp_P \Big( s_{J,m}s_{J,n} p_m^{\delta_{q,1}^{(m)} /2} p_n^{\delta_{q,1}^{(n)} /2} \Big) .
\end{equation}
Now plugging \eqref{eqn: S_Q_mn_simplify} in $\bOmega_{Q,m}$, we have
\begin{align}
    \bOmega_{Q,m} &= 
    \frac{1}{M-1} \sum_{n=1;n\neq m}^M w_{mn}^2 \Q_m \Q_n' \Q_n \Q_m' \notag \\
    &=
    \Q_m \left\{ \frac{1}{M-1} \sum_{n=1;n\neq m}^M w_{mn}^2 \Q_n' \Q_n \right\} \Q_m' 
    =:
    \Q_m \G_{Q,m} \Q_m' .
    \label{eqn: Omega_Q_m_simplify}
\end{align}
We next derive the order of the $k_r$-th eigenvalue of $\bOmega_{Q,m}$.
For any nonzero vector $\x \in \bb{R}^{k_r}$,
\begin{align}
    \x'\G_{Q,m}\x
    &=
    \frac{1}{M-1} \sum_{n=1;n\neq m}^{M} w_{mn}^{2} \x'\Q_n'\Q_n\x
    =
    \frac{1}{M-1} \sum_{n=1;n\neq m}^{M} w_{mn}^{2} \|\Q_n\x\|^{2}
    \notag \\
    &\ge
    \frac{1}{M-1} \sum_{n=1;n\neq m}^{M} w_{mn}^{2} \lambda_{\min}(\Q_n'\Q_n) \|\x\|^{2} .
    \label{eqn: G_Q_m_lower_rayleigh}
\end{align}
By Assumption~(L1), $\lambda_{\min}(\Q_n'\Q_n) \asymp p_n^{\delta_{q,k_r}^{(n)}}$ for each $n\in[M]$. Combining this with \eqref{eqn: w_mn_order} and \eqref{eqn: G_Q_m_lower_rayleigh} gives
\begin{align}
    \lambda_{\min}(\G_{Q,m})
    \gtrsim
    \frac{s_{J,m}^{2}}{M-1}
    \sum_{n=1;n\neq m}^{M} s_{J,n}^{2} p_n^{\delta_{q,k_r}^{(n)}} 
    \asymp
    s_{J,m}^{2} \max_{n\in[M] \setminus\{m\}} s_{J,n}^{2} p_n^{\delta_{q,k_r}^{(n)}} ,
    \label{eqn: lambda_min_G_Q_m}
\end{align}
where also used that $M$ is fixed.
As the right-hand side of \eqref{eqn: lambda_min_G_Q_m} is bounded away from zero, the matrix $\G_{Q,m}$ is positive definite with all eigenvalues bounded away from zero.
Then from \eqref{eqn: Omega_Q_m_simplify},
the nonzero eigenvalues of $\bOmega_{Q,m}$ are the same as those of
$\G_{Q,m}^{1/2} \Q_m'\Q_m \G_{Q,m}^{1/2}$. Therefore, we may conclude that
\begin{align}
    \lambda_{k_r}(\bOmega_{Q,m})
    &\geq
    \lambda_{\min}(\G_{Q,m}) \lambda_{\min}(\Q_m'\Q_m)
    \asymp
    \lambda_{\min}(\G_{Q,m}) p_m^{\delta_{q,k_r}^{(m)}} 
    \gtrsim
    p_m^{\delta_{q,k_r}^{(m)}} 
    s_{J,m}^{2} \max_{n\in[M] \setminus\{m\}} s_{J,n}^{2} p_n^{\delta_{q,k_r}^{(n)}} ,
    \label{eqn: Omega_Q_m_krth_eigenvalue_lower}
\end{align}
by Assumption~(L1) and \eqref{eqn: lambda_min_G_Q_m},
which particularly implies $\bOmega_{Q,m}$ has rank $k_r$.

Next, note that by Assumption~(R2), we may further simplify $\| \bDelta_{Q,mn} \|_F^2$ as
\begin{align*}
    \|\bDelta_{Q,mn}\|_F^2 
    &= 
    O_P\Big( R(T,d_{\text{fac},\star}) p_m^{\delta_{q,1}^{(m)}} p_n^{\delta_{q,1}^{(n)}} s_{J,m}^2 s_{J,n}^2 \Big)
    = 
    O_P\Big( R(T,d_{\text{fac},\star}) \, \|\S_{Q,mn}\|_F^2 \Big) ,
\end{align*}
where the last equality used \eqref{eqn: S_Q_mn_bound}.
Hence by \eqref{eqn:X_A} and the triangle inequality,
\begin{align*}
    &\;\quad
    \Big\| \X_{Q,m} - \bOmega_{Q,m} \Big\| = \left\| \frac{1}{T^2 (M-1)} \sum_{n=1;n\neq m}^M \left\{ \sum_{t=1}^T\check\Y_t^{(m)} \v_b^{(m)} \v_b^{(n)'} \check\Y_t^{(n)'} \right\}^{\otimes 2} - \bOmega_{Q,m} \right\| \\
    &=
    \left\| \left(\frac{1}{M-1} \sum_{n=1;n\neq m}^M \wh\S_{Q,mn}^{\otimes 2} \right) - \bOmega_{Q,m} \right\| 
    \leq
    \frac{1}{M-1} \sum_{n=1;n\neq m}^M \left\| \wh\S_{Q,mn} \wh\S_{Q,mn}' - \S_{Q,mn} \S_{Q,mn}' \right\| \\
    &\leq
    \frac{1}{M-1} \sum_{n=1;n\neq m}^M \left\| \bDelta_{Q,mn} \wh\S_{Q,mn}' + \S_{Q,mn} \bDelta_{Q,mn}' \right\| \\
    &=
    \frac{1}{M-1} \sum_{n=1;n\neq m}^M \left\| \bDelta_{Q,mn} \bDelta_{Q,mn}' 
    + \bDelta_{Q,mn} \S_{Q,mn}'
    + \S_{Q,mn} \bDelta_{Q,mn}' \right\| \\
    &\lesssim
    \max_{n\in[M] \setminus \{m\}} \left( \|\bDelta_{Q,mn}\|^2 + \|\S_{Q,mn}\| \|\bDelta_{Q,mn}\| \right) \\
    &\lesssim
    R(T,d_{\text{fac},\star})^{1/2} \max_{n\in[M] \setminus \{m\}}  \|\S_{Q,mn}\|_F^2
    \asymp
    R(T,d_{\text{fac},\star})^{1/2}
    p_m^{\delta_{q,1}^{(m)}} s_{J,m}^2 
    \max_{n\in[M] \setminus \{m\}}
    s_{J,n}^2 p_n^{\delta_{q,1}^{(n)}} ,
\end{align*}
which, combined with \eqref{eqn: Omega_Q_m_krth_eigenvalue_lower}, applying the Davis--Kahan theorem \citep[Theorem~2]{yu2015useful} gives the desired result on $\wh\Q_m$, and thus concludes this proof.
$\square$

\begin{lemma}\label{lemma: aligned_projected_cores}
Let all assumptions in Theorem~\ref{thm:common_comp_global_local_init} hold.
Fix any $m\in[M]$.
Recall $\check\Q_m = \Q_m (\Q_m'\Q_m)^{-1/2}$ from Lemma~\ref{lemma: common_comp_global_local_init}, and similarly let $\check\J_m := \J_m (\J_m'\J_m)^{-1/2}$. 
Also, recall $\H_{Q,m}$ and $\H_{J,m}$ from Lemma~\ref{lemma: common_comp_global_local_init}. 
For each $m\in[M]$, with
$\check\Y_{G,t}^{(m)} = \check\Y_t^{(m)} - \wh\R_m\wh\F_t^{(m)} \wh\C_m'$ from Step~3 of Section~\ref{subsec:estimation},
define
\[
\check\G_{t,m} := \check\Q_m' \Q_m \G_t\J_m' \check\J_m ,\quad
\wh\G_{t,m} := \wh\Q_m' \check\Y_{G,t}^{(m)} \wh\J_m .
\]
Then for every $t\in[T]$, $\wh\G_{t,m}$ satisfies that
\begin{align*}
    &\;\quad
    \Big\| \wh\G_{t,m} - \H_{Q,m}' \check\G_{t,m} \H_{J,m}
    \Big\|_F^2
    \notag \\
    &=
    O_P\Bigg(
    \min(p_m,q_m)
    + p_m^{\delta_{r,1}^{(m)}}
    q_m^{\delta_{c,1}^{(m)}}
    \big( \Delta_{R,m} + \Delta_{C,m} + \Delta_{F,m} \big)
    + \big( \eta_{Q,m} + \eta_{J,m} \big) p_m q_m
    \Bigg),
\end{align*}
where $\Delta_{R,m}$, $\Delta_{C,m}$, $\Delta_{F,m}$, $\eta_{Q,m}$, and $\eta_{J,m}$ are defined in the statement of Theorem~\ref{thm:common_comp_global_local_init}.
\end{lemma}
\textbf{\textit{Proof of Lemma~\ref{lemma: aligned_projected_cores}.}}
By construction, we may decompose
\begin{align}
    \wh\G_{t,m}
    &=
    \H_{Q,m}' \check\Q_m' \Q_m \G_t \J_m' \check\J_m \H_{J,m}
    + \cR_{Q,m}
    + \cR_{J,m}
    + \cR_{E,m} + \cR_{\text{local},m},
    \label{eqn: projected_core_decomp} 
\end{align}
where
\begin{align*}
    \cR_{Q,m} &:=
    \big( \wh{\Q}_m - \check{\Q}_m \H_{Q,m} \big)'
    \big( \Q_m \G_t\J_m' + \M_{p_m} \E_t^{(m)} \M_{q_m} 
    + \R_m\F_t^{(m)} \C_m'
    - \wh{\R}_m \wh{\F}_t^{(m)} \wh{\C}_m'
    \big) \wh{\J}_m , \notag 
    \\
    \cR_{J,m} &:=
    \H_{Q,m}' \check{\Q}_m' \big( \Q_m \G_t \J_m'
    + \M_{p_m} \E_t^{(m)} \M_{q_m}
    + \R_m \F_t^{(m)} \C_m'
    - \wh{\R}_m \wh{\F}_t^{(m)} \wh{\C}_m' \big)
    \big( \wh{\J}_m - \check{\J}_m \H_{J,m} \big) , \notag
    \\
    \cR_{E,m} &:=
    \H_{Q,m}' \check{\Q}_m' \M_{p_m} \E_t^{(m)} \M_{q_m} \check{\J}_m \H_{J,m} , \notag
    \\
    \cR_{\text{local},m} &:=
    \H_{Q,m}' \check{\Q}_m'
    \big( \R_m \F_t^{(m)} \C_m'
    - \wh{\R}_m \wh{\F}_t^{(m)} \wh{\C}_m' \big)
    \check{\J}_m \H_{J,m} . \notag
\end{align*}

Note that $\check{\Q}_m$ and $\check{\J}_m$ have orthonormal columns, $\H_{Q,m}$ and $\H_{J,m}$ are asymptotically orthogonal, and $\wh{\J}_m$ has orthonormal columns.
Hence we have
\begin{align}
    \big\| \check{\Q}_m \big\| =
    \big\| \check{\J}_m \big\| =
    \big\| \wh{\J}_m \big\| = 1,
    \quad
    \big\| \H_{Q,m} \big\| =
    \big\| \H_{J,m} \big\| = O_P(1).
    \label{eqn: aligned_core_norm_properties}
\end{align}
Then consider $\cR_{E,m}$. As the columns of $\check{\J}_m\H_{J,m}$ are asymptotically orthonormal, by \eqref{eqn: unit_vector_error_bound}, it holds that
\begin{align}
    \big\| \cR_{E,m} \big\|_F^2
    &\leq
    \big\| \H_{Q,m}' \check{\Q}_m' \big\|^2
    \cdot  \big\| \M_{p_m} \E_t^{(m)} \M_{q_m} \check{\J}_m \H_{J,m} \big\|_F^2
    \lesssim_P
    \min(p_m,q_m) ,
    \label{eqn: aligned_core_RE_rate}
\end{align}
where the equality follows from \eqref{eqn: aligned_core_norm_properties} and the fact that $\big\| \M_{p_m} \E_t^{(m)} \M_{q_m} \check{\J}_m \H_{J,m} \big\|_F^2  = O_P(p_m)$ as in the proof of Theorem~\ref{thm:common_comp_global_local_init}, and a symmetrical argument which also gives the rate in $O_P(q_m)$.

Next, consider $\cR_{\text{local},m}$.
By Theorem~\ref{thm:common_comp_global_local_init}, we have
\begin{align}
    &\;\quad
    \big\| \R_m \F_t^{(m)} \C_m' - \wh\R_m \wh\F_t^{(m)} \wh\C_m' \big\|_F^2  \notag \\
    &\lesssim 
    \big\| \R_m \F_t^{(m)} \C_m' -
    \wh\R_m  \H_{R,m}^{-1} \check{\F}_t^{(m)} \H_{C,m}  \wh\C_m' \big\|_F^2  
    + \big\| \wh\R_m \big( \wh\F_t^{(m)} - \H_{R,m}^{-1} \check{\F}_t^{(m)} \H_{C,m} \big) \wh\C_m' \big\|_F^2  \notag \\
    &=
    \big\| \R_m \F_t^{(m)} \C_m' -
    \wh\R_m  \H_{R,m}^{-1} \big( \R_m'\R_m \big)^{1/2} \F_t^{(m)} \big( \C_m'\C_m \big)^{1/2} \H_{C,m}  \wh\C_m' \big\|_F^2  \notag \\
    &\;\quad
    + \big\| \wh\R_m \big( \wh\F_t^{(m)} - \H_{R,m}^{-1} \check{\F}_t^{(m)} \H_{C,m} \big) \wh\C_m' \big\|_F^2  \notag \\
    &\lesssim_P
    \big\| \R_m \F_t^{(m)} \C_m' -
    \R_m (\R_m'\R_m)^{-1/2} \big( \R_m'\R_m \big)^{1/2} \F_t^{(m)} \big( \C_m'\C_m \big)^{1/2} \big( \C_m'\C_m \big)^{-1/2}  \C_m' \big\|_F^2  \notag \\
    &\;\quad
    + \big\| \wh\R_m \big( \wh\F_t^{(m)} - \H_{R,m}^{-1} \check{\F}_t^{(m)} \H_{C,m} \big) \wh\C_m' \big\|_F^2  \notag \\
    &\lesssim_P
    \big\| \wh\R_m \big( \wh\F_t^{(m)} - \H_{R,m}^{-1} \check{\F}_t^{(m)} \H_{C,m} \big) \wh\C_m' \big\|_F^2  
    \lesssim_P
    p_m^{\delta_{r,1}^{(m)}}
    q_m^{\delta_{c,1}^{(m)}}
    \big( \Delta_{R,m} + \Delta_{C,m} + \Delta_{F,m} \big) ,
    \label{eqn: aligned_core_local_error_bound}
\end{align}
where $\Delta_{R,m}$, $\Delta_{C,m}$ and $\Delta_{F,m}$ are defined in the statement of Theorem~\ref{thm:common_comp_global_local_init}.
Combining \eqref{eqn: aligned_core_norm_properties} and \eqref{eqn: aligned_core_local_error_bound},
we thus have
\begin{align}
    \big\| \cR_{\text{local},m} \big\|_F^2
    &\leq
    \big\| \H_{Q,m}' \check{\Q}_m' \big\|^2 
    \cdot \big\| \R_m \F_t^{(m)} \C_m'
    - \wh\R_m \wh\F_t^{(m)} \wh\C_m'
    \big\|_F^2 
    \cdot \big\| \check{\J}_m \H_{J,m} \big\|^2 \notag \\
    &\lesssim_P
    p_m^{\delta_{r,1}^{(m)}}
    q_m^{\delta_{c,1}^{(m)}}
    \big( \Delta_{R,m} + \Delta_{C,m} + \Delta_{F,m} \big) .
    \label{eqn: aligned_core_Rlocal_rate}
\end{align}

It remains to consider $\cR_{Q,m}$ and $\cR_{J,m}$.
For $\cR_{Q,m}$, by Assumptions~(F1), (L1), \eqref{eqn:Em11Em}, and \eqref{eqn: aligned_core_local_error_bound}, 
we have
\begin{align*}
    &\;\quad
    \big\| \Q_m \G_t \J_m'
    + \M_{p_m} \E_t^{(m)} \M_{q_m}
    + \R_m \F_t^{(m)} \C_m'
    - \wh\R_m \wh\F_t^{(m)} \wh\C_m' \big\|_F^2 \notag \\
    &\lesssim_P
    p_m^{\delta_{q,1}^{(m)}}
    q_m^{\delta_{j,1}^{(m)}}
    + p_m q_m
    + p_m^{\delta_{r,1}^{(m)}} q_m^{\delta_{c,1}^{(m)}}
    \big( \Delta_{R,m} + \Delta_{C,m} + \Delta_{F,m} \big) ,
\end{align*}
which, combined with Lemma~\ref{lemma: common_comp_global_local_init}
and \eqref{eqn: aligned_core_norm_properties},
give
\begin{align}
    \big\| \cR_{Q,m} \big\|_F^2
    &\lesssim_P
    \eta_{Q,m} \,
    \Big\{
    p_m^{\delta_{q,1}^{(m)}}
    q_m^{\delta_{j,1}^{(m)}}
    + p_m q_m
    + p_m^{\delta_{r,1}^{(m)}} q_m^{\delta_{c,1}^{(m)}}
    \big( \Delta_{R,m} + \Delta_{C,m} + \Delta_{F,m} \big)
    \Big\} .
    \label{eqn: aligned_core_RQ_rate}
\end{align}
Similarly, $\cR_{J,m}$ satisfies that
\begin{align}
    \big\| \cR_{J,m} \big\|_F^2
    &\lesssim_P
    \eta_{J,m} \,
    \Big\{
    p_m^{\delta_{q,1}^{(m)}}
    q_m^{\delta_{j,1}^{(m)}}
    + p_m q_m
    + p_m^{\delta_{r,1}^{(m)}} q_m^{\delta_{c,1}^{(m)}}
    \big( \Delta_{R,m} + \Delta_{C,m} + \Delta_{F,m} \big) \Big\} .
    \label{eqn: aligned_core_RJ_rate}
\end{align}

Finally, combining
\eqref{eqn: projected_core_decomp},
\eqref{eqn: aligned_core_RE_rate}, 
\eqref{eqn: aligned_core_Rlocal_rate}, 
\eqref{eqn: aligned_core_RQ_rate}, and \eqref{eqn: aligned_core_RJ_rate}, we conclude that
\begin{align*}
    &\;\quad
    \Big\| \wh\G_{t,m} - \H_{Q,m}' \check\Q_m' \Q_m \G_t \J_m' \check\J_m \H_{J,m} \Big\|_F^2 \\
    &\lesssim_P
    \min(p_m, q_m)
    + p_m^{\delta_{r,1}^{(m)}}
    q_m^{\delta_{c,1}^{(m)}}
    \big( \Delta_{R,m} + \Delta_{C,m} +\Delta_{F,m} \big) \\
    &\;\quad
    + \big( \eta_{Q,m} + \eta_{J,m} \big) \Big\{
    p_m^{\delta_{q,1}^{(m)}}
    q_m^{\delta_{j,1}^{(m)}}
    + p_m q_m
    + p_m^{\delta_{r,1}^{(m)}}
    q_m^{\delta_{c,1}^{(m)}}
    \big( \Delta_{R,m} + \Delta_{C,m} + \Delta_{F,m} \big)
    \Big\} \\
    &\lesssim_P
    \min(p_m, q_m)
    + p_m^{\delta_{r,1}^{(m)}}
    q_m^{\delta_{c,1}^{(m)}}
    \big( \Delta_{R,m} + \Delta_{C,m} +\Delta_{F,m} \big) 
    + \big( \eta_{Q,m} + \eta_{J,m} \big) p_m q_m ,
\end{align*}
which completes the proof of Lemma~\ref{lemma: aligned_projected_cores}.
$\square$

\begin{lemma}\label{lemma: alignment_equality}
By construction in Step~3 of Section~\ref{subsec:estimation}, it holds that
\[
\cR\big( \wh\bXi_{J,m} \otimes \wh\bXi_{Q,m} \big)
= 
\wh\sigma_{\Xi,m} \wh\u_{\Xi,m} \wh\v_{\Xi,m}' .
\]
\end{lemma}
\textbf{\textit{Proof of Lemma~\ref{lemma: alignment_equality}.}}
Recall from Step~3 of Section~\ref{subsec:estimation} the definition of the notation $\cR(\W)$ for a block matrix $\W = (\W_{ij})_{i\in[k_c], j\in[k_c]}$, with each sub-matrix $\W_{ij}$ of size $k_r\times k_r$, such that
\begin{align*}
    \cR(\W) = (\vec(\W_{11}),\ldots,\vec(\W_{k_c1}),\ldots,\vec(\W_{1k_c}),\ldots,\vec(\W_{k_ck_c})) .
\end{align*}
Then note that each sub-matrix in $\wh\bXi_{J,m} \otimes \wh\bXi_{Q,m}$ is exactly the corresponding element in $\wh\bXi_{J,m}$ multiplied by the matrix $\wh\bXi_{Q,m}$;
hence the vectorization of such sub-matrix would be the corresponding element in $\wh\bXi_{J,m}$ multiplied by $\vec(\wh\bXi_{Q,m}) = \wh\sigma_{\Xi,m}^{1/2} \wh\u_{\Xi,m}$ by construction.
Together with the construction of $\wh\bXi_{J,m}$, we complete the proof of this lemma.
$\square$

\begin{lemma}\label{lemma: aligned_projected_cores_rot_rate}
Let all assumptions in Lemma~\ref{lemma: common_comp_global_local_init} hold,
and recall the matrices $\H_{Q,m}$ and $\H_{J,m}$ within.
Then 
\begin{align*}
    \H_{Q,m}' \H_{Q,m} &=
    \I_{k_r}
    + O_P\big( \eta_{Q,m}^{1/2} \big) , \quad
    \H_{J,m}' \H_{J,m} =
    \I_{k_c}
    + O_P\big( \eta_{J,m}^{1/2} \big)  ,
\end{align*}
where $\eta_{Q,m}$ and $\eta_{J,m}$ are defined in Theorem~\ref{thm:common_comp_global_local_init}.
\end{lemma}
\textbf{\textit{Proof of Lemma~\ref{lemma: aligned_projected_cores_rot_rate}.}}
For $\H_{Q,m}$, note that
\begin{align*}
    \I_{k_r} &= \wh\Q_m' \wh\Q_m
    = (\wh\Q_m - \check{\Q}_m \H_{Q,m})' \wh\Q_m
    + \H_{Q,m}' \check{\Q}_m' \wh\Q_m \\
    &=
    (\wh\Q_m - \check{\Q}_m \H_{Q,m})' \wh\Q_m
    + \H_{Q,m}' \check{\Q}_m' (\wh\Q_m - \check{\Q}_m \H_{Q,m})
    + \H_{Q,m}' \check{\Q}_m' \check{\Q}_m \H_{Q,m} \\
    &=
    (\wh\Q_m - \check{\Q}_m \H_{Q,m})' \wh\Q_m
    + \H_{Q,m}' \check{\Q}_m' (\wh\Q_m - \check{\Q}_m \H_{Q,m}) \\
    &\;\quad
    + \H_{Q,m}' (\Q_m'\Q_m)^{-1/2} \Q_m'\Q_m (\Q_m'\Q_m)^{-1/2} \H_{Q,m} \\
    &=
    \H_{Q,m}' \H_{Q,m} + 
    O_P\left\{
R(T,d_{\text{fac},\star})^{1/2} \;
p_m^{ (\delta_{q,1}^{(m)} -\delta_{q,k_r}^{(m)})}
\left( \frac{  \max_{n\in[M] \setminus \{m\}} s_{J,n}^2 p_n^{\delta_{q,1}^{(n)}}}{\max_{n\in[M] \setminus\{m\}} s_{J,n}^2 p_n^{\delta_{q,k_r}^{(n)}}} \right)
\right\}  \\
    &=
    \H_{Q,m}' \H_{Q,m} 
    + O_P\big( \eta_{Q,m}^{1/2} \big),
\end{align*}
where the last equality used Lemma~\ref{lemma: common_comp_global_local_init},
and the $O_P(\cdot)$ bound is in terms of the Frobenius norm.
Analogous arguments hold for $\H_{J,m}$. This thus completes the proof of this lemma.
$\square$

\clearpage

\section{Additional Numerical Results}\label{app:additional_numerical_results}
\subsection{Simulation}\label{app:additional_simulation_results}
\subsubsection{Heavy-tailed robustness}\label{app:heavy_tail_robustness}

We repeat the main effect settings (Ia)--(Id) and common component settings (IIa)--(IId) under heavy-tailed innovations. The DGP is identical to that in Section~\ref{subsec: simulation}, except that the Gaussian innovations in the latent time-series processes are replaced by Student $t(6)$ innovations. Each setting is crossed with $T\in\{100,400\}$, with $500$ Monte Carlo replications.

The main effect results in Table~\ref{tab:main-effect-accuracy-t6} remain close to their Gaussian counterparts. The direct moment estimators are essentially unchanged, while the loading-space and reconstructed-component errors retain the same improvement with $T$; heavy tails mainly increase finite-sample dispersion in some entries.

\begin{table}[ht!]
\centering
\caption{Main effect estimation accuracy. All entries are Monte Carlo means multiplied by $10^2$, with standard deviations in parentheses. The latent-process innovations follow a Student $t(6)$ distribution.}
\label{tab:main-effect-accuracy-t6}
\resizebox{\textwidth}{!}{%
\begin{tabular}{l|cc|cc|cc|cc}
\toprule
Settings & \multicolumn{2}{c}{Ia} & \multicolumn{2}{c}{Ib} & \multicolumn{2}{c}{Ic} & \multicolumn{2}{c}{Id}\\
 & $T=100$ & $T=400$ & $T=100$ & $T=400$ & $T=100$ & $T=400$ & $T=100$ & $T=400$\\
\midrule
$\MSE(\wh c_t^{(m)})$ & 0.05(0.01) & 0.05(0.00) & 0.04(0.01) & 0.04(0.00) & 0.05(0.01) & 0.05(0.00) & 0.02(0.00) & 0.02(0.00)\\
$\MSE(\wh\a_t^{(m)})$ & 0.60(0.08) & 0.58(0.06) & 0.49(0.07) & 0.47(0.05) & 1.19(0.17) & 1.14(0.12) & 0.30(0.04) & 0.29(0.03)\\
$\MSE(\wh\b_t^{(m)})$ & 0.60(0.08) & 0.58(0.06) & 0.49(0.07) & 0.47(0.05) & 1.19(0.17) & 1.16(0.12) & 0.60(0.07) & 0.58(0.04)\\
$\MSE(\wh\mu_t)$ & 0.03(0.01) & 0.03(0.00) & 0.02(0.01) & 0.02(0.00) & 0.03(0.01) & 0.03(0.00) & 0.01(0.00) & 0.01(0.00)\\
$\MSE(\wh\mu_t^{(m)})$ & 0.08(0.02) & 0.08(0.01) & 0.06(0.01) & 0.06(0.01) & 0.08(0.02) & 0.08(0.01) & 0.04(0.01) & 0.04(0.00)\\
\addlinespace[0.45em]
$\cD(\A_{1m},\wh\A_{1m})$ & 7.04(5.98) & 2.80(1.13) & 9.81(10.06) & 3.22(1.73) & 7.37(5.11) & 2.97(1.01) & 7.38(6.06) & 2.70(1.15)\\
$\cD(\A_{1m},\wt\A_{1m})$ & 7.03(6.01) & 2.79(1.13) & 9.80(10.07) & 3.21(1.73) & 7.34(5.10) & 2.96(1.00) & 7.37(6.06) & 2.69(1.15)\\
$\cD(\A_{2m},\wh\A_{2m})$ & 6.33(4.15) & 2.74(0.92) & 8.23(6.34) & 3.15(1.16) & 6.68(2.95) & 2.94(0.82) & 6.41(3.80) & 2.58(0.87)\\
$\cD(\B_{1m},\wh\B_{1m})$ & 6.95(5.47) & 2.74(1.10) & 9.76(10.80) & 3.26(1.57) & 7.73(5.95) & 2.89(1.04) & 5.12(3.82) & 2.03(0.65)\\
$\cD(\B_{1m},\wt\B_{1m})$ & 6.93(5.47) & 2.73(1.10) & 9.74(10.80) & 3.26(1.57) & 7.70(5.94) & 2.87(1.04) & 5.11(3.82) & 2.02(0.64)\\
$\cD(\B_{2m},\wh\B_{2m})$ & 6.19(3.70) & 2.70(0.96) & 8.25(7.19) & 3.19(1.23) & 6.90(3.66) & 2.89(0.88) & 4.57(2.53) & 1.99(0.58)\\
\addlinespace[0.45em]
$\MSE(\wt\A_{1m}\wh\balpha_t)$ & 0.79(5.71) & 0.10(0.07) & 1.58(6.01) & 0.12(0.15) & 0.62(1.22) & 0.14(0.05) & 0.73(3.14) & 0.08(0.06)\\
$\MSE(\wh\A_{2m}\wh\balpha_t^{(m)})$ & 0.83(2.89) & 0.19(0.08) & 1.56(4.20) & 0.20(0.13) & 0.84(1.12) & 0.28(0.07) & 0.78(2.17) & 0.13(0.07)\\
$\MSE(\wt\B_{1m}\wh\bbeta_t)$ & 0.65(2.98) & 0.10(0.06) & 1.60(5.63) & 0.12(0.10) & 0.83(3.69) & 0.13(0.06) & 0.29(0.94) & 0.05(0.02)\\
$\MSE(\wh\B_{2m}\wh\bbeta_t^{(m)})$ & 0.77(2.69) & 0.18(0.08) & 1.75(5.34) & 0.21(0.13) & 0.98(2.72) & 0.28(0.08) & 0.38(1.30) & 0.09(0.03)\\
\bottomrule
\end{tabular}%
}
\end{table}

The common component results in Table~\ref{tab:common-component-accuracy-t6} show the same robustness. Refitting remains effective and the estimation errors continue to decrease with $T$, while some small-sample entries exhibit greater Monte Carlo variation.

\begin{table}[ht!]
\centering
\caption{Common component estimation accuracy. All entries are Monte Carlo means multiplied by $10^2$, with standard deviations in parentheses. The latent-process innovations follow a Student $t(6)$ distribution.}
\label{tab:common-component-accuracy-t6}
\resizebox{\textwidth}{!}{%
\begin{tabular}{l|cc|cc|cc|cc}
\toprule
Settings & \multicolumn{2}{c}{IIa} & \multicolumn{2}{c}{IIb} & \multicolumn{2}{c}{IIc} & \multicolumn{2}{c}{IId}\\
 & $T=100$ & $T=400$ & $T=100$ & $T=400$ & $T=100$ & $T=400$ & $T=100$ & $T=400$\\
\midrule
$\cD(\Q_m,\wh\Q_m)$ & 8.14(3.45) & 3.72(1.32) & 9.66(3.98) & 4.69(1.89) & 9.28(3.41) & 4.45(1.44) & 5.85(2.47) & 2.75(1.01)\\
$\cD(\Q_m,\wt\Q_m)$ & 1.84(0.46) & 0.79(0.14) & 1.97(1.54) & 0.81(0.21) & 3.22(0.64) & 1.45(0.19) & 1.16(0.20) & 0.50(0.07)\\
$\cD(\J_m,\wh\J_m)$ & 8.41(3.41) & 3.85(1.45) & 9.92(4.38) & 4.61(1.76) & 9.43(3.51) & 4.33(1.41) & 7.92(3.03) & 3.87(1.43)\\
$\cD(\J_m,\wt\J_m)$ & 1.60(0.23) & 0.71(0.08) & 1.67(1.42) & 0.67(0.08) & 3.08(0.71) & 1.40(0.17) & 1.48(0.16) & 0.67(0.06)\\
\addlinespace[0.45em]
$\cD(\R_m,\wh\R_m)$ & 2.60(0.75) & 1.01(0.17) & 3.23(1.89) & 1.04(0.22) & 4.44(1.87) & 2.43(0.57) & 2.20(0.70) & 0.78(0.15)\\
$\cD(\C_m,\wh\C_m)$ & 2.57(0.71) & 1.00(0.17) & 3.14(1.78) & 1.05(0.21) & 4.33(0.96) & 2.41(0.57) & 2.13(0.51) & 0.85(0.11)\\
\addlinespace[0.45em]
$\MSE(\wt\Q_m\wh\G_t\wt\J_m')$ & 0.07(0.03) & 0.04(0.01) & 0.19(2.72) & 0.03(0.01) & 0.26(0.31) & 0.16(0.02) & 0.04(0.01) & 0.02(0.00)\\
$\MSE(\wh\R_m\wh\F_t^{(m)}\wh\C_m')$ & 0.27(0.15) & 0.13(0.03) & 0.42(2.05) & 0.12(0.03) & 0.76(0.60) & 0.48(0.12) & 0.14(0.06) & 0.06(0.01)\\
\bottomrule
\end{tabular}%
}
\end{table}

\begin{figure}[htbp!]
\centering
\includegraphics[width=0.5\textwidth]{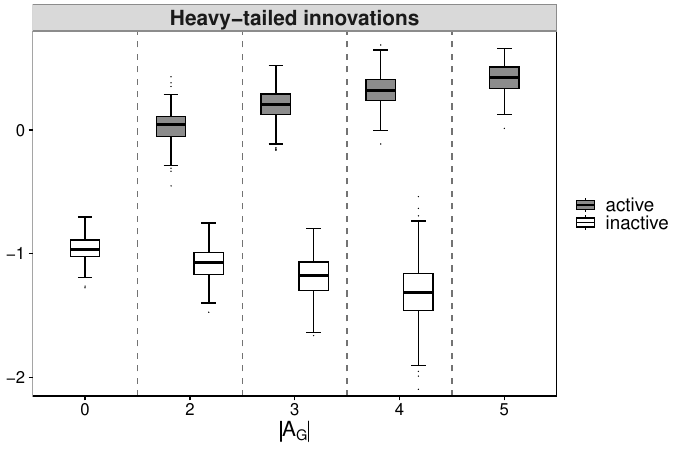}
\caption{Within-replication averages of $S_{G,m}$ over the active and inactive groups at $T=200$ under the baseline design with Student $t(6)$ innovations. Each boxplot is formed across Monte Carlo replications. The y-axis reports the base-10 logarithm of the group average.}
\label{fig:sgm_boxplot_t6}
\end{figure}

\begin{figure}[htbp!]
\centering
\includegraphics[width=0.45\textwidth]{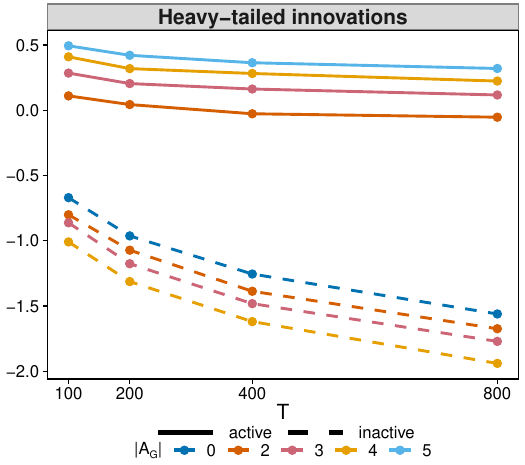}
\caption{Monte Carlo medians of the within-replication averages of $S_{G,m}$ over the active and inactive groups, respectively, over $T\in\{100,200,400,800\}$ under Student $t(6)$ innovations. The y-axis reports the statistics on the $\log_{10}$ scale. Solid and dashed lines correspond to active and inactive groups, respectively.}
\label{fig:sgm_curve_t6}
\end{figure}

We also repeat the $S_{G,m}$ behavior experiment in Section~\ref{subsubsec:sgm_behavior} under the baseline design (IIIa), using 400 Monte Carlo replications and replacing the Gaussian innovations by Student $t(6)$ innovations. Figure~\ref{fig:sgm_boxplot_t6} shows that the active and inactive group averages remain clearly separated at $T=200$, with a wider gap as more groups become active. Figure~\ref{fig:sgm_curve_t6} shows the same pattern over $T$: the inactive averages decrease, while the active averages remain comparatively stable. These results closely match the Gaussian baseline.

\subsubsection{Order-statistic calibration for detection and screening}\label{app:orderstat_comparison}

We further examine the choice of order statistic in the bootstrap procedures of Section~\ref{sec:practical_global_screening}. For existence testing, we compare tests based on $S_{G,(1)}$ and $S_{G,(2)}$, each calibrated by its corresponding bootstrap distribution. For screening, we compare two procedures indexed by $j\in\{1,2\}$. The $S_{G,(j)}$-based procedure sets both $\tau_1^{(j)}$ and $\tau_2^{(j)}$ equal to the empirical $95\%$ quantile of $S_{G,(j)}^*$ and applies the corresponding existence gate and groupwise screening rule. Since $\tau_1^{(j)}=\tau_2^{(j)}$, the existence gate is automatically satisfied whenever at least two groups exceed the screening cutoff. All results below use the baseline setting (IIIa), 400 Monte Carlo replications, and $B=199$ bootstrap replications.

Table~\ref{tab:orderstat-existence-comparison} shows that the two existence tests have very similar finite-sample performance. Both maintain size close to the nominal $5\%$ level and have nearly identical power across the alternatives.

\begin{table}[ht!]
\centering
\caption{Bootstrap existence tests based on $S_{G,(1)}$ and $S_{G,(2)}$ in the baseline setting (IIIa). Each statistic is calibrated by its corresponding bootstrap distribution. Entries are empirical rejection rates, with $|\cA_G|=0$ representing size and $|\cA_G|\geq2$ representing power.}
\label{tab:orderstat-existence-comparison}
\begin{tabular}{@{}llccccc@{}}
\toprule
$T$ & Statistic & $|\cA_G|=0$ & $|\cA_G|=2$ & $|\cA_G|=3$ & $|\cA_G|=4$ & $|\cA_G|=5$ \\
\midrule
\multirow{2}{*}{100}
& $S_{G,(1)}$ & 0.048 & 0.800 & 0.993 & 1.000 & 1.000 \\
& $S_{G,(2)}$ & 0.043 & 0.828 & 0.993 & 1.000 & 1.000 \\
\addlinespace[0.45em]
\multirow{2}{*}{200}
& $S_{G,(1)}$ & 0.058 & 0.998 & 1.000 & 1.000 & 1.000 \\
& $S_{G,(2)}$ & 0.058 & 0.995 & 1.000 & 1.000 & 1.000 \\
\addlinespace[0.45em]
\multirow{2}{*}{400}
& $S_{G,(1)}$ & 0.058 & 0.998 & 1.000 & 1.000 & 1.000 \\
& $S_{G,(2)}$ & 0.050 & 0.998 & 1.000 & 1.000 & 1.000 \\
\bottomrule
\end{tabular}
\end{table}

The difference is more pronounced for screening. As shown in Table~\ref{tab:orderstat-screening-comparison}, the $S_{G,(1)}$-based procedure is substantially more conservative: it further reduces false positives under the null but yields noticeably lower true positive and exact-selection rates, especially when $T$ is small or few groups are active. The gap narrows as $T$ and the number of active groups increase. For $|\cA_G|=3$ and $4$, false positives are zero under both procedures in the baseline design, so the differences in exact recovery are driven by missed active groups.

\begin{table}[ht!]
\centering
\caption{Bootstrap screening procedures based on $S_{G,(1)}$ and $S_{G,(2)}$ in the baseline setting (IIIa). For each procedure, the same bootstrap order statistic is used to calibrate both the existence and screening cutoffs.}
\label{tab:orderstat-screening-comparison}
\begin{tabular}{@{}llcccccc@{}}
\toprule
&& \multicolumn{2}{c}{$T=100$} & \multicolumn{2}{c}{$T=200$} & \multicolumn{2}{c}{$T=400$} \\
\cmidrule(lr){3-4}\cmidrule(lr){5-6}\cmidrule(l){7-8}
$|\cA_G|$ & Metric & $S_{G,(1)}$ & $S_{G,(2)}$ & $S_{G,(1)}$ & $S_{G,(2)}$ & $S_{G,(1)}$ & $S_{G,(2)}$ \\
\midrule
\multirow{3}{*}{$0$}
& TPR   & --    & --    & --    & --    & --    & --    \\
& FPR   & 0.001 & 0.019 & 0.000 & 0.026 & 0.002 & 0.021 \\
& Exact & 0.998 & 0.958 & 1.000 & 0.943 & 0.995 & 0.950 \\
\addlinespace[0.45em]
\multirow{3}{*}{$2$}
& TPR   & 0.255 & 0.826 & 0.755 & 0.995 & 0.973 & 0.998 \\
& FPR   & 0.000 & 0.001 & 0.000 & 0.002 & 0.000 & 0.000 \\
& Exact & 0.255 & 0.825 & 0.755 & 0.990 & 0.973 & 0.998 \\
\addlinespace[0.45em]
\multirow{3}{*}{$3$}
& TPR   & 0.729 & 0.951 & 0.947 & 0.998 & 0.994 & 1.000 \\
& FPR   & 0.000 & 0.000 & 0.000 & 0.000 & 0.000 & 0.000 \\
& Exact & 0.497 & 0.867 & 0.870 & 0.993 & 0.983 & 1.000 \\
\addlinespace[0.45em]
\multirow{3}{*}{$4$}
& TPR   & 0.860 & 0.971 & 0.959 & 0.996 & 0.998 & 0.999 \\
& FPR   & 0.000 & 0.000 & 0.000 & 0.000 & 0.000 & 0.000 \\
& Exact & 0.570 & 0.887 & 0.855 & 0.983 & 0.990 & 0.998 \\
\addlinespace[0.45em]
\multirow{3}{*}{$5$}
& TPR   & 0.896 & 0.985 & 0.971 & 0.995 & 0.999 & 1.000 \\
& FPR   & --    & --    & --    & --    & --    & --    \\
& Exact & 0.595 & 0.927 & 0.858 & 0.975 & 0.993 & 1.000 \\
\bottomrule
\end{tabular}
\end{table}

Taken together, the results support using the second-largest statistic throughout the proposed procedure. It performs similarly to the largest statistic for existence testing, while providing substantially higher screening recovery with low false positive rates.

\clearpage

\subsection{Real data}\label{app:additional_real_data_results}
\subsubsection{Data construction and preprocessing}\label{app:data_preprocessing}

Table~\ref{tab:stock_indicator_construction} summarizes the construction of the nine indicators used in the empirical analysis. The last indicator is a rolling version of the illiquidity measure of \cite{AMIHUD200231}, based on the ratio of absolute return to dollar trading volume. Rolling indicators are constructed stock by stock using November--December 2022 as warm-up history. ADF tests do not reveal systematic unit-root behavior in the constructed series, so no differencing is applied; we also do not winsorize the data. 

For each indicator, the mean and standard deviation are first computed from all finite observed cells in the 2023--2024 primary sample, before any imputation, and used for standardization. The panel is then standardized using these values. Ordinary missing cells are imputed using the same-day sector median, with the same-day market median as a fallback; observations after structural exits are left missing. Finally, each stock-indicator series is demeaned using its 2023-2024 time mean.

\begin{table}[ht!]
\centering
\caption{Construction of the equity indicators.}
\label{tab:stock_indicator_construction}
\small
\begin{tabularx}{\textwidth}{l X l}
\toprule
Indicator & Construction & Window \\
\midrule
Daily return
& CRSP daily total return, $r_t=\texttt{DlyRet}_t$
& Daily \\
20-day cumulative log return
& $\sum_{\ell=0}^{19}\log(1+r_{t-\ell})$
& 20 days \\
Log dollar volume
& $\log(1+\texttt{DlyPrcVol}_t)$
& Daily \\
Log turnover
& $\log\{\texttt{DlyPrcVol}_t/(1000\,\texttt{DlyCap}_t)\}$
& Daily \\
20-day relative turnover
& Current log turnover minus its mean over the previous 20 trading days
& 20 lagged days \\
20-day log realized volatility
& Log sample standard deviation of daily returns over the latest 20 trading days
& 20 days \\
High-low range
& $\log(\texttt{DlyHigh}_t/\texttt{DlyLow}_t)$
& Daily \\
Log bid-ask spread
& $\log[1+10^4\{2(\texttt{DlyAsk}_t-\texttt{DlyBid}_t)/
(\texttt{DlyAsk}_t+\texttt{DlyBid}_t)\}]$
& Daily \\
20-day log Amihud illiquidity
& $\log\!\left[1+\frac{1}{20}\sum_{\ell=0}^{19}
\frac{10^6|r_{t-\ell}|}{\texttt{DlyPrcVol}_{t-\ell}}\right]$
& 20 days \\
\bottomrule
\end{tabularx}
\end{table}

\subsubsection{Additional real-data results}

\begin{figure}[ht!]
\centering
\includegraphics[width=.8\textwidth]{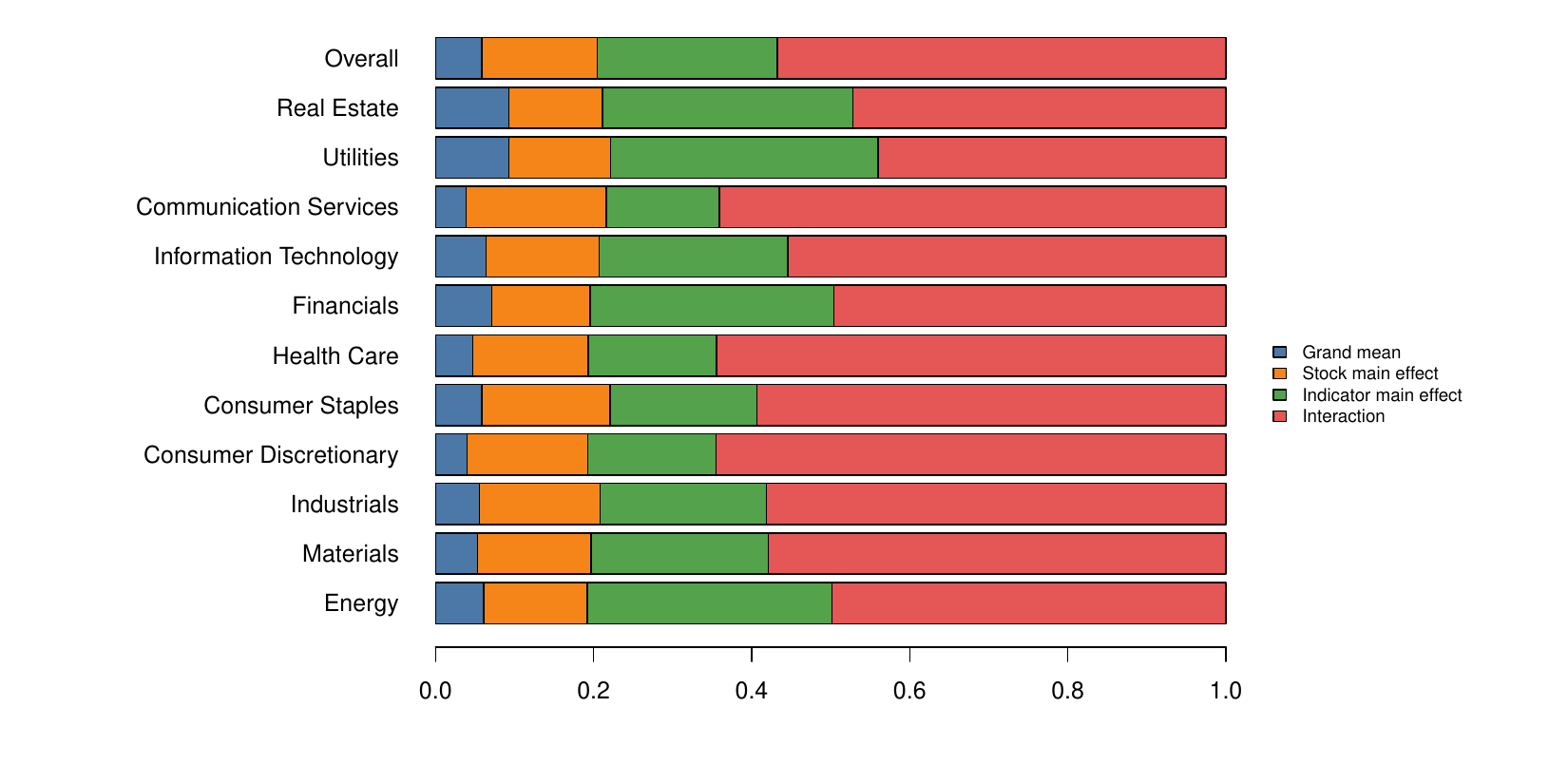}
\caption{Orthogonal decomposition of time-varying squared Frobenius norms across sectors.}
\label{fig:stock_data_geometry}
\end{figure}

\paragraph{More evidence of using MMEFM structure.}
Under (IC1), the four components, grand mean, row main effect, column main effect, and the rest interaction component are orthogonal, so for each sector $m$ and date $t$, $\norm{\Y_t^{(m)}}_F^2=p_mq_m(\wh c_t^{(m)})^2+q_m\norm{\wh\a_t^{(m)}}^2+p_m\norm{\wh\b_t^{(m)}}^2+\norm{\check\Y_t^{(m)}}_F^2$. After aggregation over sectors and dates, the additive components account for $43.23\%$ of the total squared magnitude. Figure~\ref{fig:stock_data_geometry} shows that the interaction component is the largest part in every sector, but the main effects remain substantial and vary across sectors. The grand-mean contribution is small throughout.

\begin{figure}[ht!]
\centering
\includegraphics[width=\textwidth]{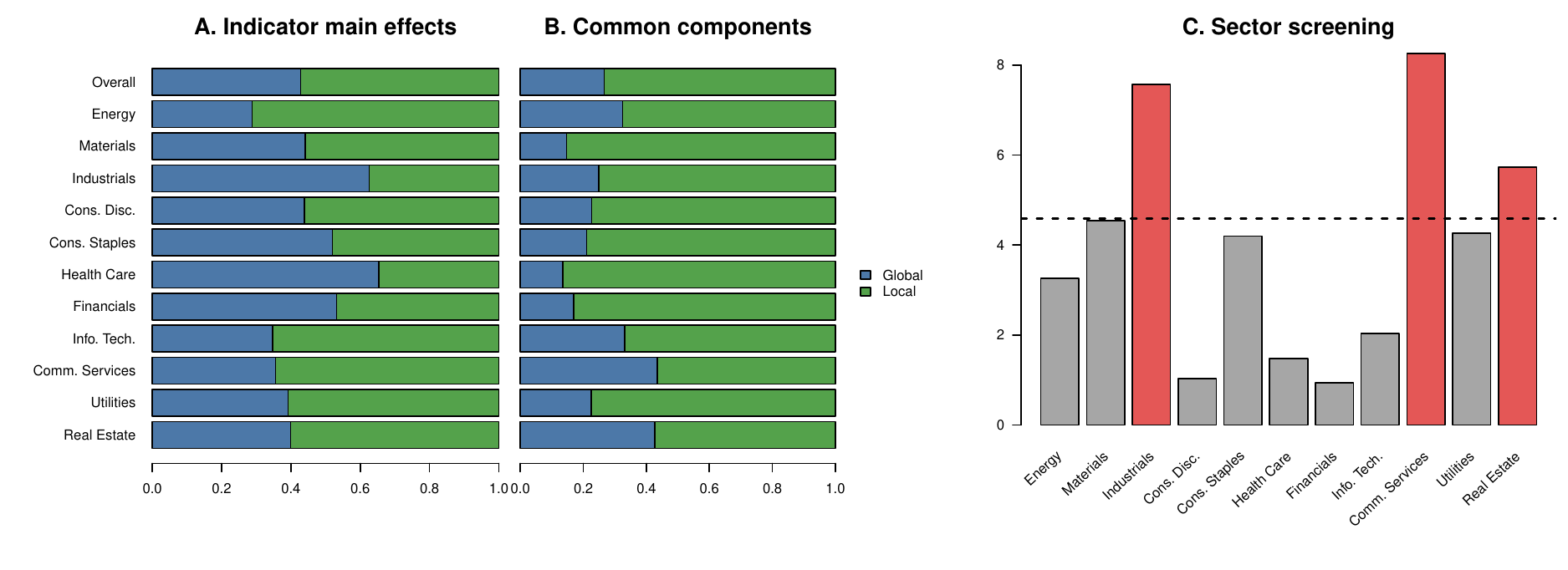}
\caption{Evidence for a multilevel structure. Panels A and B report global--local squared Frobenius magnitude ratios for the indicator main effects and common components, with Overall formed after aggregation across sectors. Panel C reports $S_{G,m}$; the dashed line is the 95\% cutoff calibrated from $B=199$ circular-shift bootstrap replications, and red bars denote screened sectors.}
\label{fig:stock_data_multilevel_structure}
\end{figure}

Since the fitted global and local components are not exactly orthogonal, we compare their squared magnitudes ratios. After aggregation over sectors and dates, the global magnitude ratios are
\begin{align*}
    \text{Indicator main effect:}\quad
    &\frac{\sum_{m=1}^M\sum_{t=1}^T\norm{\1_{p_m}(\wt\B_{1m}\wh\bbeta_t)'}_F^2}
    {\sum_{m=1}^M\sum_{t=1}^T
    \left\{\norm{\1_{p_m}(\wt\B_{1m}\wh\bbeta_t)'}_F^2+\norm{\1_{p_m}(\wh\B_{2m}\wh\bbeta_t^{(m)})'}_F^2\right\}}=42.81\%, \\
    \text{Common component:}\quad
    &\frac{\sum_{m=1}^M\sum_{t=1}^T
    \norm{\wt\Q_m\wh\G_t\wt\J_m'}_F^2}
    {\sum_{m=1}^M\sum_{t=1}^T
    \left\{\norm{\wt\Q_m\wh\G_t\wt\J_m'}_F^2+\norm{\wh\R_m\wh\F_t^{(m)}\wh\C_m'}_F^2\right\}}=26.63\%.
\end{align*}
The sector-level ratios are computed in the same way by restricting the sums to each sector. Figure~\ref{fig:stock_data_multilevel_structure} reports the corresponding sector-level ratios together with the screening statistics.

\begin{figure}[ht!]
\centering
\includegraphics[height=0.82\textheight,keepaspectratio]{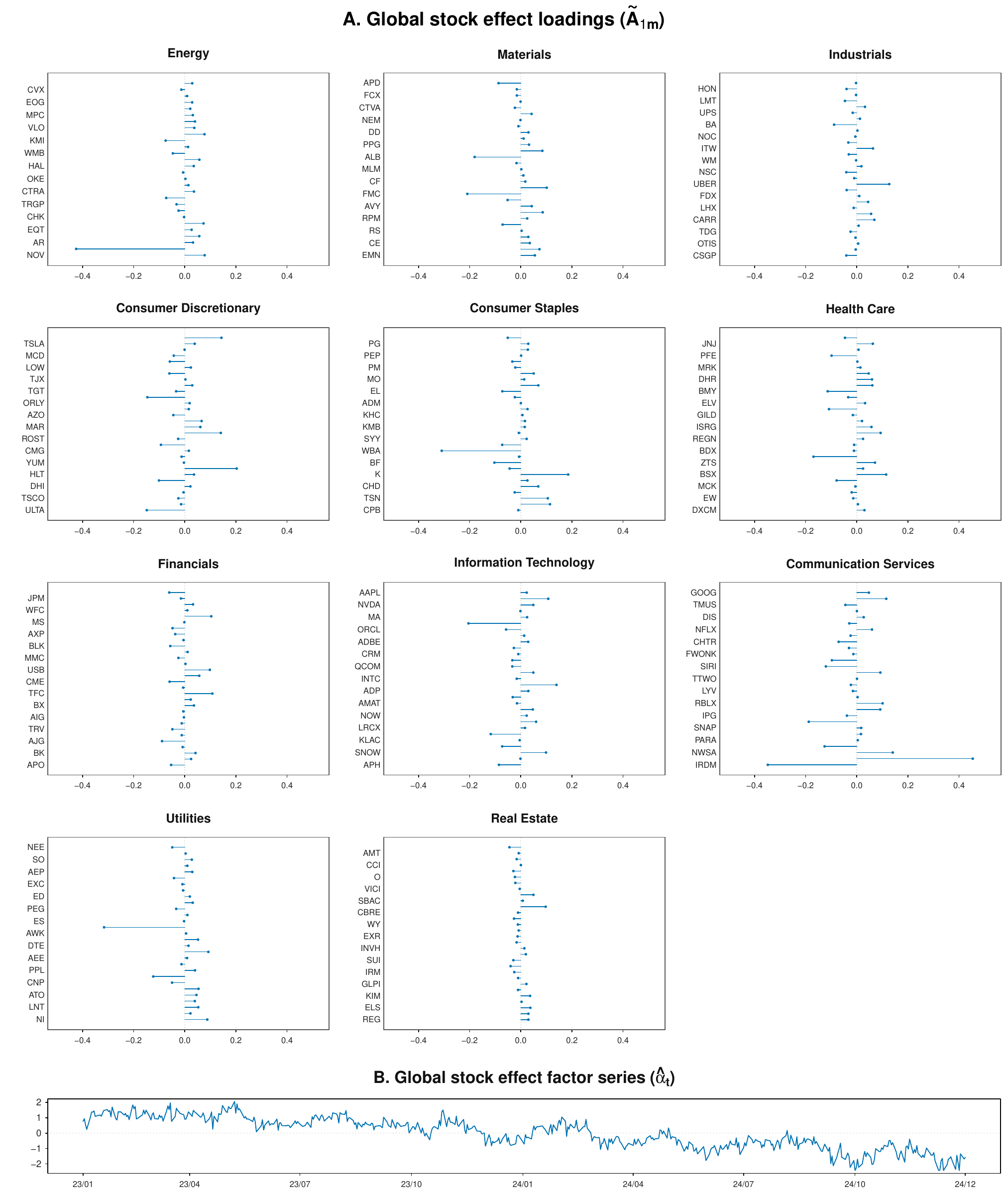}
\caption{Estimated global stock main effects.}
\label{fig:stock_global_stock_main_supp}
\end{figure}

\begin{figure}[ht!]
\centering
\includegraphics[width=\textwidth]{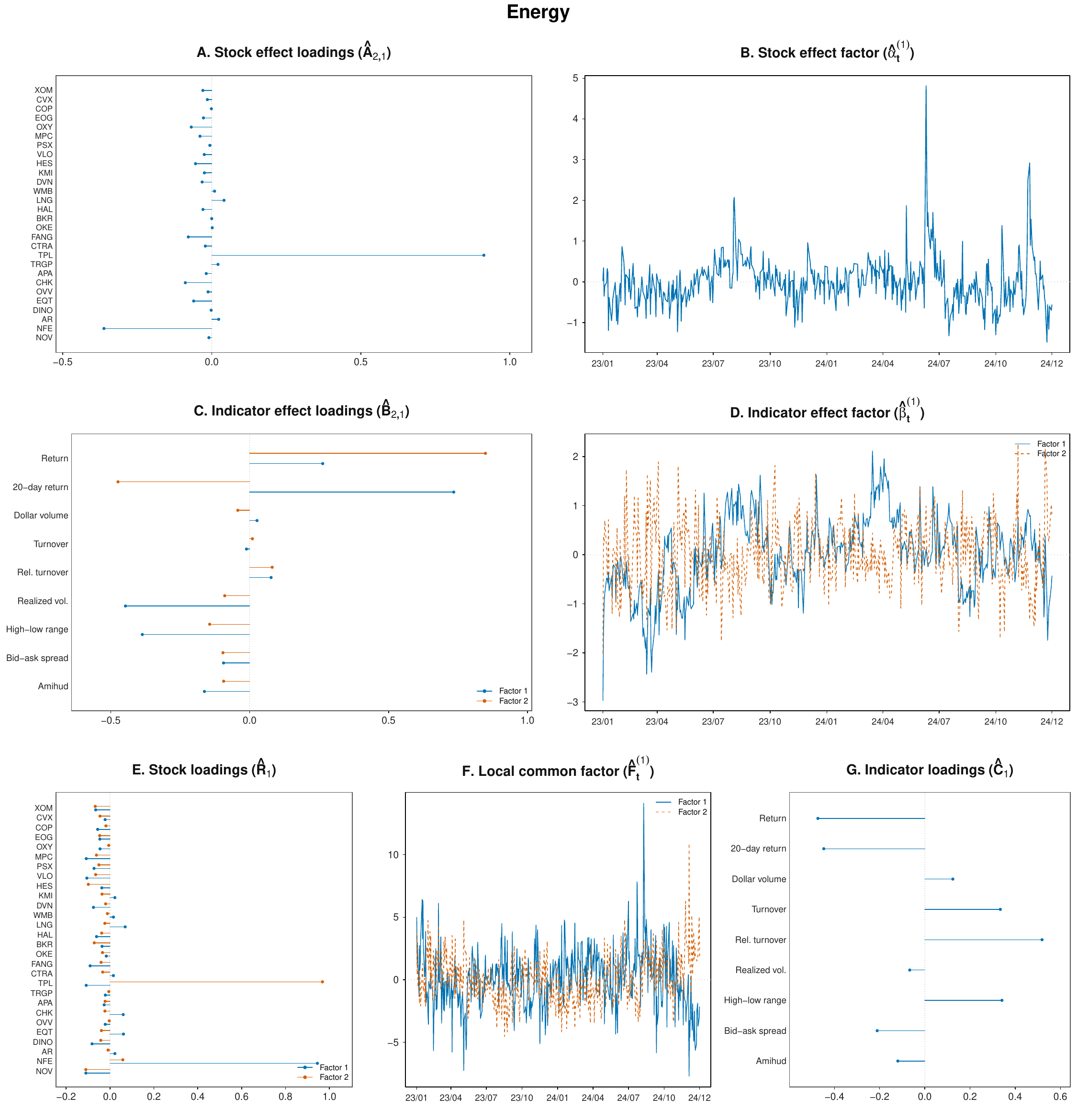}
\caption{Estimated local MMEFM components for Energy.}
\label{fig:stock_energy_local_components}
\end{figure}

\begin{figure}[ht!]
\centering
\includegraphics[width=\textwidth]{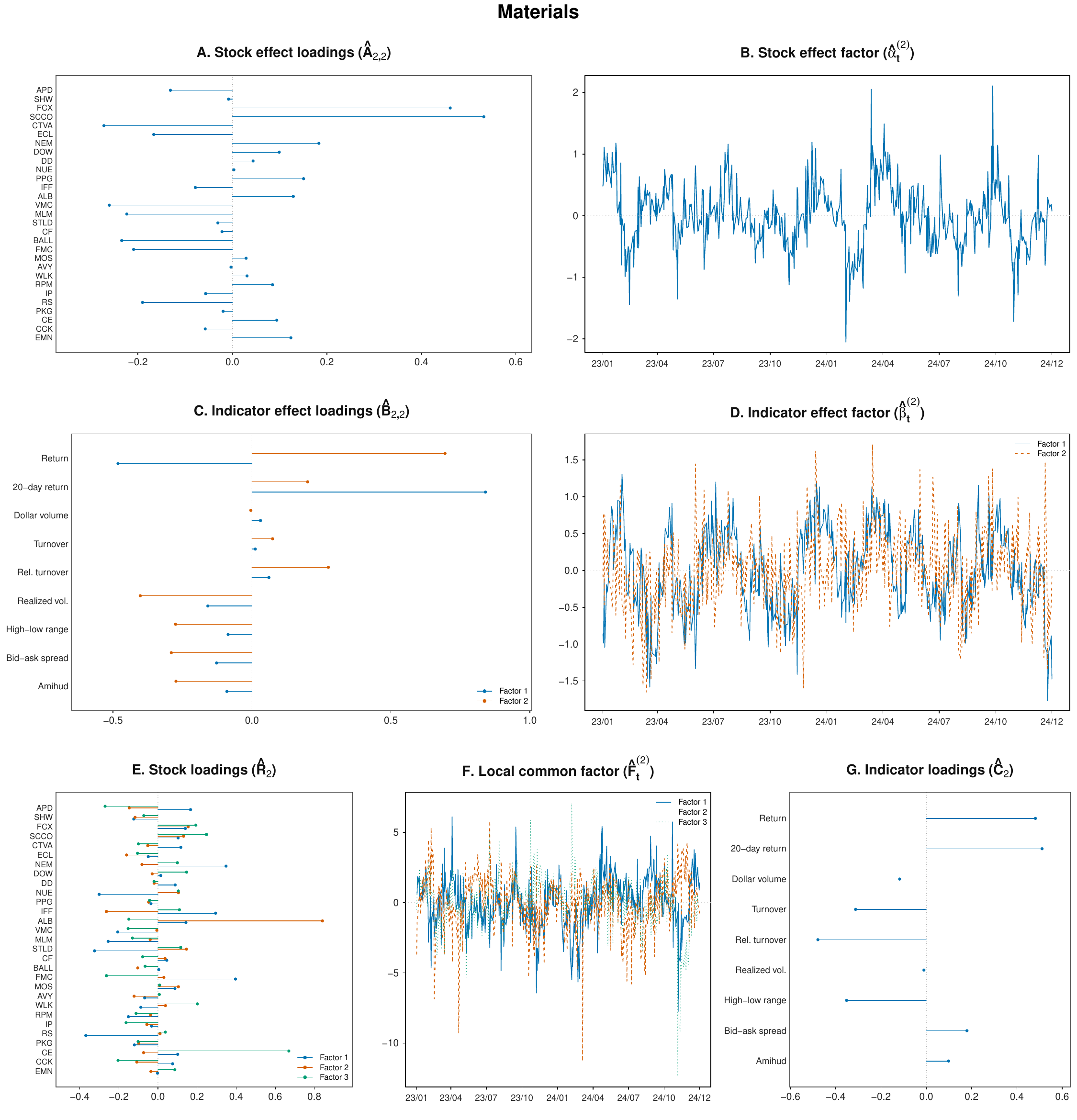}
\caption{Estimated local MMEFM components for Materials.}
\label{fig:stock_materials_local_components}
\end{figure}

\begin{figure}[ht!]
\centering
\includegraphics[width=\textwidth]{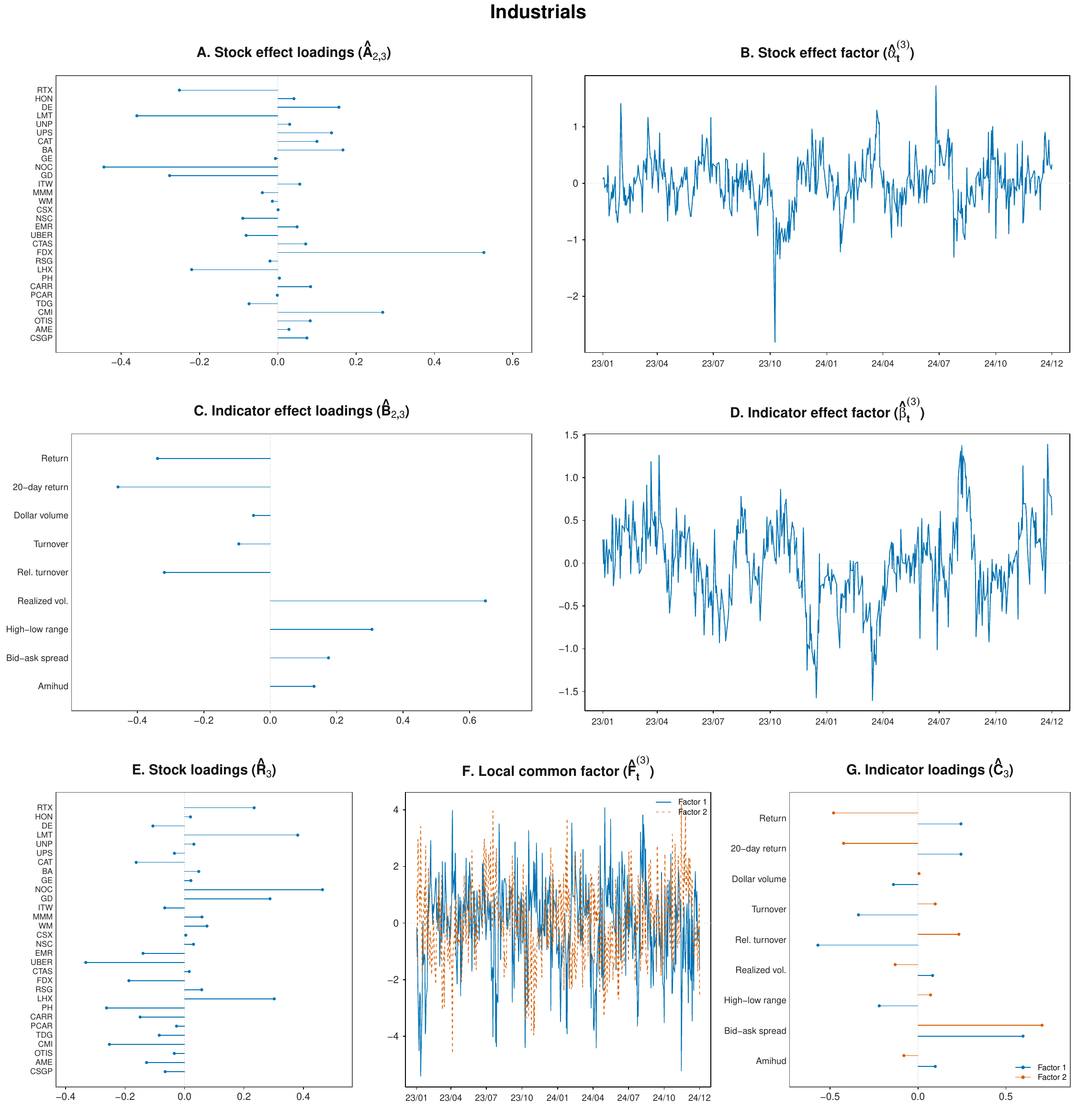}
\caption{Estimated local MMEFM components for Industrials.}
\label{fig:stock_industrials_local_components}
\end{figure}

\begin{figure}[ht!]
\centering
\includegraphics[width=\textwidth]{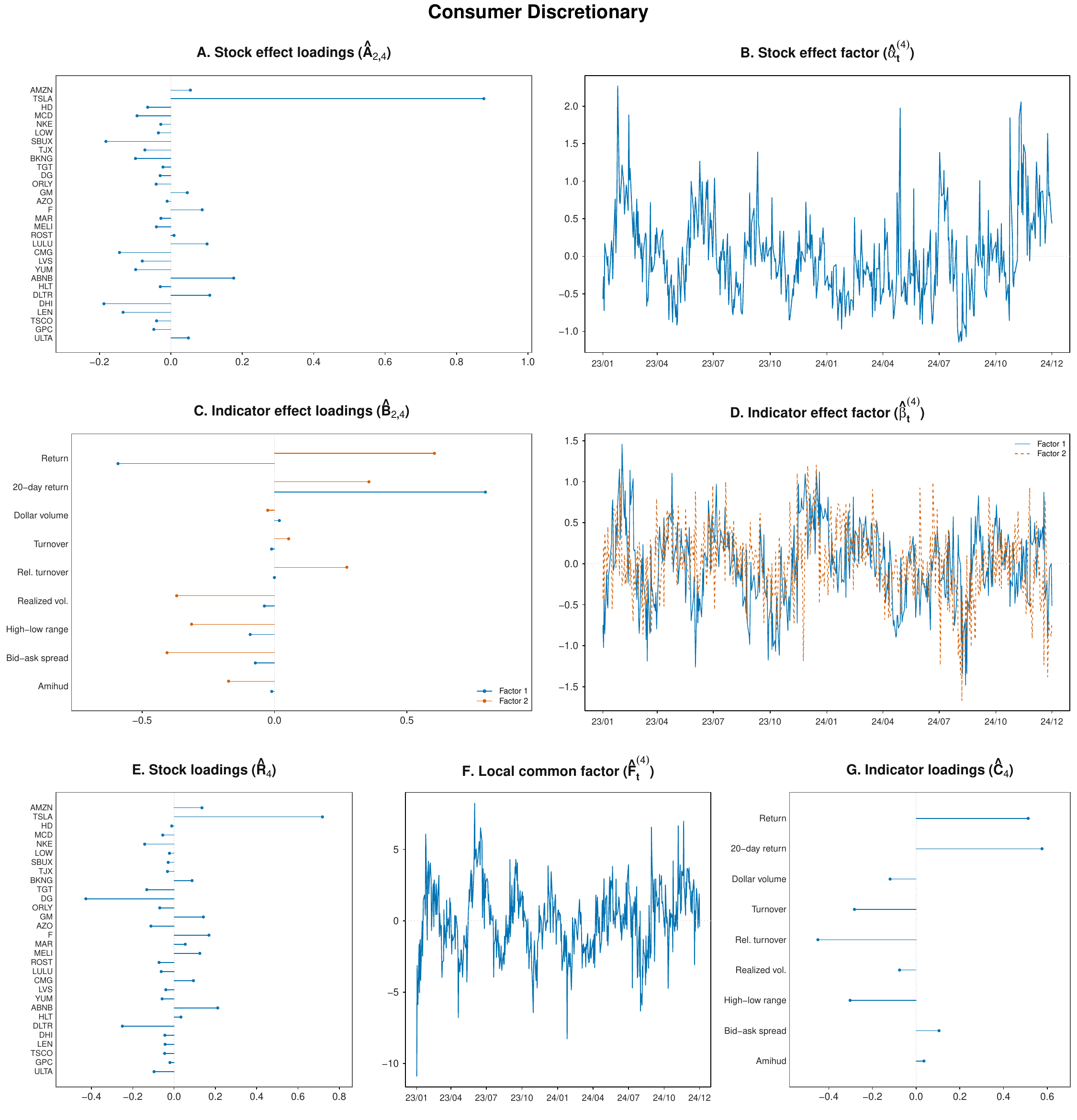}
\caption{Estimated local MMEFM components for Consumer Discretionary.}
\label{fig:stock_consumer_discretionary_local_components}
\end{figure}

\begin{figure}[ht!]
\centering
\includegraphics[width=\textwidth]{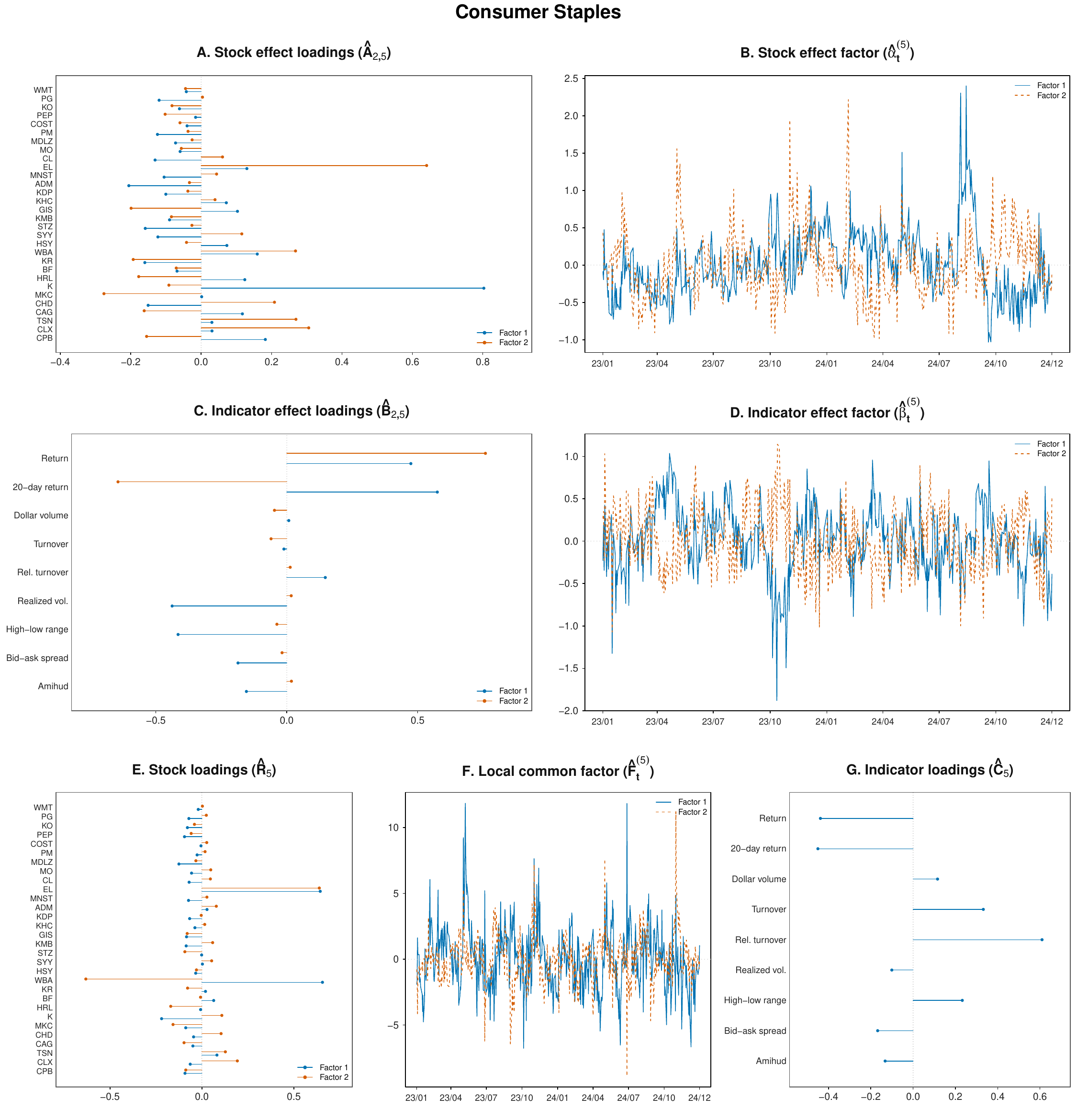}
\caption{Estimated local MMEFM components for Consumer Staples.}
\label{fig:stock_consumer_staples_local_components}
\end{figure}

\begin{figure}[ht!]
\centering
\includegraphics[width=\textwidth]{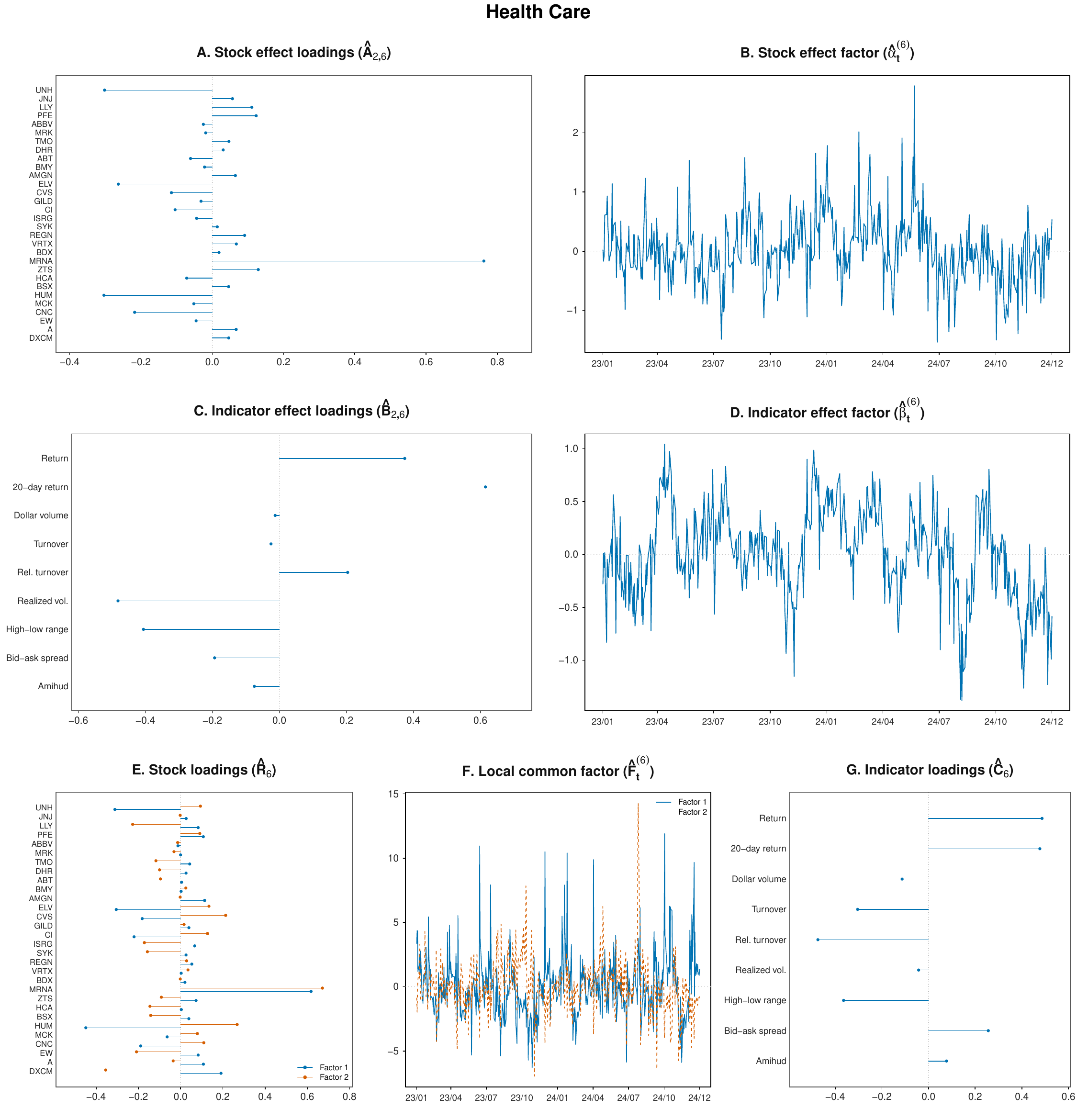}
\caption{Estimated local MMEFM components for Health Care.}
\label{fig:stock_health_care_local_components}
\end{figure}

\begin{figure}[ht!]
\centering
\includegraphics[width=\textwidth]{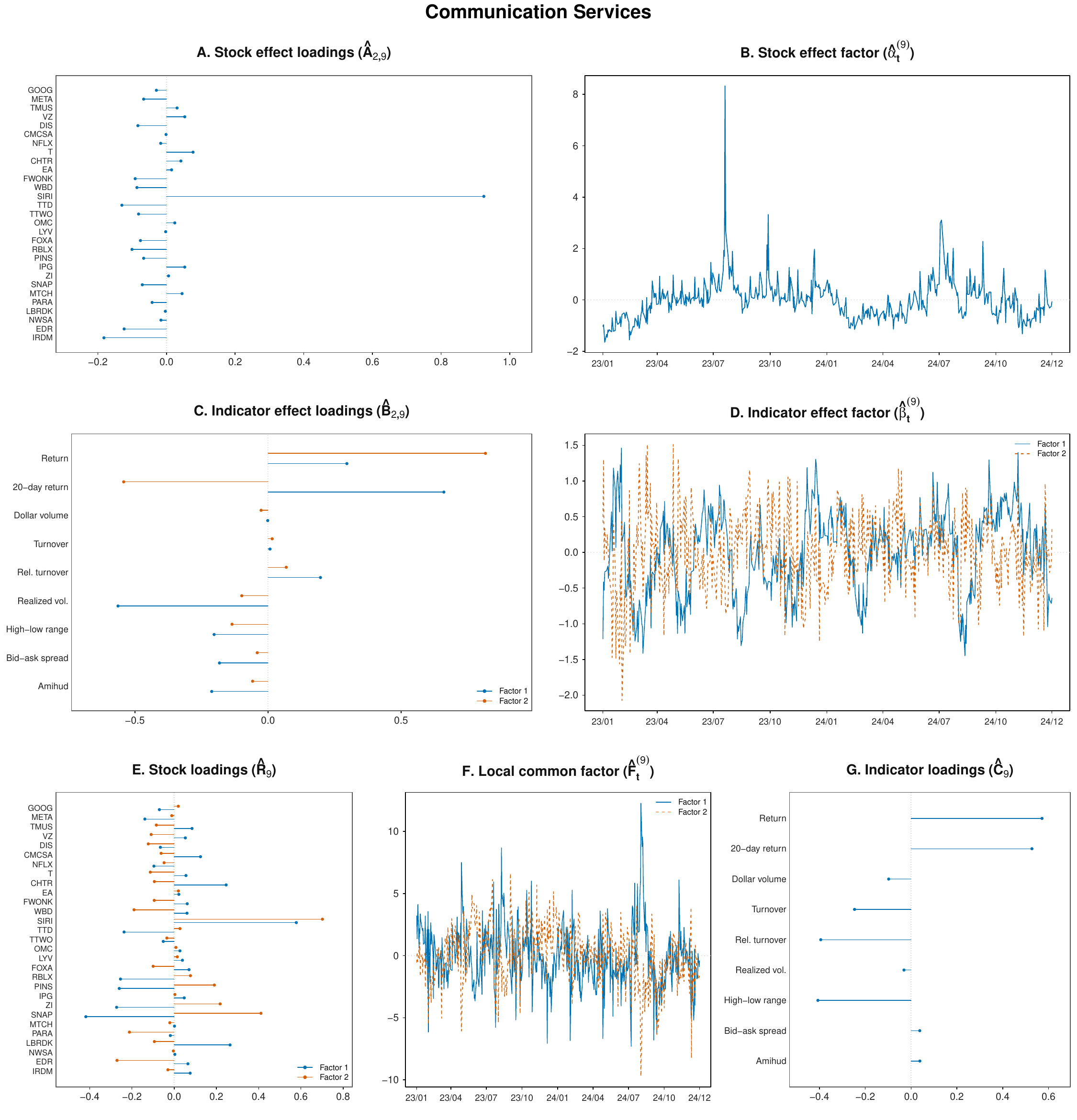}
\caption{Estimated local MMEFM components for Communication Services.}
\label{fig:stock_communication_services_local_components}
\end{figure}

\begin{figure}[ht!]
\centering
\includegraphics[width=\textwidth]{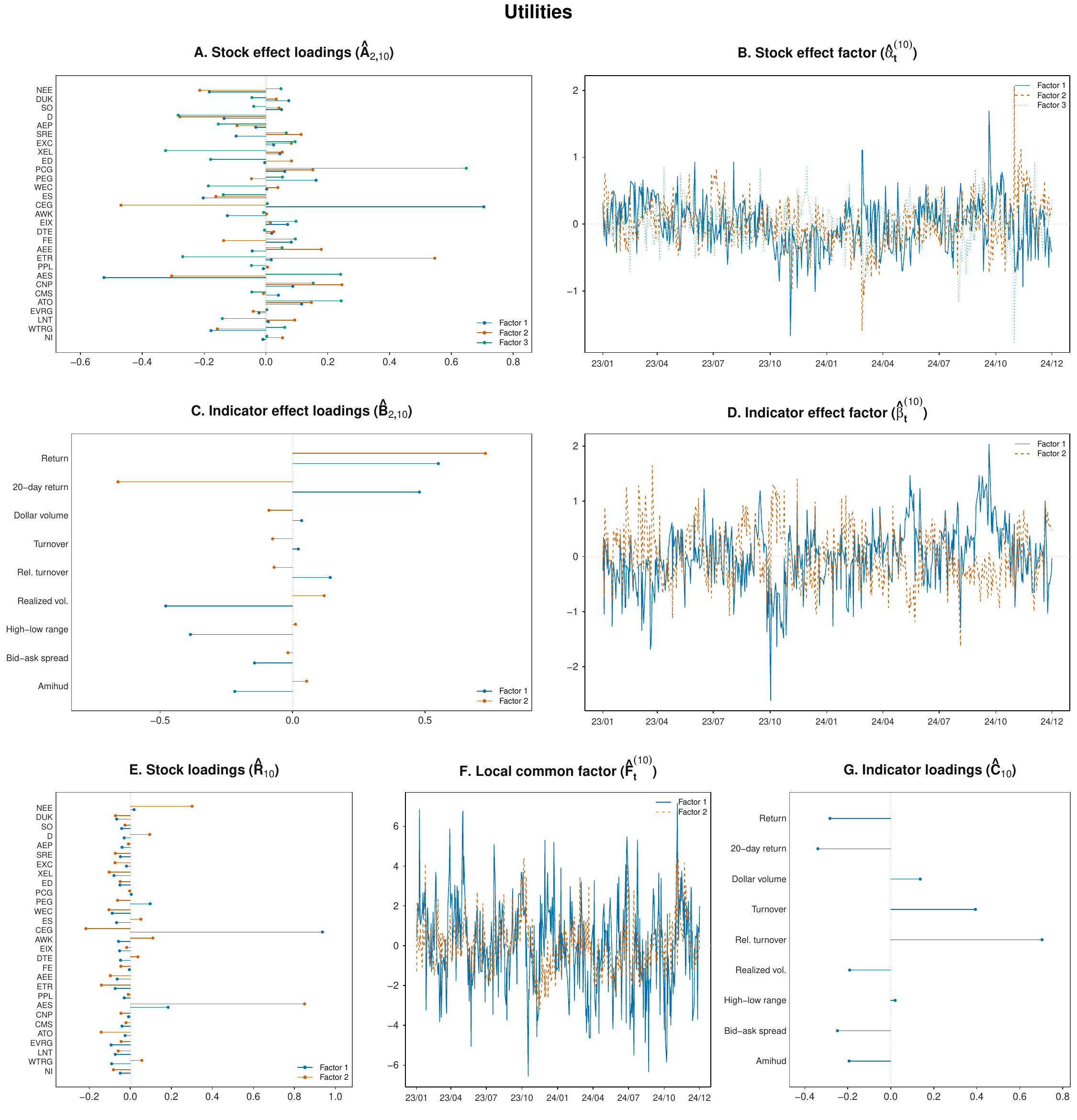}
\caption{Estimated local MMEFM components for Utilities.}
\label{fig:stock_utilities_local_components}
\end{figure}

\begin{figure}[ht!]
\centering
\includegraphics[width=\textwidth]{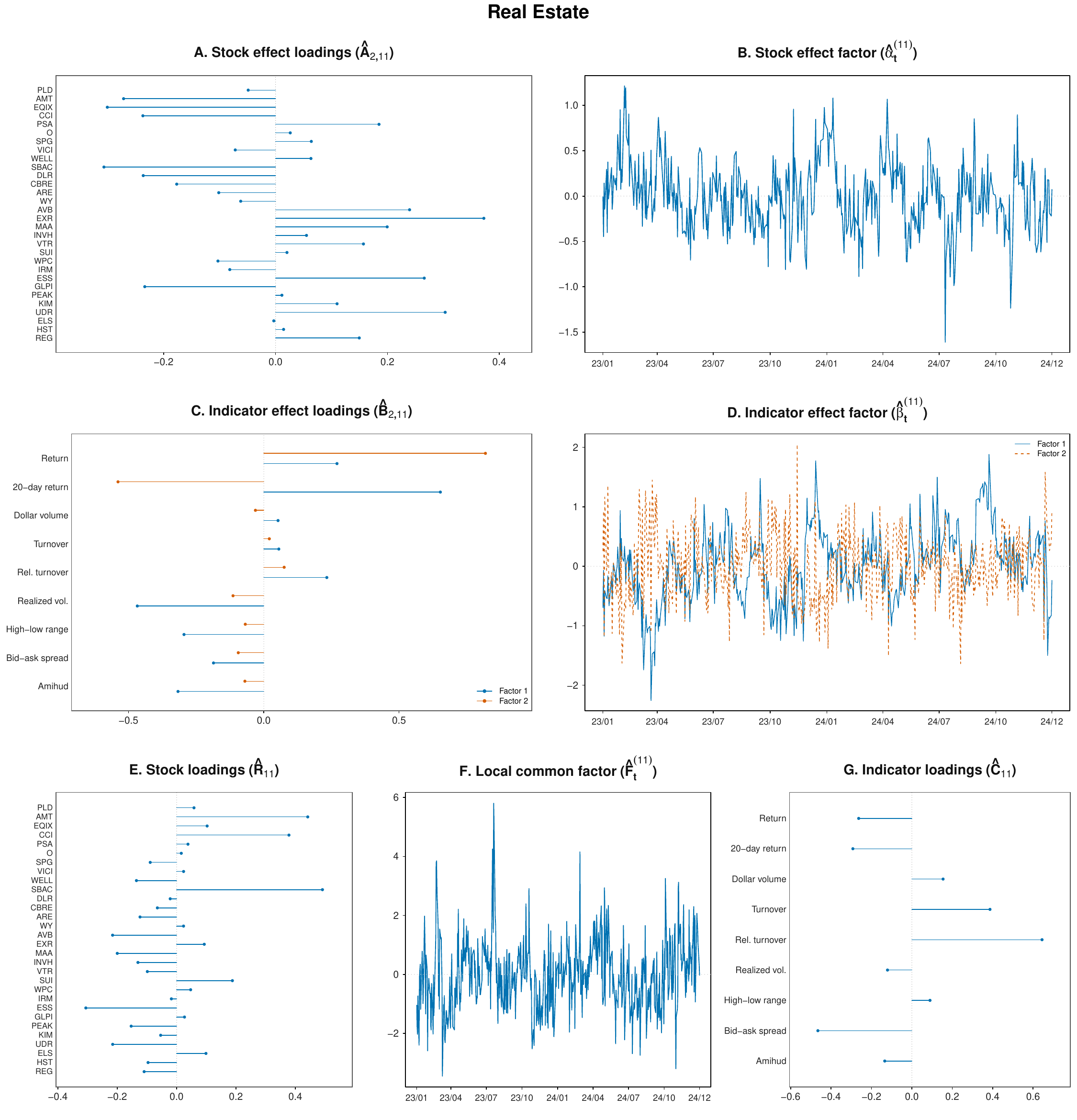}
\caption{Estimated local MMEFM components for Real Estate.}
\label{fig:stock_real_estate_local_components}
\end{figure}

\paragraph{More results.}
Table~\ref{tab:stock_rank_selection} reports the selected factor dimensions. The main text reports the global indicator main effect and illustrative local estimates for Financials and Information Technology. For completeness, Figure~\ref{fig:stock_global_stock_main_supp} reports the global stock main effect, and Figures~\ref{fig:stock_energy_local_components}--\ref{fig:stock_real_estate_local_components} report local MMEFM estimates for the remaining nine sectors. In each sector-specific figure, Panels A--B show local stock-effect loadings and factor series, Panels C--D show local indicator-effect loadings and factor series, and Panels E--G show local common-component stock loadings, factor series, and indicator loadings.

\begin{table}[ht!]
\centering
\caption{Selected factor dimensions for the MMEFM: $(r_1,\ell_1)$ and $(r_2^{(m)},\ell_2^{(m)})$ for the main effects, and $(k_r,k_c)$ and $(k_r^{(m)},k_c^{(m)})$ for the common component.}
\label{tab:stock_rank_selection}
\begin{tabular}{lcc}
\toprule
& Main effects & Common component \\
\midrule
Global (all sectors) & $(1,1)$ & $(1,1)$ \\
\midrule
\multicolumn{3}{l}{\textit{Local ranks by sector}} \\
Energy & $(1,2)$ & $(2,1)$ \\
Materials & $(1,2)$ & $(3,1)$ \\
Industrials & $(1,1)$ & $(1,2)$ \\
Consumer Discretionary & $(1,2)$ & $(1,1)$ \\
Consumer Staples & $(2,2)$ & $(2,1)$ \\
Health Care & $(1,1)$ & $(2,1)$ \\
Financials & $(1,1)$ & $(1,2)$ \\
Information Technology & $(1,2)$ & $(1,1)$ \\
Communication Services & $(1,2)$ & $(2,1)$ \\
Utilities & $(3,2)$ & $(2,1)$ \\
Real Estate & $(1,2)$ & $(1,1)$ \\
\bottomrule
\end{tabular}
\end{table}

\end{document}